\documentclass[12pt, a4paper, twoside, openright]{book}

\usepackage{msor-thesis}

\usepackage{palatino} % sets palatino as the default font

\usepackage{url} % for typesetting urls

\usepackage{xcolor} %for text colors
\usepackage[T1]{fontenc}
\usepackage{amssymb} 

\usepackage{amsmath}

\usepackage{graphicx}

\definecolor{applegreen}{rgb}{0.55, 0.71, 0.0}

\begin{document}

\frontmatter
% Book style knows about front matter
% Report style doesn't so you need to set roman numbering etc yourself :-(

%%%%%%%%%%%%%%%%%%%%%%%%%%%%%%%%%%%%%%%%%%%%%%%%%%%%%%%

\title{Traversable Wormholes, Regular Black Holes, and Black-Bounces}
\author{Alex Simpson}

\subject{Mathematics}
\abstract{\addcontentsline{toc}{chapter}{Abstract}Various spacetime candidates for traversable wormholes, regular black holes, and `black-bounces' are presented and thoroughly explored in the context of the gravitational theory of general relativity. All candidate spacetimes belong to the mathematically simple class of spherically symmetric geometries; the majority are static (time-independent as well as non-rotational), with a single dynamical (time-dependent) geometry explored. To the extent possible, the candidates are presented through the use of a global coordinate patch -- some of the prior literature (especially concerning traversable wormholes) has often proposed coordinate systems for desirable solutions to the Einstein equations requiring a multi-patch atlas. The most interesting cases include the so-called `exponential metric' -- well-favoured by proponents of alternative theories of gravity but which actually has a standard classical interpretation, and the `black-bounce' to traversable wormhole case -- where a metric is explored which represents either a traversable wormhole or a regular black hole, depending on the value of the newly introduced scalar parameter $a$. This notion of `black-bounce' is defined as the case where the spherical boundary of a regular black hole forces one to travel towards a one-way traversable `bounce' into a future reincarnation of our own universe. The metric of interest is then explored further in the context of a time-dependent spacetime, where the line element is rephrased with a Vaidya-like time-dependence imposed on the mass of the object, and in terms of outgoing\-/ingoing Eddington-Finkelstein coordinates. Analysing these candidate spacetimes extends the pre-existing discussion concerning the viability of non-singular black hole solutions in the context of general relativity, as well as contributing to the dialogue on whether an arbitrarily advanced civilization would be able to construct a traversable wormhole.}
% Books don't normally have abstracts, and this is a bit of a hack

\ack{\addcontentsline{toc}{chapter}{Acknowledgements}First and foremost I would like to thank Professor Matt Visser for his time and support as my supervisor on this project. Thanks go to him for many enlightening conversations, as well as priceless tutelage on the machinations of the world of academia; lessons which have enabled me to take my first strides toward being a fully fledged and contributing member of the scientific community. I look forward to a long-lasting collaborative relationship.

Next, thanks goes to my wonderful parents Liana and Glenn -- your perpetual love and support has been the cornerstone to my academic development through the years; specifically over the past twelve months it has given me the gift of a  robust platform from which to explore the farthest corners of our mysterious universe. Thank you for your time, your patience, and most of all, thank you for listening.

To my wonderful partner Harri I can not thank you enough for the support you have shown me over recent times. You've been a rock for me, lending me your ear when needed, your wisdom when sought, and your unconditional love throughout. This piece of work would not have been possible without you. You're the best.

To my grandmother, Flora, thank you for being my inspiration for all these years. I have always been highly motivated by your perpetual love and strength of character, and thoroughly look forward to learning more from you in the years to come.

Special thanks must go to my grandparents, Ian and Pat, for providing me with the capacity to work on this while in Whangamata -- and I do hope to have shed some light on the dark for you grandpa!

To my gravitational compatriots, Sebastian, Jessica, and Del, thank you for many an entertaining and thought provoking conversation over the past year. As colleagues one couldn't ask for scholars of a more excellent and admirable character, and as friends I am most grateful for the memories.

I would also like to acknowledge and thank Prado Mart\'in-Moruno, a Postdoctoral Fellow of the Departamento de F\'isica Te\'orica at the Universidad Complutense de Madrid. Particular credit goes to Prado for her role in the production of the figures used in \S\ref{C:Vaidya}.

Above all I hope to maintain a productive and healthy relationship with all of the above as we look to the future... it has been thoroughly enjoyable getting to know you all and I sincerely hope we maintain the connection. I suppose time will tell, as it tells all things -- except within an event horizon.}

% Uncomment the appropriate degree
%\phd
\mscthesisonly
%\mscwithhonours
%\mscbothparts
% \otherdegree{DEGREE OR DIPLOMA NAME}

%%%%%%%%%%%%%%%%%%%%%%%%%%%%%%%%%%%%%%%%%%%%%%%%%%%%%%%

\maketitle

{\addcontentsline{toc}{chapter}{\textbf{Contents\hfill}}}
\tableofcontents

{\addcontentsline{toc}{chapter}{\textbf{List of figures\hfill}}}
\listoffigures

%%%%%%%%%%%%%%%%%%%%%%%%%%%%%%%%%%%%%%%%%%%%%%%%%%%%%%%

% book style knows about mainmatter
% if you are using report style you will have to rest page numbering etc.
\mainmatter

%%%%%%%%%%%%%%%%%%%%%%%%%%%%%%%%%%%%%%%%%%%%%%%%%%%%%%%

% individual chapters included here

\chapter{Introduction}\label{C:intro}

%%%%%%

In February 2016, LIGO, (the Laser Interferometer Gravitational-Wave Observatory) confirmed detection of astrophysical gravitational waves in a groundbreaking experimental achievement~\cite{LIGO, LIGO2}. One century prior, in 1916, Albert Einstein predicted their existence as a means of transporting energy via gravitational radiation as a consequence of his theory of general relativity~\cite{speedofthought, universeofGR, pathways}.\footnote{ The birth of the idea is in fact attributed to Henri Poincar\'e as early as 1905~\cite{Poincare}, however general relativity is most certainly the foundational theory and Einstein's accreditation is contextually appropriate.} The fact that a century of ongoing work was required between the conception of this idea and its scientific verification is one of many examples that are testament to the subtle and difficult nature of gravitational physics.

 At a classical level, general relativity offers a complete theory of gravity. To date there are numerous aspects of the theory which have been experimentally verified (the detection by LIGO being one of the most significant). While it is certainly a far cry from the `holy grail' of a complete theory of quantum gravity, the combination of thoroughly reviewed theoretical justification along with experimental verification informs the conclusion that it is the `best' theory of gravitation we currently possess -- an opinion shared by a large number of the members of the global physics community. Accordingly, general relativity is the author's preferred theory of gravity, and all analyses within this thesis are conducted using the framework it provides. As such, \S\ref{C:Spacetime} and \S\ref{C:GRBackground} introduce the necessary mathematical framework from differential geometry required for tractable analysis in general relativity, as well as presenting several useful techniques which will be utilised for aspects of the subsequent analyses of specific candidate spacetimes.

Nowhere is the subtlety of gravitational physics more prevalent than in discussions pertaining to the numerous different solutions to the Einstein field equations. Throughout the decades we have discovered solutions to these equations which model many qualitatively different astrophysical objects. These range from stars, to black holes, to wormholes, each of which may be stationary, rotating, equipped with an electrical charge, or some combination of these. To this day even the most mathematically simple geometric environment of spherical symmetry provokes intensive and non-trivial discussion at conferences around the world. Hence it is still of notable scientific relevance and value to analyse proposed candidate spacetimes phenomenologically using general relativity, and to draw conclusions as to their physical nature.

This thesis focuses on the analysis of two categories of spacetime -- traversable wormholes and regular black holes. All candidate spacetimes analysed possess spherically symmetric geometries, with one dynamical (time-dependent) candidate and the remainder static (both non-rotational and time-independent). As astrophysical objects wormholes have a rather diverse history, having become significant in the popular culture surrounding science-fiction during the latter half of the twentieth century. It turns out that they also have a colourful scientific lineage, with valid solutions to the Einstein field equations corresponding to wormhole geometries being presented as early as 1935 (with the Einstein-Rosen bridge~\cite{einsteinrosen}). It should be noted that not all wormholes are traversable however -- a mathematically rigorous definition of the subset which are is presented in \S\ref{C:Morris}. Several metric candidates representing traversable wormholes are then dissected, with one particular metric of note being the so-called `exponential metric' -- a favourite for proponents of alternative theories of gravity (alternatives to general relativity). An effort is made wherever possible when examining these solutions to maximally extend all coordinates in the chosen coordinate system in the hopes of examining traversable wormhole geometries with a global coordinate patch. Historically some members of the gravitational community have proposed that wormhole geometries require a two-patch atlas,\footnote{Names suppressed to protect the guilty.} and by presenting traversable wormhole geometries in global coordinate patches it is hoped this notion will be dispelled.

Regular black holes have been objects of significant interest since their initial proposal by Bardeen in 1968~\cite{Bardeen1968}. They are intuitively attractive due to their non-singular nature, and their definition is presented in \S\ref{C:Bardeen-Hayward-Model2}. Analyses of several elementary metrics representing regular black holes are undertaken, before a metric is presented in \S \ref{C:Black-bounce} which neatly interpolates between being a traversable wormhole solution and that of a regular black hole, depending on the value of the newly-introduced scalar parameter $a$. It is with this particular candidate that the notion of `black-bounce' is introduced -- when the parameter $a<2m$ (where $m$ is the mass of our centralised object) the solution corresponds to a regular black hole geometry; this geometry possesses a spacelike hypersurface at the spherical shell $r=0$ which permits one-way travel into a future reincarnation of our own universe.\newpage

After conducting the analysis of the static case for this metric candidate, a Vaidya-like time-dependence is imposed on the metric in \S \ref{C:Vaidya}. This is achieved by allowing the mass of the object $m$ to depend on the outgoing/\-ingoing null time coordinate $w$ (which represents retarded/\-advanced time in identical fashion to the $u$ or $v$ coordinates utilised in Eddington-Finkelstein coordinates~\cite{EddFink, telebook}). This dynamical case describes several physical situations of significant interest, including a `black-bounce' geometry whose regular black hole region grows with time, the transition of a traversable wormhole geometry into a regular black hole, and the converse transition of a regular black hole leaving a wormhole remnant.

Both traversable wormhole and regular black hole geometries have canonically required violations of the energy conditions imposed on the stress-energy-momentum tensor for their construction~\cite{LorentzianWormholes}, which has informed the pre-existing discussion concerning the viability of non-singular black hole and wormhole solutions to the Einstein equations (see for example reference~\cite{viability}). Holistically, the analyses in this thesis contribute to this discussion, and the key findings are then presented in \S \ref{C:con}.

%%%%%%%%%%%%%%%%%%%%%%%%%%%%%%%%%%%%%%%%%%%%%%%%%%%%%%%%%%%%%%%%%

%%%%%%%%%%%%%%%%%%%%%%%%%%%%%%%%%%%%%%%%%%%%%%%%%%%%%%%

\chapter{The concept of `spacetime'}\label{C:Spacetime}

%%%%%%

Spacetime can be intuitively thought of as the stage on which the `play' of the universe is set. Upon this intuition, we must impose mathematically rigorous constraints on which characteristics such a background must exhibit in order to have a physical interpretation in the framework of general relativity. Since our human faculties interpret the universe as having three-dimensional space, and we think of time as being one-dimen\-sional, we construct spacetime by imagining a four-dimensional backdrop consisting of three dimensions of space and one dimension of time. Coupled with this notion is the idea that in the presence of an object equipped with a mass and/or a momentum, the four-dimensional spacetime exhibits curvature. John Archibald Wheeler puts it rather succinctly~\cite{telebook}: ``Space acts on matter, telling it how to move. In turn, matter reacts back on space, telling it how to curve.'' Spacetime is therefore malleable and responsive to the presence of matter. Specifically, Frederic P. Schuller defines a generic spacetime as follows~\cite{Schuller}: ``Spacetime is a four-dimensional topological manifold with a smooth atlas carrying a torsion-free connection compatible with a Lorentzian metric and a time orientation satisfying the Einstein equations.'' Encoded in this statement is much of the mathematical and physical machinery required to interpret a spacetime in the context of general relativity. For the sake of developing a thorough background, let us rigorously unpack these concepts individually before progressing and examining various candidate spacetimes of interest.

%%%%%%

\section{Four-dimensional topological manifold}

This is the type of topological space our so-called `background of the universe' inhabits. A topological manifold $\mathcal{M}$ of dimension $d$ is a mathematical object characterised by the following~\cite{Introtopology, MATH465}:
\begin{itemize}
    \item It is a locally Euclidean topological space -- that is a topological space $(\mathcal{E}, \mathcal{T})$ consisting of a set $\mathcal{E}$ together with a topology of open sets $\mathcal{T}$ which satisfies the following axiom:
    \begin{eqnarray}\label{Euclideanspace}
        && \forall \ x\in\mathcal{E} \ \exists \ \mathcal{O}\in\mathcal{T} \ \mbox{and} \ n\in\mathbb{Z}^{+}: \nonumber \\
        && x\in\mathcal{O} \ \mbox{and} \ \exists \ \mathcal{X}\subset\mathbb{R}^{n} \ \mbox{and} \ \exists \ \mbox{homeomorphism} \ f:\mathcal{O}\leftrightarrow\mathcal{X} \ . \quad
    \end{eqnarray}
    \emph{i.e.} There is a region surrounding each point in the manifold that is homeomorphic to a `chunk' of Euclidean space.
    \item The dimensionality of the space, $d$, is the same everywhere -- this means that with regard to the previous statement (Eq.~\ref{Euclideanspace}) the number $n=d \ \forall \ \mbox{such} \ \mathcal{X}\subset\mathbb{R}^{n}$, so the aforementioned `chunks' of Euclidean space all come from the \emph{same} Euclidean space. In terms of terminology we may call such a space a `$d$-manifold'. In the context of spacetime, since we have `$3+1$'-dimensions corresponding to space and time respectively, the space is a four-manifold.
    \item The manifold is Hausdorff: 
    \begin{equation}
        \forall \ x_{1}, x_{2}\in\mathcal{E}, \ \exists \ \mathcal{O}_{1}, \mathcal{O}_{2}\in\mathcal{T}: x_{1}\in \mathcal{O}_{1}, x_{2}\in \mathcal{O}_{2} \ \mbox{and} \ \mathcal{O}_{1}\cap \mathcal{O}_{2}=\emptyset \ .
    \end{equation}
    \emph{i.e.} All points in the manifold can be `housed off' from each other by open sets of the topology.
    \item The manifold has at least one countable atlas -- a set is `countable' if all its elements can be put in injective correspondence with the natural numbers, $\mathbb{N}$. This statement also encodes the requirement for the topological space to have a countable basis. To see what an atlas is, we may move on to the second mathematical concept in Schuller's definition for spacetime.\footnote{For further elaboration on these very basic concepts from elementary topology please see reference~\cite{Introtopology}; \emph{e.g.} rigorous definition of `homeomorphism', \emph{etc.}}
\end{itemize}

%%%%%%

\section{Smooth atlas}

We may separate `smoothness' from `atlas' for the sake of clarity of definition; first define an `atlas'~\cite{MATH465}:
\begin{itemize}
\item An atlas is a collection of charts (also called `patches') which cover the entire locally Euclidean space $\mathcal{E}$.
\item A chart $(\mathcal{O}, f, U)$ on a member of the topology $\mathcal{O}\in\mathcal{T}$ is a subset $U\subseteq\mathbb{R}^{d}$ together with a homeomorphism $f:\mathcal{O}\leftrightarrow U=f(\mathcal{O})$ (where $d$ is the dimensionality of the manifold).
\item We may therefore think of a countable atlas as being a set of charts $\mathcal{A}=\left\lbrace\left(\mathcal{O}_{i}, f_{i}, U_{i}\right)\right\rbrace: \ f_{i}:\mathcal{O}_{i}\leftrightarrow U_{i}=f_{i}\left(\mathcal{O}_{i}\right)\subseteq\mathbb{R}^{d}$, with $i\in\mathcal{I}$ (arbitrarily indexed by some countable indexing set $\mathcal{I}$), such that $\bigcup_{i\in\mathcal{I}} \ {\mathcal{O}_{i}}=\mathcal{E}$.
\end{itemize}

\noindent Now for `smoothness'~\cite{MATH465}:
\begin{itemize}
    \item An atlas is smooth if all transition maps in the atlas are smooth maps.
    \item A transition map can be defined by the following: let $(\mathcal{O}_{1}, f_{1}, U_{1})$ and $(\mathcal{O}_{2}, f_{2}, U_{2})$ be charts within our topological manifold such that $\mathcal{O}_{1}\cap\mathcal{O}_{2}$ is non-empty. Then the map defined by $g:=f_{2}\circ f_{1}^{-1}$ is the transition map $g:f_{1}(\mathcal{O}_{1}\cap \mathcal{O}_{2})\rightarrow f_{2}(\mathcal{O}_{1}\cap \mathcal{O}_{2})$.
    \item A smooth map is a map for which derivatives of all orders are defined everywhere in its domain. This concept of a `smooth' manifold for which all such transition maps are `smooth' is terminologically interchangeable with what we call a $C^{\infty}$-manifold.
\end{itemize}

Hence we have outlined the specific class of topological spaces which spacetimes must inhabit to have a physical interpretation within the context of general relativity. However, this is still an exceptionally broad class of spaces that could correspond to many qualitatively different manifolds. To draw physical conclusions concerning specific manifolds we must impose additional structure.

%%%%%%

\section{Torsion-free connection}

The additional structure we require is that of a geometry -- we eventually desire a means of making accurate statements concerning distance, angles and curvature within the spacetime. Let there exist some arbitrary smooth $d$-manifold $\mathcal{M}$. The first crucial step towards establishing a geometry on $\mathcal{M}$ is to impose a coordinate system such that we may uniquely identify each point in our manifold. In general, we must establish this coordinate system in a `chart-wise' fashion. Given a countable atlas for $\mathcal{M}$, some $\mathcal{A}=\left\lbrace\left(\mathcal{O}_{i}, f_{i}, U_{i}\right)\right\rbrace: \ \bigcup_{i\in\mathcal{I}}\mathcal{O}_{i}=\mathcal{M}$, we have the following:

\begin{eqnarray}\label{coordinatesystem}
    && \forall \ \textbf{p}\in\mathcal{M}, \exists \ i\in\mathcal{I}, \ \mbox{and} \ \left(\mathcal{O}_{i}, f_{i}, U_{i}\right)\in\mathcal{A}: \ \textbf{p}\in\mathcal{O}_{i} \ , \nonumber \\
    && \nonumber \\
    \Longrightarrow \ && \exists \ \textbf{x}=\left(x_{1},\cdots,x_{d}\right)\in U_{i}\subseteq\mathbb{R}^{d}: \ \textbf{x}=f_{i}(\textbf{p}) \ , \nonumber \\
    && \nonumber \\
    \Longrightarrow && \textbf{p}=f_{i}^{-1}\left(\textbf{x}\right) \ .
\end{eqnarray}
We may then state that point $\textbf{p}\in\mathcal{M}$ has coordinate location $\left(x_{1},\cdots,x_{d}\right)$ with respect to the chosen coordinate patch $\left(\mathcal{O}_{i}, f_{i}, U_{i}\right)$. This is a generalised process for establishing a coordinate system on a given manifold.\footnote{Restricting the domain of each $x_{1},\cdots,x_{d}$ to the reals is a standard practice for the majority of spacetime candidates; all spacetimes explored in this thesis have real-valued domains for their respective coordinate patches.}

\begin{figure}[!htb]
\begin{center}
\includegraphics[scale=0.35]{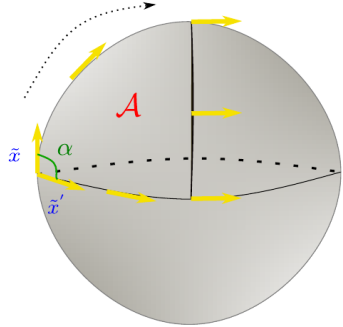}\qquad
\end{center}
{\caption[Parallel transport on the two-sphere]{The vector $\tilde{x}$ is parallel transported clockwise along the curve enclosing the region $\mathcal{A}$ back to its original location in the manifold to yield $\tilde{x}^{'}=T\left(\tilde{x}\right)$. The transport process rotates the vector by an angle $\alpha$.
}\label{paralleltransport}}
\end{figure}

Other than ensuring global coverage of $\mathcal{M}$, our choice of coordinate system is at this stage completely arbitrary. Say we impose a chosen coordinate system on $\mathcal{M}$, and $\exists \ x, y \in \mathcal{M}: \ x\neq y$ .\footnote{A slight point on notation; when defining the assignation of a coordinate system to $\mathcal{M}$ as in Eq.~\ref{coordinatesystem}, `points' in $\mathcal{M}$ were denoted using bold font to highlight their multi-dimensionality. This is a standard practice, however the convention is henceforth abandoned as the dimensionality of mathematical objects shall be obvious from context.} This construction ultimately gives rise to the existence of the respective tangent spaces, $T_{x}$ and $T_{y}$, and we may now speak freely of the behaviour of tangent vectors on $\mathcal{M}$. In order to see the role the `connection' plays in spacetime we must first define the parallel transport of these vectors. Parallel transport is the process by which a vector transports along smooth curves within our manifold. Say we have a smooth curve connecting $x,y\in\mathcal{M}$, and this curve is parameterised by an arbitrary scalar parameter, $\gamma$ (\emph{i.e.} different values of $\gamma$ simply inform placement along the curve). Parallel transport is then defined by a `transport' function, let us call it $T_{\left[x\rightarrow y;\gamma\right]}: T_{x}\rightarrow T_{y}$. This function as a mathematical object is in fact a $T^{1}_{1}$ bi-tensor; for definitions of tangent vectors/tangent spaces/tangent bundles/tensors please see references~\cite{MATH465, tensors} -- this background material is not covered here for conciseness. This function must possess several properties:

\begin{itemize}
    \item The null path $\gamma_{0}$ (where a vector remains stationary within the manifold) must define the identity of the function; $T_{\left[x\rightarrow x;\gamma_{0}\right]}=I:T_{x}\rightarrow T_{x}$.
    \item $T_{\left[x\rightarrow y;\gamma\right]}$ should be an everywhere-invertible mapping; if we are able to propagate smoothly from $x$ to $y$ within our manifold the return journey must also be possible.
    \item Reversing a path should correspond to the inverse of the transport function; $T_{\left[x\rightarrow y;\gamma\right]}=\left(T_{\left[y\rightarrow x;\tilde{\gamma}\right]}\right)^{-1}$. \footnote{Note that $\tilde{\gamma}$ is just some arbitrary scalar-valued re-parameterisation of the reverse path; we have a degree of freedom in choosing these parameters.}
    \item The transport operator ought to be a linear operator between vector spaces. If two vectors $V_{1}, V_{2} \in T_{x}$ are \emph{both} propagated along a smooth curve from $x$ to $y$ in our manifold, then $T_{\left[x\rightarrow y;\gamma\right]}\left(V_{1}+V_{2}\right)=T_{\left[x\rightarrow y;\gamma\right]}\left(V_{1}\right)+T_{\left[x\rightarrow y;\gamma\right]}\left(V_{2}\right)$.
\end{itemize}
These properties are enough to define an adequate transport operator for the parallel transport of vectors in our manifold. For a specific example of parallel transport, see Fig.~\ref{paralleltransport} where a vector is parallel transported along a smooth curve on the two-sphere from an arbitrary starting point back to its original location. It is important to note that the example of the two-sphere is purely used for intuition. It is a two-manifold rather than a four-manifold as we require for a spacetime; however the construction of the transport function extends very naturally to finitely many dimensions with identical conditions imposed on the transport operator and the corresponding number of components comprising the tangent vectors.
\enlargethispage{20pt}
%%%%

%%%%

Ultimately we wish to be able to apply this process of assigning a coordinate system and constructing a transport operator to \emph{any} $d$-manifold with a smooth atlas. We may now understand the role of the `connection'. Let us examine the simple example of the topological space $\mathbb{R}^{2}$ with respect to polar coordinates (the same concept extends naturally to arbitrary four-manifolds with arbitrary coordinate systems). In this example we have a two-manifold we observe to be `flat' (the terms `flat' and `Euclidean' will be used interchangeably henceforth), but have arbitrarily imposed a coordinate system we observe to be `curved'. The standard orthonormal basis for our system is $\left\lbrace \hat{e}_{r}, \hat{e}_{\theta}\right\rbrace$, with each point $x\in\mathbb{R}^{2}=\left(r,\theta\right)$, and with natural domains for our coordinates $r\in\left[0, +\infty\right), \theta\in\left[0,2\pi\right)$. Let $r_{1}, r_{2}\in\mathbb{R}^{+}; \ r_{2}>r_{1}$, and $\theta_{1}, \theta_{2}, \theta_{3}\in\left(\frac{3\pi}{2}, 2\pi\right); \ \theta_{3}>\theta_{2}>\theta_{1}$. We can construct the following figure (Fig.~\ref{Connexion}):
\enlargethispage{40pt}

%%%%

\begin{figure}[!htb]
\begin{center}
\includegraphics[scale=0.40]{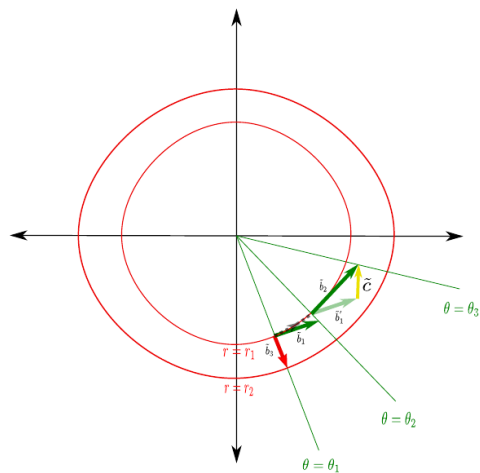}\qquad
\end{center}
{\caption[The connection in $\mathbb{R}^{2}$ with respect to polar coordinates]{Demonstrating the role of the connection in the context of $\mathbb{R}^{2}$ with respect to polar coordinates.
}\label{Connexion}}
\end{figure}

%%%%

The red and green vectors $\tilde{b}_{1}, \tilde{b}_{2}, \ \mbox{and} \ \tilde{b}_{3}$ show the direction in which position in the manifold changes when one of the coordinates is incremented by an infinitesimal amount at each location. These are the basis vectors; specifically $\tilde{b}_{1}$ and $\tilde{b}_{3}$ are the $\hat{e}_{\theta}$ and $\hat{e}_{r}$ basis vectors at the point $\left(r_{1}, \theta_{1}\right)$ respectively, and $\tilde{b}_{2}$ is the $\hat{e}_{\theta}$ basis vector at the point $\left(r_{1}, \theta_{2}\right)$. If the basis vector $\tilde{b}_{1}$ undergoes parallel transport along the smooth curve $r=r_{1}$ from $\left(r_{1}, \theta_{1}\right)$ to $\left(r_{1}, \theta_{2}\right)$, we obtain the vector $\tilde{b}_{1}^{'}$. However, $\tilde{b}_{1}^{'}$ is not a basis vector at that point in the manifold; a rotational correction (and potential scale factor) must be made to yield the correct basis vector $\tilde{b}_{2}$. The difference between vectors $\tilde{b}_{1}^{'}$ and $\tilde{b}_{2}$ is denoted by the vector $\tilde{c}$ in yellow. Note: when speaking of the `difference' between these two vectors or indeed the notion of `scale' one assumes the existence of some nonsingular parameterisation along the curves in the manifold; for now this is a sufficiently adequate notion of `distance'. By definition the vector $\tilde{c}$ must be some linear combination of the basis vectors for our manifold; \emph{i.e.} for some $\alpha_{1}, \alpha_{2} \in \mathbb{R}$, $\tilde{c}=\alpha_{1}\hat{e}_{r}+\alpha_{2}\hat{e}_{\theta}$. We may encode all of the information comprising $\tilde{c}$ in the triple-indexed object $\Gamma^{i}{}_{jk}$ which we define to be the `connection'. This object has the following properties~\cite{telebook, Hartle, Carroll}:

\begin{itemize}
    \item $i, j, k \in \left\lbrace r, \theta\right\rbrace$ ;
    \item The index $i$ indicates the basis vector one is stretching/multiplying;
    \item The index $j$ indicates which basis vector is moving;
    \item The index $k$ indicates which basis vector defines the direction of the motion.
\end{itemize}
The vector $\tilde{c}$ can then be written as: $\tilde{c}=\Gamma^{r}{}_{\theta\theta} \, \hat{e}_{r}+\Gamma^{\theta}{}_{\theta\theta} \, \hat{e}_{\theta}$. We see that the amount a basis vector is altered when propagating along smooth curves in the manifold via parallel transport is encoded by the connection. It follows that the connection is a rate of change of our previously defined transport function $T$, and in fact defines the covariant derivative of this bi-tensor. There is significant mathematical subtlety in defining tensor derivatives; for more detail on the covariant derivative, and methods of tensorial differentiation in general, please see references~\cite{telebook, Hartle, Carroll}.

The example in Fig.~\ref{Connexion} is one where a curved coordinate system has been imposed upon flat space; the very same construction may extend to a curved manifold that has been bestowed with a coordinate system as `straight' as possible. In the context of spacetime, the connection also extends naturally to arbitrary four-manifolds. The difference in such cases is that the domain of the indices is now $i, j, k \in \left\lbrace 0, 1, 2, 3\right\rbrace$ (where $0, 1, 2, 3$ are general labels for the four coordinates comprising whichever chosen coordinate system is employed). $\Gamma^{i}{}_{jk}$ is hence our first tool in the general relativity arsenal that enables us to understand the curvature of a given spacetime; by informing us how much the basis vectors themselves must rotate as they parallel propagate smoothly through the manifold.

Before defining the explicit mathematical form of $\Gamma^{i}{}_{jk}$ by discussing the metric tensor, note that a \emph{torsion-free} connection is a connection which obeys the following condition~\cite{telebook, MATH465}:

\begin{equation}
    \Gamma^{a}{}_{bc} = \Gamma^{a}{}_{cb} \ .
\end{equation}
Setting torsion to zero is non-controversial and can greatly simplify the resulting mathematics -- we do not require non-zero torsion to conduct standard analyses within general relativity. It is also a standard practice reinforced by empirical data; experiments to date are yet to demonstrate a requirement for non-zero torsion (see \emph{e.g.} the discussion in reference~\cite{marchtorsion}).

%%%%%%

\section{Lorentzian metric}\label{metric}

Given an arbitrary coordinate system on some manifold $\mathcal{M}$, the notion of distance between any two points is defined by an object called the `metric'. The form that the metric takes is context-dependent. In mathematics a general metric on a set $\mathcal{E}$ can be characterised by the following definition~\cite{Introtopology} (where $a, b, c \in \mathcal{E}$ are points expressed as coordinate locations with respect to the chosen coordinate system):

The usual notion of a mathematical metric on a set $\mathcal{E}$ is a function $ \ \ \ \ \ \ \ \ \ \ g:\mathcal{E} \ \mbox{x} \ \mathcal{E} \rightarrow \mathbb{R}$ with the following properties:
\begin{eqnarray}\label{metricproperties}
    &\bullet& \ g(a,b)\geq 0 \ \forall \ a,b\in\mathcal{E}; \ \mbox{and} \ g(a,b)=0 \ \mbox{iff} \ a=b \ , \nonumber \\
    &\bullet& \ g(a,b)=g(b,a) \ \forall \ a,b\in\mathcal{E} \ , \nonumber \\
    &\bullet& \ g(a,b)+g(b,c)\geq g(a,c) \ \forall \ a,b,c\in\mathcal{E} \ .
\end{eqnarray}
We call $g(a,b)$ the distance between points $a$ and $b$, and the pair $(\mathcal{E}, g)$, consisting of the set $\mathcal{E}$ and the metric $g$, a metric space (this mathematical notion of a metric will need modification in Lorentzian spacetimes; more on this below). It should be noted that the function $g$ is defining a sense of `direct distance' between two points in the manifold; \emph{i.e.} the distance between them is minimised (this is a corollary of the third property listed in Eq.~\ref{metricproperties}, also known as the triangle inequality). As such, $g$ is informed by an object called the `line element', which is constructed from the length traversed along the `straightest possible curve' between two points in the manifold.

Let us look at the mathematically simple environment of the Euclidean three-manifold with respect to Cartesian coordinates. In this environment, the `straightest possible curve' between two points, quite trivially, is our intuitive notion of a straight line.\footnote{To generalise these `straight curves' to arbitrary curved manifolds we define `geodesic curves'; for more information on geodesic curves see \S\ref{einsteineqs}.} In Cartesian coordinates we have an orthogonal basis. This permits a method of easily foliating the space with right-angled triangles whose shorter two sides can be expressed solely in terms of the displacements of our basis components. As such, we impose a definition of distance between two points in accordance with the Pythagorean theorem. Given the canonical basis representation for $\mathbb{R}^3$ in Cartesian coordinates, $\left(x, y, z\right)$, this yields the following mathematical form of the line element:\footnote{Use of the Einstein summation convention is employed here, where repeated `up-down' indices are summed over, for more details see references~\cite{Hartle, MATH465}. Henceforth the specific Latin indices ($i, j, k$) will index the three dimensions of space whilst any Greek indices (\emph{e.g.} $\mu, \nu$) will index the four dimensions of spacetime. The summation convention is assumed throughout the remainder of the thesis whenever pairs of `up-down' indices appear in the same expression unless otherwise stated.}
\enlargethispage{30pt}

\begin{eqnarray}
    ds^2 &=& dx^2+dy^2+dz^2 \ \nonumber \\
    &=& g_{xx}dxdx+g_{yy}dydy+g_{zz}dzdz \ \nonumber \\
    &=& g_{ij}dx^{i}dx^{j} \ .
\end{eqnarray}
We see immediately that the line element informs a diagonal metric environment for $g$, with components $g_{xx}=g_{yy}=g_{zz}=1$. Hence the Euclidean three-manifold has been equipped with the diagonal metric $g_{ij}= \ \mbox{diag}\left(1, 1, 1\right)=\delta_{ij}$, and this construction satisfies the properties of Eq.~\ref{metricproperties}. This then allows us to define the length of arbitrary smooth curves in $\mathbb{R}^3$ via the following: take the limit as each $dx^{i}\rightarrow 0$, then integrate the line element with respect to some scalar parameterisation of the desired curve (a common parameter choice is a scalar-valued temporal parameter $t$). The length of the curve is then:

\begin{equation}\label{flatdistance}
    L_{\mathbb{R}^{3}} = \int_{t_{0}}^{t_{1}}\sqrt{\delta_{ij}\frac{dx^{i}}{dt}\frac{dx^{j}}{dt}} \ dt \ .
\end{equation}
Here the curve begins at $t=t_{0}$ and ends at $t=t_{1}$. It is worth noting that as a mathematical object, the metric $g$ has a unique matrix representation. This is the case for all manifolds and contexts for which we wish to define a metric.

In order to migrate from the purely mathematical definition of the metric to the realm of physics, we relax the three properties of Eq.~\ref{metricproperties} so that we may attribute physical meaning to our notion of distance. This difference in definition is a direct result of the important role that time plays in the universe; the physical motivation behind it is to be able to clearly separate between events\footnote{To this point precise coordinate locations within a manifold with respect to some coordinate system have been dubbed `points'. `Events' are simply `points' within a manifold which are being expressed with respect to a coordinate system which has a defined notion of time. Consequently, in the context of spacetime the two terms are interchangeable.} which are timelike-separated, null-separated, and spacelike-separated. Before discussing the metric in the context of relativity it pays to clearly define what each of these terms physically mean~\cite{telebook}:

\begin{itemize}
    \item `Timelike-separated' events in a spacetime are events that can be reached from each other by traveling strictly within either the future or past-directed light cones on a spacetime diagram. This means that any `timelike-separated' events may inform each other, in the sense that it is possible for both massive and massless objects to travel between the two events without violating the condition that nothing propagate faster than the speed of light in a vacuum, $c$.
    \item `Null-separated' events in a spacetime are events that can only be reached from  one another by particles traveling at the speed of light $c$, \emph{i.e.} along the boundaries of the future and past-directed light cones in a spacetime diagram. This ultimately restricts the types of particles that may inform two null-separated events to those that have both zero rest mass and propagate with velocity $c$, \emph{e.g.} photons or gluons.
    \item `Space-like separated' events in a spacetime are events that lie strictly outside of both the future and past-directed light cones emanating from one another on a spacetime diagram. If we ignore the possibilities of superluminal velocities due to quantum phenomena (which for elementary general relativity we certainly do), these events may never inform each other, as to do so would imply particles are traveling at a velocity greater than $c$.
\end{itemize}

%%%%

\begin{figure}[!htb]
\begin{center}
\includegraphics[scale=0.30]{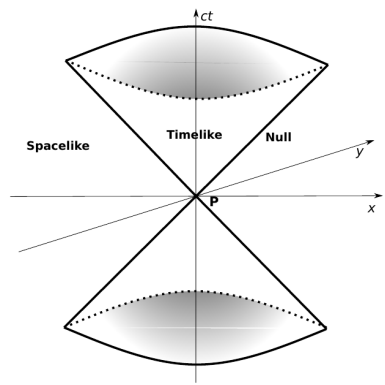}\qquad
\end{center}
{\caption[Spacetime diagram of the local light cone for $P\in\mathcal{M}$]{A spacetime diagram showing the regions which are timelike, null, and spacelike separated from the central event $P$. One of the spatial dimensions has been suppressed for ease of diagrammatic representation; the construction extends naturally to `$3+1$'-dimensions. Note that the light cone inhabits the tangent space of $P$, $T_{P}$, as such defining the separation of events in a locally flat region `near' $P$. We may parallel transport the light cone in much the same way as the basis vectors through arbitrary curved manifolds to see the generalised separation of events from $P$.
}\label{Lightcone}}
\end{figure}

%%%%

The desired physics inherent in the construction of spacetime therefore requires we have a metric which allows us to define a notion of distance within the manifold whilst still being able to clearly differentiate between these three physical scenarios. To see how this is done, let us introduce the notion of metric signature. The signature of a metric is determined by the number of both positive and negative eigenvalues that arise from its matrix representation~\cite{Carroll}. A metric with Riemannian signature has exclusively positive eigenvalues; this yields a positive definite metric tensor and the notion of distance is simply the length of the \emph{shortest} possible curve in the manifold. For the prior example of $\mathbb{R}^3$ with respect to Cartesian coordinates, the metric $g$ has Riemannian signature. The distance between two points can then naturally be found by minimising Eq.~\ref{flatdistance}. Alternatively, a `Lorentzian metric' is a metric of Lorentzian signature; defined to be a metric whose matrix representation possesses one single negative eigenvalue while the remainder are positive. In this case the notion of distance corresponds to the principle of least action in the manifold; this is an `extremal distance' defined by the principles of variational calculus (for details see reference~\cite{variational}).

In the context of general relativity, the separation between timelike/\-null/\-spacelike-separated events is achieved by imposing a metric of Lor\-entzian signature on the four-manifold. To see this, let us suppose we provide the prior example of three-dimensional Euclidean space with an additional dimension for time, parameterised by some scalar-valued temporal variable $t$, treating the temporal dimension identically to the three spatial dimensions. Our manifold is now the four-manifold $\mathbb{R}^4$. One would naturally extend the Pythagorean notion of distance imposed on $\mathbb{R}^3$ to apply to the new scenario, yielding the new line element:

\begin{equation}
    ds^2 = dt^2+dx^2+dy^2+dz^2 = \delta_{\mu\nu} dx^{\mu} dx^{\nu} \ .
\end{equation}
The metric's matrix representation is the Kronecker delta function in four dimensions; this is positive definite and therefore Riemannian. However, notice that if one considers events that are space-like separated \emph{versus} those that are time-like separated, the `distance' in each instance between the events is strictly positive, and there is no distinct value which marks the transition between the two. Consequently one receives no information about the underlying physicality of events in spacetime by using this metric. Let us now impose a metric of Lorentzian signature (note -- there are several ways to do this as we can technically choose any of the four basis variables to correspond to the negative eigenvalue which will be present in the metric's matrix representation; this is a matter of taste). The author's preferred convention is to enforce the signature `-,+,+,+', such that the temporal metric coefficient inherits the negative eigenvalue. This convention will be employed for every spacetime candidate analysed in this thesis.
\newpage
\noindent We may then define a line element:

\begin{equation}
    ds^2 = -dt^2+dx^2+dy^2+dz^2 \ .
\end{equation}
The metric is now $g_{\mu\nu}=\eta_{\mu\nu}=\mbox{diag}\left(-1, 1, 1, 1\right)$, and one can easily see that when adopting geometric units (where the speed of light $c=1$), all null curves are characterised by the property that $ds^2 = 0$. This is due to the fact that given a null curve in the manifold, \emph{i.e.} the worldline of a particle traveling at the speed of light, then the spatial and temporal displacements must be equal for all values of $t$. Hence $dt=\sqrt{dx^2+dy^2+dz^2}$, and the result follows. Furthermore, if events are timelike-separated then particles traveling between them must be propagating at a velocity less than the speed of light, which implies $ds^2<0$ as the temporal term provides the dominant balance. Conversely, spacelike-separated events are characterised by a positive line element, $ds^2>0$. Equipping the four-manifold with a Lorentzian metric has enabled us to easily differentiate between the three cases as desired.

This specific example is in fact the flat-space limit to general relativity of special relativity, also known as Minkowski space (notice that for a specific $t=$constant time-slice the induced three-metric is strictly Euclidean). This extends to a generalised definition for the metric in general relativity as follows~\cite{Hartle, ONeill}:

The metric is defined as a symmetric, non-degenerate, and position-depend\-ent matrix that is a rank-two tensor, denoted canonically by $g_{\mu\nu}$, and which has constant Lorentzian signature. Its form is governed by the `line element'; the distance in the manifold $\mathcal{M}$ between points which are infinitesimally displaced with respect to the chosen coordinate basis. The generalised line element is given by:

\begin{equation}
    ds^2 = g_{\mu\nu}dx^{\mu}dx^{\nu} \ ,
\end{equation}
where $dx^{\mu}$ are the infinitesimal displacements of each of the coordinates. The generalised notion of distance between two points is found by extremising the action of the following integral:

\begin{equation}\label{arclengthgen}
    L_{\mathcal{M}} = \int_{\gamma_{0}}^{\gamma_{1}}\sqrt{g_{\mu\nu}\frac{dx^{\mu}}{d\gamma}\frac{dx^{\nu}}{d\gamma}} \ d\gamma \ ,
\end{equation}
where a geodesic curve connecting the points is arbitrarily parameterised by some scalar parameter $\gamma$ (and the points are located on the curve at $\gamma_{0}$ and $\gamma_{1}$ respectively).

Having established what a `Lorentzian metric' is and why we equip our manifold with such a structure we may now explicitly define the connection's mathematical form in terms of the metric. Recalling that the connection is torsion-free, we may specialise it to the case of Christoffel symbols of the second kind -- simply a torsion free, affine, metric connection (for details see references~\cite{MATH465, Hartle, Carroll}). For some metric tensor $g_{\mu\nu}$, these Christoffel symbols are defined by:

\begin{equation}\label{Christoffel}
    \Gamma^{\mu}{}_{\alpha\beta} = \frac{1}{2}g^{\mu\nu}\left(\partial_{\alpha}g_{\nu\beta}+\partial_{\beta}g_{\nu\alpha}-\partial_{\nu}g_{\alpha\beta}\right) \ ,
\end{equation}
where the index $\nu$ is contracted over in accordance with the Einstein summation convention. Hence we have equipped our four-manifold with a means of measuring both distance and curvature -- a geometry on the space as desired. This concludes the mathematical aspects from differential geometry required for the construction of spacetime; the remaining conditions of imposing a time orientation and satisfying the Einstein equations at all coordinate locations represent physical constraints we wish to place on candidate spacetimes in order to satisfy desirable laws of physics.

%%%%%%

\section{Time orientation}

Suppose we have a construction $\left(\mathcal{M}, g_{\mu\nu}\right)$; that is a generic four-manifold equipped with a Lorentzian metric (and a corresponding torsion free metric connection as defined by Eq.~\ref{Christoffel}). The purpose of imposing a time orientation on $\left(\mathcal{M}, g_{\mu\nu}\right)$ is to preserve global causal structure. Locally, the causal structure is akin to that of special relativity -- this ensues due to the fact that we may always approximate local flatness in general relativity. Globally however, topology may not be so trivial, and objects such as manifold singularities or self-identified `twists' within $\mathcal{M}$ may result in a confused notion of causality. This is something we wish to avoid for standard general relativity. To make this concept precise, we may refer back to the light cone construction in Fig.~\ref{Lightcone}. Arbitrarily we wish to assign half of the cone to be the `future' and the other half to therefore be the `past' -- the events lying within the `future' half of the light cone from $P$ are then deemed to belong to the `chronological future' of $P$, whilst events lying within the past half belong to the `chronological past' of $P$ (note that all events in both the chronological future and past of $P$ are strictly timelike-separated from $P$). Keeping in mind that the light cone belongs to the tangent space of $P$, $T_{P}$, a time orientable spacetime $\left(\mathcal{M}, g_{\mu\nu}\right)$ is defined as a spacetime in which it is possible to make a continuous designation of the chronological future of $P$ and the chronological past of $P$ as $P$ varies over $\mathcal{M}$~\cite{Wald}. Physically this encodes a requirement that as we move through the manifold we must always know in which direction through time we are propagating; without this the base causality assumption of `cause' and then subsequent `effect' collapses. To better understand this statement, let us examine an example of a spacetime which is \emph{not} time orientable (Fig.~\ref{timeorientable}):
\newpage

%%%%

\begin{figure}[!htb]
\begin{center}
\includegraphics[scale=0.45]{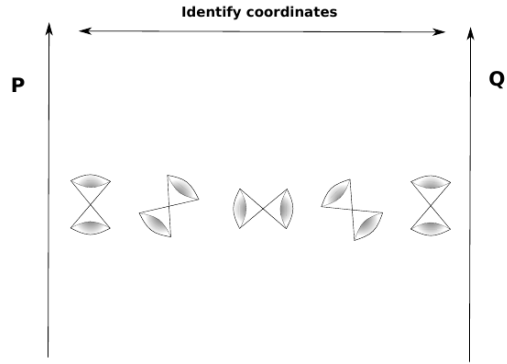}\qquad
\end{center}
{\caption{{A temporally non-orientable spacetime.
}}\label{timeorientable}}
\end{figure}

%%%%

In this example, the coordinate location of event $P$ is identified with event $Q$ (via imposing some periodicity on the time coordinate, $t$). Correspondingly their respective light cones in their respective tangent spaces are also identified. Through a periodic cycle of $t$, the topology has permitted the light cones to `tip' sufficiently such that they rotate completely; what was the chronological future of event $P$ becomes the chronological past of event $Q$. This occurs via smooth propagation through the manifold. As such we have no clear notion of a temporal direction -- the same coordinate location within the spacetime has a contradicting chronological future and past. A continuous designation of chronological future and past can not be made, and by definition the spacetime therefore does not possess a valid time orientation.

To codify the notion of time orientable spacetimes we may make the following mathematical statement~\cite{delmatt, dewitt}: A spacetime $\left(\mathcal{M}, g_{\mu\nu}\right)$ is time orientable if and only if there exists a globally defined timelike vector field on $\mathcal{M}$. All candidate spacetimes examined in this thesis are time-orientable.

%%%%%%

\section{Einstein's general relativity}\label{einsteineqs}

\subsection{Einstein's motivation}

Before presenting and explaining the Einstein field equations which underpin all of general relativity, it pays to provide historical context to fully understand their origin. In 1686, Isaac Newton developed his famous equation for the force felt due to gravity between two objects of respective masses $m_{1}$ and $m_{2}$, separated by some distance $r$~\cite{Newton}:

\begin{equation}\label{Newton}
    F = G_{N}\frac{m_{1}m_{2}}{r^2} \ .
\end{equation}
Newton's Law (Eq.~\ref{Newton}) was the pillar of all gravitational physics for over two hundred years. Specifically, in the context of the gravitational field of the Earth it was remarkably accurate and sufficient to satisfy essentially all practical/engineering requirements that were faced pre-twentieth century. Meanwhile, a very similar equation was developed in 1785 by Charles-Augustin de Coulomb describing the electrostatic force felt between two charged particles of respective charges $q_{1}$ and $q_{2}$, separated by some distance $r$~\cite{Griffiths}:

\begin{equation}\label{Coulomb}
    F = \frac{1}{4\pi \epsilon_{0}}\frac{q_{1}q_{2}}{r^2} \ .
\end{equation}

One can immediately see that other than the constants of proportionality, both Coulomb's and Newton's Laws (Eq.~\ref{Coulomb} and Eq.~\ref{Newton} respectively) are essentially identical. Gravitational/electrostatic force between two objects equipped with mass/charge are directly proportional to the product of their masses/charges and inversely proportional to the square of the distance between them. However physicists quickly saw a problem with these equations. They noticed that if one arbitrarily changes the distance $r$ to some new value, the force felt between the particles will change in accordance with the formula; the problem here is that time has not been factored into account -- action at a distance is instantaneous. Permitting instantaneous action at a distance disagreed with all relevant experimental evidence at the time, as it was well known that all propagating objects take measurable quantities of time to move through space and for their subsequent effects to be felt. Specifically pertaining to Coulomb's Law, a new school of thought was required to explain electrostatics. The consensus was reached that there must exist some as yet `undiscovered' physical objects which permit the propagation of electrostatic force between particles. These objects were deemed to be `electrical fields', and field theory in the context of electromagnetism was born. A large number of notable physicists worked together to fully describe the nature of these fields, and eventually James Clerk Maxwell published the unification of electromagnetism in the form of Maxwell's equations (this particular form using vector calculus was presented in 1884 by Oliver Heaviside, based on the initial publication by Maxwell \emph{circa} 1862~\cite{Hunt}):

\begin{eqnarray}
\nabla\cdot\textbf{E} &=& \frac{\rho}{\epsilon_{0}} \ ; \nonumber \\
\nabla\cdot\textbf{B} &=& 0 \ ; \nonumber \\
\nabla  \times  \textbf{E} &=& -\frac{\partial\textbf{B}}{\partial t} \ ; \nonumber \\
\nabla  \times  \textbf{B} &=& \mu_{0}\left(\textbf{J}+\epsilon_{0}\frac{\partial\textbf{E}}{\partial t}\right) \ .
\end{eqnarray}
These field equations are some of the most celebrated equations in all of physics, and encode all information required for classical electromagnetism (we now know that quantum electrodynamics is a more fundamental theory, and that Maxwell's equations are the classical limit of this theory). The problem of instantaneous action at a distance with respect to Coulomb's Law was resolved. However Newton's Law (Eq.~\ref{Newton}) was still the only tool we had to describe the gravitational force, and still permitted instantaneous action at a distance. The objective therefore was to alter gravitational theory in a very similar way to what was done for electromagnetism; by supposing there exist gravitational fields which encode and transmit information pertaining to the gravitational force between massive objects.
\enlargethispage{10pt}

While many physicists attempted to revolutionise the gravitational theory, it was Albert Einstein who almost single-handedly did so with the formulation of general relativity. This theory was predicated on the Einstein equivalence principle (also known as the `strong' equivalence principle). In order to formulate this principle Einstein began with an observationally well established and fundamental tenet of gravity; that the trajectory of an object which is exclusively under the influence of a gravitational force is independent of the object's mass.\footnote{Knowledge of this fact is often attributed to Galileo Galilei, who is said to have dropped spherical objects of different masses from the leaning tower of Pisa (\emph{circa}~1590) and observed that they hit the ground at identical times. The factual accuracy of this anecdote is commonly doubted, however beginning \emph{circa} 1885, Lor\'and E\"otv\"os conducted a highly accurate experiment which concluded that the inertial and gravitational masses of objects are indeed equal~\cite{Eotvos}.} This phenomenon is known as the `universality of free fall' and is often referred to as the `weak' equivalence principle. The intriguing thing about this phenomenon is that it is different from all of the other fundamental forces in physics; electromagnetic forces for example are certainly dependent on the charges of the objects involved in the system, and objects with different charges will have a fundamentally different response when propagated through an electrical field. Equipped with this knowledge, Einstein made an incredible insight -- that the `fields' through which gravitational force propagates must in fact be the fabric of space and time itself.

It follows naturally that in much the same way as a charged particle alters the electrical fields around it, an object equipped with a mass must alter the fabric of space and time in its vicinity as well. This was the first notion that space and time could be `curved'. He posited that all objects in a gravitational field simply move along the `straightest possible' paths through the curved space and time, and will continue to do so perpetually until some other non-gravitational force acts on them. In the context of our construction of the four-manifold for spacetime, these `straight' paths are called geodesic curves, and they are defined mathematically by the fact that tangent vectors will always remain parallel to each other when undergoing parallel transport along the curve. As we have already seen, the connection $\Gamma^{\mu}{}_{\alpha\beta}$ encodes the measure of parallelism within a manifold, and as such we may formulate the strong equivalence principle.

\subsubsection{Einstein equivalence principle:}

Gravity is encoded by the Christoffel connection $\Gamma^{\mu}{}_{\alpha\beta}$ on a topological four-manifold equipped with an associated metric tensor $g_{\mu\nu}$ of Lorentzian signature such that:

\begin{itemize}
    \item The universality of free fall is defined by the geodesic equations of motion of the Christoffel connection with respect to the chosen coordinate system. This means that a curve in the manifold parameterised by some arbitrary scalar parameter $\gamma$ is a geodesic if (and only if) the tangent vectors to the curve, given by $t^{\mu}=\frac{dX^{\mu}}{d\gamma}$, satisfy the following differential equation (known as the geodesic equation):
    \begin{equation}\label{geodesic1}
        \frac{d^2X^{\mu}}{d\gamma^2}+\Gamma^{\mu}{}_{\alpha\beta}\frac{dX^{\alpha}}{d\gamma}\frac{dX^{\beta}}{d\gamma} = f(\gamma)\frac{dX^{\mu}}{d\gamma} \ ,
    \end{equation}
    for the full derivation of this equation please see \S\ref{geodesic}. The parameter $\gamma$ is deemed to be `affine' if $f(\gamma)=0$ for the function $f$ in Eq.~\ref{geodesic1}, hence the geodesic equations for an affine parameter are given by:
    \begin{equation}\label{affinegeodesic}
        \frac{d^2X^{\mu}}{d\gamma^2}+\Gamma^{\mu}{}_{\alpha\beta}\frac{dX^{\alpha}}{d\gamma}\frac{dX^{\beta}}{d\gamma} = 0 \ .
    \end{equation}
    \item The flat space limit of spacetime recovers the theory of special relativity, where the metric tensor becomes the Minkowski metric; $g_{\mu\nu} = \eta_{\mu\nu} = \ \mbox{diag}\left[-1, 1, 1, 1\right]$.
\end{itemize}

From this Einstein had a framework for how objects propagate through gravitational `fields', \emph{i.e.} curved spacetime. He still required a codifying set of `field' equations that would relate the curvature of the manifold to the source of that curvature; the specific distribution of mass and energy informed by objects within the manifold itself.

\subsection{Geodesic equations of motion}\label{geodesic}

Before presenting the Einstein field equations, let us demonstrate that geo\-desic motion does in fact correspond to Eq.~\ref{geodesic1}. In accordance with fundamental geometry we know that the shortest distance between two points in Euclidean space is a straight line. The length of a curve in Euclidean space parameterised by an arbitrary scalar parameter $\gamma$ is given by the arc length formula as follows:

\begin{equation}
    L = \int\sqrt{\delta_{ij}\frac{dX^{i}}{d\gamma}\frac{dX^{j}}{d\gamma}} \ d\gamma \ ,
\end{equation}
we extend this definition naturally to the flat space limit of general relativity (the Minkowski space environment of special relativity) by equipping the four-manifold with the metric tensor $\eta_{\mu\nu}=\mbox{diag}\left(-1, 1, 1, 1\right)$. This yields the following definition for arc length in special relativity:

\begin{equation}
    L = \int\sqrt{\eta_{\mu\nu}\frac{dX^{\mu}}{d\gamma}\frac{dX^{\nu}}{d\gamma}} \ d\gamma \ ,
\end{equation}
and finally we may analogously define generalised arc length for general relativity in an arbitrary four-manifold equipped with some Lorentzian metric $g_{\mu\nu}$ by:

\begin{equation}\label{arclengthgeo}
    L = \int\frac{dL}{d\gamma} \ d\gamma = \int\sqrt{g_{\mu\nu}\frac{dX^{\mu}}{d\gamma}\frac{dX^{\nu}}{d\gamma}} \ d\gamma \ .
\end{equation}

We defined the geodesic equation previously as the equation which governs universal free fall; hence in accordance with Einstein's suppositions it should define a curve in the manifold which is `as straight as possible'. Keeping analogous with Euclidean space, this would imply that a geodesic curve between two points on the manifold (say $x, y\in\mathcal{M}$) must be such that the arc length of the curve extremises the distance between $x$ and $y$. Therefore a curve on $\mathcal{M}$ is a geodesic if (and only if) its tangent vectors extremise Eq.~\ref{arclengthgeo}. Realising that the expression within the integrand is a Lagrangian $\mathcal{L}(\gamma, \textbf{q}(\gamma), \textbf{q}'{\gamma})$, we see that the path between points which extremises arclength $L$ is characterised by the Euler-Lagrange equations from elementary variational calculus~\cite{variational}. As such a geodesic curve must satisfy the following differential equation:

\begin{equation}
    \frac{d}{d\gamma}\left\lbrace\frac{\partial}{\partial\left(dX^{\alpha}/d\gamma\right)}\left(\sqrt{g_{\mu\nu}\frac{dX^{\mu}}{d\gamma}\frac{dX^{\nu}}{d\gamma}}\right)\right\rbrace-\frac{\partial}{\partial X^{\alpha}}\left(\sqrt{g_{\mu\nu}\frac{dX^{\mu}}{d\gamma}\frac{dX^{\nu}}{d\gamma}}\right) = 0 \ .
\end{equation}

We can now show that this differential equation is in fact equivalent to the equation for geodesic motion presented in Eq.~\ref{geodesic1}. Evaluating:

\begin{eqnarray}
\frac{d}{d\gamma}\left\lbrace \frac{1}{\sqrt{g_{\mu\nu}\frac{dX^{\mu}}{d\gamma}\frac{dX^{\nu}}{d\gamma}}}g_{\alpha\beta}\frac{dX^{\beta}}{d\gamma}\right\rbrace &-& \frac{1}{2\sqrt{g_{\mu\nu}\frac{dX^{\mu}}{d\gamma}\frac{dX^{\nu}}{d\gamma}}}\frac{\partial g_{\beta\zeta}}{\partial X^{\alpha}}\frac{dX^{\beta}}{d\gamma}\frac{dX^{\zeta}}{d\gamma} = 0 \ ; \nonumber \\
&& \nonumber \\
\frac{d}{d\gamma}\left(\frac{1}{\sqrt{g_{\mu\nu}\frac{dX^{\mu}}{d\gamma}\frac{dX^{\nu}}{d\gamma}}}\right)g_{\alpha\beta}\frac{dX^{\beta}}{d\gamma} &+& \frac{1}{\sqrt{g_{\mu\nu}\frac{dX^{\mu}}{d\gamma}\frac{dX^{\nu}}{d\gamma}}}\left\lbrace g_{\alpha\beta}\frac{d^2X^{\beta}}{d\gamma^2}+\frac{\partial g_{\alpha\beta}}{\partial\gamma}\frac{dX^{\beta}}{d\gamma}\right\rbrace \nonumber \\
&& \nonumber \\
&-& \frac{1}{2\sqrt{g_{\mu\nu}\frac{dX^{\mu}}{d\gamma}\frac{dX^{\nu}}{d\gamma}}}\frac{\partial g_{\beta\zeta}}{\partial X^{\alpha}}\frac{dX^{\beta}}{d\gamma}\frac{dX^{\zeta}}{d\gamma} = 0 \ , \nonumber
\end{eqnarray}\newpage
\noindent and if we define a function:

\begin{eqnarray}
   f(\gamma) &=& -\sqrt{g_{\mu\nu}\frac{dX^{\mu}}{d\gamma}\frac{dX^{\nu}}{d\gamma}}\frac{d}{d\gamma}\left(\frac{1}{\sqrt{g_{\mu\nu}\frac{dX^{\mu}}{d\gamma}\frac{dX^{\nu}}{d\gamma}}}\right) \nonumber \\
   && \nonumber \\
   &=& \frac{1}{2} \ \frac{d}{d\gamma}\left[ln\left(g_{\mu\nu}\frac{dX^{\mu}}{d\gamma}\frac{dX^{\nu}}{d\gamma}\right)\right] \ ,
\end{eqnarray}
we have:

\begin{eqnarray}\label{geodesicderivation}
g_{\alpha\beta}\frac{d^2X^{\beta}}{d\gamma^2}+\frac{\partial g_{\alpha\beta}}{\partial\gamma}\frac{dX^{\beta}}{d\gamma}-\frac{1}{2}\frac{\partial g_{\beta\zeta}}{\partial X^{\alpha}}\frac{dX^{\beta}}{d\gamma}\frac{dX^{\zeta}}{d\gamma} &=& f(\gamma)g_{\alpha\beta}\frac{dX^{\beta}}{d\gamma} \ ; \nonumber \\
&& \nonumber \\
g_{\alpha\beta}\frac{d^2X^{\beta}}{d\gamma^2}+\left\lbrace \frac{\partial g_{\alpha\beta}}{\partial X^{\zeta}}-\frac{1}{2}\frac{\partial g_{\beta\zeta}}{\partial X^{\alpha}}\right\rbrace\frac{dX^{\beta}}{d\gamma}\frac{dX^{\zeta}}{d\gamma} &=& f(\gamma)g_{\alpha\beta}\frac{dX^{\beta}}{d\gamma} \ . \nonumber \\
\end{eqnarray}

In Eq.~\ref{Christoffel} we defined the Christoffel symbols of the second kind; the Christoffel symbols of the first kind are obtained via lowering indices using the metric tensor (for details on this process see references~\cite{MATH465, telebook}) and are as follows:

\begin{equation}
    \Gamma_{\alpha\beta\zeta} = \frac{1}{2}\left(\partial_{\beta}g_{\alpha\zeta}+\partial_{\zeta}g_{\alpha\beta}-\partial_{\alpha}g_{\beta\zeta}\right) \ .
\end{equation}

Looking specifically at the second term on the left hand side of Eq.~\ref{geodesicderivation}, in view of the fact that there is a contraction on the indices $\beta$ and $\zeta$, we may rewrite it as follows:

\begin{eqnarray}
    \left\lbrace \frac{\partial g_{\alpha\beta}}{\partial X^{\zeta}}-\frac{1}{2}\frac{\partial g_{\beta\zeta}}{\partial X^{\alpha}}\right\rbrace\frac{dX^{\beta}}{d\gamma}\frac{dX^{\zeta}}{d\gamma} &=& \left\lbrace \frac{1}{2}\frac{\partial g_{\alpha\zeta}}{\partial X^{\beta}}+\frac{1}{2}\frac{\partial g_{\alpha\beta}}{\partial X^{\zeta}}-\frac{1}{2}\frac{\partial g_{\beta\zeta}}{\partial X^{\alpha}}\right\rbrace \frac{dX^{\beta}}{d\gamma}\frac{dX^{\zeta}}{d\gamma} \nonumber \\
    && \nonumber \\
    &=& \Gamma_{\alpha\beta\zeta}\frac{dX^{\beta}}{d\gamma}\frac{dX^{\zeta}}{d\gamma} \ ,
\end{eqnarray}
hence the geodesic equation of motion is given by:

\begin{equation}
    g_{\alpha\beta}\frac{d^2X^{\beta}}{d\gamma^2}+\Gamma_{\alpha\beta\zeta}\frac{dX^{\beta}}{d\gamma}\frac{dX^{\zeta}}{d\gamma} = f(\gamma)g_{\alpha\beta}\frac{dX^{\beta}}{d\gamma} \ ,
\end{equation}
and if we use the inverse metric to raise the appropriate index on both sides of the equation, we return Eq.~\ref{geodesic1} as required.

\subsection{Einstein field equations}\label{energyconditions}

The final step to establishing the framework of general relativity is to examine the Einstein field equations. They are canonically presented as~\cite{telebook, Hartle, Carroll}:\footnote{An alternative form includes a term containing the cosmological constant $\Lambda$; approximating $\Lambda\approx 0$ is common practice however since $\Lambda$ has been observed to be extremely small.}

\begin{equation}
    R_{\mu\nu} - \frac{1}{2}R g_{\mu\nu} = \frac{8\pi G_{N}}{c^4} T_{\mu\nu} \ .
\end{equation}
The left-hand side of the equations describes the `fields' themselves, \emph{i.e} the curvature of spacetime. The right-hand side describes the source of the curvature -- the distribution of stress-energy, energy density and momentum throughout the spacetime. Let us unpack these individually.

\subsubsection{The quasi-local curvature of spacetime}\footnote{There is a distinction being made here between 'quasi-local' and 'local' due to the fact that all finite regions in all spacetimes are \emph{locally} Minkowski in general relativity.}

This is encoded in the expression: $R_{\mu\nu}-\frac{1}{2}Rg_{\mu\nu}$.\footnote{This is only the Ricci curvature. To fully express global curvature we must also factor into account the Weyl curvature. Qualitatively, Ricci curvature contains information pertaining to how volumes of objects are distorted in the presence of tidal forces, while Weyl curvature encodes the changes in shape. For more information on the nature of Weyl and Ricci curvature, please see~\cite{Weylcurv}.} We have developed all of the mathematical objects required to fully understand this expression. To understand what these symbols mean we can observe the following:

\begin{itemize}
    \item $g_{\mu\nu}$ -- The metric tensor (see \S\ref{metric}). The desired physics enforces that this has Lorentzian signature.
    \item $R_{\mu\nu}$ -- The Ricci curvature tensor. This is constructed via the following steps; first the connection specialised to the Christoffel symbols of the second kind:
    \begin{equation}
        \Gamma^{\mu}{}_{\alpha\beta} = \frac{1}{2}g^{\mu\nu}\left(\partial_{\alpha}g_{\nu\beta}+\partial_{\beta}g_{\nu\alpha}-\partial_{\nu}g_{\alpha\beta}\right) \ ,
    \end{equation}
    we may then construct the four-indexed Riemann curvature tensor as follows~\cite{telebook}:
    \begin{equation}
        R^{\mu}{}_{\nu\alpha\beta} = \partial_{\alpha}\Gamma^{\mu}{}_{\nu\beta}-\partial_{\beta}\Gamma^{\mu}{}_{\nu\alpha}+\Gamma^{\mu}{}_{\sigma\alpha}\Gamma^{\sigma}{}_{\nu\beta}-\Gamma^{\mu}{}_{\sigma\beta}\Gamma^{\sigma}{}_{\nu\alpha} \ ,
    \end{equation}
    and the Ricci curvature tensor $R_{\mu\nu}$ is obtained via contraction of the Riemann tensor on the first and third indices~\cite{telebook}:
    \begin{equation}
        R_{\mu\nu} = R^{\sigma}{}_{\mu\sigma\nu} \ .
    \end{equation}
    \item The Ricci scalar $R$. This is obtained by contracting the inverse metric with the Ricci tensor~\cite{telebook}:
    \begin{equation}
        R = g^{\mu\nu}R_{\mu\nu} \ .
    \end{equation}
\end{itemize}
Gathering all of these terms defines the Einstein curvature tensor, $G_{\mu\nu}=R_{\mu\nu}-\frac{1}{2}R g_{\mu\nu}$. Note that this implies that there are in total ten equations in the Einstein field equations at each point in spacetime; this comes from the fact that both $R_{\mu\nu}$ and $g_{\mu\nu}$ have a four-by-four matrix representation which is strictly symmetric. Now we fully comprehend the left-hand side of the Einstein field equations; it is important to note that all information encoded in the curvature tensors and scalar invariants ultimately stems directly from the metric tensor, $g_{\mu\nu}$. It therefore follows that defining a line element (and hence a metric environment) is all one requires in order to characterise all the relevant information pertaining to the curvature of a particular spacetime. We therefore conduct all analyses of candidate spacetimes beginning solely with the metric as our starting point. It is worth noting that there are numerous other curvature tensors and curvature invariants which provide geometric information about the four-manifold but are not explicitly part of the Einstein field equations; specifically the tensors and invariants which will form additional parts of the analyses in this thesis include: the Weyl tensor $C_{\mu\nu\alpha\beta}$, the Ricci contraction $R_{\mu\nu}R^{\mu\nu}$, the Kretschmann scalar $R_{\mu\nu\alpha\beta}R^{\mu\nu\alpha\beta}$, and the Weyl contraction $C_{\mu\nu\alpha\beta}C^{\mu\nu\alpha\beta}$ (for details on these mathematical objects see references~\cite{MATH465, telebook}).

\subsubsection{The stress-energy-momentum tensor}

The right-hand side of the Einstein field equations encodes the source of the curvature in spacetime. Before proceeding note the following -- until now all expressions unless otherwise stated have assumed the international system of units (SI-units); henceforth geometric units shall be used to simplify calculation as is conventional in physics. As such, the speed of light in a vacuum $c=1$, and Newton's gravitational constant $G_{N}=1$. Multiplication of combinations of $c$ and $G_{N}$ where appropriate will return SI-units if desired. Hence we rephrase the Einstein field equations:

\begin{equation}\label{einsteineqsfinal}
    G_{\mu\nu} = R_{\mu\nu}-\frac{1}{2}Rg_{\mu\nu} = 8\pi T_{\mu\nu} \ .
\end{equation}
The general form of $T_{\mu\nu}$ is as follows~\cite{Hartle}:

\begin{equation}
    T_{\mu\nu} = \begin{bmatrix}
    \rho & F_{i} \\
    F_{j} & \pi_{ij}
    \end{bmatrix} \ ,
\end{equation}
(recall latin indices $i,j\in\left\lbrace 1, 2, 3\right\rbrace$).\newpage These objects are the following:

\begin{itemize}
    \item $\rho$ is the energy density of the relativistic masses present in the spacetime,
    \item $F_{i}$ and $F_{j}$ represent the directional energy flux of relativistic mass across each $x_{i}$ surface (these are analogous to the Poynting vectors from electromagnetism) - this encodes momentum density,
    \item $\pi_{ij}$ represents the spatial shear stress tensor, with the diagonal entries (those independent of direction) defining radial and transverse pressure.
\end{itemize}

One of the simplest examples for the stress-energy-momentum tensor is in the case when the matter distribution is that of a perfect fluid, \emph{i.e} we have spherical symmetry with uniform pressure. In this case:

\begin{equation}
    T_{\mu\nu} = \begin{bmatrix}
    \rho & 0 & 0 & 0 \\
    0 & p & 0 & 0 \\
    0 & 0 & p & 0 \\
    0 & 0 & 0 & p
    \end{bmatrix} \ ,
\end{equation}
and in general for spherically symmetric matter distributions:

\begin{equation}\label{stress}
    T_{\mu\nu} = \begin{bmatrix}
    \rho & 0 & 0 & 0 \\
    0 & p_{\parallel} & 0 & 0 \\
    0 & 0 & p_{\perp} & 0 \\
    0 & 0 & 0 & p_{\perp}
    \end{bmatrix} \ .
\end{equation}
All candidate spacetimes analysed in this thesis have a spherically symmetric matter distribution and as such the corresponding stress-energy-momentum tensor will take the form of Eq.~\ref{stress}.

As well as the form of the components of $T_{\mu\nu}$ being dictated by the Einstein field equations, there are numerous (at least seven) different energy conditions in the context of classical general relativity which imply mathematical constraints on $T_{\mu\nu}$~\cite{LorentzianWormholes}. The most fundamental of these conditions is the null energy condition (NEC), the satisfaction of which is mathematically represented by the following~\cite{LorentzianWormholes}:

\begin{equation}
    \mbox{NEC} \quad \Longleftrightarrow \quad T_{\mu\nu}t^{\mu}t^{\nu}\geq 0 \ ,
\end{equation}
where $t^{\mu}$ is any arbitrary null vector and the inequality must hold globally. Given a stress-energy-momentum tensor of the form in Eq.~\ref{stress}, this assertion translates to the following in terms of the pressures within the matter distribution:

\begin{equation}
    \mbox{NEC} \quad \Longleftrightarrow \quad \rho+p_{\parallel}\geq 0, \ \mbox{\emph{and}} \ \rho+p_{\perp}\geq 0 \ .
\end{equation}
The primary energy conditions of interest are the null, weak, strong, and dominant energy conditions, and if the null energy condition is violated it mathematically implies direct violation of these remaining three conditions also; this fact is used in the subsequent analyses of the various candidate spacetimes (for details pertaining to the corollaries of the violation of the NEC, please see reference~\cite{LorentzianWormholes}). Canonically, traversable wormhole geometries have required a violation of the null energy condition~\cite{MorrisThorne}, and as such represent what are known as `exotic' solutions to the Einstein equations.

Regular black hole geometries typically require a violation of \emph{some} of the energy conditions, however the NEC might be satisfied in some regular black hole geometries. This is due to the fact that the violation of the radial NEC is directly related to the `flare-out' condition pertaining to wormhole throats~\cite{LorentzianWormholes} (this condition is defined in \S\ref{C:Morris}). For regular black hole spacetimes, one instead expects a violation of the strong energy condition (SEC). This is a consequence of the fact that the lack of curvature singularities implies geodesic completeness on the manifold; and geodesic completeness implies the SEC will not be satisfied (for details on the corollaries of geodesic completeness, please see reference~\cite{largescale}). In terms of the principal pressures from the form of the stress-energy-momentum tensor as presented in Eq.~\ref{stress}, satisfaction of the SEC amounts to the following:

\begin{equation}
    \mbox{SEC} \qquad \Longleftrightarrow \qquad \rho+p_{\parallel}+2p_{\perp}\geq 0 \ .
\end{equation}
This condition will be utilised as part of the analysis of regular black hole spacetimes in \S\ref{C:Bardeen-Hayward-Model2}.

We have established the framework of general relativity; every term in Schuller's definition for a generic spacetime has been thoroughly unpacked. There remain various mathematical and physical subtleties to be discussed before proceeding to analyse various spacetimes of interest using these foundational building blocks -- these are presented in \S\ref{C:GRBackground}.

%%%%%%%%%%%%%%%%%%%%%%%%%%%%%%%%%%%%%%%%%%%%%%%%%%%%%%

%%%%%%%%%%%%%%%%%%%%%%%%%%%%%%%%%%%%%%%%%%%%%%%%%%%%%%%

\chapter{Fundamentals: Analysing candidate spacetimes}\label{C:GRBackground}

%%%%%%

Let us introduce some terminology and techniques which are essential in performing the level of analysis desired for each of the prospective candidate spacetimes.

\section{Spherical symmetry}\label{sphericalsym}

All candidate spacetimes analysed in this thesis possess a spherically symmetric matter distribution. This implies that the four-manifolds corresponding to the left-hand side of the Einstein field equations are all spherically symmetrical geometries. It is worth noting that each metric presented models a spacetime with one `centralised' (with respect to the chosen coordinate patch) massive object which is controlling the curvature of that specific spacetime. Surfaces which correspond to a fixed distance (with respect to the metric) from this object are then spatial two-spheres. Usually this fixed distance simply corresponds to a specific designation of our $r$-coordinate, however this is subject to our chosen coordinate patch. A corollary of spherical symmetry is that we may fix an angular coordinate arbitrarily when discussing the worldlines of particles and simplify any physical problems to a reduced equatorial state -- for example when calculating innermost stable circular orbits (ISCOs) and photon spheres\footnote{Discussion and definition of these objects is presented in \S\ref{ISCOintro}.} this greatly reduces the complexity of calculations. We may also discuss the curvature of the spacetime with respect to the changing areas of the induced spatial two-spheres as we vary distance from the centralised mass; a powerful tool for verifying certain qualitative aspects for each candidate spacetime (for dynamical spacetimes this discussion requires that we fix a time slice, however the majority of metrics analysed henceforth are time-independent and we have the freedom to vary time arbitrarily without affecting the nature of the curvature).

%%%%%%

\section{Horizons}\label{Horizon}

Horizons in the context of general relativity are subtle physical objects, and there are multiple qualitatively different types of horizon corresponding to disparate technical definitions. Fundamentally all classes of horizon are characterised as a physical surface within the four-manifold permitting the passage of massive and massless particles in one direction only, and whose location is such that the notion of time for all external observers comes to a stop at the horizon~\cite{telebook, Wald, largescale, LorentzianWormholes}.

The most commonly encountered class of horizon is the event horizon, or absolute horizon -- this is primarily due to the popularity of the event horizon in science fiction, although some physicists and applied mathematicians also advocate using the definition of the event horizon as it enables an easier environment in which to prove mathematical theorems~\cite{Visser:14}. In the book Lorentzian Wormholes, the following definition for the event horizon is provided~\cite{LorentzianWormholes}: ``For each asymptotically flat region the associated future/past event horizon is defined as the boundary of the region from which causal curves (that is, null or timelike curves) can reach asymptotic future/past null infinity.''\footnote{Future/past null infinity is defined as~\cite{Hartle}: the spatial surface corresponding to the set of all coordinate locations which outgoing/ingoing null curves (\emph{e.g.} light rays) are able to asymptotically approach as $\vert t\vert\rightarrow+\infty$. This boundary is only defined in an asymptotically flat region of spacetime, and is one of the `conformal infinities' used in constructing the Carter-Penrose diagrams which diagrammatically represent the global causal structure of specific spacetimes (the others being future/past timelike infinity and spacelike infinity -- these objects are utlilised for the construction of the diagrams in \S\ref{C:Black-bounce} and \S\ref{C:Vaidya}). For details on conformal infinities, timelike and spacelike infinity, and their uses in general relativity, please see reference~\cite{conformal}.} It turns out that the event horizon may not in fact be the most well-informed category of horizon when describing physical reality -- one runs into many technical issues, one of which concerns black hole evaporation across large timescales and the potential recovery of information deemed to be strictly `lost' by the definition of these event/absolute horizons (for details see reference~\cite{Visser:14}).

Another type of horizon is the apparent horizon~\cite{LorentzianWormholes}: ``defined locally in terms of trapped surfaces. Pick any closed, spacelike, two-dimensional surface (two-surface). At any point on the two-surface there are two null geodesics that are orthogonal to the surface. They can be used to define inward and outward propagating wavefronts. If the area of both inward and outward propagating wavefronts decrease as a function of time, then the original two-surface is a trapped surface and one is inside the apparent horizon. More precisely, if the expansion of both sets of orthogonal null geodesics is negative, then the two-surface is a trapped surface.'' So the apparent horizon is the boundary between these trapped and untrapped surfaces, characterised by conditions on the focussing/defocussing of null geodesics. Note that in a time-independent metric environment these two horizon definitions are one and the same. Given the fact the geometry remains unchanged with time, any apparent horizon is also an absolute horizon, however this is not the case with a dynamical metric environment and it is important to distinguish between them.

In view of the fact that all metric candidates analysed henceforth model spherically symmetric geometries, we may simplify the mathematical definition of the apparent horizon. When specialised to spherical symmetry, the apparent horizon simply corresponds to the locus of coordinate locations such that radially propagating light rays have zero coordinate velocity (in most coordinate patches this will be when $d\theta=0, d\phi=0, \ \mbox{and} \ \frac{dr}{dt}=0$).\footnote{In the special case when the chosen coordinate system informs a diagonal metric environment \emph{and} we have spherical symmetry, this definition implies that the location of the horizon is simply defined by the surface which forces $g_{tt}=0$; this simplification is utilised henceforth where appropriate.} Accordingly the analyses in this thesis adopt this simplified definition of the apparent horizon, which for all time-independent candidate spacetimes acts as \emph{both} an event and an apparent horizon. The exception of the dynamical spacetime is in \S\ref{C:Vaidya}, where a careful distinction is made between the two qualitatively different horizons. It follows that we naturally define a black hole region within a spacetime as the region of the geometry which lies strictly within the apparent horizon.

%%%%%%

\section{Singularity}\label{singularity}

When discussing the presence/location of a singularity within a specific geometry, it is important to differentiate between a gravitational singularity and a coordinate singularity. Both are mathematically characterised by coordinate locations which correspond to poles of a coefficient function or functions in the metric; their qualitative difference is that a gravitational singularity is representative of a physical source of infinite curvature in the spacetime (a tear in the topological manifold), whilst coordinate singularities represent nothing physically special at all and may always be removed through an alternative choice of coordinate patch. We therefore need a way of mathematically separating the two so as to draw meaningful physical conclusions. There are multiple methods of doing this, for example via geodesic incompleteness or via analysis of the Riemann curvature tensor with respect to an orthonormal tetrad~\cite{largescale}.\footnote{This is the method utilised in \S\ref{sec:curvature}; there is no particularly special reason for this; the calculation was contextually convenient.} For our purposes however a gravitational singularity will generally be defined as a coordinate location which forces one or more of the scalar curvature invariants $R, R_{\mu\nu}R^{\mu\nu}, R_{\mu\nu\alpha\beta}R^{\mu\nu\alpha\beta}$, and $C_{\mu\nu\alpha\beta}C^{\mu\nu\alpha\beta}$ to have infinite magnitude. By imposing the conditions on the scalar invariants, which are coordinate-independent quantities, we remove the possibility of mistaking a coordinate artefact for a genuine gravitational singularity. For the cases of interest, \emph{i.e.} traversable wormhole and regular black hole geometries, we are in fact looking for an absence of gravitational singularities and wish to demonstrate that no such coordinate locations exist -- this therefore corresponds to ensuring that the curvature invariants remain finite over the entire domain of our chosen coordinate patch.

%%%%%%

\section{Killing symmetries}\label{Killing}

\subsection{Killing vector fields}

A Killing vector field in the context of spacetime is defined as a vector field which preserves the metric~\cite{telebook}. This means that $\xi^{\mu}$ is a Killing vector if and only if any set of points displaced by some $\xi^{\mu}dx_{\mu}$ leaves all distance relationships are unchanged. It follows that displacements along Killing vector fields are isometries; bijective maps $f:\mathbb{R}^{n}\rightarrow\mathbb{R}^{n}$ such that $g_{\mu\nu}\left(f(x), f(y)\right)=g_{\mu\nu}\left(x, y\right)$. A physical corollary of this displacement being an isometric map is that worldlines of particles displaced infinitesimally in an arbitrary direction along the Killing vector field are congruent. From this definition, one can derive Killing's equation: $\mathcal{L}_{X} \ g_{\mu\nu}=\nabla_{\mu}X_{\nu}+\nabla_{\nu}X_{\mu}=0$; that the Lie derivative of the metric tensor with respect to a given Killing vector field $X$ is manifestly zero.\footnote{For details on the Lie derivative and its relationship with the Killing vector, please see references~\cite{MATH465, telebook}.}

\subsection{Conserved physical quantities}

Given some arbitrary spacetime, often there are symmetries in the spacetime geometry, and as such associated Killing vectors which yield conserved physical quantities in accordance with the same conservation laws that arise from classical analytic mechanics~\cite{telebook}. To see that these symmetries exist, without loss of generality we may use the example of a metric expressed with respect to some coordinate patch $\left(t, r, \theta, \phi\right)$ such that the metric is both diagonal and time-independent, \emph{i.e.} independent of the $t$-coordinate (this argument can be generalised to arbitrary coordinate patches in various domains with ease). It therefore follows that $\partial g_{\mu\nu}/{\partial t}=0$. The geometric interpretation of this relation is that any curve in the manifold can be shifted by some $\Delta t=\epsilon, \ \epsilon\in\mathbb{R}$, to form a congruent curve; \emph{i.e.} the transformation $t\rightarrow t+\epsilon$ preserves the metric. We may conclude that $\xi^{\mu}=\partial_{t}=\left(1, 0, 0, 0\right)=\delta^{\mu}{}_{t}$ is a Killing vector. Let us derive Killing's equation by taking the covariant derivative of the Killing vector field in the following manner:

\begin{eqnarray}\label{Killingeq}
\nabla_{\nu}\xi_{\mu} &=& g_{\mu\sigma}\nabla_{\nu}\xi^{\sigma} \nonumber \\
&=& g_{\mu\sigma}\left(\frac{\partial{\xi^{\sigma}}}{\partial{x^{\nu}}}+\Gamma^{\sigma}{}_{\nu\alpha}\xi^{\alpha}\right) \qquad \mbox{(definition of covariant derivative)} \nonumber \\
&=& g_{\mu\sigma}\Gamma^{\sigma}{}_{\nu t} \ = \Gamma_{\mu\nu t} \qquad \qquad \qquad \quad \mbox{(using $\frac{\partial{\xi^{\sigma}}}{\partial{x^{\nu}}}=0, \ \xi^{\alpha}=\delta^{\alpha}{}_{t}$)} \nonumber \\
&& \nonumber \\
&=& \frac{1}{2}\left(\frac{\partial{g_{\mu t}}}{\partial{x^{\nu}}}+\frac{\partial{g_{\mu\nu}}}{\partial{t}}-\frac{\partial{g_{\nu t}}}{\partial{x^{\mu}}}\right) \qquad \qquad \mbox{(definition of $\Gamma_{\mu\nu t}$)} \nonumber \\
&=&\frac{1}{2}\left(\partial_{\nu}g_{\mu t}-\partial_{\mu}g_{\nu t}\right) \qquad \qquad \qquad \quad \quad \ \ \mbox{(using $\frac{\partial{g_{\mu\nu}}}{\partial{t}}=0$)} \nonumber \\
&=& -\frac{1}{2}\left(\partial_{\mu}g_{\nu t}-\partial_{\nu}g_{\mu t}\right) \nonumber \\
&=& -\nabla_{\mu}\xi_{\nu} \ . \ \ \ \qquad \qquad \qquad \qquad \qquad \qquad \mbox{(by symmetry)} \ .
\end{eqnarray}
So we have returned to Killing's equation, $\nabla_{\mu}\xi_{\nu}+\nabla_{\nu}\xi_{\mu}=0$.

If we have an affinely parameterised geodesic curve in our spacetime, parameterised by some affine scalar parameter $\gamma$, then for the affinely parameterised tangent vector to the curve, $X^{\nu}$, we have the following result as a corollary from Eq.~\ref{affinegeodesic}: $X^{\mu}\nabla_{\mu}X^{\nu}=0$. As a consequence of our connection being a metric connection, we also must satisfy the metricity condition~\cite{telebook, MATH465}, that $\nabla_{\sigma}g_{\mu\nu}=0$.

Combining these results with Killing's equation, we can prove the following assertion: for any Killing vector field $\xi^{\mu}$ on a spacetime $\left(\mathcal{M}, g_{\mu\nu}\right)$, and some affinely parameterised tangent vector field $X^{\nu}$ to an affinely parameterised geodesic curve, $\xi^{\mu}g_{\mu\nu}X^{\nu}=K$, where $K$ is some scalar-valued constant. To prove this we show that $\frac{d}{d\gamma}\left(\xi^{\mu}g_{\mu\nu}X^{\nu}\right)=0$, by employing the product rule for the covariant derivative:

\begin{eqnarray}
\frac{d}{d\gamma}\left(\xi^{\mu}g_{\mu\nu}X^{\nu}\right) &=& \left(X^{\sigma}\nabla_{\sigma}\xi^{\mu}\right)g_{\mu\nu}X^{\nu} \nonumber \\
&& + \xi^{\mu}\left(X^{\sigma}\nabla_{\sigma}g_{\mu\nu}\right)X^{\nu} \quad \mbox{(zero due to metricity condition)} \nonumber \\
&& + \xi^{\mu}g_{\mu\nu}\left(X^{\sigma}\nabla_{\sigma}X^{\nu}\right) \quad \ \ \mbox{(zero as a corollary of Eq.~\ref{affinegeodesic})} \nonumber \\
&& \nonumber \\
&=& \left(X^{\sigma}\nabla_{\sigma}\xi^{\mu}\right)g_{\mu\nu}X^{\nu} \nonumber \\
&=& X^{\sigma}\nabla_{\sigma}\xi_{\nu}X^{\nu} \qquad \quad \ \ \mbox{(using metric tensor to lower index)} \nonumber \\
&=& X^{\sigma}\nabla_{(\sigma}\xi_{\nu)}X^{\nu} \qquad \quad \mbox{(since $\sigma$ and $\nu$ are dummy indices)} \nonumber \\
&& \nonumber \\
&=& X^{\sigma}\left[\frac{1}{2}\left(\nabla_{\sigma}\xi_{\nu}+\nabla_{\nu}\xi_{\sigma}\right)\right]X^{\nu} \qquad \ \mbox{(zero due to Eq.~\ref{Killingeq})} \nonumber \\
&=& 0 \ .
\end{eqnarray}
As such, for the specific example where $\xi^{\mu}=\partial_{t}$, and if we specify the tangent vector field $X^{\nu}$ to be the four-momentum of some test particle along its worldline, \emph{i.e.} $P^{\nu}=m_{0}V^{\nu}$ (where $V^{\nu}$ is the four-velocity of the test particle), we have the following:

\begin{eqnarray}
    \xi^{\mu}g_{\mu\nu}P^{\nu} &=& \delta^{\mu}{}_{t}g_{\mu\nu}P^{\nu} \nonumber \\
    &=& g_{t\nu}P^{\nu} \nonumber \\
    &=& g_{t\nu}\delta^{\nu}{}_{t}P^{\nu} \qquad \mbox{(due to diagonal metric environment)} \nonumber \\
    &=& g_{tt}P^{t} \nonumber \\
    &=& E \ .
    %&=& \sqrt{-g_{tt}}P^{\hat{t}} \nonumber \\
    %&=& \sqrt{-g_{tt}} \, E \ .
\end{eqnarray}
It follows that the symmetry of the spacetime in the $t$-coordinate ultimately yields the conservation of energy, $E$. Hence the Killing vector implies the conservation of energy along our geodesic curves of interest. This specific conserved quantity is relied upon heavily in the subsequent analyses of candidate spacetimes; also utilised (where appropriate) is the metric independence of the azimuthal $\phi$-coordinate, which implies the conservation of the quantity $g_{\phi\phi}P^{\phi}$. This quantity in turn implies the conservation of angular momentum, $L$.

%%%%%%

\section{Static spacetimes}\label{static}

A spacetime is mathematically characterised as static if it admits a hypersurface orthogonal timelike Killing vector field~\cite{largescale}. Physically this means that the geometry is time-independent and also non-rotational (note that spherical symmetry implies lack of rotation). This is a special case of the more general stationary spacetime -- one which is time-independent but permits rotation (hence mathematically only requires that it admits a timelike Killing vector field). For example the Kerr geometry is a spacetime which is stationary without being static~\cite{Kerr}, however every metric analysed in this thesis is non-rotational due to the relative tractability of the mathematics, and all bar one are time-independent (and hence static). The exception has a dynamical metric environment which is time-dependent; neither static nor stationary.

%%%%%%

\section{ISCO and photon sphere}\label{ISCOintro}

The innermost stable circular orbit (ISCO) and the photon sphere correspond to specific locations within a given spacetime which are of significant observational interest. The ISCO is defined as the innermost stable circular orbit which a massive particle is able to maintain around some massive object~\cite{telebook}, and physically corresponds to the innermost edge of the accretion disc; an important astrophysical object for empirical observation. The photon sphere corresponds to the locus of coordinate locations sufficiently near the centralised massive object such that photons (or any massless particle) are forced to propagate in circular geodesic orbits (which may be stable or unstable)~\cite{telebook}. Given an appropriate test particle for each case, both of these mathematical concepts are characterised by coordinate locations corresponding to stationary points of the effective energy potentials of the test particles -- this is a corollary of the desired physics inherent in orbital mechanics; that a test particle orbiting a massive body be in mechanical equilibrium (in the classical sense; please see reference~\cite{mechequilibrium} for details). To find the coordinate location of the ISCO we re-parameterise the line element using tangent vectors to a timelike worldline, whilst for the photon sphere we re-parameterise the line element with respect to the tangent vectors of a null worldline.

Given the fact that both the ISCO and the photon sphere are strictly \emph{circular} orbits, they are typically only discussed in the context of spherically symmetrical geometries. There are much messier generalisations, ISCOs and photon rings, for the equatorial plane of axially symmetric geometries such as Kerr~\cite{Kerr}; these are not explored in this thesis. In view of this, without loss of generality we may always (through an appropriate choice of coordinate patch) define the effective energy potentials of arbitrary test particles as some function $V(r)$, the form of which is found via analysis of the conserved quantities implied by the Killing symmetries of each candidate spacetime. This implies that the coordinate locations of the circular orbits are found at the $r$-values which satisfy $V^{'}(r)=0$. The stability of each circular orbit is determined by the sign of $V^{''}(r)$ in the following manner~\cite{mechequilibrium, corben}:

\begin{eqnarray}
V^{''}(r)<0 \qquad &\Longrightarrow& \qquad \mbox{unstable orbit} \ ; \nonumber \\
V^{''}(r)=0 \qquad &\Longrightarrow& \qquad \mbox{marginally stable} \ ; \nonumber \\
V^{''}(r)>0 \qquad &\Longrightarrow& \qquad \mbox{stable orbit} \ .
\end{eqnarray}
We define the notion of stability to be whether small peturbations orthogonal to the geodesic orbit on either side cause the particle to remain in the circular orbit (stable), or cause it to follow some altogether qualitatively different worldline (unstable). This raises a notable point: terminologically `ISCO' is standard, but one would assume that use of the word `stable' in the term ISCO implies that the corresponding orbit must always be stable. In general, this is not the case. Quite often the ISCO is at best one-sided stable, and in the specific case for the Morris-Thorne traversable wormhole spacetime analysed in \S\ref{C:Morris}, the ISCO is in fact two-sided \emph{unstable}. This is merely an oddity in use of language; there are no significant ramifications pertaining to whether the ISCO is stable/unstable in the context of the desired astronomical analysis.

The motivation for this section of analysis is that given a specific solution to the Einstein equations, astronomers will be informed as to where to point their telescopes (with respect to a specified coordinate patch) in order to gain as much pertinent information regarding the behaviour of both massless and massive particles as they near the region of spacetime with the highest curvature. Due to the fact that the ISCO corresponds to the innermost edge of the accretion disc, it is inherently crucial for the astrophysical imaging of black hole regions. The photon sphere is also of significance for astronomers; for extensive discussion on the importance of the photon sphere for astrophysical imaging see references~\cite{Virbhadra:1999, Virbhadra:2002, Virbhadra:1998, Virbhadra:2007,Claudel:2000, Virbhadra:2008}.

It should be noted that in Newtonian gravity, there is no concept of an ISCO, as one may easily stabilise the orbits of test particles which are arbitrarily close to a mass source. It follows that ISCO locations are intrinsically general relativistic.

%%%%%%

\section{Regge-Wheeler equation}

In order to use the Regge-Wheeler equation to conduct tractable analysis, we must first rewrite the metric environment in terms of a naturally defined tortoise coordinate. In a spherically symmetric geometry with a diagonal metric environment, the tortoise coordinate $r_{*}$ is constructed such that it must satisfy the first-order differential equation~\cite{ReggeWheeler1}:

\begin{eqnarray}
    d r_{*} &=& \sqrt{-\frac{g_{rr}}{g_{tt}}} \, d r \ ; \nonumber \\
    && \nonumber \\
    \Longrightarrow \quad r_{*} &=& \int \sqrt{-\frac{g_{rr}}{g_{tt}}} \, d r \ .
\end{eqnarray}
For the specialised case where $g_{rr}g_{tt}=-1$, as is the case for all spacetime candidates analysed in this thesis, we have the simplified tortoise coordinate given by:

\begin{equation}
    d r_{*} = g_{rr} \, d r \ ; \qquad \Longrightarrow \qquad r_{*} = \int g_{rr} \, d r \ .
\end{equation}
In a spherically symmetric \emph{and} static environment, without loss of generality, this enables us to rewrite the metric as follows:

\begin{equation}
    ds^{2} = A(r)\left\lbrace -dt^{2}+dr_{*}^{2}\right\rbrace+g_{\theta\theta}\,d\Omega^{2} \ .
\end{equation}
The use of the tortoise coordinate normalises the relation between $dt^{2}$ and $dr^{2}$, such that radially propagating test particles (\emph{i.e.} $d\Omega^{2}=0$) have worldlines which correspond to $\pi/4$-radian lines on a spacetime diagram. This is a useful feature, however it comes at the cost of some hitherto unknown conformal factor in the form of $A(r)$ (for radial null propagation use of the tortoise coordinate is particularly nice as we may simply ignore the effect of the conformal factor in view of the fact that $d\hat{s}^{2}=0$).

Having rewritten the metric in terms of $r_{*}$, the Regge-Wheeler equation enables one to draw conclusions pertaining to the energy potentials of the following objects, subject to linear peturbations induced by greybody factors~\cite{ReggeWheeler1}:

\begin{itemize}
    \item The spin zero massless scalar field minimally coupled to gravity,
    \item The spin one Maxwell vector field,
    \item The spin two axial peturbation mode.
\end{itemize}
The general form of the Regge-Wheeler equation is given by~\cite{ReggeWheeler1}:

\begin{equation}
    \partial_{r_{*}}^{2}\hat{\phi}+\lbrace \omega^2-\mathcal{V}\rbrace\hat\phi = 0 \ ,
\end{equation}
where $\hat\phi$ is the scalar or vector field of interest, $\mathcal{V}$ is the spin-dependent Regge-Wheeler potential for our particle, and $\omega$ is a temporal frequency component in the Fourier domain. The formalism of all subsequent Regge-Wheeler analyses in this thesis closely follows that of reference~\cite{ReggeWheeler1}, and extracting Regge-Wheeler potentials forms part of the standard analysis for each candidate spacetime.

%%%%%%

\section{Surface gravity and Hawking temperature}

Canonical surface gravity in general relativity requires the existence of a static Killing horizon, defined to be a null hypersurface at a coordinate location where the norm of the Killing vector field goes to zero~\cite{telebook, Wald}. Given a static Killing horizon in a spacetime, and some suitably normalised Killing vector field $\xi^{\mu}$, the surface gravity $\kappa$ is then calculated by evaluating the following equation at the coordinate location of the Killing horizon~\cite{telebook, Wald}:

\begin{equation}
    \xi^{\mu}\nabla_{\nu}\xi_{\mu} = -\kappa \; \xi_{\nu} \ ,
\end{equation}
and via Killing's equation and appropriate contraction with the metric tensor we may rewrite this as:

\begin{equation}
    \xi^{\mu}\nabla_{\mu}\xi_{\nu} = \kappa \; \xi_{\nu} \ .
\end{equation}
Surface gravity $\kappa$ is then directly related to the Hawking temperature $T_{H}$ as a consequence of Hawking evaporation by the following formula~\cite{largescale}:

\begin{equation}
    T_{H} = \frac{\hbar\kappa}{2\pi k_{B}} \ .
\end{equation}
In \S\ref{sec:3+1} a calculation is performed to obtain the surface gravity $\kappa$ at the horizon for the regular black hole geometry. This calculation is not repeated for other sections as it is either physically uninteresting (\emph{e.g.} for the Bardeen and Hayward metrics in \S\ref{C:Bardeen-Hayward-Model2}), or it is not a well-defined process (\emph{e.g.} in the time-dependent environment presented in \S\ref{C:Vaidya}, the lack of a suitable Killing horizon prevents classical calculation of $\kappa$). For traversable wormhole spacetimes the notion of surface gravity is at best unusual~\cite{LorentzianWormholes}, and discussion on this point is omitted from this thesis.

We are now sufficiently armed with the necessary tools to conduct thorough analysis in the context of general relativity. Accordingly, let us proceed to analysing various candidate spacetimes of interest.
%%%%%%%%%%%%%%%%%%%%%%%%%%%%%%%%%%%%%%%%%%%%%%%%%%%%%

%%%%%%%%%%%%%%%%%%%%%%%%%%%%%%%%%%%%%%%%%%%%%%%%%%%%%%%

\chapter{Introducing the traversable wormhole}\label{C:Morris}

%%%%%%

Science fiction is littered with examples of `wormholes' which permit all sorts of miraculous travel through space and time; as such the concept has blossomed in popularity in contemporary times -- even to the layperson. Often less well known however is that there are specific and rigorous ways of defining these objects within the context of general relativity, and that there exist model spacetimes of interest whose geometric curvature informs a traversable wormhole environment and subsequent matter distribution in accordance with the Einstein field equations. First let us define what a traversable wormhole is as a physical object.\footnote{Here we are defining a wormhole geometry which is merely traversable in principle; in order for a real traveller to pass through the geometry there are numerous engineering concerns also, \emph{e.g.} tidal forces due to gravity mustn't tear a would-be traveller to pieces.} At its most elementary, a "wormhole is a short-cut through space and time" \cite{LorentzianWormholes}, and fundamentally traversable wormholes can be characterised by the following criteria \cite{MorrisThorne}:

\begin{itemize}
    \item We are working within the framework of general relativity. As such we require that any wormhole solution obeys the Einstein field equations at all coordinate locations, hence defining the form of the stress-energy-momentum tensor, $T_{\mu\nu}$.
    \item The geometry should possess a specific coordinate location called the `throat' which connects two asymptotically flat regions of spacetime -- these flat regions may either be from the same or different universes; the geometry is still a traversable wormhole provided their asymptotic limits model Minkowski space (in most coordinate patches this occurs as $\vert r\vert \rightarrow +\infty$). Without loss of generality, we can always choose a coordinate system where the throat is located at $r=0$ (and in fact it is usually prudent to do so for simplicity of calculations near the throat). In the local area near the throat, the geometry must satisfy the `flare-out' condition -- that the areas of the induced spatial hypersurfaces on either side of the throat are strictly increasing as a function of distance from the throat. This is often mathematically characterised by the condition that $A''(r_{throat})>0$; \emph{i.e.} the throat locally minimises the area function of the spatial hypersurfaces.
    \item There must be no horizons in the geometry as the wormhole must be two-way traversable -- naturally the existence of a horizon would permit one-way travel, but in the opposite direction travel would require one to violate the condition from special relativity that nothing propagate faster than the speed of light, $c$.
    \item The geometry should be gravitationally non-singular (or at the very least one should be able to isolate any gravitational singularities in small, easily avoided regions); a gravitational singularity in the vicinity of a particle's worldline would nullify any need to discuss engineering concerns as any object experiencing travel through a region of infinite curvature will most certainly be destroyed. This is also of mathematical convenience, as a gravitationally non-singular geometry enforces that all curvature tensor components and curvature invariants remain everywhere-finite with appropriate coordinate selection.\footnote{Note that there certainly are wormhole geometries possessing singularities -- \emph{e.g.} the Einstein-Rosen bridge. However this geometry is strictly not traversable~\cite{LorentzianWormholes}.} In a spherically symmetrical geometry with a traversable wormhole throat the combination of the `flare-out' condition and the requirement for asymptotic flatness informs that there be strictly no gravitational singularity \emph{anywhere} in the geometry.
\end{itemize}

These criteria are sufficient to ensure that a wormhole geometry is traversable in principle. We may now begin analysing specific candidates for traversable wormhole solutions phenomenologically, and draw conclusions as to their nature. Specifically we are interested in traversable wormhole solutions which correspond to global (or near-global) coordinate patches in order to extend the pre-existing discussion, where a global coordinate patch is a coordinate patch defined on the manifold $\mathcal{M}$ such that every point in $\mathcal{M}$ has a corresponding coordinate location with respect to the chosen coordinate patch (\emph{i.e.} a one-chart atlas completely covering $\mathcal{M}$).

%%%%%%

\section{Morris-Thorne wormhole}\label{Morristhorne}

Morris and Thorne provide what is most likely the simplest metric representing a traversable wormhole geometry as follows~\cite{MorrisThorne}:\footnote{The subsequent analysis of this metric is not at all new, rather it is intended to act as a straightforward template for which all remaining analyses of other candidate spacetimes adheres to. Analysis which extends the pre-existing discussion begins from \S\ref{C:Exponential}.}

\begin{equation}\label{Morristhorne}
    ds^2 = -dt^2+dr^2+\left(r^2+a^2\right)\left(d\theta^2+\sin^2\theta \ d\phi^2\right) \ . 
\end{equation}
Here $a$ is simply a scalar parameter and we may enforce $a\neq 0$. If $a=0$ then we have the metric for Minkowski space with respect to spherical polar coordinates -- clearly not the traversable wormhole candidate we desire. Note that the metric is time-independent and spherically symmetric in view of the diagonal metric environment; therefore it is non-rotational. Hence $\xi=\partial_{t}$ is a Killing vector and the metric is static (see \S\ref{static}). Furthermore, the areas of spherical symmetry are given by:

\begin{eqnarray}
A(r) &=& 4\pi\left(r^2+a^2\right) \quad
\Longrightarrow \quad A'(r) = 8\pi r \ ; \nonumber \\
A'(r) &=& 0 \quad \Longrightarrow \quad r=0 \ ,
\end{eqnarray}
and we have a stationary point at $r=0$, which is a minimum for the area of spherical surfaces in view of the fact that: $A''(r)=8\pi$, $+8\pi>0 \ \forall \  r$. As $\vert r\vert\rightarrow +\infty$, $a^2\ll r^2$, so $r^2+a^2 \sim r^2$ and the metric becomes asymptotic to that of Minkowski space. We may therefore conclude that the geometry described by Eq.~\ref{Morristhorne} has a throat at coordinate location $r=0$ which connects two asymptotically flat regions of spacetime, and satisfies the `flare-out' condition for a traversable wormhole. Note that at the throat when $r=0$, $A(r)=4\pi a^2$, hence we conclude that the scalar parameter $a$ is in fact informing the radial width of the throat of our geometry.

By definition a traversable wormhole geometry must also have no horizons for all coordinate locations in the spacetime. We may observe immediately that $g_{tt}=-1$; this is clearly non-zero irrespective of any parameter values. As such there are no horizons as required. Furthermore the radial null curves arising from this metric (\emph{i.e.} $ds^2=0$;  $\theta, \phi=\mbox{constant}$) are given by $\frac{dr}{dt}=1$. This defines a coordinate speed of light for the metric that is equal to the speed of light in a vacuum, and a corresponding refractive index of $n=1$. Also note the metric permits a global coordinate patch, where $t\in(-\infty. +\infty), r\in(-\infty,+\infty), \ \theta\in[0,\pi], \ \phi\in[-\pi, +\pi)$, and the entire manifold is covered. It remains to check that the geometry is gravitationally non-singular.

%%%%%%

\subsection{Curvature tensors and invariants analysis}
\vfil
The Ricci scalar:

\begin{equation}
    R = \frac{-2a^2}{\left(r^2+a^2\right)^2} \ .
\end{equation}\clearpage
\noindent Ricci tensor non-zero components:\footnote{Non-zero curvature tensor components for this analysis are all presented using the mixed components with one or two indices raised (depending on whether it is a rank two or rank four tensor) via contraction with the contravariant metric. This is done for simplification of the resulting algebraic expressions, and this form is utilised where appropriate for all other candidate spacetimes.}

\begin{eqnarray}
    R^{r}{}_{r} = \frac{-2a^2}{\left(r^2+a^2\right)^2} = R \ &;& \nonumber \\
    \mbox{as} \ \vert r\vert\rightarrow 0, \quad R^{r}{}_{r} \rightarrow \frac{-2}{a^2} \ &;& \nonumber \\
    \mbox{as} \ \vert r\vert\rightarrow+\infty, \quad  R^{r}{}_{r} \rightarrow 0 \ &.&
\end{eqnarray}
Riemann tensor non-zero components:

\begin{eqnarray}
    R^{r\theta}{}_{r\theta} = R^{r\phi}{}_{r\phi} = -R^{\theta\phi}{}_{\theta\phi} = \frac{-a^2}{\left(r^2+a^2\right)^2} = \frac{1}{2}R \ &;& \nonumber \\
    \mbox{as} \ \vert r\vert\rightarrow 0, \quad R^{r\theta}{}_{r\theta} \rightarrow -\frac{1}{a^2} \ &;& \nonumber \\
    \mbox{as} \ \vert r\vert\rightarrow+\infty, \quad R^{r\theta}{}_{r\theta} \rightarrow 0 \ &.&
\end{eqnarray}
Einstein tensor non-zero components:

\begin{eqnarray}\label{einsteinmorristhorne}
    G^{t}{}_{t} = -G^{r}{}_{r} = G^{\theta}{}_{\theta} = G^{\phi}{}_{\phi} = \frac{a^2}{\left(r^2+a^2\right)^2} = -\frac{1}{2}R \ &;& \nonumber \\
    \mbox{as} \ \vert r\vert\rightarrow 0, \quad G^{t}{}_{t} \rightarrow \frac{1}{a^2} \ &;& \nonumber \\
    \mbox{as} \ \vert r\vert\rightarrow+\infty, \quad G^{t}{}_{t} \rightarrow 0 \ &.&
\end{eqnarray}
Weyl tensor non-zero components:

\begin{eqnarray}
    -2C^{tr}{}_{tr} = C^{t\theta}{}_{t\theta} = C^{t\phi}{}_{t\phi} &=& C^{r\theta}{}_{r\theta} = C^{r\phi}{}_{r\phi} \nonumber \\
    && \nonumber \\
    &=& -2C^{\theta\phi}{}_{\theta\phi} = \frac{-a^2}{3\left(r^2+a^2\right)^2} = \frac{1}{6}R \ ; \nonumber \\
    && \mbox{as} \ \vert r\vert\rightarrow 0, \quad C^{t\theta}{}_{t\theta} \rightarrow -\frac{1}{3a^2} \ ; \nonumber \\
    && \mbox{as} \ \vert r\vert\rightarrow+\infty, \quad C^{t\theta}{}_{t\theta} \rightarrow 0 \ .
\end{eqnarray}\clearpage
\noindent The Ricci contraction $R_{\mu\nu}R^{\mu\nu}$:

\begin{eqnarray}
    R_{\mu\nu}R^{\mu\nu} = \frac{4a^4}{\left(r^2+a^2\right)^4} = R^{2} \ &;& \nonumber \\
    \mbox{as} \ \vert r\vert\rightarrow 0, \quad R_{\mu\nu}R^{\mu\nu} \rightarrow \frac{4}{a^4} \ &;& \nonumber \\
    \mbox{as} \ \vert r\vert\rightarrow+\infty, \quad R_{\mu\nu}R^{\mu\nu} \rightarrow 0 \ &.&
\end{eqnarray}
The Kretschmann scalar:

\begin{eqnarray}
    R_{\mu\nu\alpha\beta}R^{\mu\nu\alpha\beta} = \frac{12a^4}{\left(r^2+a^2\right)^4} = 3R^{2} \ &;& \nonumber \\
    \mbox{as} \ \vert r\vert\rightarrow 0, \quad R_{\mu\nu\alpha\beta}R^{\mu\nu\alpha\beta} \rightarrow \frac{12}{a^4} \ &;& \nonumber \\
    \mbox{as} \ \vert r\vert\rightarrow+\infty, \quad R_{\mu\nu\alpha\beta}R^{\mu\nu\alpha\beta} \rightarrow 0 \ &.&
\end{eqnarray}
The Weyl contraction $C_{\mu\nu\alpha\beta}C^{\mu\nu\alpha\beta}$:

\begin{eqnarray}
    C_{\mu\nu\alpha\beta}C^{\mu\nu\alpha\beta} = \frac{16a^4}{3\left(r^2+a^2\right)^4} = \frac{4}{3}R^{2} \ &;& \nonumber \\
    \mbox{as} \ \vert r\vert\rightarrow 0, \quad C_{\mu\nu\alpha\beta}C^{\mu\nu\alpha\beta} \rightarrow \frac{16}{3a^4} \ &;& \nonumber \\
    \mbox{as} \ \vert r\vert\rightarrow+\infty, \quad C_{\mu\nu\alpha\beta}C^{\mu\nu\alpha\beta} \rightarrow 0 \ &.&
\end{eqnarray}

It is therefore clear that all curvature tensor components and invariants are strictly finite at all coordinate locations in the domain for the Morris-Thorne metric. Characterising a gravitational singularity as a coordinate location corresponding to infinite curvature as in \S\ref{singularity}, we may conclude that the geometry possesses no gravitational singularities. Furthermore the components and scalars are very simple near the throat of the wormhole geometry; of particular interest is the fact that all non-zero tensor components and curvature invariants can be expressed as simple functions of the Ricci scalar, $R$.  The Morris-Thorne metric is therefore singularity-free, has no horizons, has a throat connecting two asymptotically flat regions of spacetime which also satisfies the `flare-out' condition, and as such is indeed a traversable wormhole as presupposed.

%%%%%%

\subsection{ISCO and photon sphere analysis}

Let us find the coordinate locations of the photon sphere for massless particles and the ISCO for massive particles as functions of $r$ and $a$, emphasising that the current calculation is essentially a template for future re-use with respect to other metric environments.

Consider the tangent vector to the worldline of a massive or massless particle, parameterised by some arbitrary affine parameter, $\lambda$:

\begin{equation}
    g_{\mu\nu}\frac{dx^{\mu}}{d\lambda}\frac{dx^{\nu}}{d\lambda}=-\left(\frac{dt}{d\lambda}\right)^{2}+\left(\frac{dr}{d\lambda}\right)^{2}+\left(r^{2}+a^{2}\right)\left\lbrace\left(\frac{d\theta}{d\lambda}\right)^{2}+\sin^{2}\theta \left(\frac{d\phi}{d\lambda}\right)^{2}\right\rbrace \ .
\end{equation}
Since we have used an affine parameter here, and we are certainly not dealing with a spacelike separation in either the massive or massless case, we may, without loss of generality, separate the two cases by defining a scalar-valued object as follows:

\begin{equation}
    \epsilon = \left\{
    \begin{array}{rl}
    -1 & \qquad\mbox{Massive particle, \emph{i.e.} timelike worldline} \ ; \\
     0 & \qquad\mbox{Massless particle, \emph{i.e.} null geodesic} \ .
    \end{array}\right. 
\end{equation}
That is, $g_{\mu\nu}\frac{dx^{\mu}}{d\lambda}\frac{dx^{\nu}}{d\lambda}=\epsilon$, and due to the metric being spherically symmetric we may fix $\theta=\frac{\pi}{2}$ arbitrarily and view the reduced equatorial problem:
\begin{equation}
    g_{\mu\nu}\frac{dx^{\mu}}{d\lambda}\frac{dx^{\nu}}{d\lambda}=-\left(\frac{dt}{d\lambda}\right)^{2}+\left(\frac{dr}{d\lambda}\right)^{2}+\left(r^{2}+a^{2}\right)\left(\frac{d\phi}{d\lambda}\right)^{2}=\epsilon \ .
\end{equation}

At this stage we must note that there are symmetries in the spacetime geometry, and as such associated Killing vectors which yield conserved physical quantities in accordance with the same conservation laws that arise from classical analytic mechanics.\footnote{For a more rigorous insight into the mathematics here, please see \S\ref{Killing}.} The metric is independent of both time, $t$, and azimuthal angle, $\phi$; this yields the following expressions for the conservation of energy $E$, and angular momentum $L$:

\begin{equation}
    \left(\frac{dt}{d\lambda}\right)=E \ ; \qquad\quad \left(r^{2}+a^{2}\right)\left(\frac{d\phi}{d\lambda}\right)=L \ .
\end{equation}
Hence:

\begin{equation}
    -E^2+\left(\frac{dr}{d\lambda}\right)^2+\frac{L^{2}}{r^{2}+a^{2}}=\epsilon \ ,
\end{equation}
\\
\begin{equation}
    \Longrightarrow\quad\left(\frac{dr}{d\lambda}\right)^{2}=E^2+ \epsilon-\frac{L^2}{r^2+a^2} \ .
\end{equation}
Noting that we may assume both the photon sphere and ISCO locations to correspond to geodesic orbits, and in the context of spherical symmetry this corresponds to a fixed $r$-coordinate (\emph {i.e.} $\frac{dr}{d\lambda}=0$), this gives `effective potentials' for geodesic orbits as follows (`potentials' are proportional to $E^{2}$):

\begin{equation}
    V_{\epsilon}(r)=-\epsilon+\frac{L^{2}}{r^{2}+a^{2}} \ .
\end{equation}

\begin{itemize}
    \item For a photon orbit we have the massless particle case $\epsilon=0$. Since we are in a spherically symmetric environment, solving for the locations of such orbits amounts to finding the coordinate location of the 'photon sphere'; \emph {i.e.} the value of the $r$-coordinate sufficiently close to our mass such that photons are forced to propagate in circular geodesic orbits. These circular orbits occur at $V_{0}^{'}(r)=0$, as such:
    \begin{equation}
        V_{0}(r)=\frac{L^{2}}{r^{2}+a^{2}} \ ,
    \end{equation}
    leading to:
    \begin{equation}
        V_{0}^{'}(r)=\frac{-2r L^2}{\left(r^2+a^2\right)^2} \ .
    \end{equation}
    Evaluating where $V_{0}^{'}(r)=0$, this yields the location of these circular orbits to be the coordinate location $r=0$, at the wormhole throat.
    
    \noindent To verify stability, check the sign of $V_{0}^{''}(r)$:
    
    \begin{equation}
        V_{0}^{''}(r) = \frac{-2L^2}{\left(r^2+a^2\right)^2}\left[1-\frac{4r^2}{r^2+a^2}\right] \ ,
    \end{equation}
    
    evaluating this at $r=0$ we have:
    
    \begin{equation}
        V_{0}^{''}(r=0) = \frac{-2L^2}{a^4} \ ,
    \end{equation}
    
    which is strictly less than zero in view of $L$ being an angular momentum (\emph{i.e.} $L>0$), and $a\in(0, +\infty)$. Therefore this corresponds to an unstable photon orbit.

    \item For massive particles the geodesic orbit corresponds to a timelike worldline and we have the case that $\epsilon=-1$. Therefore:
    \begin{equation}
        V_{-1}(r)=1+\frac{L^{2}}{r^{2}+a^{2}} \ , 
    \end{equation}
    and it is easily verified that this leads to:
    \begin{equation}
        V_{-1}^{'}(r)= V_{0}^{'}(r) = \frac{-2rL^2}{\left(r^2+a^2\right)^2} \ ,
    \end{equation}
    and we similarly have a coordinate location for our ISCO at $r=0$, right at the wormhole throat. Similarly to the photon sphere, this orbit will be unstable -- please see \S\ref{ISCOintro} for details on the stability of ISCOs in general.
\end{itemize}

%%%%%%

\subsection{Regge-Wheeler analysis}\label{MorristhorneRW}

Considering the Regge-Wheeler equation in view of the formalism developed in~\cite{ReggeWheeler1}, we may explicitly evaluate the Regge-Wheeler potentials for particles of spin $S\in\lbrace 0,1\rbrace$ in our spacetime. Firstly note that the metric Eq.~\ref{Morristhorne} can be written as:

\begin{equation}
    ds^2 = \bigg\lbrace -dt^2+dr_{*}^2\bigg\rbrace+\left(r^2+a^2\right)\left(d\theta^2+\sin^2\theta\; d\phi^2\right) \ ,
\end{equation}
where $dr_{*}=dr$ is the naturally defined tortoise coordinate, and we can express $\left(r^2+a^2\right)$ as a function of $r_{*}$, $B(r_{*})$ (the only reason for re-expressing the metric in terms of the tortoise coordinate in this case is to remain consistent with the general formalism surrounding the Regge-Wheeler equation as developed in~\cite{ReggeWheeler1}; for this particular analysis $g_{rr}=1$, so the tortoise coordinate and our standard $r$-coordinate are in fact identical), yielding:

\begin{equation}
    ds^2 = \bigg\lbrace -dt^2+dr_{*}^2\bigg\rbrace+B(r_*)^2\left(d\theta^2+\sin^2\theta \; d\phi^2\right) \ .
\end{equation}

The general Regge--Wheeler equation is~\cite{ReggeWheeler1}:

\begin{equation}
    \partial_{r_{*}}^{2}\hat{\phi}+\lbrace \omega^2-\mathcal{V}\rbrace\hat\phi = 0 \ .
\end{equation}
For a scalar field ($S=0$) examination of the d'Alembertian equation quickly yields~\cite{ReggeWheeler1}:

\begin{equation}
\mathcal{V}_{S=0} =   \left\lbrace{1 \over B^2} \right\rbrace \ell(\ell+1)
+ {\partial_{r_{*}}^2 B \over B} \ .
\end{equation}
For a vector field ($S=1$) conformal invariance in `3+1'-dimensions guarantees that the Regge-Wheeler potential can depend only on the ratio $1/B$, whence normalizing to known results implies~\cite{ReggeWheeler1}:

\begin{equation}
\mathcal{V}_{S=1} =   \left\lbrace{1 \over B^2} \right\rbrace \ell(\ell+1).
\end{equation}
Collecting results, for $S\in\{0,1\}$ we have:

\begin{equation}
\mathcal{V}_{S} =   \left\lbrace{1 \over B^2} \right\rbrace \ell(\ell+1)
+ (1-S) {\partial_{r_{*}}^2 B \over B} \ .
\end{equation}
The spin 2 axial mode ($S=2$) is somewhat messier in this particular case, and not of immediate interest.

Noting that for our metric $\partial_{r_{*}}=\partial_{r}$ and $B=\sqrt{r^2+a^2}$  we have:

\begin{equation}
    \frac{\partial_{r_{*}}^2 B}{B} = \frac{a^2}{\left(r^2+a^2\right)^{2}} \ ,
\end{equation}\clearpage
\noindent therefore:

\begin{equation}
\mathcal{V}_{S\in\{0,1\}} = \frac{1}{\left(r^2+a^2\right)}\left\lbrace \ell\left(\ell+1\right)+\left(1-S\right)\frac{a^2}{\left(r^2+a^2\right)}\right\rbrace \ .
\end{equation}
This has the correct behaviour as $\vert r\vert\rightarrow +\infty$, since $\mathcal{V}_{S\in\left\lbrace 0, 1\right\rbrace}\rightarrow 0$. At the throat we observe the following:

\begin{equation}
    \mbox{As} \ \vert r\vert\rightarrow 0, \qquad \mathcal{V}_{S\in\left\lbrace 0, 1\right\rbrace}\rightarrow \frac{\ell\left(\ell+1\right)+1-S}{a^2} \ .
\end{equation}
For the specific spin cases we have:

\subsubsection{Spin zero}

Let $S=0$:

\begin{equation}
    \mathcal{V}_{S=0} = \frac{1}{\left(r^2+a^2\right)}\left\lbrace \ell\left(\ell+1\right)+\frac{a^2}{\left(r^2+a^2\right)}\right\rbrace \ .
\end{equation}
For scalars the $s$-wave ($\ell=0$) is particularly important:

\begin{equation}
    \mathcal{V}_{0, \ell=0} = \frac{\partial_{r_{*}}^{2}r}{r} = \frac{a^2}{\left(r^2+a^2\right)^{2}} \ . 
\end{equation}

\subsubsection{Spin one}

Let $S=1$:

\begin{equation}
    \mathcal{V}_{S=1} = \frac{\ell\left(\ell+1\right)}{\left(r^2+a^2\right)} \ .
\end{equation}

%%%%%%

\subsection{Stress-energy-momentum tensor}

Let us examine the resulting Einstein field equations for this spacetime, and subsequently analyse the various energy conditions. In view of the form of the energy conditions presented in \S\ref{energyconditions}, we shall first use the metric tensor to lower one index on the non-zero Einstein tensor components from Eq.~\ref{einsteinmorristhorne}. Combining this with the form of the Einstein field equations presented in Eq.~\ref{einsteineqsfinal}, we have: $g_{\mu\sigma}G^{\sigma}{}_{\nu}=G_{\mu\nu}=8\pi T_{\mu\nu}$. This yields the following specific form of the stress-energy-momentum tensor:

\begin{equation}
    \rho = p_{\parallel} = -p_{\perp} = \frac{-a^2}{8\pi\left(r^2+a^2\right)^2} \ .
\end{equation}
We may now conduct analysis of the various energy conditions and see whether they are violated in our spacetime.

\subsubsection{Null energy condition}

We require that $\forall \ r, a, m$, both $\rho+p_{\parallel}\geq 0$, \emph{and} $\rho+p_{\perp}\geq 0$. Firstly looking at $\rho+p_{\parallel}$:

\begin{equation}
    \rho = p_{\parallel} = \frac{-a^2}{8\pi\left(r^2+a^2\right)} \quad \Longrightarrow \quad \rho+p_{\parallel} = \frac{-2a^2}{8\pi\left(r^2+a^2\right)} < 0 \ .
\end{equation}
In view of the fact that this is manifestly negative for all coordinate locations in our domain, it is clear the null energy condition is strictly violated in this spacetime. It is a corollary of this that the remaining weak, strong, and dominant energy conditions will be similarly violated (see \S\ref{energyconditions} for details). We therefore have an intriguing spacetime which in the framework of general relativity requires us to thread the throat with some exotic mass source possessing a negative energy density (\emph{i.e.} $\rho<0$); not consistent with the energy conditions.

%%%%%%%%%%%%%%%%%%%%%%%%%%%%%%%%%%%%%%%%%%%%%%%%%%%%%%%%

\section{The exponential metric}\label{C:Exponential}

\noindent A specific metric candidate of interest, the so-called `exponential metric', has been favoured by certain members of the community for some time. Emphasis has been placed on the lack of horizons present in the subsequent geometry (which implies the metric is certainly not modeling a black hole), although rarely has it been noted that instead one is dealing with a traversable wormhole, in the sense of Morris and Thorne~\cite{MorrisThorne}. Furthermore it is of specific interest in this context as it permits a global coordinate patch. Some of the proponents of this metric are also in favour of pursuing alternative theories of gravity to that of general relativity, and believe that this metric supports this pursuit. As such, let us undertake a standard analysis of the metric and see whether it yields a straightforward interpretation through the lens of general relativity.

The `exponential metric' is described by the line element:

\begin{equation}
ds^2 = - e^{-2m/r} dt^2 + e^{+2m/r}\{dr^2 + r^2(d\theta^2+\sin^2\theta \, d\phi^2)\} \ ,
\end{equation}
 and has now been in circulation for some sixty years~\cite{Yilmaz:1958, Yilmaz:1971, Yilmaz:1973, Clapp:1973, Rastall:1975, Fennelly:1976, Misner:1995, Alley:1995, MECO:1999, Robertson:1999, Ibison:2006a, Ibison:2006b, BenAmots:2007, Svidzinsky:2009, Martinis:2010, BenAmots:2011, Svidzinsky:2015, Aldama:2015, MECO:2016};
at least since 1958. In weak fields, ($\frac{2m}{r}\ll 1$), one has:

\begin{equation}
ds^2 = \{- dt^2 +dr^2 + r^2(d\theta^2+\sin^2\theta \, d\phi^2)\} +
{2m\over r} \{dt^2 +dr^2 + r^2(d\theta^2+\sin^2\theta \, d\phi^2)\} \ ,
\end{equation}\clearpage
\noindent that is:

\begin{equation}
g_{ab} = \eta_{ab} + {2m\over r} \delta_{ab} \ .
\end{equation}
This exactly matches the lowest-order weak-field expansion of the can\-onical Schwarzschild solution, and so this exponential metric will automatically pass all of the standard lowest-order weak-field tests of general relativity. However strong-field behaviour,  ($\frac{2m}{r}\gg 1$),  and even medium-field behaviour,  ($\frac{2m}{r} \sim 1$), is rather different.

The exponential metric has no horizons, $g_{tt}\neq0$, and so is clearly not a black hole. On the other hand, it does not seem to have been previously remarked that the exponential metric describes a traversable wormhole in the sense of Morris and Thorne~\cite{MorrisThorne,MTY,Visser:1989a,Visser:1989b,LorentzianWormholes, Cramer:1994, Poisson:1995, Hochberg:1997, Visser:1997, Hochberg:1998, Hochberg:1998b, Barcelo:1999, Barcelo:2000, Dadhich:2001, Visser:2003,Lemos:2003,Kar:2004,Lobo:2005,Sushkov:2005,Garcia:2011, Bhawal, Arias, Gao, Maldacena, Willenborg, Sahoo}. In fact, the exponential metric has a wormhole throat at $r=m$, with the region $r<m$ corresponding to an infinite-volume `other universe' where time runs slower on the other side of the wormhole throat. Note also that metric is both static and spherically symmetric.

%%%%%%

\subsection{Traversable wormhole throat}\label{S:throat}

Consider the area of the spherical surfaces of constant $r$ coordinate:

\begin{equation}
A(r) = 4\pi r^2 e^{2m/r} \ .
\end{equation}
Then:

\begin{equation}
{dA(r)\over dr} = 8\pi (r-m) e^{2m/r} \ ,
\end{equation}
and:

\begin{equation}
{d^2A(r)\over dr^2} = 8\pi  e^{2m/r} \left(1-{2m\over r} +{2m^2\over r^2}\right)
=
8\pi  e^{2m/r} \left\{\left(1-{m\over r}\right)^2  +{m^2\over r^2}\right\} > 0 \ .
\end{equation}
That is: the area is a concave function of the $r$ coordinate, and has a minimum at $r=m$, where it satisfies the `flare-out' condition  $A''|_{r=m} = +8\pi e^2 > 0$.
Furthermore, all metric components are finite at $r=m$, and the diagonal components are non-zero. This is sufficient to guarantee that the surface $r=m$ is a traversable wormhole throat,  in the sense of Morris and Thorne~\cite{MorrisThorne,MTY,Visser:1989a,Visser:1989b,LorentzianWormholes, Cramer:1994, Poisson:1995, Hochberg:1997, Visser:1997, Hochberg:1998, Hochberg:1998b, Barcelo:1999, Barcelo:2000, Dadhich:2001, Visser:2003,Lemos:2003,Kar:2004,Lobo:2005,Sushkov:2005,Garcia:2011, Bhawal, Arias, Gao, Maldacena, Willenborg, Sahoo}. There is a rich phenomenology of traversable wormhole physics that has been developed over the last thirty years (since the Morris-Thorne paper~\cite{MorrisThorne}), much of which can be readily adapted (\emph{mutatis mutandi}) to the exponential metric. 

%%%%%%

\subsection{Comparison: exponential \emph{versus} Schwarzschild}\label{S:compare}

Let us briefly compare the exponential and Schwarzschild metrics.

\subsubsection{Isotropic coordinates}\label{S:isotropic}

In isotropic coordinates the Schwarzschild spacetime is:

\begin{equation}
ds_\mathrm{Sch}^2 = - \left(1-{m\over2r}\over1+{m\over2r}\right)^2 dt^2
+ \left(1+{m\over2r}\right)^4 \{dr^2 +  r^2(d\theta^2+\sin^2\theta \, d\phi^2)\} \ ,
\end{equation}
which we should compare with the exponential metric in isotropic coordinates:

\begin{equation}
ds^2 = - e^{-2m/r} dt^2 + e^{+2m/r}\{dr^2 + r^2(d\theta^2+\sin^2\theta \, d\phi^2)\} \ .
\end{equation}

It is clear that in the Schwarzschild spacetime there is a horizon present at $r=\frac{m}{2}$.
Recalling that the domain for the $r$-coordinate in the isotropic coordinate system for Schwarzschild is $r\in (0, +\infty)$, we see that the horizon also corresponds to where the area of spherical constant-$r$ surfaces is minimised:

\begin{eqnarray}
A(r) &=& 4\pi r^2\left(1+\frac{m}{2r}\right)^{4} \ ; \nonumber \\
&& \nonumber \\
\frac{dA(r)}{dr} &=& 8\pi r\left(1-\frac{m}{2r}\right)\left(1+\frac{m}{2r}\right)^{3} \ ; \nonumber \\
&& \nonumber \\
\frac{d^{2}A(r)}{dr^2} &=& 8\pi\left(1+\frac{m}{2r}\right)^{2}\left(\frac{3}{4}\left(\frac{m}{r}\right)^{2}-\frac{m}{r}+1\right) \ .
\end{eqnarray}
So for the Schwarzschild geometry in isotropic coordinates the area has a minimum at $r=\frac{m}{2}$, where $A'\vert_{r=\frac{m}{2}}=0$, and $A''\vert_{r=\frac{m}{2}}=+64\pi >0$.
While this satisfies the `flare-out' condition the corresponding wormhole (it is in fact the Einstein-Rosen bridge) is \emph{non-traversable} due to the presence of the horizon.

In contrast the geometry described by the exponential metric clearly has no horizons, since $\forall \ r\in (0,+\infty)$ we have $g_{tt}=\exp\left({\frac{-2m}{r}}\right)\neq 0$. As already demonstrated, there is a  traversable wormhole throat located at $r=m$, where the area of the spherical surfaces is minimised, and the `flare-out' condition is satisfied, in the \emph{absence} of a horizon. Thus the Schwarzschild horizon at $r=\frac{m}{2}$ in isotropic coordinates is replaced by a wormhole throat at $r=m$ in the exponential metric.

Furthermore, for the exponential metric, since $\exp\left({\frac{-2m}{r}}\right) >0$ is monotone decreasing as $r\to0$, it follows that proper time for a stationary observer evolves increasingly slowly as a function of coordinate time as one moves closer to the centre $r\to 0$.  

%%%%%%

\subsubsection{Curvature coordinates}\label{S:curvature}

To go to so-called `curvature coordinates', $r_{s}$, for the exponential metric we make the coordinate transformation:

\begin{equation}
r_s = r \, e^{m/r}; \qquad\qquad dr_s = e^{m/r}  \, (1 - m/r) \ dr \ .
\end{equation}
So for the exponential metric in curvature coordinates:

\begin{equation}
ds^2 = - e^{-2m/r} dt^2 + {dr_s^2\over (1-m/r)^2} + r_s^2(d\theta^2+\sin^2\theta \, d\phi^2) \ .
\end{equation}
Here $r$ is regarded as an implicit function of $r_s$. Note that as the isotropic coordinate $r$  ranges over the interval $(0,\infty)$, the curvature coordinate $r_s$ has a minimum at $r_s=m\,e$. In fact for the exponential metric the curvature coordinate $r_s$ double-covers the interval $r_s\in[m\, e,\infty)$, first descending from $\infty$ to $m\, e$ and then increasing again to $\infty$.
Indeed, looking for the minimum of the coordinate $r_{s}$:

\begin{equation}
\frac{dr_{s}}{dr} = e^{m/r}\left(1-\frac{m}{r}\right)
\qquad \Longrightarrow \qquad 
\left.\frac{dr_{s}}{dr}\right\vert_{r=m}=0 \ .
\end{equation}
So we have a stationary point at $r=m$, which corresponds to $r_{s}=m\,e$, and furthermore:
\begin{equation}
\frac{d^{2}r_{s}}{dr^{2}}  = \frac{m^2}{r^{3}}\;e^{m/r}
\qquad \Longrightarrow \qquad  \left.\frac{d^{2}r_{s}}{dr^{2}}\right\vert_{r=m}>0 \ .
\end{equation}
The curvature coordinate $r_{s}$ therefore has a minimum at $r_{s}=m\,e$, and in these curvature  coordinates the exponential metric exhibits a wormhole throat at $r_{s}=m\,e > 2m$.

Compare this with the Schwarzschild metric in curvature coordinates:

\begin{equation}
ds_\mathrm{Sch}^2 = - (1-2m/r_s) \, dt^2 + {dr_s^2\over 1-2m/r_s} + r_s^2(d\theta^2+\sin^2\theta \, d\phi^2) \ .
\end{equation}
By inspection it is clear that there is a horizon at $r_{s}=2m$, since at that location  $g_{tt}\vert_{r_{s}=2m}=0$. For the Schwarzschild metric the isotropic and curvature coordinates are related by $r_s = r \left(1+{m\over2r}\right)^2$. 

If for the exponential metric one really wants the fully explicit inversion of $r$ as a function of $r_s$, then observe:

\begin{equation}
r =r_s \exp\left( W(-m/r_s)\right) = - {m\over W(-m/r_s)} \ .
\end{equation}
Here $W(x)$ is the `appropriate branch' of the Lambert $W$ function -- implicitly defined by the relation $W(x)\, e^{W(x)}=x$. This function has a convoluted two hundred and fifty-year history; only recently has it become common to view it as one of the standard `special functions' of mathematics~\cite{Corless}. Applications vary~\cite{Corless,Valluri:00}, including combinatorics (enumeration of rooted trees)~\cite{Corless},  delay differential equations~\cite{Corless}, falling objects subject to linear drag~\cite{Vial:12}, evaluating the numerical constant in Wien's displacement law~\cite{Stewart:11, Stewart:12}, quantum statistics~\cite{Valluri:09},
the distribution of prime numbers~\cite{primes}, constructing the `tortoise' coordinate for Schwarzs\-child black holes~\cite{tortoise}, \emph{etcetera}.

In terms of the  Lambert $W$ function and the curvature coordinate $r_s$, the explicit version of the exponential metric becomes:

\begin{equation}
ds^2 = - e^{2W(-m/r_s)} \, dt^2 + {dr_s^2\over (1+W(-m/r_s))^2} + r_s^2(d\theta^2+\sin^2\theta \, d\phi^2) \ .
\end{equation}
The $W_0(x)$ branch corresponds to the region $r>m$ outside the wormhole throat;
whereas the $W_{-1}(x)$ branch corresponds to the region $r<m$ inside the wormhole.
The Taylor series for $W_0(x)$ for $|x| < e^{-1}$ is~\cite{Corless}:

\begin{equation}
W_0(x) = \sum_{n=1}^\infty {(-n)^{n-1} x^n\over n!} \ .
\end{equation}
A key asymptotic formula for $W_{-1}(x)$ is~\cite{Corless}:\footnote{Eq.~\ref{littleo} employs use of the `Little o' notation; this means that $o(x)$ exhibits the asymptotic behaviour $\vert o(x)\vert/\vert x\vert\rightarrow 0$~\cite{little-o}.}

\begin{equation}\label{littleo}
W_{-1}(x) =  \ln(-x) - \ln(-\ln(-x)) + o(1);  \qquad (x\to 0^-) \ .
\end{equation}
The two real branches meet at $W_0(-1/e)=W_{-1}(-1/e)=-1$, and in the vicinity of that meeting point:\footnote{Eq.~\ref{bigO} employs use of the `Big $\mathcal{O}$' notation; this means that $\mathcal{O}(x)$ is some expression which is \emph{at most} a positive constant multiple of the input, $x$~\cite{Big-O}.}

\begin{equation}\label{bigO}
W(x) =  -1 + \sqrt{2(1+ex)} - {2\over3} (1+ex) + \mathcal{O}[(1+ex)^{3/2}] \ .
\end{equation}
More details regarding the Lambert $W$ function can be found in Corless~\emph{et al}, please see reference~\cite{Corless}.

%%%%%%

\subsection{Curvature tensors and invariants analysis}\label{S:Exp-tensor-inv}

For the Riemann tensor the non-vanishing components are:

\begin{eqnarray}
R^{tr}{}_{tr} &=& -2R^{t\theta}{}_{t\theta} = -2 R^{t\phi}{}_{t\phi} = {2m(r-m)e^{-2m/r}\over r^4} \ ; \nonumber \\
R^{r\theta}{}_{r\theta} &=& R^{r\phi}{}_{r\phi} = -{m e^{-2m/r}\over r^3} \ ; \nonumber \\
R^{\theta\phi}{}_{\theta\phi} &=&  {m(2r-m)e^{-2m/r}\over r^4} \ .
\end{eqnarray}
Weyl tensor non-zero components:

\begin{eqnarray}
C^{tr}{}_{tr} &=& -2C^{t\theta}{}_{t\theta} = -2 C^{t\phi}{}_{t\phi} = -2C^{r\theta}{}_{r\theta} = -2 C^{r\phi}{}_{r\phi} \nonumber \\
&& \nonumber \\
&=& C^{\theta\phi}{}_{\theta\phi} = {2m(3r-2m)e^{-2m/r}\over 3r^4} \ .
\end{eqnarray}
For the Ricci and Einstein tensors, as well as the Ricci scalar:

\begin{eqnarray}
R^{\mu}{}_{\nu} &=& -{2m^2 e^{-2m/r}\over r^4} \; \text{diag}\{0,1,0,0\}^\mu{}_{\nu} \ ;
\\
R &=& -{2m^2 e^{-2m/r}\over r^4} \ ;
\\
G^{\mu}{}_{\nu} &=& {m^2 e^{-2m/r}\over r^4} \; \text{diag}\{1,-1,1,1\}^\mu{}_{\nu} \ .
\end{eqnarray}
For the Kretschmann and other related scalars we have:

\begin{equation}
R_{\mu\nu\alpha\beta}\, R^{\mu\nu\alpha\beta} = {4m^2(12r^2-16mr+7m^2)e^{-4m/r}\over r^8} \ ;
\end{equation}

\begin{equation}
C_{\mu\nu\alpha\beta}\, C^{\mu\nu\alpha\beta} = {16\over3} {m^2(3r-2m)^2 e^{-4m/r}\over r^8} \ ;
\end{equation}

\begin{equation}
R_{\mu\nu}\, R^{\mu\nu} = R^2 = {4m^4 e^{-4m/r}\over r^8} \ .
\end{equation}

All of the curvature components and scalar invariants exhibited above are finite everywhere in the exponential spacetime -- in particular they are finite at the throat ($r=m$) and decay to zero both as $r\to+\infty$ and as $r\to0$. They take on maximal values near the throat, where $r=\hbox{(dimensionless number)}\times m$.

%%%%%%

\subsection{Ricci convergence conditions}

In the usual framework of general relativity the standard energy conditions are useful (null energy condition, weak energy condition, \emph{etc...}) because they feed back into the Raychaudhuri equations and its generalizations, and so give information about the focussing and defocussing of geodesic congruences~\cite{Raychad1, Hochberg:1998}. In the absence of the Einstein equations were one to attempt analysis through an alternative theory of gravity, one can instead impose conditions directly on the Ricci tensor.

Specifically, a Lorentzian spacetime is said to satisfy the timelike, null, or spacelike Ricci convergence condition if for all timelike, null, or spacelike vectors $t^{\mu}$ one has:
\begin{equation}
R_{\mu\nu}\; t^{\mu}t^{\nu}\geq 0 \ .
\end{equation}
For the exponential metric one has:

\begin{equation}
R_{\mu\nu} = -{2m^2 \over r^4} \; \text{diag}\{0,1,0,0\}_{\mu\nu} \ .
\end{equation}
So the failure of the Ricci convergence condition amounts to:
\begin{equation}
R_{\mu\nu}\; t^{\mu}t^{\nu} =  -{2m^2 \over r^4} \; (t^r)^2 \ngeq 0 \ .
\end{equation}
\enlargethispage{40pt}

\noindent The Ricci convergence condition clearly will \emph{not} be satisfied for all timelike, null, or spacelike vectors $t^{\mu}$ (if $t^{r}>0$ the contraction is in fact strictly negative).
Specifically, the  violation of the null Ricci convergence condition is crucial for understanding the `flare-out' at the throat of the traversable wormhole~\cite{LorentzianWormholes}.

%%%%%%

\subsection{Effective refractive index -- lensing properties}\label{S:refractive}

The exponential metric can be written in the form:

\begin{equation}
ds^2 = e^{2m/r} \left\{ - e^{-4m/r} dt^2 + \{dr^2 + r^2(d\theta^2+\sin^2\theta \, d\phi^2)\} \right\} \ .
\end{equation}
If we are only interested in photon propagation, then the overall conformal factor is irrelevant (since $ds^{2}=0$), and we might as well work with:

\begin{equation}
d\hat s^2 =- e^{-4m/r} dt^2 + \{dr^2 + r^2(d\theta^2+\sin^2\theta \, d\phi^2)\} \ .
\end{equation}
That is:

\begin{equation}
d\hat s^2 =- e^{-4m/r} dt^2 + \{dx^2 +dy^2 +dz^2\} \ .
\end{equation}
But this metric has a very simple physical interpretation: It corresponds to a coordinate speed of light $c(r) = e^{-2m/r}$, or equivalently an effective refractive index:

\begin{equation}
n(r) = e^{2m/r} \ .
\end{equation}
This effective refractive index is well defined all the way down to $r=0$, and (via Fermat's principle of least time) completely characterizes the focussing/defocussing of null geodesics. This notion of `effective refractive index' for the gravitational field has in the weak field limit been considered in reference~\cite{refractive}, and in the strong-field limit falls naturally into the `analogue spacetime' programme~\cite{LRR,LNP}.

Compare the above with Schwarzschild spacetime in isotropic coordinates where the effective refractive index is:

\begin{equation}
n(r) =  {(1+{m\over2r})^3\over|1-{m\over2r}|} \ .
\end{equation}
The two effective refractive indices have the same large-$r$ limit, $n(r)\approx 1 + {2m\over r}$, but differ markedly once $r \lesssim \frac{m}{2}$.

%---------------------------------------------------------------------
\begin{figure}[!htb]
\begin{center}
\includegraphics[scale=0.35]{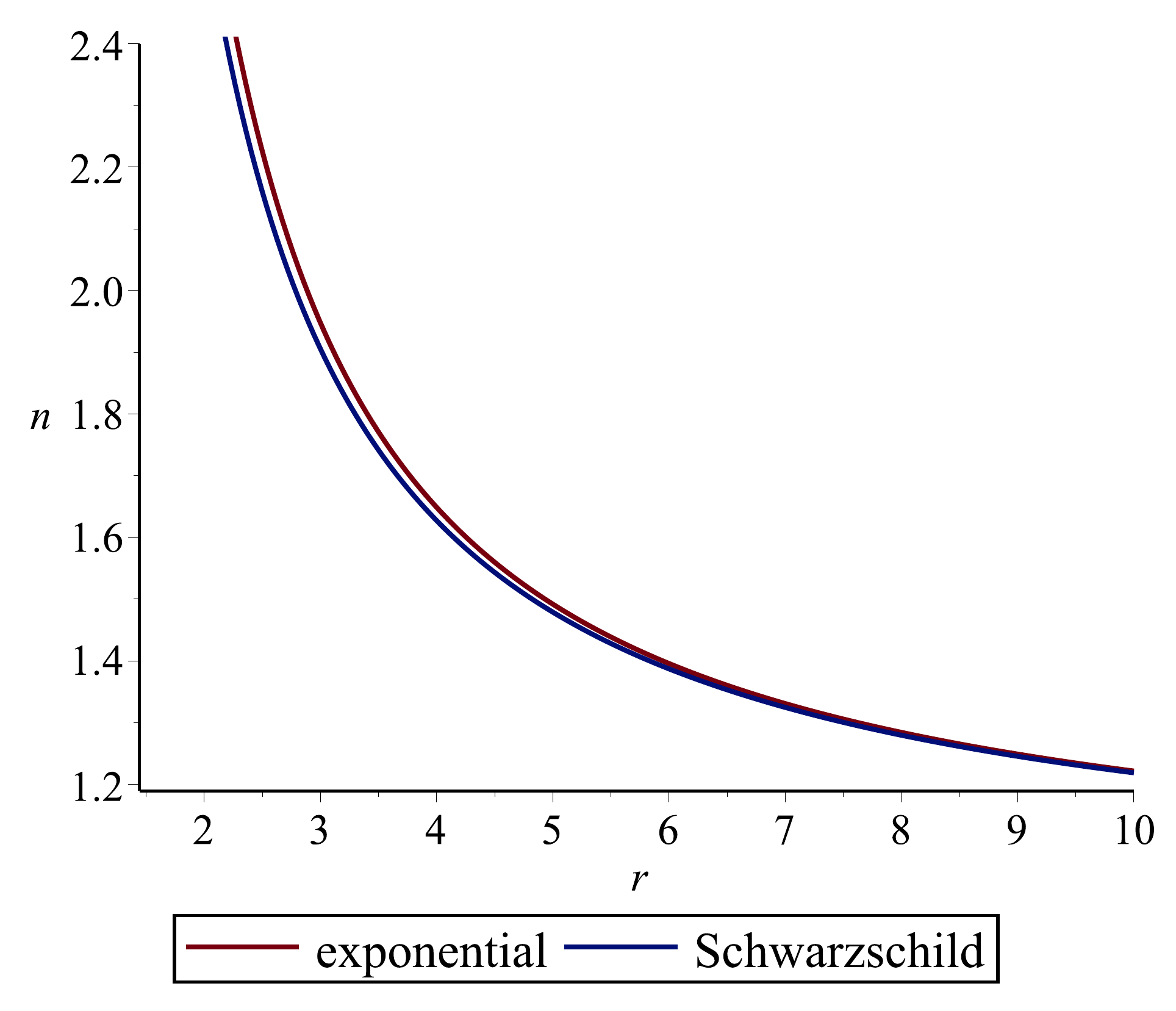}\qquad
\includegraphics[scale=0.35]{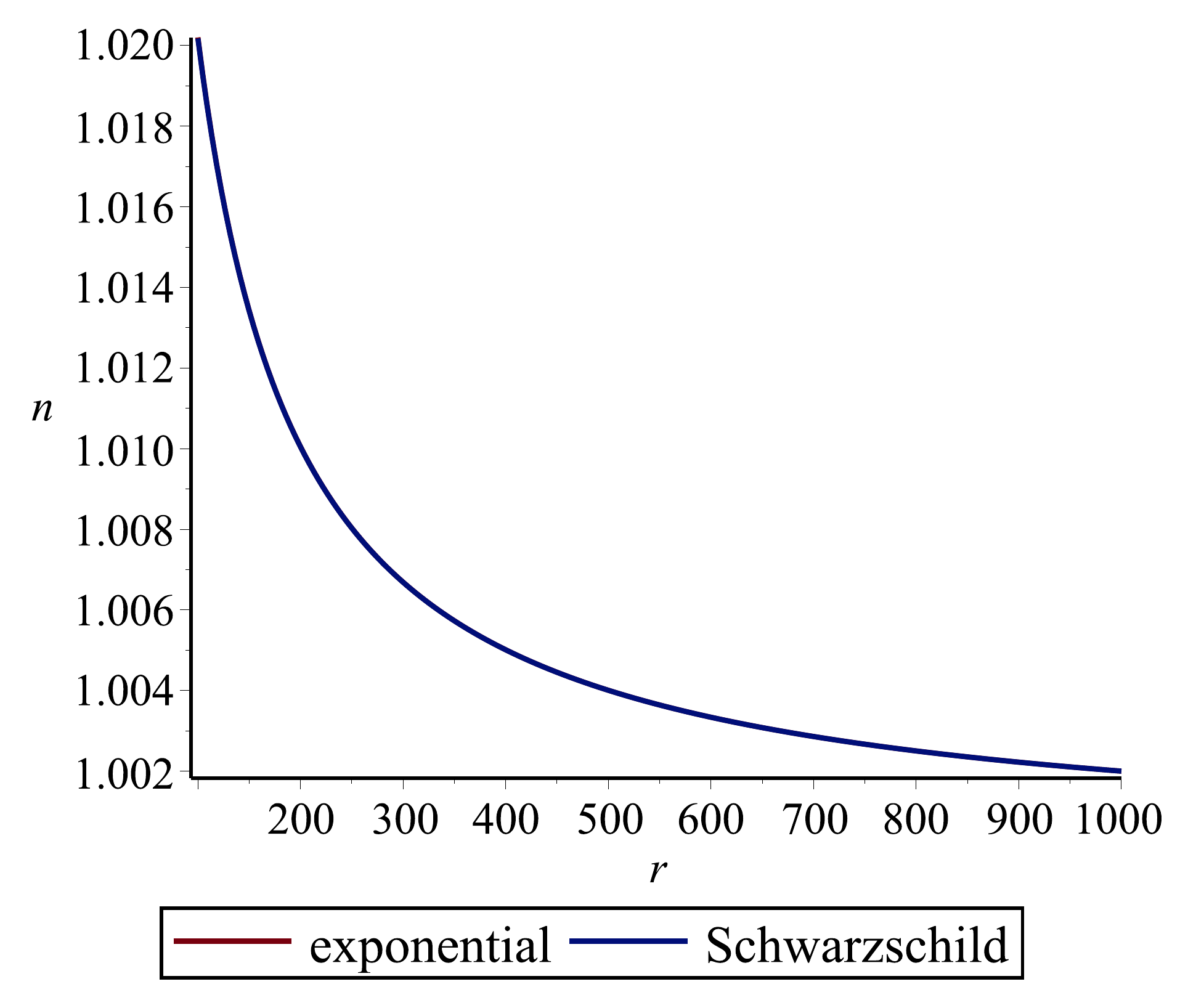}
\includegraphics[scale=0.45]{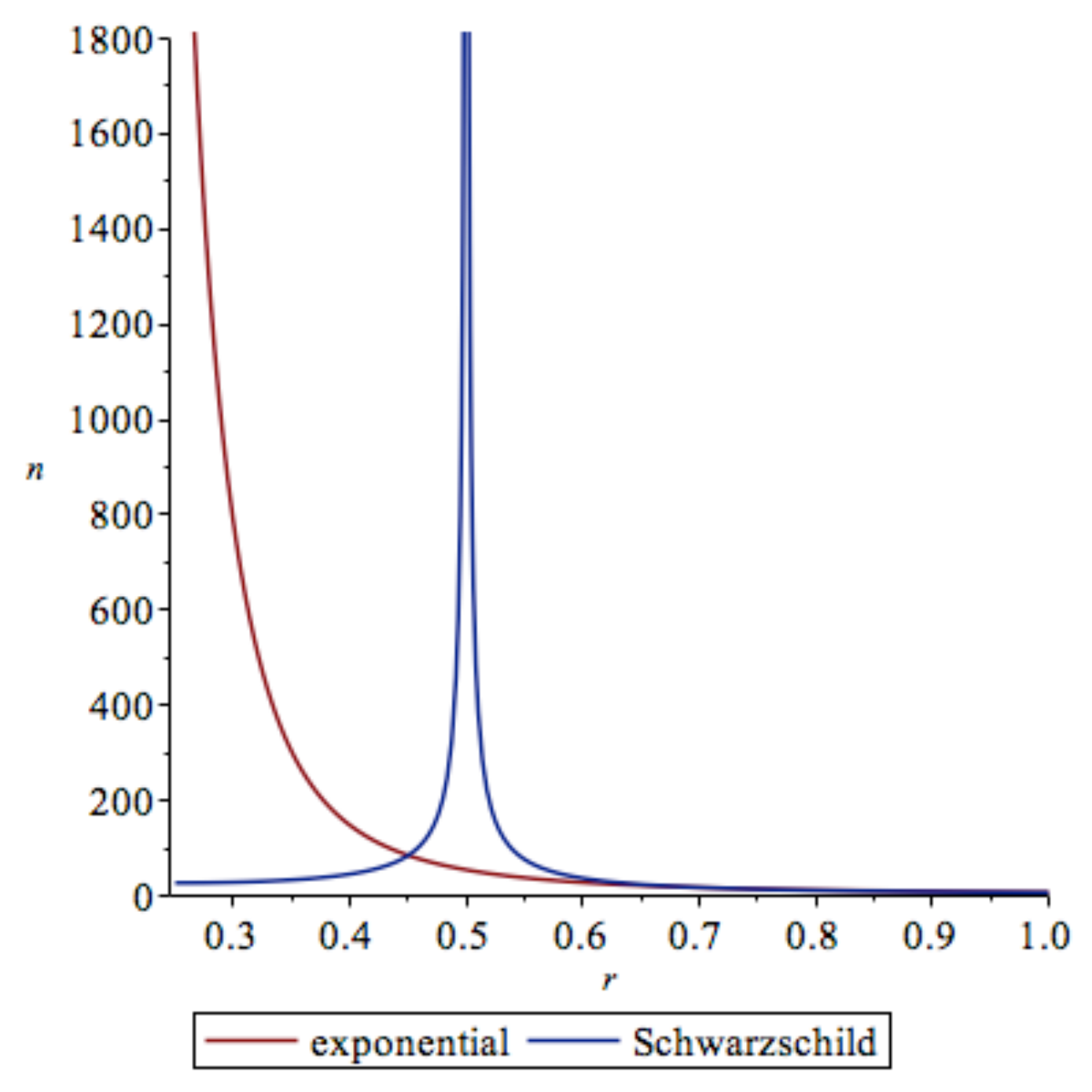}
\end{center}
{\caption[Refractive indexes for the exponential metric \emph{versus}\newline Schwarzschild]{The graph shows the refractive index for the exponential metric compared to the Schwarzschild metric in the isotropic coordinate. The parameter $m = 1$. The top panel is for relatively small $r\gtrsim2m$ and the bottom-left panel is for large $r$. The bottom-right panel is for the strong field region $r \sim \frac{m}{2}$.}\label{pott}}
\end{figure}
%--------------------------------------------------------------------

From the graphs presented in Fig.~\ref{pott}, we can see that the refractive index for the exponential metric is greater than that of the Schwarzschild metric in the isotropic coordinate at tolerably small $r\gtrsim 2m$. For large $r$, they converge to each other and hence are asymptotically equal. In the strong field region they differ radically. Observationally, once you get close enough to where you would have expected to see the Schwarzschild horizon, the lensing properties differ markedly.

%%%%%%

\subsection{ISCO and photon sphere analysis}\label{S:Exp-ISCO}

For massive particles, it is relatively easy to find the innermost stable circular orbit (ISCO) for the exponential metric,  while for massless particles such as photons there  is a unique unstable circular orbit. These can then be compared with the Schwarzschild spacetime.
It is emphasised that the notion of ISCO depends only on the geodesic equations, not on the assumed field equations chosen for setting up the spacetime. 
Since Schwarzschild ISCOs for massive particles at $r_s=6m$ have already been seen by astronomers,
this might place interesting bounds somewhat restraining the exponential-metric enthusiasts.
Additionally, the Schwarzschild unstable circular photon orbit for massless particles is at $r_s=3m$ (the photon sphere); the equivalent for the exponential metric is relatively easy to find.

\def\L{{\mathcal{L}}}
To determine the circular orbits, consider the affinely parameterised tangent vector to the worldline of a massive or massless particle:
\begin{eqnarray}
\frac{ds^2}{d\lambda^2}
&=& -e^{-2m/r} \left(dt \over d\lambda\right)^2 \nonumber \\
&& \nonumber \\
&& + e^{2m/r} \left\{ \left(dr \over d\lambda\right)^2 + r^2\left[\left(d\theta \over d\lambda\right)^2+ \sin^2\theta\left(d\phi \over d\lambda\right)^2\right] \right\} = \epsilon \ . \quad
\end{eqnarray}
Here $\epsilon \in\{-1,0\}$; with $-1$ corresponding to a timelike trajectory and $0$ corresponding to a null trajectory. 
In view of the spherical symmetry we might as well just set $\theta=\pi/2$ and work with the reduced equatorial problem:
\begin{equation}
\frac{ds^2}{d\lambda^2}
= -e^{-2m/r} \left(dt \over d\lambda\right)^2
+ e^{2m/r} \left\{ \left(dr \over d\lambda\right)^2 + r^2\left(d\phi \over d\lambda\right)^2 \right\}
= \epsilon \in\{-1,0\} \ .\quad
\end{equation}

The Killing symmetries imply two conserved quantities (energy and angular momentum):
\begin{equation}
e^{-2m/r} \left(dt \over d\lambda\right)=E; \qquad\qquad
e^{2m/r} r^2 \left(d\phi \over d\lambda\right)=L \ .
\end{equation}
Thence:
\begin{equation}
e^{2m/r} \left\{-E^2 + \left(dr \over d\lambda\right)^2 \right\} + e^{-2m/r} {L^2\over r^2}
=\epsilon \ ,
\end{equation}
that is:
\begin{equation}
\left(dr \over d\lambda\right)^2 =  E^2 + e^{-2m/r} \left\{ \epsilon - e^{-2m/r} {L^2\over r^2}\right\} \ .
\end{equation}
This defines the `effective potential' for geodesic orbits:
\begin{equation}
V_\epsilon(r) = e^{-2m/r} \left\{- \epsilon + e^{-2m/r} {L^2\over r^2}\right\} \ ,
\end{equation}
note that this is significantly more complicated than for the Morris-Thorne wormhole presented in \S\ref{Morristhorne}.

\begin{itemize}
\item
For $\epsilon=0$ (massless particles such as photons), the effective potential simplifies to:
\begin{equation}
V_0(r) = {e^{-4m/r} L^2\over r^2} \ .
\end{equation}
This has a single peak at $r=2m$ corresponding to $V_{0,max} = {L^2\over (2me)^2}$.
This is the only place where $V_0'(r)=0$, and at this point $V''(r)<0$. 
Thus there is an unstable photon sphere at $r=2m$, corresponding to the curvature coordinate $r_s = 2m \,e^{1/2} \approx 3.297442542\, m$ (this is not too far from what we would expect for Schwarzschild, where the photon sphere is at $r_s=3m$).

\item
For $\epsilon=-1$ (massive particles such as atoms, electrons, protons, or planets), the  effective potential is:
\begin{equation}
V_{-1}(r) =  e^{-2m/r} \left\{1 + e^{-2m/r} {L^2\over r^2}\right\}
=
e^{2W(m/r_s)} \left\{1 + {L^2\over r_s^2}\right\} \ .
\end{equation}
It is easy to verify that:
\begin{equation}
V_{-1}'(r) = {2 e^{-2m/r} ( L^2 e^{-2m/r} [2m-r] + m r^2) \over r^4} \ ,
\end{equation}
and that:
\begin{equation}
V_{-1}''(r) = {2 e^{-2m/r} ( L^2 e^{-2m/r} [8m^2-12mr+3r^2] + 2m^2 r^2 -2mr^3)\over r^6} \ .
\end{equation}
Circular orbits, denoted $r_c$, occur at $V_{-1}'(r)=0$, but there is no simple analytic way of determining $r_c(m,L)$ as a function of $m$ and $L$. Working more indirectly, by assuming a circular orbit ar $r=r_c$,  one can solve for the required angular momentum $L_c(r_c,m)$ as a function of $r_c$ and $m$. Explicitly:
\begin{equation}\label{expangular}
L_c(r_c,m) =  { r_c \, e^{m/r_c} \sqrt{m} \over \sqrt{{r_c}-2m}} \ .
\end{equation}
Note that at large $r_c$ we have $L_c(r_c,m)\sim \sqrt{m r_c}$ as one would expect from considering circular orbits in Newtonian gravity. This is a useful consistency check; in classical physics the angular momentum per unit mass for a particle with angular velocity $\omega$ is $L_{c}\sim\omega r_{c}$~\cite{Worthington}. Kepler's third law of planetary motion implies that $\omega^2\sim {G_{N}m}/{r_{c}}$~\cite{Kepler} (here $m$ is the mass of the centralised object, as above). It therefore follows that $L_{c}\sim\sqrt{{G_Nm}/{r_{c}}}\; r_{c}$. That is, $L_{c}\sim\sqrt{mr_{c}}$, as above. Eq.~\ref{expangular} is enough to tell you that circular orbits for massive particles do exist all the way down to $r_c=2m$, the location of the unstable photon orbit; this does not yet guarantee stability. 
Noting that:
\begin{equation}
{\partial L_c(r_c,m)\over\partial r_c} =
{e^{m/r_c} (r_c^2 -6mr_c+4m^2) \sqrt{m} \over 2 r_c (r_c-2m)^{3/2}} \ ,
\end{equation}
we observe that the curve $L_c(r_c,m)$ has a minimum at $r_c=\left(3+\sqrt{5}\right)m$ where $L_\mathrm{min} \approx 3.523216438\, m$ (see Fig.~\ref{F:angular}).

%---------------------------------------------------------------------------------------------
\begin{figure}[!htb]
\begin{center}
\includegraphics[scale=0.5]{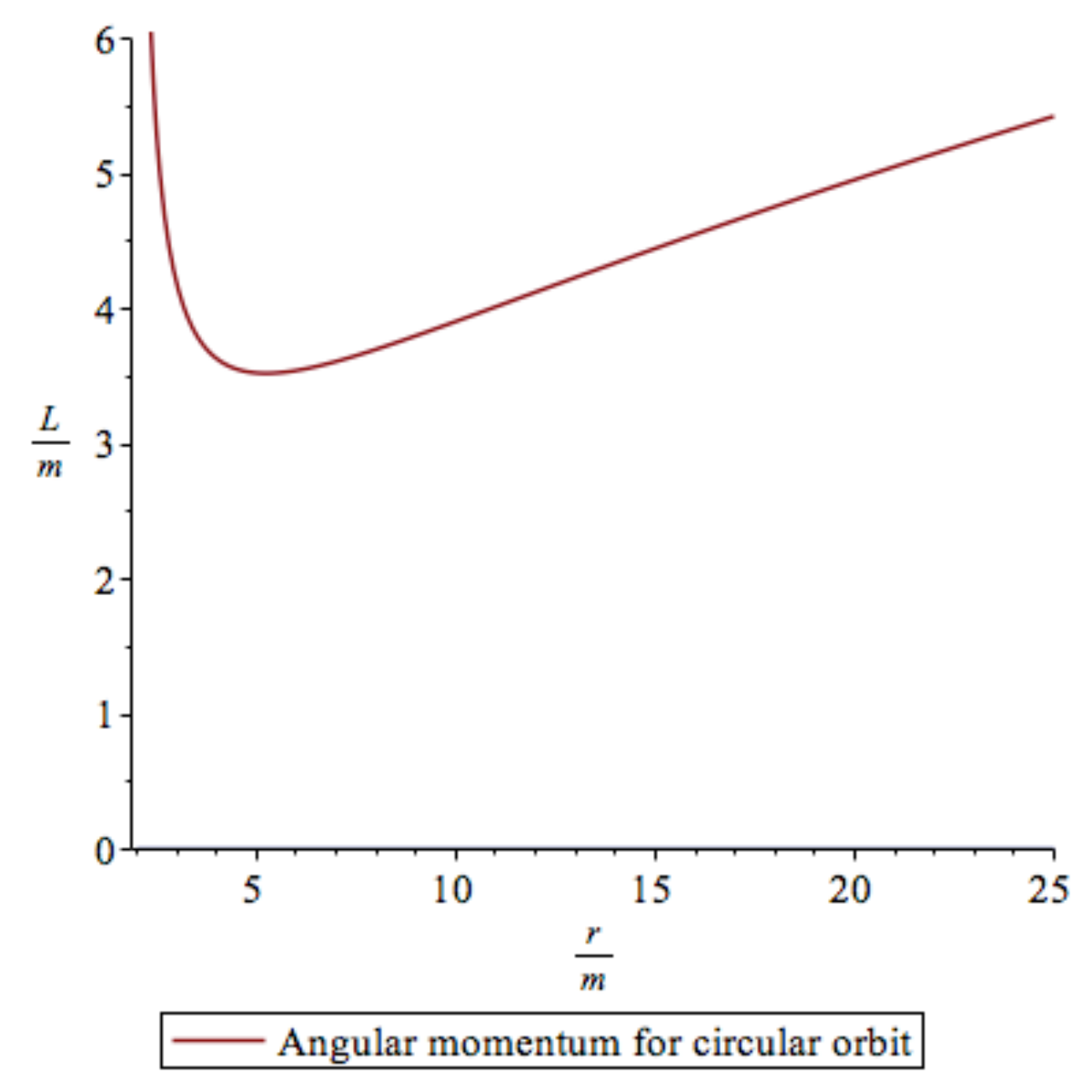}
\end{center}
{\caption[Required angular momentum to establish circular orbits in the exponential spacetime]{The graph shows the angular momentum $L/m$ required to establish a circular orbit at radius $r/m$. Note the minimum at $r= (3+\sqrt{5})m$ where $L_\mathrm{min} \approx 3.523216438\, m$. Circular orbits for $r \geq (3+\sqrt{5})m$ are stable; whereas circular orbits for 
$r < (3+\sqrt{5})m$ are unstable (and circular orbits for $r<2m$ do not exist).
}\label{F:angular}}
\end{figure}
%---------------------------------------------------------------------------------------------

To check stability substitute $L_c(r_c,m)$ into $V_{-1}''(r)$ to obtain:
\begin{equation}
V_{-1}''(r_c) =  {2m e^{-2m/r_c} (r_c^2 -6mr_c+4m^2) \over r_c^4 (r_c-2m)} \ .
\end{equation}
This changes sign when $r_c^2 -6mr_c+4m^2=0$, that is $r_c = \left(3\pm\sqrt{5}\right) m$. 
Only the positive root is relevant (the negative root lies below $r_c=2m$ where there are no circular orbits, stable or unstable).  Consequently we identify the location for the massive particle ISCO (for the exponential metric in isotropic coordinates) as:
\begin{equation}
r_{\hbox{\tiny ISCO}} = \left(3+\sqrt{5}\right)m \approx 5.236067977\, m \ .
\end{equation}
In curvature coordinates:
\begin{equation}
r_{s,\hbox{\tiny ISCO}} = \left(3+\sqrt{5}\right)\, \exp\left\{ {1\over4} \left(3-\sqrt{5}\right)\right\} \, m \approx 6.337940263\, m \ .
\end{equation}
This is not too far from what would have been expected in the standard Schwarzschild spacetime, where the Schwarzschild geometry ISCO is at  $r_{s,\hbox{\tiny ISCO}}=6m$.
\end{itemize}

%%%%%%

\subsection{Regge-Wheeler analysis}\label{S:Exp-RW}

Consider now the Regge--Wheeler equation for scalar and vector perturbations around the exponential metric spacetime. 
We will invoke the inverse Cowling approximation (wherein we keep the geometry fixed while letting the scalar and vector fields oscillate; we do this since we do not \emph{a priori} know the spacetime dynamics). The analysis closely parallels the general formalism developed in~\cite{ReggeWheeler1}.

Start from the exponential metric:
\begin{equation}
ds^2 = - e^{-2m/r} dt^2 + e^{+2m/r}\{dr^2 + r^2(d\theta^2+\sin^2\theta \, d\phi^2)\} \ .
\end{equation}
Define a tortoise coordinate by $dr_* = e^{2m/r} \; dr$, then:
\begin{equation}
ds^2 =  e^{-2m/r} (-dt^2 +dr_*^2) + e^{+2m/r}r^2(d\theta^2+\sin^2\theta \, d\phi^2) \ .
\end{equation}
Here $r$ is now implicitly a function of $r_*$.
We can also write this as:
\begin{equation}
ds^2 =  e^{-2m/r} (-dt^2 +dr_*^2) + r_s^2(d\theta^2+\sin^2\theta \, d\phi^2) \ .
\end{equation}
Using the formalism developed in~\cite{ReggeWheeler1}, the Regge--Wheeler equation can be written in the form:
\begin{equation}
 \partial_{r_{*}}^2\, \hat \phi 
+ \left\{\omega^2- \mathcal{V} \right\}
  \hat \phi = 0 \ .
\end{equation}
For a general spherically symmetric metric, specifying the metric components in curvature coordinates and with the coefficient of $d\Omega^{2}$ being $r_{s}^{2}$ (\emph{i.e.} this replaces the function $B(r_{*})$ seen in the Regge-Wheeler analysis of the Morris-Thorne spacetime in \S\ref{MorristhorneRW}), the Regge-Wheeler potential for spins $S\in\{0,1,2\}$ and angular momentum $\ell\geq S$ is given by~\cite{ReggeWheeler1}:\footnote{Note that the spin two axial mode has a far more tractable analysis in this context than for that of Morris-Thorne. This is directly related to the simplicity of the coefficient of $d\Omega^{2}$ when the metric is represented with respect to curvature coordinates.}
\begin{equation}
\mathcal{V}_S =  (-g_{tt}) \left[{\ell(\ell+1)\over r_s^2} 
+ {S(S-1) (g^{rr}-1)\over r_s^2}\right]+ (1-S) {\partial_{r_{*}}^2 r_s \over r_s} \ .
\end{equation}
For the exponential metric in curvature coordinates we have already seen that both $g_{tt} = -e^{-2m/r}$ and $g^{rr} = (1-m/r)^2$. Therefore:
\begin{equation}
\mathcal{V}_S =  e^{-2m/r} \left[{\ell(\ell+1)\over r_s^2} 
+ {S(S-1) [(1-m/r)^2-1]\over r_s^2}\right]+ (1-S) {\partial_{r_{*}}^2 r_s \over r_s} \ .
\end{equation}
It is important to realize that both $r_s$ and $r$ occur in the equation above. 
By noting that $\partial_{r_{*}} = e^{-2m/r} \partial_r$ it is possible to evaluate:
\begin{equation}
 {\partial_{r_{*}}^2 r_s \over r_s} = {e^{-4m/r} m (2r-m)\over r^4} 
 = - {e^{-4m/r} [(1-m/r)^2-1]\over r^2} \ ,
\end{equation}
and so rephrase the Regge--Wheeler potential as:
\begin{equation}
\mathcal{V}_S =  e^{-4m/r} \left[{\ell(\ell+1)\over r^2} 
+ {(S^2-1) [(1-m/r)^2-1]\over r^2}\right] \ .
\end{equation}
This is always zero at $r=0$ and $r=\infty$, with some extrema at non-trivial values of $r$. 

The corresponding result for the Schwarzschild spacetime is:
\begin{equation}
\mathcal{V}_{S,\mathrm{Sch}} =  \left(1-{2m\over r_s}\right) \left[{\ell(\ell+1)\over r_s^2} 
- {S(S-1) 2m \over r_s^3}\right]+ (1-S) {\partial_{r_{*}}^2 r_s \over r_s} \ .
\end{equation}
For the Schwarzschild metric  $\partial_{r_{*}} = (1-2m/r_s) \partial_{r_s}$, and so
 it is possible to evaluate:
\begin{equation}
 {\partial_{r_{*}}^2 r_s \over r_s} =  \left(1-{2m\over r_s}\right) {2m\over r_s^3} \ .
\end{equation}
Then:
\begin{equation}
\mathcal{V}_{S,\mathrm{Sch}} =  \left(1-{2m\over r_s}\right) 
\left[{\ell(\ell+1)\over r_s^2} 
- {(S^2-1) 2m \over r_s^3}\right] \ .
\end{equation}
Converting to isotropic coordinates, which for the Schwarzschild geometry means one is applying
$r_s = r \left(1+{m\over2r}\right)^2$, we have:
\begin{equation}
\mathcal{V}_{S,\mathrm{Sch}} =  \left(1-{m\over2r}\over1+{m\over2r}\right)^2
\left[{\ell(\ell+1)\over r^2 \left(1+{m\over2r}\right)^4 } 
- {(S^2-1) 2m \over r^3 \left(1+{m\over2r}\right)^6}\right] \ .
\end{equation}
This is always zero at the horizon $r=m/2$ and at $r=+\infty$, with some extrema at non-trivial values of $r$. 

%%%%

\subsubsection{Spin zero}

In particular for spin zero one has:
\begin{eqnarray}
\mathcal{V}_0 
&=&  e^{-2m/r} {\,\ell(\ell+1)\over r_s^2} +  {\partial_{r_{*}}^2 r_s \over r_s} 
\nonumber\\
&=&  e^{-4m/r} \,{\ell(\ell+1)\over r^2} 
+  {\partial_{r_{*}}^2 r_s \over r_s}
\nonumber\\
&=&
e^{-4m/r} \left[{\ell(\ell+1)
- [(1-m/r)^2-1]\over r^2}\right] \ .
\end{eqnarray}
This result can also be readily checked by brute force computation.

\noindent The corresponding result for the Schwarzschild spacetime is:
\begin{equation}
\mathcal{V}_{0,\mathrm{Sch}} =  \left(1-{m\over2r}\over1+{m\over2r}\right)^2
\left[{\ell(\ell+1)\over r^2 \left(1+{m\over2r}\right)^4 } 
+ {2m \over r^3 \left(1+{m\over2r}\right)^6}\right] \ .
\end{equation}

%%%%%%%%%%%%%%%%%%%%%%%%%%%%%%%%%----------------------

\begin{figure}[!htb]
\begin{center}
\includegraphics[scale=0.5]{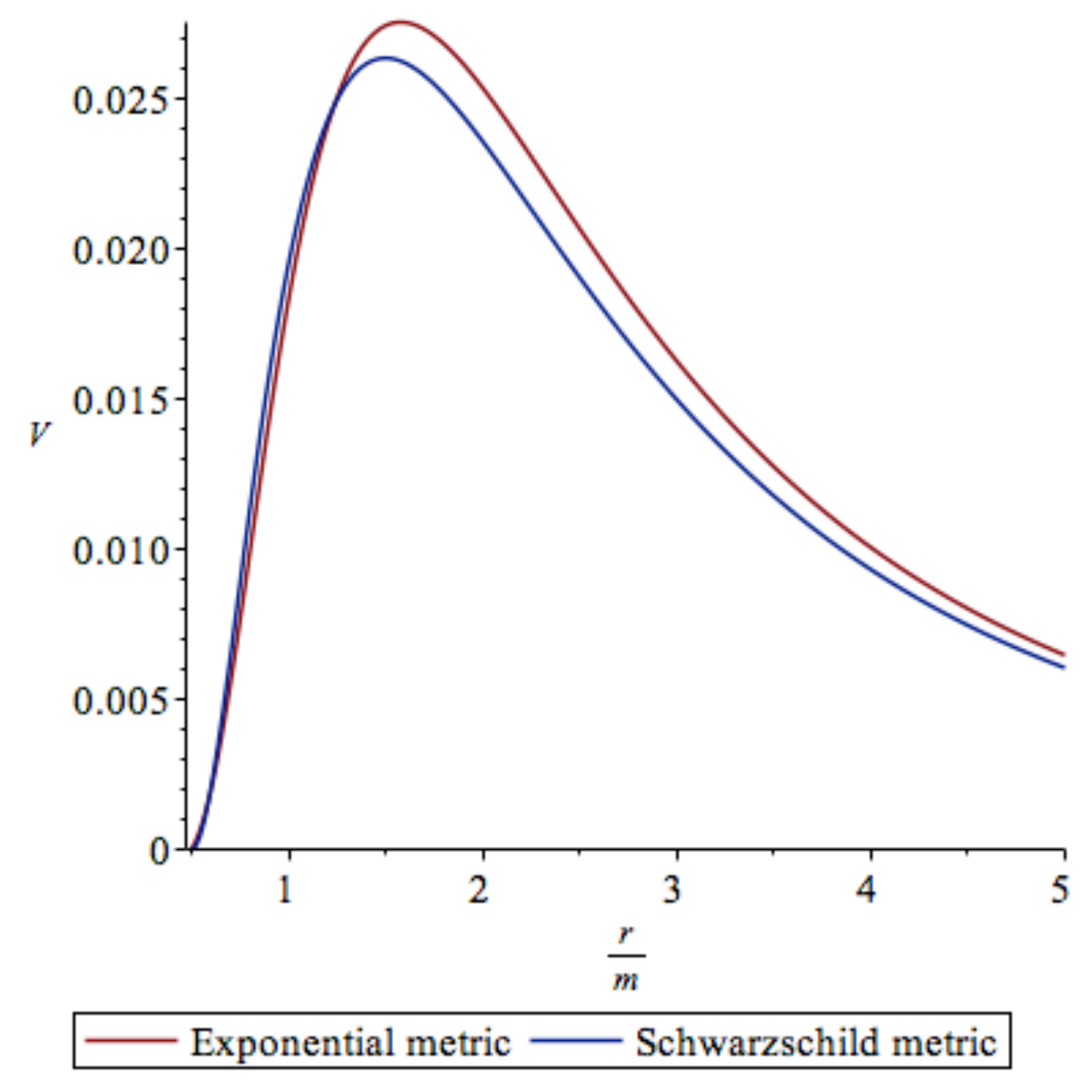}\qquad
\includegraphics[scale=0.5]{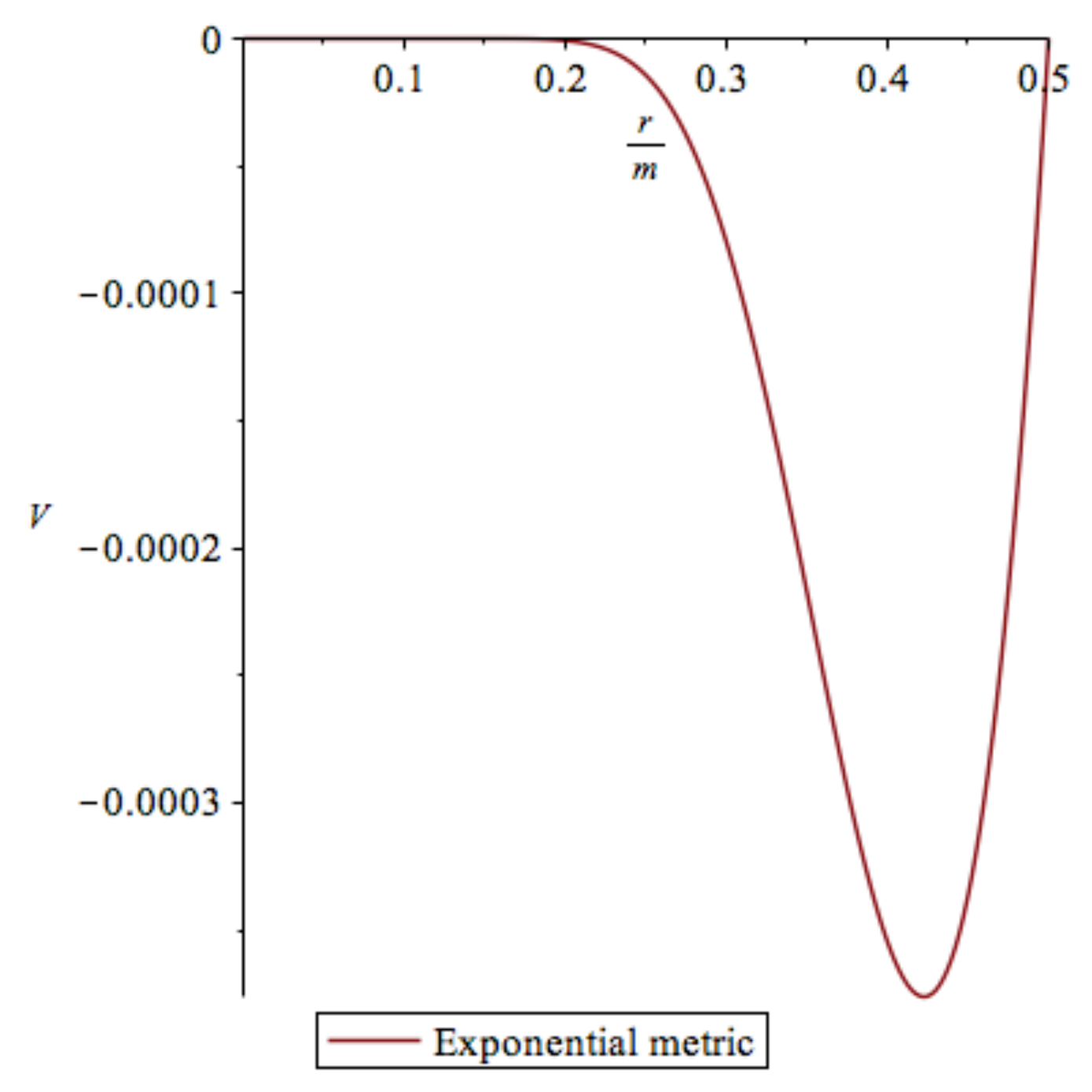}\qquad
\end{center}
{\caption[Spin zero Regge-Wheeler potentials for the exponential \emph{vs} Schwarzschild spacetimes]{The graph shows the spin zero Regge-Wheeler potential for $\ell=0$. 
While the Regge-Wheeler potentials are not dissimilar for $r>m/2$,
they are radically different once one goes to small $r<m/2$ (where the Regge-Wheeler potential for Schwarzschild is only formal since one is behind a horizon and cannot interact with the domain of outer communication). 
}\label{F:V0}}
\end{figure}

%%%%%%%%%%%%%%%%%%%%--------------------------------

\noindent For scalars the $s$-wave ($\ell=0$) is particularly important:
\begin{equation}
\mathcal{V}_{0,\ell=0}  = e^{-4m/r} \left[ {1- (1-m/r)^2\over r^2}\right]
= e^{-4m/r} \left[{2m\over r^3 } \left(1-{m\over2r} \right)\right] \ ,
\end{equation}
versus:
\begin{equation}
\mathcal{V}_{0,\ell=0,\mathrm{Sch}} =\left(1-{m\over2r}\over1+{m\over2r}\right)^2
\left[ {2m \over r^3 \left(1+{m\over2r}\right)^6}\right] \ .
\end{equation}
Note that these potentials both have zeroes at $r=m/2$ and that for $r<m/2$ only the exponential Regge-Wheeler potential is of physical interest (thanks to the horizon at $r=m/2$ in the Schwarzschild metric). See Fig.~\ref{F:V0} for qualitative features of the potential. The potential peaks are at $r=\left(1+{1\over\sqrt{3}}\right)m$ and $r={3m\over2}$ respectively.  For the exponential metric there is also a trough (a local minimum) at $r=\left(1-{1\over\sqrt{3}}\right)m$. 

%%%%

\subsubsection{Spin one}

For the spin one vector field the $\left\{ {r_s^{-1}} \partial_{*}^2{ r_s} \right\} $ term drops out; this can ultimately be traced back to the conformal invariance of massless spin one particles in `3+1'-dimensions. We are left with the particularly simple result ($\ell\geq 1$):
\begin{equation}
\mathcal{V}_1 =  {e^{-4m/r} \ell(\ell+1)\over r^2} \ .
\end{equation}
See related brief comments regarding conformal invariance in reference~\cite{ReggeWheeler1}.
Note that this rises from zero ($r\to0$) to some maximum at $r=2m$, where $\mathcal{V}_1\to {\ell(\ell+1)\over(2me)^2}$, and then decays back to zero (as $r\to\infty$).

\begin{figure}[!htb]
\begin{center}
\includegraphics[scale=0.6]{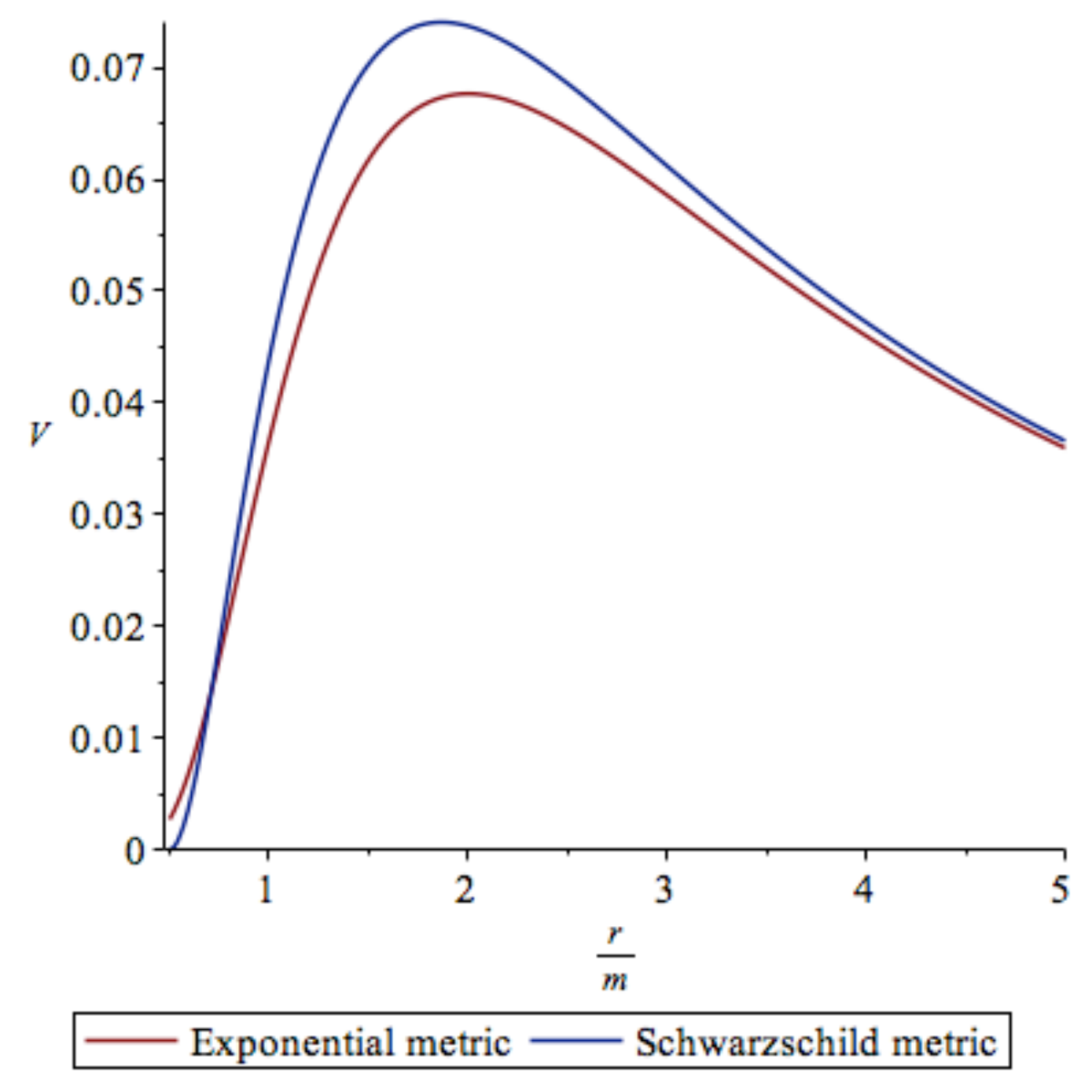}\qquad
\includegraphics[scale=0.6]{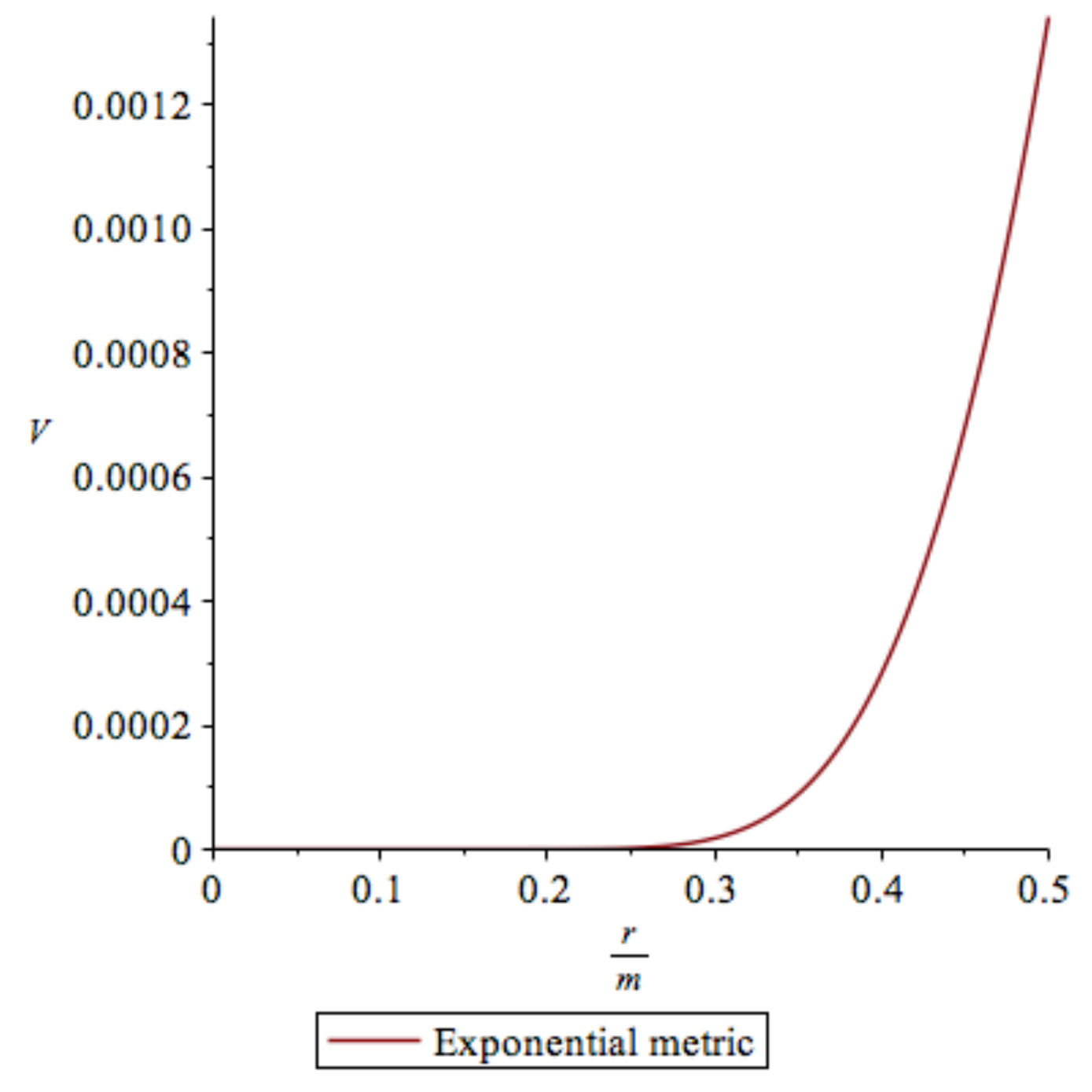}\qquad
\end{center}
{\caption[Spin one Regge-Wheeler potentials for the exponential \emph{vs} Schwarzschild spacetimes]{The graph shows the spin one Regge-Wheeler potential for $\ell=1$. 
While the Regge-Wheeler potentials are not dissimilar for $r>m/2$,
they are radically different once one goes to small $r<m/2$ (where the Regge-Wheeler potential for Schwarzschild is only formal since one is behind a horizon and cannot interact with the domain of outer communication). 
}\label{F:V1}}
\end{figure}

\noindent The corresponding result for the Schwarzschild spacetime is:
\begin{equation}
\mathcal{V}_{1,\mathrm{Sch}} =  {\left(1-{m\over2r}\right)^2\over\left(1+{m\over2r}\right)^6} \; 
{\ell(\ell+1)\over r^2} \ .
\end{equation}\\
Note that this rises from zero (at $r= m/2$) to some maximum at $r=\left(1+{\sqrt{3}\over2}\right) m$, where $\mathcal{V}_1\to {2\ell(\ell+1)\over27m^2}$, and then decays back to zero (as $r\to\infty$).
See Fig.~\ref{F:V1} for qualitative features of the potential.

%---------------------------------------------------------
%-----------------------------------------------------------

%%%%

\subsubsection{Spin two}

\begin{figure}[!htb]
\begin{center}
\includegraphics[scale=0.6]{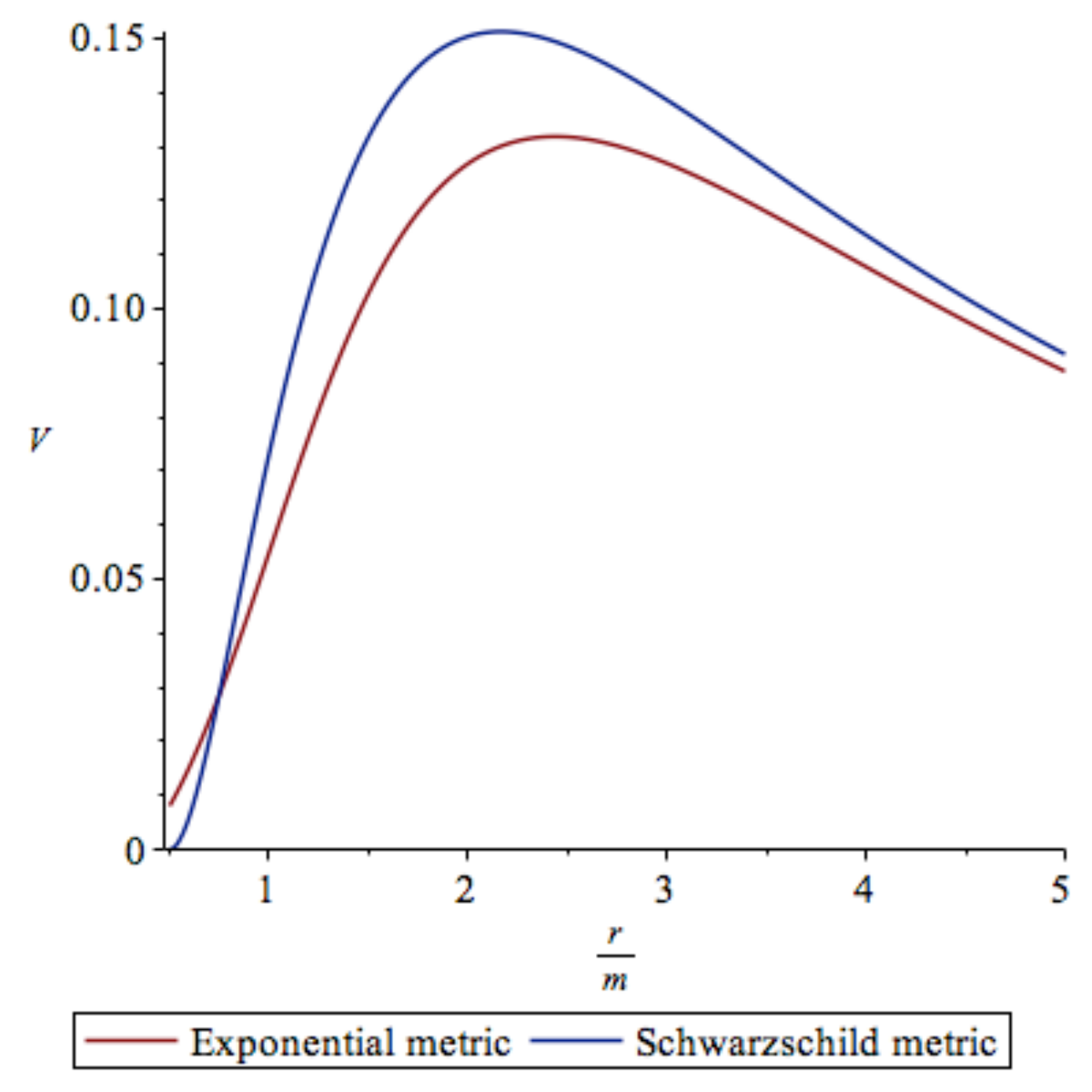}\qquad
\includegraphics[scale=0.6]{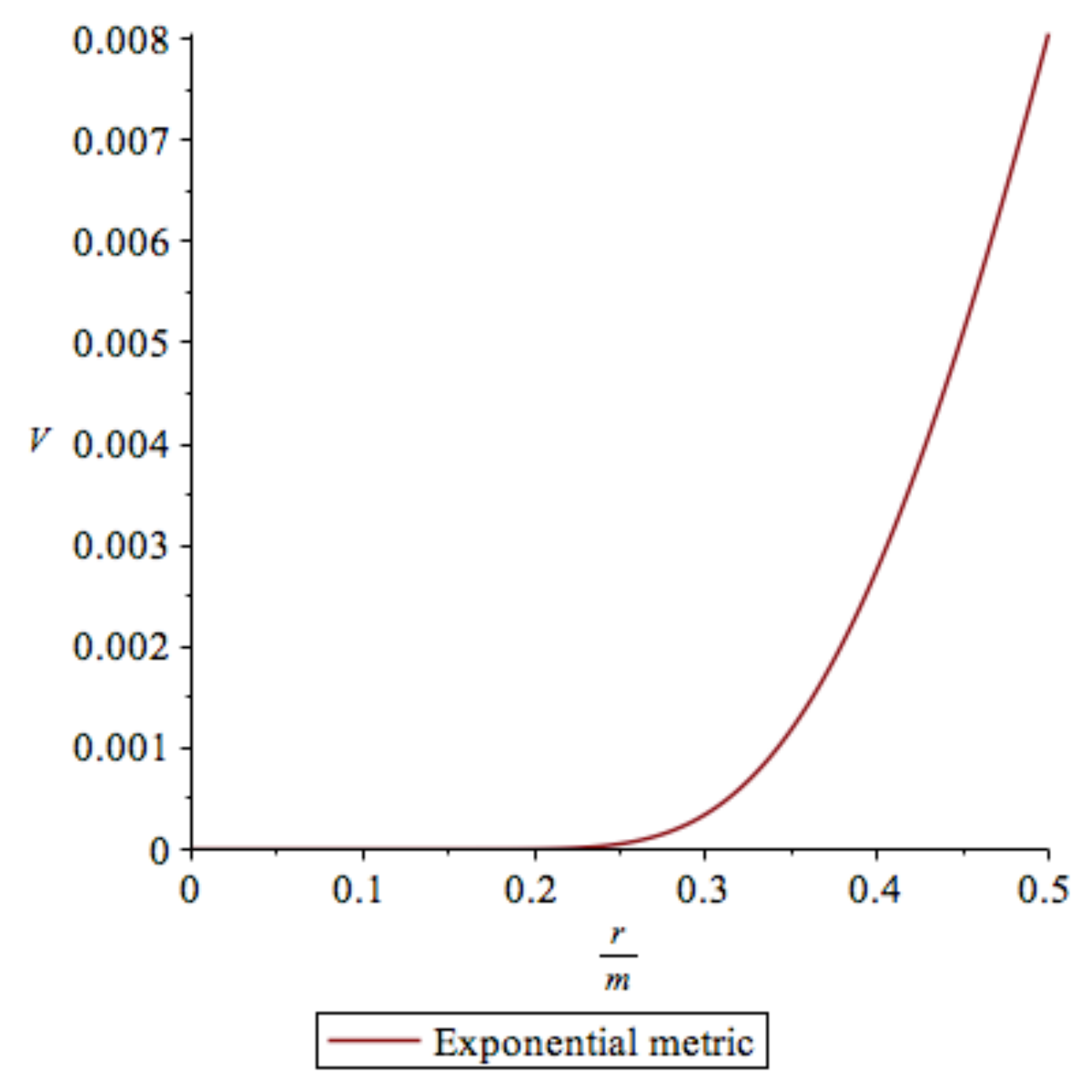}\qquad
\end{center}
{\caption[Spin two Regge-Wheeler potentials for the exponential \emph{vs} Schwarzschild spacetimes]{The graph shows the spin two (axial) Regge-Wheeler potential for $\ell=2$. 
The Regge-Wheeler potentials are somewhat dissimilar for $r>m/2$,
and are radically different once one goes to small $r<m/2$ (where the Regge-Wheeler potential for Schwarzschild is only formal since one is behind a horizon and cannot interact with the domain of outer communication). 
}\label{F:V2}}
\end{figure}

For spin two, more precisely for spin 2 axial perturbations (see reference~\cite{ReggeWheeler1}), one has ($\ell\geq 2$):
\begin{eqnarray}
\mathcal{V}_2 
&=&  e^{-2m/r} {\,\ell(\ell+1)\over r_s^2} -3  \,{\partial_{r_{*}}^2 r_s \over r_s} 
\nonumber\\
&=&  e^{-4m/r} \,{\ell(\ell+1)\over r^2} 
-3 \, {\partial_{r_{*}}^2 r_s \over r_s}
\nonumber\\
&=&
e^{-4m/r} \left[{\ell(\ell+1)
+3 [(1-m/r)^2-1]\over r^2}\right] \ .
\end{eqnarray}
The corresponding result for Schwarzschild spacetime is:
\begin{equation}
\mathcal{V}_{2,\mathrm{Sch}} =  \left(1-{m\over2r}\over1+{m\over2r}\right)^2
\left[{\ell(\ell+1)\over r^2 \left(1+{m\over2r}\right)^4 } 
- {6m \over r^3 \left(1+{m\over2r}\right)^6}\right] \ .
\end{equation}
See Fig.~\ref{F:V2} for qualitative features of the potential.

%-----------------------------------------------------------------

%-----------------------------------------------------------------

%%%%%%
\clearpage
\subsection[General relativistic interpretation]{General relativistic interpretation for the exponential metric}\label{S:Exp-GR}

\enlargethispage{10pt}
While many of the proponents of the exponential metric have for one reason or another been trying to build `alternatives' to standard general relativity, there is nevertheless a relatively simple interpretation of the exponential metric within the framework of standard general relativity and the standard Einstein equations, albeit with an `exotic' matter source. The key starting point is to note:
\begin{equation}
R_{\mu\nu} 
= -{2m^2 \over r^4} \; \text{diag}\{0,1,0,0\}_{\mu\nu} 
= - {1\over2} \nabla_{\mu}\left(2m\over r\right) \nabla_{\nu}\left(2m\over r\right) 
= - {1\over2} \nabla_{\mu} \Phi \, \nabla_{\nu} \Phi \ .
\end{equation}
Equivalently:
\begin{equation}
G_{\mu\nu} 
= - {1\over2} \left\{ \nabla_{\mu} \Phi \, \nabla_{\nu} \Phi - {1\over 2}  g_{\mu\nu} \, (g^{\alpha\beta} \nabla_{\alpha} \Phi \nabla_{\beta} \Phi) \right\} \ .
\end{equation}
This is just the usual Einstein equation for a \emph{negative kinetic energy massless scalar field}, a `ghost' or `phantom' field. The contracted Bianchi identity $G^{\mu\nu}{}_{;\nu}$ then automatically yields the scalar field equations of motion $(g^{\mu\nu} \nabla_{\mu}\nabla_{\nu}) \Phi=0$. That the scalar field has negative kinetic energy is intimately related to the fact that the exponential metric describes a traversable wormhole (and so must violate the null energy condition~\cite{MorrisThorne, LorentzianWormholes}). 

So, perhaps ironically, despite the fact that many of the proponents of the exponential metric for one reason or another reject general relativity, the exponential metric they advocate has a straightforward if somewhat exotic general relativistic interpretation.\footnote{It is also possible to interpret the exponential metric as a special sub-case of the Brans class IV solution of Brans-Dicke theory, which in turn is a special case of the general spherical, asymptotically flat, vacuum solution~\cite{Faraoni:2016,Faraoni:2018}; in this context it is indeed known that some solutions admit a wormhole throat, but that message seems not to have reached the wider community.}

%%%%%%

\subsection{Overview}

Phenomenologically, the exponential metric has a number of interesting features:

\begin{itemize}
\item It is a traversable wormhole, with time slowed down for stationary observers on the other side of the wormhole throat. 
\item Strong field lensing phenomena are markedly different from Schwarzs\-child.
\item ISCOs and unstable photon orbits still exist, and are moderately shifted from where they would be located in Schwarzschild spacetime.
\item  Regge-Wheeler potentials can still be extracted,  and are moderately different from what they would be in Schwarzschild spacetime.
\end{itemize}
Notably, the exponential metric has a natural interpretation in terms of general relativity coupled to a phantom scalar field.

%%%%%%%%%%%%%%%%%%%%%%%%%%%%%%%%%%%%%%%%%%%%%%%%%%%%%%%%

%%%%%%%%%%%%%%%%%%%%%%%%%%%%%%%%%%%%%%%%%%%%%%%%%%%%%%%

%\include{05-Generalised-TW}

%%%%%%%%%%%%%%%%%%%%%%%%%%%%%%%%%%%%%%%%%%%%%%%%%%%%%%%

\chapter{Introducing the regular black hole}\label{C:Bardeen-Hayward-Model2}

%%%%%%

So-called `regular black holes' are a topic currently of considerable interest in the general relativity and astrophysics communities. Ever since James Bardeen initially proposed the concept of a regular black hole over fifty years ago in 1968~\cite{Bardeen1968}, see also the more recent references~\cite{Bergmann-Roman,Hayward:2005, Bardeen:2014, Frolov:2014, Frolov:2014b, Frolov:2016, Frolov:2017, Frolov:2018, Cano:2018, Bardeen:2018, viability, beyond}, the notion has been intuitively attractive due to its non-singular nature. Rodrigues, Marcos and Silva give the following general definition \cite{Rodrigues2018Bardeen}: ``If some energy conditions on the stress-energy tensor are violated, [it] is possible [to] construct regular black holes in General Relativity and in alternative theories of gravity. This type of solution has horizons but does not present singularities.'' When referring to a regular black hole one therefore requires a spacetime that possesses a horizon but no singularity. Similarly to the analysis on traversable wormhole geometries, it pays to explore various metric candidates for regular black hole geometries within the framework of general relativity, and provide thorough analyses of their phenomenological properties~\cite{viability, beyond}.

%%%%%%

\section{Bardeen regular black hole}\label{BardeenRBH}

One metric which models a regular black hole is the aptly named `Bardeen metric', presented by Bardeen in the initial proposal for a regular black hole geometry, defined by the following line element~\cite{Bardeen1968}:

\begin{eqnarray}
    ds^2=-\left(1-\frac{2mr^2}{\left[r^2+\left(2ml^2\right)^{\frac{2}{3}}\right]^{\frac{3}{2}}}\right)dt^2 &+& \frac{dr^2}{\left(1-\frac{2mr^2}{\left[r^2+\left(2ml^2\right)^{\frac{2}{3}}\right]^{\frac{3}{2}}}\right)} \nonumber \\
    && \nonumber \\
    &+& r^2\left(d\theta^2+\sin^2\theta d\phi^2\right) \ .
\end{eqnarray}
Here $l$ is a length scale, typically associated with the Planck length~\cite{viability}.

We may note immediately that the metric is static and spherically symmetric. The areas of spherical symmetry of constant-$r$ coordinate are trivially modelled by the area function $A(r)=4\pi r^{2}$, clearly minimised when $r=0$. We may conclude that $r=0$ marks the coordinate location of the mass controlling the spacetime curvature.
 
 Note that the pole of $g_{tt}$ is complex-valued, found at coordinate location $r=\pm\left(2ml^{2}\right)^{\frac{1}{3}}i$ ; as such this does not affect the domain of the coordinate patch in view of the fact we are not modelling a complex spacetime (by construction our topological environment is homeomorphic to $\mathbb{R}^{4}$). However, there will be a coordinate singularity at the horizon location,\footnote{Demonstration that this is merely a coordinate singularity and not a curvature singularity is a consequence of the analysis in \S\ref{bardeentensors}. Note that this is also qualitatively similar to the Schwarzschild solution, in the sense that with respect to Schwarzschild curvature coordinates the Schwarzschild solution has a coordinate artefact at the horizon location, $r_{s}$.} where $\left[r^{2}+\left(2ml^{2}\right)^{\frac{2}{3}}\right]^{\frac{3}{2}}=2mr^{2}$ (these $r$-values are the poles of $g_{rr}$; two of the three roots will be eliminated as candidates for the horizon location in the subsequent analysis); without loss of generality let us call the location of the remaining root $r_{H}$. The metric therefore admits an `almost-global' coordinate patch, with the specific coordinate domains expressed further below.

Bardeen's objective when introducing this form of the metric was to present a non-singular black hole solution which was a minimal perversion of the Schwarzschild solution. The complex pole of $g_{tt}$ is directly related to this motivation; Bardeen's desire was to remove the gravitational singularity which is present at $r=0$ in the Schwarzschild solution. Note that if we make the convenient mathematical approximation $r^{2}\gg \left(2ml^{2}\right)^{\frac{2}{3}}$, we return the Schwarzschild solution in quite straightforward fashion via the following:

\begin{eqnarray}
    r^{2}\gg \left(2ml^{2}\right)^{\frac{2}{3}} \quad &\Longrightarrow& \quad r^{2}+\left(2ml^{2}\right)^{\frac{2}{3}}\approx r^{2} \nonumber \\
    && \nonumber \\
    &\Longrightarrow& \quad \frac{2mr^2}{\left[r^2+\left(2ml^{2}\right)^{\frac{2}{3}}\right]^{\frac{3}{2}}}\approx\frac{2m}{r} = \frac{r_{s}}{r} \ .
\end{eqnarray}
Making the approximation $r^2\gg\left(2ml^{2}\right)^{\frac{2}{3}}$ is uncontroversial for all regions of spacetime that are not \emph{extremely} close to the centralised massive object with respect to the chosen coordinate patch.
\clearpage

\noindent To see this, we have:

\begin{eqnarray}\label{scales}
    r^{2}\gg\left(2ml^{2}\right)^{\frac{2}{3}} \quad &\Longrightarrow& \quad r^{3}\gg 2ml^{2} \nonumber \\
    && \nonumber \\
    &\Longrightarrow& \quad r\gg \left(2ml^{2}\right)^{\frac{1}{3}} \ . \nonumber \\
\end{eqnarray}
\emph{A la} Bardeen, we associate $l$ as a length scale parameter to be the Planck length. This means that with respect to SI-units, $l\approx 1.6 \times 10^{-35}$ m~\cite{plancklength}. $l^{2}$, therefore, is of an order of magnitude of $10^{-70} \ \mbox{m}^{2}$. The solar mass, typically used as a unit of measurement when discussing the masses at the centres of black hole regions, is $M_{\odot}\approx 2\times 10^{30}$ kg~\cite{solarmass}. Converting these units to metres for consistency, \emph{i.e.} multiplying by $G_{N}/c^{2}$, yields the solar mass in metres to be $M_{\odot}\approx 1.5\times 10^{3}$ m. It therefore follows that even when dealing with the largest estimates of the masses of supermassive black holes in our universe (which are of the order of several billion solar masses~\cite{supermassive}, so we may approximate $m$ from Eq.~\ref{scales} to be of an order of magnitude of $10^{13}$ m), we observe that $\left(2ml^{2}\right)^{\frac{1}{3}}\varpropto 10^{-19}$ m. Therefore we resolve that $r\gg\left(2ml^{2}\right)^{\frac{1}{3}}$ is equivalent to stating that we are significantly further from the centralised mass than a distance proportional to $10^{-19}$ m. In order to conduct general analysis concerning the standard physical objects in this spacetime, which are typically \emph{much} further from the centralised mass than this, the separation of scales is sufficient to ensure the approximation is wholly uncontroversial. Accordingly we expect both mathematical and physical results from the analysis to look very similar to those from the analysis of the Schwarszchild solution, subject to very minor peturbations.

Hence we find the following for the horizon location in the Bardeen metric (employing the approximation $r^{2}\gg\left(2ml^{2}\right)^{\frac{2}{3}}$ as above):\footnote{Eq.~\ref{bardeenhorizon} employs use of the `Big $\mathcal{O}$' notation; this means that $\mathcal{O}(l)$ is some expression which is \emph{at most} a positive constant multiple of the Planck length, $l$~\cite{Big-O}. Contextually we may assume whichever constant of proportionality to $l$ is the genuine `tight' upper bound is sufficiently small that our $\mathcal{O}(l)$ is negligible (this assumption is again predicated on the minuteness of the Planck length informing the separation of scales argument).}

\begin{eqnarray}\label{bardeenhorizon}
    g_{tt}=0 \quad &\Longrightarrow& \quad \left[r^2+\left(2ml^2\right)^{\frac{2}{3}}\right]^{\frac{3}{2}}=2mr^2 \nonumber \\
    && \nonumber \\
    &\Longrightarrow& \quad r^{3}\approx 2mr^{2} \nonumber \\
    && \nonumber \\
    &\Longrightarrow& \quad r\in\left\lbrace +\mathcal{O}(l), 2m+ \mathcal{O}(l)\right\rbrace \ .
\end{eqnarray}
We may immediately discount the solutions near $r=0$ (note that $+\mathcal{O}(l)$ is a repeated root) in view of the mathematical approximation. Therefore we have a horizon at coordinate location $r_{H}=2m + \mathcal{O}(l)$, \emph{i.e.} $r_{H}\approx r_{s}$, and we may conclude the geometry contains a black hole region with a horizon location which is a small peturbation to that of the Schwarzschild solution (consistent with the aforementioned expectation). The `almost-global' coordinate patch therefore has domains: $t\in(-\infty, +\infty), \ r\in\mathbb{R}^{+}-\left\lbrace +\mathcal{O}(l), 2m+\mathcal{O}(l)\right\rbrace, \theta\in [0, \pi], \ \mbox{and} \ \phi\in[-\pi, \pi)$; note that while the metric mathematically permits negative $r$-values we restrict the domain of the $r$-coordinate to locations that lie within our universe, as any structure lying on the `other side' of the curvature controlling mass at $r=0$ is grossly unphysical. It remains to demonstrate that the geometry is gravitationally non-singular in order to show that this black hole region is `regular' in the sense of Bardeen.

%%%%

\subsection{Curvature tensors and invariants analysis}\label{bardeentensors}

The Ricci scalar:

\begin{eqnarray}
    R &=& \frac{6m\left(2ml^2\right)^{\frac{2}{3}}\left(4\left(2ml^2\right)^{\frac{2}{3}}-r^2\right)}{\left(r^2+\left(2ml^2\right)^{\frac{2}{3}}\right)^{\frac{7}{2}}} \ ; \nonumber \\
    && \nonumber \\
    && \mbox{as} \ \vert r\vert\rightarrow 0, \quad R\rightarrow \ \frac{12}{l^{2}} \ .
\end{eqnarray}
\noindent Ricci tensor non-zero components:

\begin{eqnarray}
    R^{t}{}_{t} &=& R^{r}{}_{r} = \frac{3m\left(2ml^2\right)^{\frac{2}{3}}\left(2\left(2ml^2\right)^{\frac{2}{3}}-3r^2\right)}{\left(r^2+\left(2ml^2\right)^{\frac{2}{3}}\right)^{\frac{7}{2}}} \ , \nonumber \\
    && \nonumber \\
    R^{\theta}{}_{\theta} &=& R^{\phi}{}_{\phi} = \frac{6m\left(2ml^2\right)^{\frac{2}{3}}}{\left(r^2+\left(2ml^2\right)^{\frac{2}{3}}\right)^{\frac{5}{2}}} \ ; \nonumber \\
    && \nonumber \\
    && \mbox{as} \ \vert r\vert\rightarrow 0, \quad R^{\mu}{}_{\nu}\rightarrow \ \frac{3}{l^{2}} \ .
\end{eqnarray}
Riemann tensor non-zero components:

\begin{eqnarray}
    R^{tr}{}_{tr} &=& \frac{m\left(2\left(r^2-\left(2ml^2\right)^{\frac{2}{3}}\right)^{2}-7r^2\left(2ml^2\right)^{\frac{2}{3}}\right)}{\left(r^2+\left(2ml^2\right)^{\frac{2}{3}}\right)^{\frac{7}{2}}} \ , \nonumber
\end{eqnarray}\vfill\newpage
\begin{eqnarray}
    R^{t\theta}{}_{t\theta} &=& R^{t\phi}{}_{t\phi} = R^{r\theta}{}_{r\theta} = R^{r\phi}{}_{r\phi} = \frac{m\left(2\left(2ml^2\right)^{\frac{2}{3}}-r^2\right)}{\left(r^2+\left(2ml^2\right)^{\frac{2}{3}}\right)^{\frac{5}{2}}} \ , \nonumber \\
    && \nonumber \\
    R^{\theta\phi}{}_{\theta\phi} &=& \frac{2m}{\left(r^2+\left(2ml^2\right)^{\frac{2}{3}}\right)^{\frac{3}{2}}} \ ; \nonumber \\
    && \nonumber \\
    && \mbox{as} \ \vert r\vert\rightarrow 0, \quad R^{\mu\nu}{}_{\alpha\beta}\rightarrow \ \frac{1}{l^{2}} \ .
\end{eqnarray}
Einstein tensor has non-zero components:

\begin{eqnarray}\label{bardeeneinstein}
    G^{t}{}_{t} &=& G^{r}{}_{r} = \frac{-6m\left(2ml^2\right)^{\frac{2}{3}}}{\left(r^2+\left(2ml^2\right)^{\frac{2}{3}}\right)^{\frac{11}{2}}}\Bigg\lbrace r^6+3r^4\left(2ml^2\right)^{\frac{2}{3}}+3r^2\left(2ml^2\right)^{\frac{4}{3}} \nonumber \\
    && \nonumber \\
    && \qquad \qquad \qquad \qquad \qquad \qquad \qquad \qquad \qquad +\left(2ml^2\right)^{2}\Bigg\rbrace \ , \nonumber \\
    && \nonumber \\
    G^{\theta}{}_{\theta} &=& G^{\phi}{}_{\phi} = \frac{3m\left(2ml^2\right)^{\frac{2}{3}}\left(3r^2-2\left(2ml^2\right)^{\frac{2}{3}}\right)}{\left(r^2+\left(2ml^2\right)^{\frac{2}{3}}\right)^{\frac{7}{2}}} \ ; \nonumber \\
    && \nonumber \\
    && \mbox{as} \ \vert r\vert\rightarrow 0, \quad G^{\mu}{}_{\nu}\rightarrow \ -\frac{3}{l^{2}} \ .
\end{eqnarray}
Weyl tensor non-zero components:

\begin{eqnarray}
    && -\frac{1}{2}C^{tr}{}_{tr} = C^{t\theta}{}_{t\theta} = C^{t\phi}{}_{t\phi} \nonumber \\
    && \nonumber \\
    && = \frac{-r^2m}{2\left(r^2+\left(2ml^2\right)^{\frac{2}{3}}\right)^{\frac{17}{2}}}\Bigg\lbrace r^2\left(r^2+\left(2ml^2\right)^{\frac{2}{3}}\right)\Bigg[r^6\left(2r^2+5\left(2ml^2\right)^{\frac{2}{3}}\right) \nonumber \\
    && \nonumber \\
    && -\left(2ml^2\right)^{2}\left(10r^2+13\left(2ml^2\right)^{\frac{2}{3}}\right)\Bigg]+3\left(2ml^2\right)^{\frac{8}{3}}\left(r^4-\left(2ml^2\right)^{\frac{4}{3}}\right)\Bigg\rbrace \ , \nonumber
\end{eqnarray}\vfill\newpage
\begin{eqnarray}
    && C^{r\theta}{}_{r\theta} = C^{r\phi}{}_{r\phi} = \frac{r^2m}{2\left(r^2+\left(2ml^2\right)^{\frac{2}{3}}\right)^{\frac{11}{2}}}\Bigg\lbrace 2r^6+r^4\left(2ml^2\right)^{\frac{2}{3}} \nonumber \\
    && \nonumber \\
    && \qquad \qquad \qquad \qquad \qquad \ \ -4r^2\left(2ml^2\right)^{\frac{4}{3}}-3\left(2ml^2\right)^{2}\Bigg\rbrace \ , \nonumber \\
    && \nonumber \\
    && C^{\theta\phi}{}_{\theta\phi} = \frac{r^2m\left(3\left(2ml^2\right)^{\frac{2}{3}}-2r^2\right)}{\left(r^2+\left(2ml^2\right)^{\frac{2}{3}}\right)^{\frac{7}{2}}} \ ; \nonumber \\
    && \nonumber \\
    && \mbox{as} \ \vert r\vert\rightarrow 0, \quad C^{\mu\nu}{}_{\alpha\beta}\rightarrow 0 \ .
\end{eqnarray}
The Ricci contraction $R_{\mu\nu}R^{\mu\nu}$:

\begin{eqnarray}
    R_{\mu\nu}R^{\mu\nu} &=& \frac{18m^2\left(2ml^2\right)^{\frac{4}{3}}\left\lbrace13r^{4}-4r^{2}\left(2ml^{2}\right)^{\frac{2}{3}}+8\left(2ml^{2}\right)^{\frac{4}{3}}\right\rbrace}{\left(r^2+\left(2ml^2\right)^{\frac{2}{3}}\right)^{7}} \ ; \nonumber \\
    && \nonumber \\
    && \mbox{as} \ \vert r\vert\rightarrow 0, \quad R_{\mu\nu}R^{\mu\nu}\rightarrow \ \frac{36}{l^{4}} \ .
\end{eqnarray}
The Kretschmann scalar:

\begin{eqnarray}
    R_{\mu\nu\alpha\beta}R^{\mu\nu\alpha\beta} &=& \frac{24m^2}{\left(r^2+\left(2ml^2\right)^{\frac{2}{3}}\right)^{7}}\Bigg\lbrace 2r^8-6r^6\left(2ml^2\right)^{\frac{2}{3}}+47r^4ml^2\left(2ml^2\right)^{\frac{1}{3}} \nonumber \\
    && \nonumber \\
    && \qquad \qquad \qquad \qquad -2r^2\left(2ml^2\right)^{2}+4\left(2ml^2\right)^{\frac{8}{3}}\Bigg\rbrace \ ; \nonumber \\
    && \nonumber \\
    && \mbox{as} \ \vert r\vert\rightarrow 0, \quad R_{\mu\nu\alpha\beta}R^{\mu\nu\alpha\beta}\rightarrow \ \frac{24}{l^{4}} \ .
\end{eqnarray}
The Weyl contraction $C_{\mu\nu\alpha\beta}C^{\mu\nu\alpha\beta}$:

\begin{eqnarray}
    C_{\mu\nu\alpha\beta}C^{\mu\nu\alpha\beta} &=& \frac{768r^4m^2}{\left(r^2+\left(2ml^2\right)^{\frac{2}{3}}\right)^{17}}\Bigg\lbrace 289r^4\left(2ml^2\right)^{\frac{2}{3}}\Bigg[\left(ml^2\right)^{6}+\frac{57}{289}r^6\left(ml^2\right)^{4} \nonumber \\
    && \nonumber \\
    && -\frac{195}{4624}r^{12}\left(ml^2\right)^{2}+\frac{7}{4624}r^{18}\Bigg]+78r^2\left(2ml^2\right)^{\frac{4}{3}}\Bigg[\left(ml^2\right)^{6} \nonumber
\end{eqnarray}\vfill\clearpage
\begin{eqnarray}
    && +\frac{105}{52}r^{6}\left(ml^2\right)^{4}-\frac{9}{26}r^{12}\left(ml^2\right)^{2}+\frac{23}{1664}r^{18}\Bigg]+36\left(ml^2\right)^{8} \nonumber \\
    && \nonumber \\
    && +580r^{6}\left(ml^2\right)^{6}-\frac{147}{2}r^{12}\left(ml^2\right)^{4}+\frac{45}{24}r^{18}\left(ml^2\right)^{2}+\frac{1}{16}r^{24}\Bigg\rbrace \ ; \nonumber \\
    && \nonumber \\
    && \mbox{as} \ \vert r\vert\rightarrow 0, \quad C_{\mu\nu\alpha\beta}C^{\mu\nu\alpha\beta}\rightarrow 0 \ .
\end{eqnarray}
All non-zero curvature tensor components and all scalar curvature invariants exhibit the correct behaviour as $\vert r\vert\rightarrow+\infty$, asymptotically tending towards zero (indicative of the fact that the spacetime is asymptotically Minkowski as we move further from the centralised mass at $r=0$). Furthermore, all components and invariants exhibit finite behaviour as they tend toward the region of highest curvature; we may conclude that they are everywhere-finite within the spacetime. As such the spacetime possesses no gravitational singularities, and the singularity present at the horizon is indeed a coordinate artefact, removable through an appropriate change of coordinate patch (this is also true for the repeated pole of $g_{rr}$ at $r=+\mathcal{O}(l)$ near the centralised mass). We have verified that this spacetime models a regular black hole in the sense of Bardeen.

%%%%%%

\subsection{ISCO and photon sphere analysis}

Let us examine the locations of the ISCO for massive particles and the photon sphere for massless particles as functions of $m$ and $l$.

Consider the tangent vector to the worldline of a massive or massless particle, parameterized by some arbitrary affine parameter, $\lambda$:

\begin{equation}
    g_{\mu\nu}\frac{dx^{\mu}}{d\lambda}\frac{dx^{\nu}}{d\lambda}=-g_{tt}\left(\frac{dt}{d\lambda}\right)^{2}+g_{rr}\left(\frac{dr}{d\lambda}\right)^{2}+r^{2}\left\lbrace\left(\frac{d\theta}{d\lambda}\right)^{2}+\sin^{2}\theta \left(\frac{d\phi}{d\lambda}\right)^{2}\right\rbrace \ .
\end{equation}

\noindent We may, without loss of generality, define a scalar-valued object as follows:

\begin{equation}
    \epsilon = \left\{
    \begin{array}{rl}
    -1 & \qquad\mbox{Massive particle, \emph{i.e.} timelike worldline} \ ; \\
     0 & \qquad\mbox{Massless particle, \emph{i.e.} null geodesic} \ .
    \end{array}\right. 
\end{equation}
That is, $g_{\mu\nu}\frac{dx^{\mu}}{d\lambda}\frac{dx^{\nu}}{d\lambda}=\epsilon$, and due to the metric being spherically symmetric we may fix $\theta=\frac{\pi}{2}$ arbitrarily and view the reduced equatorial problem:
\begin{equation}
    g_{\mu\nu}\frac{dx^{\mu}}{d\lambda}\frac{dx^{\nu}}{d\lambda}=-g_{tt}\left(\frac{dt}{d\lambda}\right)^{2}+g_{rr}\left(\frac{dr}{d\lambda}\right)^{2}+r^{2}\left(\frac{d\phi}{d\lambda}\right)^{2}=\epsilon \ .
\end{equation}

Observe that the vectors $\xi^{t}$ and $\xi^{\phi}$ are Killing vectors, as all metric components $g_{\mu\nu}$ are independent of $t$ and $\phi$ respectively. In accordance with the conserved quantities associated with each Killing vector, this yields the following expressions for the conservation of energy $E$, and angular momentum $L$:

\begin{equation}
    \left(1-\frac{2mr^2}{\left[r^2+\left(2ml^2\right)^{\frac{2}{3}}\right]^{\frac{3}{2}}}\right)\left(\frac{dt}{d\lambda}\right)=E \ ; \qquad\quad r^{2}\left(\frac{d\phi}{d\lambda}\right)=L \ .
\end{equation}
Hence:

\begin{equation}
    \left(1-\frac{2mr^2}{\left[r^2+\left(2ml^2\right)^{\frac{2}{3}}\right]^{\frac{3}{2}}}\right)^{-1}\left\lbrace -E^{2}+\left(\frac{dr}{d\lambda}\right)^{2}\right\rbrace+\frac{L^{2}}{r^{2}}=\epsilon \ ,
\end{equation}
\\
\begin{equation}
    \Longrightarrow\quad\left(\frac{dr}{d\lambda}\right)^{2}=E^{2}+\left(1-\frac{2mr^2}{\left[r^2+\left(2ml^2\right)^{\frac{2}{3}}\right]^{\frac{3}{2}}}\right)\left\lbrace\epsilon-\frac{L^{2}}{r^{2}}\right\rbrace \ .
\end{equation}
This gives the `effective potentials' for geodesic orbits as follows:

\begin{equation}
    V_{\epsilon}(r)=\left(1-\frac{2mr^2}{\left[r^2+\left(2ml^2\right)^{\frac{2}{3}}\right]^{\frac{3}{2}}}\right)\left\lbrace -\epsilon+\frac{L^{2}}{r^{2}}\right\rbrace \ .
\end{equation}

\begin{itemize}
    \item For a photon orbit we have the massless particle case $\epsilon=0$. Since we are in a spherically symmetric environment, solving for the locations of such orbits amounts to finding the coordinate location of the 'photon sphere'; \emph {i.e.} the value of the $r$-coordinate sufficiently close to our mass such that photons are forced to propogate in circular geodesic orbits. These circular orbits occur at $V_{0}^{'}(r)=0$, as such:
    \begin{equation}
        V_{0}(r)=\left(1-\frac{2mr^2}{\left[r^2+\left(2ml^2\right)^{\frac{2}{3}}\right]^{\frac{3}{2}}}\right)\left(\frac{L^{2}}{r^{2}}\right) \ ,
    \end{equation}
    leading to:
    \begin{equation}
        V_{0}^{'}(r)=\frac{2rL^2}{\left[r^2+\left(2ml^2\right)^{\frac{2}{3}}\right]^{\frac{5}{2}}}\left\lbrace 3m-\frac{\left[r^2+\left(2ml^2\right)^{\frac{2}{3}}\right]^{\frac{5}{2}}}{r^4}\right\rbrace \ .
    \end{equation}
    When $V_{0}^{'}(r)=0$, if we discount the solution $r=0$ (in view of the fact that $r=0$ lies within the horizon; certainly not a location in which one may observe photons), one obtains: $3mr^4=\left[r^2+\left(2ml^2\right)^{\frac{2}{3}}\right]^{\frac{5}{2}}$. Using the approximation $r^{2}\gg\left(2ml^{2}\right)^{\frac{2}{3}}$, as we employed previously, this yields a photon sphere location of $r=3m+\mathcal{O}(l)$; a slight peturbation on the expected $r=3m$ result for Schwarzschild.
    
    To verify stability, we check the sign of $V_{0}^{''}(r)$; it can be easily shown that:
    \begin{equation}
        V_{0}^{''}(r)=6L^{2}\left\lbrace\frac{1}{r^4}+\frac{m}{\left(r^2+\left(2ml^2\right)^{\frac{2}{3}}\right)^{\frac{7}{2}}}\left[\left(2ml^2\right)^{\frac{2}{3}}-4r^2\right]\right\rbrace \ .
    \end{equation}
    In view of the fact that the photon sphere is very near $r=3m$, let us examine behaviour of $V_{0}^{''}(r)$ at $r=3m$:
    
    \begin{equation}
        V_{0}^{''}\vert_{r=3m} = \frac{6L^{2}}{\left(3m\right)^{4}}+\frac{6L^{2}m}{\left(\left(3m\right)^{2}+\left(2ml^{2}\right)^{\frac{2}{3}}\right)^{\frac{7}{2}}}\left[\left(2ml^{2}\right)^{\frac{2}{3}}-4\left(3m\right)^{2}\right] \ ,
    \end{equation}
    and making the subsequent approximations $36m^{2}, 9m^{2}\gg\left(2ml^{2}\right)^{\frac{2}{3}}$:
    
    \begin{eqnarray}
        V_{0}^{''}\vert_{r=3m} &\approx& 6L^{2}\left\lbrace\frac{1}{\left(3m\right)^{4}}-\frac{4}{3\left(3m\right)^{4}}\right\rbrace \nonumber \\
        && \nonumber \\
        &\approx& -\frac{2L^{2}}{\left(3m\right)^{4}} < 0 \ .
    \end{eqnarray}
    We may conclude (in view of the approximations above) that the circular orbits for massless particles in the `local' area near $r=3m$ are unstable, hence $r=3m+\mathcal{O}(l)$ corresponds to a photon sphere with an unstable circular orbit. This is consistent with expectations.

    \item For massive particles the geodesic orbit corresponds to a timelike worldline and we have the case that $\epsilon=-1$. Therefore:
    \begin{equation}
        V_{-1}(r)=\left(1-\frac{2mr^2}{\left[r^{2}+\left(2ml^2\right)^{\frac{2}{3}}\right]^{\frac{3}{2}}}\right)\left(1+\frac{L^{2}}{r^{2}}\right) \ ,
    \end{equation}
    and it is easily verified that this leads to:
    \begin{eqnarray}
        V_{-1}^{'}(r) &=& \frac{2mr}{\left[r^2+\left(2ml^2\right)^{\frac{2}{3}}\right]^{\frac{5}{2}}}\Bigg\lbrace 3L^2+r^2-2\left(2ml^2\right)^{\frac{2}{3}} \nonumber \\
        && \nonumber \\
        && \qquad\qquad\qquad \ \ -\frac{L^2\left(r^2+\left(2ml^2\right)^{\frac{2}{3}}\right)^{\frac{5}{2}}}{mr^4}\Bigg\rbrace \ .
    \end{eqnarray}
    There is no straightforward analytic way of equating $V_{-1}'(r)$ to zero and solving for $r$; it is again preferable to assume a circular orbit at some $r_c$ and rearrange for the required angular momentum $L_c$ at that orbital radius. It then follows that the ISCO for a massive particle will lie at the $r$-coordinate for which that angular momentum is minimised. Therefore, when $V_{-1}'(r)=0$, discounting the solution $r=0$ (as this lies within the horizon), it follows that:
    
    \begin{equation}
        3L^2+r^2-2\left(2ml^2\right)^{\frac{2}{3}}-\frac{L^2\left(r^2+\left(2ml^2\right)^{\frac{2}{3}}\right)^{\frac{5}{2}}}{mr^4}=0 \ .
    \end{equation}
    Assuming fixed circular orbits at values $r_c$ and rearranging for $L_c$ it can be shown that:
    
    \begin{equation}
        L_c=\frac{\sqrt{m}r_c^2\sqrt{2\left(2ml^2\right)^{\frac{2}{3}}-r_{c}^2}}{\sqrt{3mr_c^4-\left[r_c^2+\left(2ml^2\right)^{\frac{2}{3}}\right]^{\frac{5}{2}}}} \ .
    \end{equation}
    Here we take the positive square root of $L_c^2$ to keep solutions physical, since we desire a positive angular momentum. Let us check that for large $r$ we recover $L_{c}\sim\sqrt{mr_{c}}$ in accordance with classical mechanics:
    
    \begin{eqnarray}
        \mbox{as} \ \vert r\vert\rightarrow+\infty, \quad L_{c} &\rightarrow& \frac{\sqrt{m}r_{c}^{2}\sqrt{-r_{c}^{2}}}{\sqrt{3mr_{c}^{4}-r_{c}^{5}}} \ , \nonumber \\
        && \nonumber \\
        &\rightarrow& \frac{\sqrt{m}r_{c}^{2}\sqrt{-r_{c}^{2}}}{r_{c}\sqrt{-r_{c}^{2}}\sqrt{r_{c}-3m}} \ , \nonumber \\
        && \nonumber \\
        &\sim& \frac{\sqrt{m}r_{c}}{\sqrt{r_{c}}} \ \sim \sqrt{mr_{c}} \ ,
    \end{eqnarray}
    as desired.\vfill\clearpage
    
    \noindent It can then be easily shown that:
    
    \begin{eqnarray}
        \frac{\partial L_c}{\partial r_{c}} &=& \frac{\sqrt{m}r_{c}\sqrt{2\left(2ml^2\right)^{\frac{2}{3}}-r_{c}^2}}{\sqrt{3mr_{c}^4-\left(r_{c}^2+\left(2ml^2\right)^{\frac{2}{3}}\right)^{\frac{5}{2}}}}\Bigg\lbrace\frac{4\left(2ml^2\right)^{\frac{2}{3}}-3r_{c}^2}{2\left(2ml^2\right)^{\frac{2}{3}}-r_{c}^2} \nonumber \\
        && \nonumber \\
        && \qquad \qquad \quad -\frac{12mr_{c}^4-5r_{c}^2\left(r_{c}^2+\left(2ml^2\right)^{\frac{2}{3}}\right)^{\frac{3}{2}}}{6mr_{c}^4-2\left(r_{c}^2+\left(2ml^2\right)^{\frac{2}{3}}\right)^{\frac{5}{2}}}\Bigg\rbrace \ .
    \end{eqnarray}
    Equating $\frac{\partial L_{c}}{\partial{r_{c}}}=0$ yields (and ignoring the solutions at $r_{c}=0$, and $r_{c}=\pm\sqrt{2}\left(2ml^{2}\right)^{\frac{1}{3}}$, as these lie either within the photon sphere location or outside of the domain for our $r$-coordinate; clearly not valid for an ISCO location):
    
    \begin{equation}\label{bardeenISCO}
        \frac{4\left(2ml^2\right)^{\frac{2}{3}}-3r_{c}^2}{2\left(2ml^2\right)^{\frac{2}{3}}-r_{c}^2} = \frac{12mr_{c}^4-5r_{c}^2\left(r_{c}^2+\left(2ml^2\right)^{\frac{2}{3}}\right)^{\frac{3}{2}}}{6mr_{c}^4-2\left(r_{c}^2+\left(2ml^2\right)^{\frac{2}{3}}\right)^{\frac{5}{2}}} \ .
    \end{equation}
    Now we make the following approximation in view of the separation of scales: $r_{c}^{2}\gg 2\left(2ml^{2}\right)^{\frac{2}{3}}$. This allows the following substitutions to approximate solutions for $r_{c}$ to Eq.~\ref{bardeenISCO}:
    
    \begin{eqnarray}
        &\bullet& \ 4\left(2ml^{2}\right)^{\frac{2}{3}}-3r_{c}^{2} \approx -3r_{c}^{2} \ ; \nonumber \\
        && \nonumber \\
        &\bullet& \ 2\left(2ml^{2}\right)^{\frac{2}{3}}-r_{c}^{2} \approx -r_{c}^{2} \ ; \nonumber \\
        && \nonumber \\
        &\bullet& \ r_{c}^{2} + \left(2ml^{2}\right)^{\frac{2}{3}} \approx r_{c}^{2} \ .
    \end{eqnarray}
    Eq.~\ref{bardeenISCO} can then be approximated by:
    
    \begin{eqnarray}
        3-\frac{12mr_{c}^{4}-5r_{c}^{5}}{6mr_{c}^{4}-2r_{c}^{5}} &\approx& 0 \ , \nonumber \\
        && \nonumber \\
        \Longrightarrow \quad 6mr_{c}^{4}-r_{c}^{5} &\approx& 0 \ , \nonumber \\
        && \nonumber \\
        \Longrightarrow \quad r_{c}^{4}\left[6m-r_{c}\right] &\approx& 0 \ .
    \end{eqnarray}
    Discounting the solution at $r_{c}=0$, we therefore have an ISCO location at $r_{c}\approx 6m$, or $r_{c}=6m+\mathcal{O}(l)$. This once again is a small peturbation to the expected ISCO location for the Scwarzschild solution, as expected.
    \end{itemize}
    
    Denoting $r_{H}$ as the location of the horizon, $r_{Ph}$ as the location of the photon sphere, and $r_{ISCO}$ as the location of the ISCO, we have the following summary:
    
    \begin{itemize}
        \item $r_{H}=2m+\mathcal{O}(l) \ ;$
        \item $r_{Ph}=3m+\mathcal{O}(l) \ ;$
        \item $r_{ISCO}=6m+\mathcal{O}(l) \ .$
    \end{itemize}
    All locations are very near those of the Schwarzschild solution and we see that the Bardeen metric is indeed a good choice for a geometry modelling a regular black hole with minimal perversion to Schwarzschild.
    
%%%%

\subsection{Regge-Wheeler analysis}

Consider now the Regge-Wheeler equation for scalar and vector perturbations around this spacetime. The analysis closely parallels the general formalism developed in~\cite{ReggeWheeler1}. We begin with the Bardeen metric:

\begin{eqnarray}
    ds^2=-\left(1-\frac{2mr^2}{\left[r^2+\left(2ml^2\right)^{\frac{2}{3}}\right]^{\frac{3}{2}}}\right)dt^2 &+& \frac{dr^2}{\left(1-\frac{2mr^2}{\left[r^2+\left(2ml^2\right)^{\frac{2}{3}}\right]^{\frac{3}{2}}}\right)} \nonumber \\
    && \nonumber \\
    &+& r^2\left(d\theta^2+\sin^2\theta d\phi^2\right) \ .
\end{eqnarray}
Define a tortoise coordinate by:

\begin{equation}
    dr_{*} = \left(1-\frac{2mr^2}{\left[r^2+\left(2ml^2\right)^{\frac{2}{3}}\right]^{\frac{3}{2}}}\right)^{-1}dr \ ,
\end{equation}
then the metric can be rewritten as:

\begin{equation}
    ds^2 = \left(1-\frac{2mr^2}{\left[r^2+\left(2ml^2\right)^{\frac{2}{3}}\right]^{\frac{3}{2}}}\right)\left\lbrace -dt^2+dr_{*}^{2}\right\rbrace +r^2\left(d\theta^2+\sin^2\theta d\phi^2\right) \ .
\end{equation}
Here $r$ is now implicitly a function of $r_*$. The Regge-Wheeler equation can be written as~\cite{ReggeWheeler1}:
\begin{equation}
 \partial_{r_{*}}^2\, \hat \phi 
+ \left\{\omega^2- \mathcal{V} \right\}
  \hat \phi = 0 \ .
\end{equation}\clearpage
\noindent For a general spherically symmetric metric with respect to curvature coordinates, the Regge-Wheeler potential for spins $S\in\lbrace 0, 1, 2\rbrace$ and angular momentum $\ell\geq S$ is~\cite{ReggeWheeler1}:

\begin{equation}
    \mathcal{V}_{S}=\left(-g_{tt}\right)\Bigg\lbrace\frac{\ell\left(\ell+1\right)}{r^2}+\frac{S\left(S-1\right)\left(g^{rr}-1\right)}{r^{2}}\Bigg\rbrace+\left(1-S\right)\frac{\partial_{r_{*}}^{2}r}{r} \ .
\end{equation}
For the Bardeen metric we therefore have the following Regge-Wheeler potential:

\begin{eqnarray}
    \mathcal{V}_{S} &=& \left(1-\frac{2mr^2}{\left[r^2+\left(2ml^2\right)^{\frac{2}{3}}\right]^{\frac{3}{2}}}\right)\Bigg\lbrace\frac{\ell\left(\ell+1\right)}{r^{2}}-\frac{2mS\left(S-1\right)}{\left(r^{2}+\left(2ml^{2}\right)^{\frac{2}{3}}\right)^{\frac{3}{2}}}\Bigg\rbrace \nonumber \\
    && \nonumber \\
    && \nonumber \\
    && \qquad \qquad \qquad \qquad \qquad \qquad \qquad \qquad +\left(1-S\right)\frac{\partial_{r_{*}}^{2}r}{r} \ .
\end{eqnarray}
It can be readily shown that:

\begin{equation}
    \frac{\partial_{r_{*}}^{2}r}{r} = \frac{2m\left\lbrace\left[r^{2}+\left(2ml^{2}\right)^{\frac{2}{3}}\right]^{\frac{3}{2}}-2mr^{2}\right\rbrace\left\lbrace r^{2}-2\left(2ml^{2}\right)^{\frac{2}{3}}\right\rbrace}{\left[r^{2}+\left(2ml^{2}\right)^{\frac{2}{3}}\right]^{4}} \ ,
\end{equation}
and so we may rephrase the Regge-Wheeler potential as:

\begin{eqnarray}
\mathcal{V}_S &=& \left(1-\frac{2mr^2}{\left[r^2+\left(2ml^2\right)^{\frac{2}{3}}\right]^{\frac{3}{2}}}\right)\Bigg\lbrace\frac{\ell\left(\ell+1\right)}{r^{2}}-\frac{2mS\left(S-1\right)}{\left(r^{2}+\left(2ml^{2}\right)^{\frac{2}{3}}\right)^{\frac{3}{2}}}\Bigg\rbrace \nonumber \\
&& \nonumber \\
&& \nonumber \\
&& +\left(1-S\right)\frac{2m\left\lbrace\left[r^{2}+\left(2ml^{2}\right)^{\frac{2}{3}}\right]^{\frac{3}{2}}-2mr^{2}\right\rbrace\left\lbrace r^{2}-2\left(2ml^{2}\right)^{\frac{2}{3}}\right\rbrace}{\left[r^{2}+\left(2ml^{2}\right)^{\frac{2}{3}}\right]^{4}} \ . \nonumber \\
&& \nonumber \\
&& 
\end{eqnarray}\vfill\newpage

%%%%

\subsubsection{Spin zero}

In particular for spin zero one has:
\begin{eqnarray}
\mathcal{V}_0
&=& \left(1-\frac{2mr^2}{\left[r^2+\left(2ml^2\right)^{\frac{2}{3}}\right]^{\frac{3}{2}}}\right)\left\lbrace\frac{\ell\left(\ell+1\right)}{r^2}\right\rbrace \nonumber \\
&& \nonumber \\
&& +\frac{2m\left\lbrace\left[r^{2}+\left(2ml^{2}\right)^{\frac{2}{3}}\right]^{\frac{3}{2}}-2mr^{2}\right\rbrace\left\lbrace r^{2}-2\left(2ml^{2}\right)^{\frac{2}{3}}\right\rbrace}{\left[r^{2}+\left(2ml^{2}\right)^{\frac{2}{3}}\right]^{4}} \ .
\end{eqnarray}
This result can also be readily checked by brute force computation. For scalars the $s$-wave ($\ell=0$) is particularly important:
\begin{equation}
\mathcal{V}_{0,\ell=0}  = \frac{\partial_{r_{*}}^{2}r}{r} = \frac{2m\left\lbrace\left[r^{2}+\left(2ml^{2}\right)^{\frac{2}{3}}\right]^{\frac{3}{2}}-2mr^{2}\right\rbrace\left\lbrace r^{2}-2\left(2ml^{2}\right)^{\frac{2}{3}}\right\rbrace}{\left[r^{2}+\left(2ml^{2}\right)^{\frac{2}{3}}\right]^{4}} \ .
\end{equation}

\subsubsection{Spin one}

For the spin one vector field the $\left\{ {r^{-1}} \partial_{r_{*}}^2{ r} \right\} $ term drops out; this can ultimately be traced back to the conformal invariance of massless spin one particles in `3+1'-dimensions. We are left with the particularly simple result ($\ell\geq 1$):
\begin{equation}
\mathcal{V}_1 = \left(1-\frac{2mr^2}{\left[r^2+\left(2ml^2\right)^{\frac{2}{3}}\right]^{\frac{3}{2}}}\right)\left\lbrace\frac{\ell\left(\ell+1\right)}{r^2}\right\rbrace \ .
\end{equation}

%%%%

\subsubsection{Spin two}

For the spin two axial mode (\emph{i.e.} $S=2$) we have the following ($\ell\geq 2$):

\begin{eqnarray}
\mathcal{V}_2 &=& \left(1-\frac{2mr^2}{\left[r^2+\left(2ml^2\right)^{\frac{2}{3}}\right]^{\frac{3}{2}}}\right)\Bigg\lbrace\frac{\ell\left(\ell+1\right)}{r^{2}}-\frac{4m}{\left(r^{2}+\left(2ml^{2}\right)^{\frac{2}{3}}\right)^{\frac{3}{2}}}\Bigg\rbrace \nonumber \\
&& \nonumber \\
&& \nonumber \\
&& -\frac{2m\left\lbrace\left[r^{2}+\left(2ml^{2}\right)^{\frac{2}{3}}\right]^{\frac{3}{2}}-2mr^{2}\right\rbrace\left\lbrace r^{2}-2\left(2ml^{2}\right)^{\frac{2}{3}}\right\rbrace}{\left[r^{2}+\left(2ml^{2}\right)^{\frac{2}{3}}\right]^{4}} \ . \nonumber \\
&& \nonumber \\
&& 
\end{eqnarray}

%%%%%%

\subsection{Stress-energy-momentum tensor}

Let us examine the Einstein field equations for this spacetime. In this instance, due to the complexity of the algebraic expressions for the non-zero Einstein tensor components as presented in Eq.~\ref{bardeeneinstein}, instead of using the metric tensor to lower the upper index on the mixed components, it is preferable to raise an index on either side of the Einstein field equations. As such, the form of the equations becomes: $G^{\mu}{}_{\nu}=8\pi T^{\mu}{}_{\nu}$. Due to the negative eigenvalue corresponding to the temporal metric coefficient as a consequence of the preferred type of Lorentzian signature (`-,+,+,+'), this process will yield the following general form of the stress-energy-momentum tensor:

\begin{equation}\label{stressbar}
    T^{\mu}{}_{\nu} = \begin{bmatrix}
    -\rho & 0 & 0 & 0 \\
    0 & p_{\parallel} & 0 & 0 \\
    0 & 0 & p_{\perp} & 0 \\
    0 & 0 & 0 & p_{\perp}
    \end{bmatrix} \ .
\end{equation}
\emph{i.e.} $\rho$ has switched sign due to the contraction process. We therefore have the following specific form of the stress-energy-momentum tensor for the Bardeen metric:

\begin{eqnarray}
   \rho &=& \frac{6m\left(2ml^2\right)^{\frac{2}{3}}}{8\pi\left(r^2+\left(2ml^2\right)^{\frac{2}{3}}\right)^{\frac{11}{2}}}\Bigg\lbrace r^6+3r^4\left(2ml^2\right)^{\frac{2}{3}}+3r^2\left(2ml^2\right)^{\frac{4}{3}} \nonumber \\
    && \nonumber \\
    && \qquad \qquad \qquad \qquad \qquad \qquad \qquad \qquad \qquad +\left(2ml^2\right)^{2}\Bigg\rbrace \ , \nonumber \\
    && \nonumber \\
    p_{\parallel} &=& -\frac{6m\left(2ml^2\right)^{\frac{2}{3}}}{8\pi\left(r^2+\left(2ml^2\right)^{\frac{2}{3}}\right)^{\frac{11}{2}}}\Bigg\lbrace r^6+3r^4\left(2ml^2\right)^{\frac{2}{3}}+3r^2\left(2ml^2\right)^{\frac{4}{3}} \nonumber \\
    && \nonumber \\
    && \qquad \qquad \qquad \qquad \qquad \qquad \qquad \qquad \qquad +\left(2ml^2\right)^{2}\Bigg\rbrace \ , \nonumber \\
    && \nonumber \\
    p_{\perp} &=& \frac{3m\left(2ml^2\right)^{\frac{2}{3}}\left(3r^2-2\left(2ml^2\right)^{\frac{2}{3}}\right)}{8\pi\left(r^2+\left(2ml^2\right)^{\frac{2}{3}}\right)^{\frac{7}{2}}} \ .
\end{eqnarray}
Let us now analyse the various energy conditions and see whether they are violated in this spacetime.

\subsubsection{Null energy condition}

In order to satisfy the null energy condition (NEC), both $\rho + p_{r}\geq 0$ \emph{and} $\rho+p_\perp\geq 0$  for all $r$ and $m$.\clearpage
\noindent First let us examine $\rho + p_{\parallel}$:

\begin{eqnarray}
    \rho+p_{\parallel} &=& \frac{6m\left(2ml^2\right)^{\frac{2}{3}}}{8\pi\left(r^2+\left(2ml^2\right)^{\frac{2}{3}}\right)^{\frac{11}{2}}}\Bigg\lbrace r^6+3r^4\left(2ml^2\right)^{\frac{2}{3}}+3r^2\left(2ml^2\right)^{\frac{4}{3}} \nonumber \\
    && \nonumber \\
    && +\left(2ml^2\right)^{2}\Bigg\rbrace-\frac{6m\left(2ml^2\right)^{\frac{2}{3}}}{8\pi\left(r^2+\left(2ml^2\right)^{\frac{2}{3}}\right)^{\frac{11}{2}}}\Bigg\lbrace r^6+3r^4\left(2ml^2\right)^{\frac{2}{3}} \nonumber \\
    && \nonumber \\
    && +3r^2\left(2ml^2\right)^{\frac{4}{3}}+\left(2ml^2\right)^{2}\Bigg\rbrace \ , \nonumber \\
    && \nonumber \\
    &=& 0 \ .
\end{eqnarray}
This is manifestly zero in our manifold; we have $\rho=-p_{\parallel}$. It follows that the first inequality required for the NEC to hold will be satisfied for this spacetime. Now let us examine $\rho+p_{\perp}$:

\begin{eqnarray}
    \rho+p_{\perp} &=& \frac{6m\left(2ml^2\right)^{\frac{2}{3}}}{8\pi\left(r^2+\left(2ml^2\right)^{\frac{2}{3}}\right)^{\frac{11}{2}}}\Bigg\lbrace r^6+3r^4\left(2ml^2\right)^{\frac{2}{3}}+3r^2\left(2ml^2\right)^{\frac{4}{3}} \nonumber \\
    && \nonumber \\
    && +\left(2ml^2\right)^{2}\Bigg\rbrace+\frac{3m\left(2ml^2\right)^{\frac{2}{3}}\left(3r^2-2\left(2ml^2\right)^{\frac{2}{3}}\right)}{8\pi\left(r^2+\left(2ml^2\right)^{\frac{2}{3}}\right)^{\frac{7}{2}}} \ , \nonumber \\
    && \nonumber \\
    &=& \frac{15mr^{2}\left(2ml^{2}\right)^{\frac{2}{3}}}{\left(r^{2}+\left(2ml^{2}\right)^{\frac{2}{3}}\right)^{\frac{17}{2}}}\Bigg\lbrace r^{10}+5r^{8}\left(2ml^{2}\right)^{\frac{2}{3}}+10r^{6}\left(2ml^{2}\right)^{\frac{4}{3}} \nonumber \\
    && \nonumber \\
    && +10r^{4}\left(2ml^{2}\right)^{2}+5r^{2}\left(2ml^{2}\right)^{\frac{8}{3}}+\left(2ml^{2}\right)^{\frac{10}{3}}\Bigg\rbrace \ .
\end{eqnarray}
The conformal factor is strictly positive, and all individual terms present within the braced brackets are globally positive. It follows that this sum is manifestly positive on our manifold and we may conclude that the NEC is satisfied for the Bardeen regular black hole spacetime.

\subsubsection{Strong energy condition}

In order to satisfy the strong energy condition (SEC) the spacetime must globally satisfy the following inequality: $\rho+p_{\parallel}+2p_{\perp}\geq 0$.\clearpage
\noindent Evaluating:

\begin{eqnarray}
    \rho+p_{\parallel}+2p_{\perp} &=& 2p_{\perp} \ , \nonumber \\
    && \nonumber \\
    &=& \frac{3m\left(2ml^2\right)^{\frac{2}{3}}\left(3r^2-2\left(2ml^2\right)^{\frac{2}{3}}\right)}{4\pi\left(r^2+\left(2ml^2\right)^{\frac{2}{3}}\right)^{\frac{7}{2}}} \ .
\end{eqnarray}
This switches sign when $r=\sqrt{\frac{2}{3}}\left(2ml^{2}\right)^{\frac{1}{3}}$, and is negative in the region $r<\sqrt{\frac{2}{3}}\left(2ml^{2}\right)^{\frac{1}{3}}$. The Bardeen metric therefore models a geometry which accurately depicts a regular black hole, but clearly violates the strong energy condition associated with the stress-energy-momentum tensor. This is consistent with the statement by Rodrigues \emph{et al}.
    
%%%%%%%%%%%%%%%%%%%%%%%%%%%%%%%%%%%%%%%%%%%%%%%%%%%%%%%%%%%%%%%%%%%%%%%%%%%%%%%%%%%%%%%%%%%%%%%%%%%%%%%%%%%%%%%%%%%%%%%%%%%%%%%%

\section{Hayward regular black hole}\label{HaywardRBH}

Another metric modeling a regular black hole is the Hayward metric~\cite{Hayward:2005}:

\begin{equation}\label{hayward}
    ds^2=-\left(1-\frac{2mr^2}{r^3+2ml^2}\right)dt^2+\frac{dr^2}{\left(1-\frac{2mr^2}{r^3+2ml^2}\right)}+r^2\left(d\theta^2+\sin^2\theta d\phi^2\right) \ .
\end{equation}
Similarly to the Bardeen metric, here $l$ is a length scale typically identified with the Planck length~\cite{viability}.

Once again, we have a time-independent and non-rotational metric (static and spherically symmetric). The areas of spherical symmetry of constant-$r$ coordinate are once again trivially modelled by the area function $A(r)=4\pi r^{2}$, clearly minimised when $r=0$. We may conclude that $r=0$ marks the coordinate location of the mass controlling the spacetime curvature.

The pole of $g_{tt}$ is where $r^{3}+2ml^{2}=0$, \emph{i.e.} at $r=-\left(2ml^{2}\right)^{\frac{1}{3}}$. This is slightly different from the Bardeen metric, where the pole of $g_{tt}$ was complexified in order to avoid it affecting the coordinate patch. However the pole corresponds to a strictly negative value for our $r$-coordinate; as such the coordinate location of the singularity does not lie within our universe (as our universe corresponds to the domain $r\in\mathbb{R}^{+}$). The Hayward metric therefore replaces the curvature singularity present at $r=0$ in the Schwarzschild solution with a pole at $r=-\left(2ml^{2}\right)^{\frac{1}{3}}$. This lies outside the domain we impose on our $r$-coordinate, and as such presents no issues.

The presence of this negative pole in the \emph{absence} of a gravitational singularity is indicative of the fact that, in a very similar fashion to Bardeen, the proposal of the Hayward metric as a regular black hole spacetime was designed to be a minimal perversion of the Schwarzschild solution. Accordingly we can expect physics between the Hayward and Schwarzschild spacetimes to be fairly similar, particularly for sufficiently large $r$. In fact, if we may make the specific mathematical approximation $r\gg\left(2ml^{2}\right)^{\frac{1}{3}}$, it is easily demonstrated that we return the Schwarzschild solution when making the appropriate substitutions into the metric environment of Eq.~\ref{hayward}. The validity of this approximation is again intrinsically linked to the separation of scales between coordinate locations of interest within the spacetime and multiples of the Planck length $l$; see \S\ref{BardeenRBH} for details on this argument.

Assuming that $r\gg\left(2ml^{2}\right)^{\frac{1}{3}}$, we may examine the location of the horizon in the Hayward spacetime:\footnote{Once again use of the `Big $\mathcal{O}$' notation is employed here. For details please see reference~\cite{Big-O}.}

\begin{eqnarray}
   g_{tt} = 0 \qquad &\Longrightarrow& \qquad r^3 \approx 2mr^2 \nonumber \\
    &\Longrightarrow& \qquad r\in\lbrace  +\mathcal{O}(l), 2m +\mathcal{O}(l)\rbrace \ .
\end{eqnarray}
We may immediately discount the solutions (remembering that $+\mathcal{O}(l)$ corresponds to a repeated root) near $r=0$ in view of the mathematical approximation made, hence there is a horizon location for the Hayward metric at coordinate location $r_{H}=2m+\mathcal{O}(l)$; this is once again very similar to the expected result for the Schwarzschild solution at $r_{s}=2m$, subject to small peturbations. The geometry governed by the Hayward metric therefore certainly possesses a black hole region of some kind. We may now also state clearly that the metric permits an `almost-global' coordinate patch with domains as follows: $t\in\left(-\infty,+\infty\right),\newline r\in \mathbb{R}^{+}-\left\lbrace +\mathcal{O}(l), 2m+\mathcal{O}(l)\right\rbrace, \theta\in[0,\pi], \ \mbox{and} \ \phi\in[-\pi, \pi)$. This is identical to the coordinate patch for the Bardeen metric. It remains to demonstrate that the geometry has no curvature singularities in order to show that the black hole region is `regular' in the sense of Bardeen.

%%%%%%

\subsection{Curvature tensors and invariants analysis}\label{haywardcurv}

The Ricci scalar:

\begin{eqnarray}
    R &=& \frac{24m^2l^2\left(4ml^2-r^3\right)}{\left(2ml^2+r^3\right)^{3}} \ ; \nonumber \\
    && \nonumber \\
    && \mbox{as} \ \vert r\vert\rightarrow 0, \quad R\rightarrow\frac{12}{l^2} \ .
\end{eqnarray}\clearpage
\noindent Ricci tensor non-zero components:

\begin{eqnarray}
    R^{t}{}_{t} &=& R^{r}{}_{r} = \frac{24m^2l^2\left(ml^2-r^3\right)}{\left(2ml^2+r^3\right)^{3}} \ , \nonumber \\
    && \nonumber \\
    R^{\theta}{}_{\theta} &=& R^{\phi}{}_{\phi} = \frac{12m^2l^2}{\left(2ml^2+r^3\right)^{2}} \ ; \nonumber \\
    && \nonumber \\
    && \mbox{as} \ \vert r\vert\rightarrow 0, \quad R^{\mu}{}_{\nu}\rightarrow\frac{3}{l^{2}} \ .
\end{eqnarray}
Riemann tensor non-zero components:

\begin{eqnarray}
    R^{tr}{}_{tr} &=& \frac{2m\left(\left(2ml^2\right)^{2}-7r^3\left(2ml^2\right)+r^6\right)}{\left(2ml^2+r^3\right)^{3}} \ , \nonumber \\
    && \nonumber \\
    R^{t\theta}{}_{t\theta} &=& R^{t\phi}{}_{t\phi} = R^{r\theta}{}_{r\theta} = R^{r\phi}{}_{r\phi} = \frac{m\left(4ml^2-r^3\right)}{\left(2ml^2+r^3\right)^{2}} \ , \nonumber \\
    && \nonumber \\
    R^{\theta\phi}{}_{\theta\phi} &=& \frac{2m}{2ml^2+r^3} \ ; \nonumber \\
    && \nonumber \\
    && \mbox{as} \ \vert r\vert\rightarrow 0, \quad R^{\mu\nu}{}_{\alpha\beta}\rightarrow\frac{1}{l^{2}} \ .
\end{eqnarray}
Einstein tensor non-zero components:

\begin{eqnarray}\label{einsteinhayward}
    G^{t}{}_{t} &=& G^{r}{}_{r} = \frac{-12m^2l^2}{\left(2ml^2+r^3\right)^{2}} \ , \nonumber \\
    && \nonumber \\
    G^{\theta}{}_{\theta} &=& G^{\phi}{}_{\phi} = \frac{-24m^2l^2\left(ml^2-r^3\right)}{\left(2ml^2+r^3\right)^{3}} \ ; \nonumber \\
    && \nonumber \\
    && \mbox{as} \ \vert r\vert\rightarrow 0, \quad G^{\mu}{}_{\nu}\rightarrow -\frac{3}{l^{2}} \ .
\end{eqnarray}
The Weyl tensor non-zero components are straightforward:

\begin{eqnarray}
    -\frac{1}{2}C^{tr}{}_{tr} = -\frac{1}{2}C^{\theta\phi}{}_{\theta\phi} &=& C^{t\theta}{}_{t\theta} = C^{t\phi}{}_{t\phi}
    = C^{r\theta}{}_{r\theta} \nonumber \\
    && \nonumber \\
    &=& C^{r\phi}{}_{r\phi} = \frac{mr^3\left(4ml^2-r^3\right)}{\left(2ml^2+r^3\right)^{3}} \ ; \nonumber \\
    && \nonumber \\
    && \mbox{as} \ \vert r\vert\rightarrow 0, \quad C^{\mu\nu}{}_{\alpha\beta}\rightarrow 0 \ .
\end{eqnarray}\clearpage
\noindent For the Kretschmann scalar and other related curvature invariants:

\begin{eqnarray}
    R_{\mu\nu\alpha\beta}R^{\mu\nu\alpha\beta} &=& \frac{48m^{2}}{\left(2ml^{2}+r^{3}\right)^{6}}\Bigg\lbrace r^{12}-4r^{9}\left(2ml^{2}\right)+18r^{6}\left(2ml^{2}\right)^{2} \nonumber \\
    && \nonumber \\
    && \qquad \qquad \qquad -2r^{3}\left(2ml^{2}\right)^{3}+2\left(2ml^{2}\right)^{4}\Bigg\rbrace \ ; \nonumber \\
    && \nonumber \\
    && \mbox{as} \ \vert r\vert\rightarrow 0, \quad R_{\mu\nu\alpha\beta}R^{\mu\nu\alpha\beta}\rightarrow\frac{24}{l^4} \ . \\
    && \nonumber \\
    R_{\mu\nu}R^{\mu\nu} &=& \frac{288m^4l^4\left(2\left(2ml^2\right)^{2}-2r^3\left(2ml^2\right)+5r^6\right)}{\left(2ml^2+r^3\right)^{6}} \ ; \nonumber \\
    && \nonumber \\
    && \mbox{as} \ \vert r\vert\rightarrow 0, \quad R_{\mu\nu}R^{\mu\nu}\rightarrow\frac{36}{l^4} \ . \\
    && \nonumber \\
    C_{\mu\nu\alpha\beta}C^{\mu\nu\alpha\beta} &=& \frac{48m^{2}r^{6}\left(4ml^{2}-r^{3}\right)^{2}}{\left(2ml^{2}+r^{3}\right)^{6}} \ ; \nonumber \\
    && \nonumber \\
    && \mbox{as} \ \vert r\vert\rightarrow 0, \quad C_{\mu\nu\alpha\beta}C^{\mu\nu\alpha\beta}\rightarrow 0 \ .
\end{eqnarray}
All non-zero curvature tensor components and all scalar curvature invariants exhibit the correct behaviour as $\vert r\vert\rightarrow+\infty$, asymptotically tending towards zero (indicative of the fact that the spacetime is asymptotically Minkowski as we move further from the centralised mass at $r=0$). Furthermore, all components and invariants exhibit finite behaviour as they tend toward the region of highest curvature; we may conclude that they are everywhere-finite within the spacetime. As such the spacetime possesses no gravitational singularities, and the singularities present at locations $r\in\left\lbrace +\mathcal{O}(l), 2m+\mathcal{O}(l)\right\rbrace$ are indeed coordinate artefacts, removable through an appropriate change of coordinate patch. We have verified that this spacetime models a regular black hole in the sense of Bardeen. Notably, all algebraic expressions for the non-zero curvature tensor components and the curvature invariants are of the general form:

\begin{equation}
    \frac{F(r)}{\left(2ml^{2}+r^{3}\right)^{n}} \ ,
\end{equation}
where $F(r)$ is some polynomial function of $r$ and $n\in\mathbb{Z}^{+}$.\clearpage
%%%%%%

\subsection{ISCO and photon sphere analysis}

Let us now calculate the locations of the ISCO for massive particles and the photon sphere for massless particles as functions of $m$ and $l$.

Consider the tangent vector to the worldline of a massive or massless particle, parameterized by some arbitrary affine parameter, $\lambda$:

\begin{equation}
    g_{\mu\nu}\frac{dx^{\mu}}{d\lambda}\frac{dx^{\nu}}{d\lambda}=-g_{tt}\left(\frac{dt}{d\lambda}\right)^{2}+g_{rr}\left(\frac{dr}{d\lambda}\right)^{2}+r^{2}\left\lbrace\left(\frac{d\theta}{d\lambda}\right)^{2}+\sin^{2}\theta \left(\frac{d\phi}{d\lambda}\right)^{2}\right\rbrace \ .
\end{equation}

\noindent Since we have used an affine parameter here, and we are certainly not dealing with a spacelike separation in either the massive or massless case, we may, without loss of generality, separate the two cases by defining the following scalar-valued object:

\begin{equation}
    \epsilon = \left\{
    \begin{array}{rl}
    -1 & \qquad\mbox{Massive particle, \emph{i.e.} timelike worldline} \ ; \\
     0 & \qquad\mbox{Massless particle, \emph{i.e.} null geodesic} \ .
    \end{array}\right. 
\end{equation}
That is, $g_{\mu\nu}\frac{dx^{\mu}}{d\lambda}\frac{dX^{\nu}}{d\lambda}=\epsilon$, and due to the metric being spherically symmetric we may fix $\theta=\frac{\pi}{2}$ arbitrarily and view the reduced equatorial problem:
\begin{equation}
    g_{\mu\nu}\frac{dx^{\mu}}{d\lambda}\frac{dx^{\nu}}{d\lambda}=-g_{tt}\left(\frac{dt}{d\lambda}\right)^{2}+g_{rr}\left(\frac{dr}{d\lambda}\right)^{2}+r^{2}\left(\frac{d\phi}{d\lambda}\right)^{2}=\epsilon \ .
\end{equation}
Once again, the metric is independent of time $t$ and azimuthal coordinate $\phi$. This means that $\xi^{t}$ and $\xi^{\phi}$ are Killing vectors. In accordance with the conserved quantities associated with each Killing vector, this yields the following expressions for the conservation of energy $E$, and angular momentum $L$:

\begin{equation}
    \left(1-\frac{2mr^2}{r^3+2ml^2}\right)\left(\frac{dt}{d\lambda}\right)=E \ ; \qquad\quad r^{2}\left(\frac{d\phi}{d\lambda}\right)=L \ .
\end{equation}
Hence:

\begin{equation}
    \left(1-\frac{2mr^2}{r^3+2ml^2}\right)^{-1}\left\lbrace -E^{2}+\left(\frac{dr}{d\lambda}\right)^{2}\right\rbrace+\frac{L^{2}}{r^{2}}=\epsilon \ ,
\end{equation}
\\
\begin{equation}
    \Longrightarrow\quad\left(\frac{dr}{d\lambda}\right)^{2}=E^{2}+\left(1-\frac{2mr^2}{r^3+2ml^2}\right)\left\lbrace\epsilon-\frac{L^{2}}{r^{2}}\right\rbrace \ .
\end{equation}
This gives `effective potentials' for geodesic orbits as follows:

\begin{equation}
    V_{\epsilon}(r)=\left(1-\frac{2mr^2}{r^3+2ml^2}\right)\left\lbrace -\epsilon+\frac{L^{2}}{r^{2}}\right\rbrace \ .
\end{equation}

\begin{itemize}
    \item For a photon orbit we have the massless particle case $\epsilon=0$. Since we are in a spherically symmetric environment, solving for the locations of such orbits amounts to finding the coordinate location of the 'photon sphere'; \emph {i.e.} the value of the $r$-coordinate sufficiently close to our mass such that photons are forced to propogate in circular geodesic orbits. These circular orbits occur at $V_{0}^{'}(r)=0$, as such:
    \begin{equation}
        V_{0}(r)=\left(1-\frac{2mr^2}{r^3+2ml^2}\right)\left(\frac{L^{2}}{r^{2}}\right) \ ,
    \end{equation}
    leading to:
    \begin{equation}
        V_{0}^{'}(r)=\frac{2r^2L^2}{\left(r^3+2ml^2\right)^2}\left\lbrace 3m-\frac{\left(r^3+2ml^2\right)^2}{r^5}\right\rbrace \ .
    \end{equation}
    When $V_{0}^{'}(r)=0$, if we discount the solution $r=0$ (as $r=0$ lies within the horizon; not a location in which one may observe photons), one obtains: $3mr^{5}-\left(r^{3}+2ml^{2}\right)^{2}=0$. Making the approximation $r\gg\left(2ml^{2}\right)^{\frac{1}{3}}$, as employed previously, we have: $3mr^{5}\approx r^{6}$, implying a photon sphere location of $r=3m+\mathcal{O}(l)\approx r_{s, Photon}$. As expected, this is very near the location of the photon sphere for the Schwarzschild spacetime.
    
    \noindent To verify stability, check the sign of $V_{0}^{''}(r)$:
    \begin{equation}
        V_{0}^{''}(r)=\frac{6L^2}{r^4}+\frac{24mrL^2}{\left(r^3+2ml^2\right)^3}\left\lbrace ml^{2}-r^{3}\right\rbrace \ .
    \end{equation}
    In view of the fact that the photon sphere location is very near $r=3m$, let us examine behaviour of $V^{''}_{0}(r)$ at $r=3m$:
    
    \begin{equation}
        V_{0}^{''}\vert_{r=3m} = \frac{6L^{2}}{\left(3m\right)^{4}}+\frac{72m^{2}L^{2}}{\left(\left(3m\right)^{3}+2ml^{2}\right)^{3}}\left\lbrace ml^{2}-\left(3m\right)^{3}\right\rbrace \ ,
    \end{equation}
    and making the subsequent approximation $27m^{3}\gg 2ml^{2}$:
    
    \begin{eqnarray}
        V_{0}^{''}\vert_{r=3m} &\approx& 6L^{2}\left\lbrace\frac{1}{\left(3m\right)^{4}}-\frac{4}{3\left(3m\right)^{4}}\right\rbrace \nonumber \\
        && \nonumber \\
        &\approx& -\frac{2L^{2}}{\left(3m\right)^{4}} < 0 \ .
    \end{eqnarray}
    We may conclude (in view of the approximations above) that the circular orbits for massless particles in the `local' area near $r=3m$ are unstable, hence $r=3m+\mathcal{O}(l)$ corresponds to a photon sphere with an unstable circular orbit -- consistent with expectations.

    \item For massive particles the geodesic orbit corresponds to a timelike worldline and we have the case that $\epsilon=-1$. Therefore:
    \begin{equation}
        V_{-1}(r)=\left(1-\frac{2mr^2}{r^3+2ml^2}\right)\left(1+\frac{L^{2}}{r^{2}}\right) \ ,
    \end{equation}
    and it is easily verified that this leads to:
    
    \begin{equation}
        V_{-1}'(r)=\frac{2mr}{\left(r^3+2ml^2\right)^2}\left\lbrace r^3+3rL^2-4ml^2-\frac{L^2\left(r^3+2ml^2\right)^2}{mr^4}\right\rbrace \ .
    \end{equation}
    
    There is no straightforwad analytic way of equating $V_{-1}'(r)$ to zero and solving for $r$; it is once again preferable to assume a circular orbit at some $r_c$ and rearrange for the required angular momentum $L_c$ at that orbital radius. It then follows that the ISCO for a massive particle will lie at the $r$-coordinate for which that angular momentum is minimised. Therefore, when $V_{-1}'(r)=0$, discounting the solution $r=0$ (as this lies within the photon sphere location; not a valid candidate for an ISCO location), it follows that:
    
    \begin{equation}
        r^3+3rL^2-4ml^2-\frac{L^2\left(r^3+2ml^2\right)^2}{mr^4}=0 \ .
    \end{equation}
    Assuming a fixed circular orbit at $r_c$ and rearranging for $L_c$ yields the following (taking the positive square root of $L_{c}^{2}$ to keep solutions physical):
    
    \begin{equation}
        L_c = \sqrt{\frac{mr_{c}^{4}\left(4ml^{2}-r_{c}^{3}\right)}{3mr_{c}^{5}-\left(r_{c}^{3}+2ml^{2}\right)^{2}}} \ .
    \end{equation}
    At large $r_{c}$ we observe:
    
    \begin{eqnarray}
        \mbox{as} \ \vert r\vert\rightarrow+\infty, \quad L_{c} &\rightarrow& \frac{\sqrt{m}r_{c}^{3}\sqrt{-r_{c}}}{\sqrt{3mr_{c}^{5}-r_{c}^{6}}} \ , \nonumber \\
        && \nonumber \\
        &\rightarrow& \frac{\sqrt{m}r_{c}^{3}\sqrt{-r_{c}}}{r_{c}^{2}\sqrt{-r_{c}}\sqrt{r_{c}-3m}} \ , \nonumber \\
        && \nonumber \\
        &\sim& \sqrt{mr_{c}} \ .
    \end{eqnarray}
    This is consistent with the desired result from classical mechanics.\clearpage
    \noindent Taking the partial derivative of $L_{c}$ with respect to our orbit location $r_{c}$, we obtain the following:
    
    \begin{eqnarray}
        \frac{\partial{L_{c}}}{\partial{r_{c}}} &=& \frac{\sqrt{m}r_{c}}{\sqrt{3mr_{c}^{5}-\left(r_{c}^{3}+2ml^{2}\right)^{2}}}\Bigg\lbrace \frac{16ml^{2}-7r_{c}^{3}}{2\sqrt{4ml^{2}-r_{c}^{3}}} \nonumber \\
        && \nonumber \\
        && \quad -\frac{r_{c}\sqrt{4ml^{2}-r_{c}^{3}}\left(15mr_{c}^{4}-6r_{c}^{5}-12r_{c}^{2}ml^{2}\right)}{2\left(3mr_{c}^{5}-\left(r_{c}^{3}+2ml^{2}\right)^{2}\right)}\Bigg\rbrace \ .
    \end{eqnarray}
    Equating this to zero, and discounting the nonphysical solution at $r_{c}=0$, we obtain:
    
    \begin{eqnarray}\label{hayISCO}
        && \left(16ml^{2}-7r_{c}^{3}\right)\left(3mr_{c}^{5}-\left(r_{c}^{3}+2ml^{2}\right)^{2}\right) \nonumber \\
        && \nonumber \\ && \qquad \qquad -r_{c}^{3}\left(4ml^{2}-r_{c}^{3}\right)\left(15mr_{c}^{2}-6r_{c}^{3}-12ml^{2}\right) = 0 \ .
    \end{eqnarray}
    We may now make the following approximations in view of the separation of scales:
    
    \begin{eqnarray}
        &\bullet& 16ml^{2}-7r_{c}^{3} \approx -7r_{c}^{3} \ , \nonumber \\
        && \nonumber \\
        &\bullet& r_{c}^{3}+2ml^{2} \approx r_{c}^{3} \ , \nonumber \\
        && \nonumber \\
        &\bullet& 4ml^{2}-r_{c}^{3} \approx -r_{c}^{3} \ , \nonumber \\
        && \nonumber \\
        &\bullet& 15r_{c}^{2}-12l^{2} \approx 15r_{c}^{2} \ .
    \end{eqnarray}
    Accordingly, Eq.~\ref{hayISCO} can be approximated by:
    
    \begin{eqnarray}
        r_{c}^{6}\left(15r_{c}^{2}-6r_{c}^{3}\right)-7r_{c}^{3}\left(3mr_{c}^{5}-r_{c}^{6}\right) &\approx& 0 \ , \nonumber \\
        && \nonumber \\
        \Longrightarrow r_{c}^{9}-6mr_{c}^{8} &\approx& 0 \ , \nonumber \\
        && \nonumber \\
        \Longrightarrow r_{c}^{8}\left[r_{c}-6m\right] &\approx& 0 \ .
    \end{eqnarray}
    Once again discounting the solution at $r_{c}=0$, we obtain the ISCO location for this spacetime at $r_{c}\approx 6m$, or $r_{c}=6m+\mathcal{O}(l)$. This is a small petrubation to the expected ISCO location for the Schwarzschild solution, as expected.
    \end{itemize}
    
    Denoting $r_{H}$ as the location of the horizon, $r_{Ph}$ as the location of the photon sphere, and $r_{ISCO}$ as the location of the ISCO, we have the following summary:
    
    \begin{itemize}
        \item $r_{H}=2m+\mathcal{O}(l) \ ;$
        \item $r_{Ph}=3m+\mathcal{O}(l) \ ;$
        \item $r_{ISCO}=6m+\mathcal{O}(l) \ .$
    \end{itemize}
    All locations are very near those of the Schwarzschild solution. As such, similarly to the Bardeen metric, we can conclude that the Hayward metric is indeed a good choice for a geometry modelling a regular black hole with minimal perversion to Schwarzschild.

%%%%%%

\subsection{Regge-Wheeler analysis}

Consider now the Regge-Wheeler equation for scalar and vector perturbations around this spacetime. The analysis closely parallels the general formalism developed in~\cite{ReggeWheeler1}. We begin with the Hayward metric:

\begin{equation}\label{hayward}
    ds^2=-\left(1-\frac{2mr^2}{r^3+2ml^2}\right)dt^2+\frac{dr^2}{\left(1-\frac{2mr^2}{r^3+2ml^2}\right)}+r^2\left(d\theta^2+\sin^2\theta d\phi^2\right) \ .
\end{equation}
Define a tortoise coordinate by:

\begin{equation}
    dr_{*} = \left(1-\frac{2mr^2}{r^{3}+2ml^{2}}\right)^{-1}dr \ ,
\end{equation}
then the metric can be rewritten as:

\begin{equation}
    ds^2 = \left(1-\frac{2mr^2}{r^{3}+2ml^{2}}\right)\left\lbrace -dt^2+dr_{*}^{2}\right\rbrace +r^2\left(d\theta^2+\sin^2\theta d\phi^2\right) \ .
\end{equation}
Here $r$ is now implicitly a function of $r_*$. The Regge-Wheeler equation can be written as~\cite{ReggeWheeler1}:
\begin{equation}
 \partial_{r_{*}}^2\, \hat \phi 
+ \left\{\omega^2- \mathcal{V} \right\}
  \hat \phi = 0 \ .
\end{equation}
For a general spherically symmetric metric with respect to curvature coordinates, the Regge-Wheeler potential for spins $S\in\lbrace 0, 1, 2\rbrace$ and angular momentum $\ell\geq S$ is~\cite{ReggeWheeler1}:

\begin{equation}
    \mathcal{V}_{S}=\left(-g_{tt}\right)\Bigg\lbrace\frac{\ell\left(\ell+1\right)}{r^2}+\frac{S\left(S-1\right)\left(g^{rr}-1\right)}{r^{2}}\Bigg\rbrace+\left(1-S\right)\frac{\partial_{r_{*}}^{2}r}{r} \ .
\end{equation}
For the Hayward metric we therefore have the following Regge-Wheeler potential:

\begin{eqnarray}
    \mathcal{V}_{S} &=& \left(1-\frac{2mr^2}{r^{3}+2ml^{2}}\right)\Bigg\lbrace\frac{\ell\left(\ell+1\right)}{r^{2}}-\frac{2mS\left(S-1\right)}{2ml^{2}+r^{3}}\Bigg\rbrace \nonumber \\
    && \nonumber \\
    && \qquad \qquad \qquad \qquad \qquad \qquad \qquad +\left(1-S\right)\frac{\partial_{r_{*}}^{2}r}{r} \ .
\end{eqnarray}
It can be readily shown that:

\begin{equation}
    \frac{\partial_{r_{*}}^{2}r}{r} = \frac{2m\left(r^{3}+2ml^{2}-2mr^{2}\right)\left(r^{3}-4ml^{2}\right)}{\left(r^{3}+2ml^{2}\right)^{3}} \ ,
\end{equation}
and so we may rephrase the Regge--Wheeler potential as:

\begin{eqnarray}
    \mathcal{V}_{S} &=& \left(1-\frac{2mr^2}{r^{3}+2ml^{2}}\right)\Bigg\lbrace\frac{\ell\left(\ell+1\right)}{r^{2}}-\frac{2mS\left(S-1\right)}{r^{3}+2ml^{2}} \nonumber \\
    && \nonumber \\
    && \qquad \qquad \qquad \qquad \qquad +\frac{2m\left(1-S\right)\left(r^{3}-4ml^{2}\right)}{\left(r^{3}+2ml^{2}\right)^{2}}\Bigg\rbrace \ .
\end{eqnarray}

%%%%

\subsubsection{Spin zero}

In particular for spin zero one has:
\begin{eqnarray}
\mathcal{V}_0 &=& \left(1-\frac{2mr^{2}}{r^{3}+2ml^{2}}\right)\left\lbrace\frac{\ell\left(\ell+1\right)}{r^{2}}+\frac{2m\left(r^{3}-4ml^{2}\right)}{\left(r^{3}+2ml^{2}\right)^{2}}\right\rbrace \ .
\end{eqnarray}
This result can also be readily checked by brute force computation. For scalars the $s$-wave ($\ell=0$) is particularly important:
\begin{equation}
\mathcal{V}_{0,\ell=0}  =  \frac{\partial_{r_{*}}^{2}r}{r} = \frac{2m\left(r^{3}+2ml^{2}-2mr^{2}\right)\left(r^{3}-4ml^{2}\right)}{\left(r^{3}+2ml^{2}\right)^{3}} \ .
\end{equation}

\subsubsection{Spin one}

For the spin one vector field the $\left\{ {r^{-1}} \partial_{r_{*}}^2{ r} \right\} $ term drops out; this can ultimately be traced back to the conformal invariance of massless spin one particles in `3+1'-dimensions. We are left with the particularly simple result ($\ell\geq 1$):
\begin{equation}
\mathcal{V}_1 = \left(1-\frac{2mr^{2}}{r^{3}+2ml^{2}}\right)\left\lbrace\frac{\ell\left(\ell+1\right)}{r^2}\right\rbrace \ .
\end{equation}

%%%%

\subsubsection{Spin two}

For the spin two axial mode (\emph{i.e.} $S=2$) we have the following ($\ell\geq 2$):

\begin{eqnarray}
    \mathcal{V}_{S} &=& \left(1-\frac{2mr^2}{r^{3}+2ml^{2}}\right)\Bigg\lbrace\frac{\ell\left(\ell+1\right)}{r^{2}}-\frac{4m}{r^{3}+2ml^{2}} \nonumber \\
    && \nonumber \\
    && \qquad \qquad \qquad \qquad \qquad -\frac{2m\left(r^{3}-4ml^{2}\right)}{\left(r^{3}+2ml^{2}\right)^{2}}\Bigg\rbrace \ .
\end{eqnarray}

%%%%%%

\subsection{Stress-energy-momentum tensor}

Let us examine the Einstein field equations for this spacetime, and subsequently analyse the various energy conditions. We shall use the form of the stress-energy-momentum tensor defined by the mixed non-zero Einstein tensor components from Eq.~\ref{einsteinhayward}. This gives the following form of the Einstein field equations: $G^{\mu}{}_{\nu}=8\pi T^{\mu}{}_{\nu}$. Accordingly, this yields the following general form of the stress-energy-momentum tensor:

\begin{equation}\label{stresshay}
    T^{\mu}{}_{\nu} = \begin{bmatrix}
    -\rho & 0 & 0 & 0 \\
    0 & p_{\parallel} & 0 & 0 \\
    0 & 0 & p_{\perp} & 0 \\
    0 & 0 & 0 & p_{\perp}
    \end{bmatrix} \ ,
\end{equation}
and we have the following specific forms for the principal pressures:

\begin{eqnarray}
    \rho &=& \frac{12m^2l^2}{8\pi\left(2ml^2+r^3\right)^{2}} \ , \nonumber \\
    && \nonumber \\
    p_{\parallel} &=& -\frac{12m^2l^2}{8\pi\left(2ml^2+r^3\right)^{2}} \ , \nonumber \\
    && \nonumber \\
    p_{\perp} &=& -\frac{24m^2l^2\left(ml^2-r^3\right)}{8\pi\left(2ml^2+r^3\right)^{3}} \ .
\end{eqnarray}
We may now analyse the various energy conditions and see whether they are violated in our spacetime.

\subsubsection{Null energy condition}

In order to satisfy the null energy condition, we require that both $\rho+p_{\parallel}\geq 0$ \emph{and} $\rho+p_{\perp}\geq 0$ globally in our spacetime. Let us first consider $\rho+p_{\parallel}$:

\begin{equation}\label{nullhayward}
    \rho+p_{\parallel} = \frac{12m^2l^2}{8\pi\left(2ml^2+r^3\right)^{2}}-\frac{12m^2l^2}{8\pi\left(2ml^2+r^3\right)^{2}} = 0 \ .
\end{equation}
This is manifestly zero for our spacetime, and the condition that $\rho+p_{\parallel}\geq 0$ is satisfied. Let us now consider $\rho+p_{\perp}$:

\begin{eqnarray}
    \rho+p_{\perp} &=& \frac{12m^2l^2}{8\pi\left(2ml^2+r^3\right)^{2}}-\frac{24m^2l^2\left(ml^2-r^3\right)}{8\pi\left(2ml^2+r^3\right)^{3}} \nonumber \\
    && \nonumber \\
    &=& \frac{12m^{2}l^{2}}{8\pi\left(2ml^{2}+r^{3}\right)^{2}}\left\lbrace 1-\frac{2ml^{2}-2r^{3}}{2ml^{2}+r^{3}}\right\rbrace \nonumber \\
    && \nonumber \\
    &=& \frac{12m^{2}l^{2}}{8\pi\left(2ml^{2}+r^{3}\right)^{2}}\left\lbrace \frac{3r^{3}}{2ml^{2}+r^{3}}\right\rbrace > 0 \ .
\end{eqnarray}
Given that $r\in\mathbb{R}^{+}$, this is manifestly non-negative, and we may conclude that the NEC is satisfied for the geometry induced by the Hayward metric.

\subsubsection{Strong energy condition}

In order to satisfy the strong energy condition (SEC), we require that $\rho+p_{\parallel}+2p_{\perp}\geq 0$ globally in our spacetime. Evaluating:

\begin{eqnarray}
    \rho+p_{\parallel}+2p_{\perp} &=& 2p_{\perp} \nonumber \\
    && \nonumber \\
    &=& -\frac{48m^2l^2\left(ml^2-r^3\right)}{8\pi\left(2ml^2+r^3\right)^{3}} \ , \nonumber \\
    && \nonumber \\
    \mbox{as} \ \vert r\vert\rightarrow 0, &\quad& \rho+p_{\parallel}+2p_{\perp} \rightarrow -\frac{3}{4\pi l^{2}} \ .
\end{eqnarray}
The expression switches sign at $r=\left(ml^{2}\right)^{\frac{1}{3}}$, and is strictly negative in the region $r<\left(ml^{2}\right)^{\frac{1}{3}}$. Accordingly the inequality will not be satisfied. We may conclude that the SEC is violated for this spacetime. As such, we have a geometry which satisfies the null energy condition, whilst violating the strong energy condition -- very similar to the spacetime induced by the Bardeen metric.

%%%%%%%%%%%%%%%%%%%%%%%%%%%%%%%%%%%%%%%%%

\section[Regular black hole with suppressed mass]{Regular black hole with exponentially\\ suppressed mass}\label{C:suppressed}

Presented here is a new and rather different metric, dubbed the `exponentially suppressed mass' metric. Conducting a standard general relativistic analysis of the resulting spacetime (in a similar fashion to the analyses performed for both the Bardeen and Hayward metrics), one can demonstrate that this metric does indeed correspond to a regular black hole in the sense of Bardeen, but with the mass of the centralised object becoming exponentially suppressed as one nears the coordinate location $r=0$. The line element is as follows:

\begin{equation}
    ds^2=-\left(1-\frac{2me^{{-a/r}}}{r}\right)dt^2+\frac{dr^2}{\left(1-\frac{2me^{{-a/r}}}{r}\right)}+r^2\left(d\theta^2+\sin^2{\theta}d\phi^2\right) \ .
\end{equation}
First we note that the metric is static and spherically symmetric. The areas of spherical symmetry of constant $r$-coordinate are trivial, modelled by the area function $A(r)=4\pi r^{2}$, which we can clearly see is minimised at $r=0$. We may conclude that the mass controlling the spacetime curvature has coordinate location $r=0$.

Note that this representation of the metric corresponds to the central mass being $r$-dependent in the following manner: $m(r)=me^{-a/r}$. We may immediately enforce that $a\in\mathbb{R}^{+}$ in order to ensure the mass is being exponentially `suppressed' as $\vert r\vert\rightarrow 0$; if $a=0$ we simply have the Schwarzschild solution, and if $a<0$ we have an altogether different scenario where asymptotic behaviour for small $r$ indicates massive exponential `growth'. The exponential expression has the following properties:
\begin{equation}
    \lim_{r\rightarrow 0^{+}}e^{{-a/r}}=0 \ , \qquad \lim_{r\rightarrow 0^{-}}e^{{-a/r}}=+\infty \ .
\end{equation}
The metric is therefore not analytic at coordinate location $r=0$. Looking at behaviour as $r\rightarrow 0^{-}$ can be omitted from the analysis; the severe discontinuity at $r=0$ implies that behaviour in the negative $r$ domain is grossly unphyiscal. This does not affect our coordinate patch regardless, as $r=0$ also marks a coordinate singularity (demonstration that this is a coordinate singularity and not a curvature singularity is a corollary of analysis in \S\ref{suppressedtensors}) as it is the pole of $g_{tt}$. We may trivially avoid these `issues' by enforcing $r\in\mathbb{R}^{+}$, \emph{i.e.} strictly remaining within our universe, which is the primary region of interest. In view of the diagonal metric environment, we may now examine horizon locations for the spacetime by setting $g_{tt}=0$:

\begin{eqnarray}
    g_{tt} = 0 \quad &\Longrightarrow& \quad r-2me^{-a/r} = 0 \ , \nonumber \\
    && \nonumber \\
    &\Longrightarrow& \quad -\frac{a}{2m}=-\frac{a}{r}e^{-a/r} \ , \nonumber \\
    && \nonumber \\
    &\Longrightarrow& \quad -\frac{a}{r} = W\left(-\frac{a}{2m}\right) \ , \nonumber \\
    && \nonumber \\
    &\Longrightarrow& \quad r = -\frac{a}{W\left(-\frac{a}{2m}\right)} \ , \nonumber \\
    && \nonumber \\
    &\Longrightarrow& \quad r = +2me^{W\left(-a/2m\right)} \ .
\end{eqnarray}
We have a coordinate location of the horizon defined explicitly in terms of the real-valued branches of the Lambert $W$ function (please see \S\ref{S:curvature} for detailed discussion on the Lambert $W$ function as one of the `special functions' in mathematics). Using the convention that positive $r$-coordinate values correspond to locations in our universe, and having enforced $a>0$, we may restrict the Lambert $W$ function on the denominator to only taking negative values. This presents two possibilities:
\begin{itemize}
    \item Taking the $W_{0}\left(x\right)$ branch of the real-valued Lambert $W$ function:
    \begin{eqnarray}
        W_{0}\left(-\frac{a}{2m}\right)<0 \quad &\Longrightarrow& \quad -\frac{a}{2m}\in \ \left(-\frac{1}{e}, \ 0\right) \ , \nonumber \\
        && \nonumber \\
        &\Longrightarrow& \quad a\in \ \left(0, \ \frac{2m}{e}\right) \ .
    \end{eqnarray}
    Provided $a$ lies in this interval we will therefore have a defined $r$-coordinate location for a horizon in our universe when taking the $W_{0}(x)$ branch of the Lambert $W$ function. Keeping in mind that fixing $a$ in this interval causes $W_{0}\left(-\frac{a}{2m}\right)\in\left(-1, \ 0\right)$, the possible coordinate locations of the horizon are given by $r_{H}\in \ (a, +\infty)$.
    
    \item Taking the $W_{-1}\left(x\right)$ branch of the real-valued Lambert $W$ function:
    
    The $W_{-1}(x)$ branch only returns outputs for $x\in \ \left(-\frac{1}{e}, 0\right)$, hence we have the same restriction on $a$ as before to ensure a defined coordinate location for the horizon; that $a\in \ \left(0, \frac{2m}{e}\right)$. The range of the $W_{-1}(x)$ branch is entirely negative so all possible solutions will correspond to locations in our universe. However the difference is that fixing $a$ in the interval of interest causes $W_{-1}\left(-\frac{a}{2m}\right)\in \ (-1, -\infty)$, hence the possible coordinate locations for the horizon are given by $r_{H}\in \ (0, a)$.
\end{itemize}
It follows then that in order for our geometry to possess a horizon in our universe, we require $a\in \left(0, \frac{2m}{e}\right)$. Then depending on whether we take the $W_{-1}(x)$ or $W_{0}(x)$ branch of the Lambert $W$ function, the horizon will be located either in the region $r_{H}\in \ (0, a)$ or $r_{H}\in (a, +\infty)$ respectively. In both cases the geometry is certainly modelling a black hole region of some description; it remains to demonstrate that the spacetime is gravitationally nonsingular in order to show this is a regular black hole in the sense of Bardeen. However, first let us take a look at what happens to the geometry when $a\geq\frac{2m}{e}$:

\begin{eqnarray}
    a\geq\frac{2m}{e} \quad &\Longrightarrow& \quad -\frac{a}{2m}\leq-\frac{1}{e} \ , \nonumber \\
    && \nonumber \\
    &\Longrightarrow& W\left(-\frac{a}{2m}\right) \ \mbox{is undefined} \ .
\end{eqnarray}
Therefore no such $r$-value exists in our geometry such that $g_{tt}=0$, \emph{i.e.} there are no horizons in the geometry. It follows that when $a\geq\frac{2m}{e}$, there is no black hole of any kind; the geometry is modelling something qualitatively independent. The geometry admits a generalised `almost-global' coordinate patch with the following domains:\footnote{In view of $r_{H}$ being dependent on the choice of branch of the Lambert $W$ function, we keep the terminology for the coordinate location of the horizon general, some $r_{H}\in\mathbb{R}^{+}$.} $t\in\left(-\infty, +\infty\right), r\in\mathbb{R}^{+}-\left\lbrace r_{H}\right\rbrace, \theta\in[0,\pi], \ \mbox{and} \ \phi\in[-\pi, \pi)$. Analysis of the radial null curves leads to a radial coordinate speed of light:

\begin{equation}
    c(r) = 1-\frac{2me^{{-a/r}}}{r} \ ,
\end{equation}
\emph{i.e.} a radial refractive index of:

\begin{equation}
    n(r) = \frac{r}{r-2me^{{-a/r}}} \ .
\end{equation}
Let us now examine the non-zero components of the curvature tensors as well as the curvature invariants to show that, for $a\in\left(0, \frac{2m}{e}\right)$, this metric is indeed modelling a regular black hole geometry.

%%%%%%

\subsection{Curvature tensors and invariants analysis}\label{suppressedtensors}

Before proceeding with the standard analysis of the mixed non-zero curvature tensor components and curvature invariants, it is prudent to introduce a relevant piece of mathematical detail. For any polynomial function $f(r)$, as $\vert r\vert\rightarrow 0$, $\frac{e^{-a/r}}{f(r)}\rightarrow e^{-a/r}\rightarrow 0$; \emph{i.e.} the exponential expression provides the dominant balance for the asymptotic behaviour for small $r$. Keeping this in mind, let us examine the mixed non-zero curvature tensor components and the curvature invariants.

\noindent The Ricci scalar:

\begin{eqnarray}
    R &=& \frac{2ma^{2}e^{-a/r}}{r^{5}} \ ; \nonumber \\
    && \nonumber \\
    && \mbox{as} \ \vert r\vert\rightarrow 0, \quad R\rightarrow 0 \ .
\end{eqnarray}
Ricci tensor non-zero components:

\begin{eqnarray}
    R^{t}{}_{t} &=& R^{r}{}_{r} = \frac{ma\left(a-2r\right)e^{-a/r}}{r^{5}} \ , \nonumber \\
    && \nonumber \\
    R^{\theta}{}_{\theta} &=& R^{\phi}{}_{\phi} = \frac{2mae^{-a/r}}{r^{4}} \ ; \nonumber \\
    && \nonumber \\
    && \mbox{as} \ \vert r\vert\rightarrow 0, \quad R^{\mu}{}_{\nu}\rightarrow 0 \ .
\end{eqnarray}
Riemann tensor non-zero components:

\begin{eqnarray}
    R^{tr}{}_{tr} &=& \frac{m\left(a^{2}-4ar+2r^{2}\right)e^{-a/r}}{r^{5}} \ , \nonumber \\
    && \nonumber \\
    R^{t\theta}{}_{t\theta} &=& R^{t\phi}{}_{t\phi} = \frac{m\left(a-r\right)e^{-a/r}}{r^{4}} \ , \nonumber \\
    && \nonumber \\
    R^{r\theta}{}_{r\theta} &=& R^{r\phi}{}_{r\phi} = \frac{m\left(a-r\right)e^{-a/r}}{r^{4}} \ , \nonumber \\
    && \nonumber \\
    R^{\theta\phi}{}_{\theta\phi} &=& \frac{2me^{-a/r}}{r^{3}} \ ; \nonumber \\
    && \nonumber \\
    && \mbox{as} \ \vert r\vert\rightarrow 0, \quad R^{\mu\nu}{}_{\alpha\beta}\rightarrow 0 \ .
\end{eqnarray}
Einstein tensor non-zero components:

\begin{eqnarray}\label{einsteinsuppressed}
    G^{t}{}_{t} &=& G^{r}{}_{r} = -\frac{2mae^{-a/r}}{r^{4}} \ , \nonumber \\
    && \nonumber \\
    G^{\theta}{}_{\theta} &=& G^{\phi}{}_{\phi} = -\frac{ma\left(a-2r\right)e^{-a/r}}{r^{5}} \ ; \nonumber \\
    && \nonumber \\
    && \mbox{as} \ \vert r\vert\rightarrow 0, \quad G^{\mu}{}_{\nu}\rightarrow 0 \ .
\end{eqnarray}
Weyl tensor non-zero components:

\begin{eqnarray}
    -\frac{1}{2}C^{tr}{}_{tr} = -\frac{1}{2}C^{\theta\phi}{}_{\theta\phi} = C^{t\theta}{}_{t\theta} &=& C^{t\phi}{}_{t\phi} = C^{r\theta}{}_{r\theta} \nonumber \\
    &=& C^{r\phi}{}_{r\phi} = -\frac{m\left(a^{2}-6ar+6r^{2}\right)e^{-a/r}}{6r^{5}} \ ; \nonumber \\
    && \nonumber \\
    && \mbox{as} \ \vert r\vert\rightarrow 0, \quad C^{\mu\nu}{}_{\alpha\beta}\rightarrow 0 \ .
\end{eqnarray}
The Ricci contraction $R_{\mu\nu}R^{\mu\nu}$:

\begin{eqnarray}
    R_{\mu\nu}R^{\mu\nu} &=& \frac{2m^{2}a^{2}\left(a^{2}-4ar+8r^{2}\right)e^{-2a/r}}{r^{10}} \ ; \nonumber \\
    && \nonumber \\
    && \mbox{as} \ \vert r\vert\rightarrow 0, \quad R_{\mu\nu}R^{\mu\nu}\rightarrow 0 \ .
\end{eqnarray}\clearpage
\noindent The Kretschmann scalar:

\begin{eqnarray}
    R_{\mu\nu\alpha\beta}R^{\mu\nu\alpha\beta} &=& \frac{4m^{2}\left(a^{4}-8a^{3}r+24a^{2}r^{2}-24ar^{3}+12r^{4}\right)e^{-2a/r}}{r^{10}} \ ; \nonumber \\
    && \nonumber \\
    && \mbox{as} \ \vert r\vert\rightarrow 0, \quad R_{\mu\nu\alpha\beta}R^{\mu\nu\alpha\beta}\rightarrow 0 \ .
\end{eqnarray}
The Weyl contraction $C_{\mu\nu\alpha\beta}C^{\mu\nu\alpha\beta}$:

\begin{eqnarray}
    C_{\mu\nu\alpha\beta}C^{\mu\nu\alpha\beta} &=& \frac{4m^{2}\left(a^{2}-6ar+6r^{2}\right)^{2}e^{-2a/r}}{3r^{10}} \ ; \nonumber \\
    && \nonumber \\
    && \mbox{as} \ \vert r\vert\rightarrow 0, \quad C_{\mu\nu\alpha\beta}C^{\mu\nu\alpha\beta}\rightarrow 0 \ .
\end{eqnarray}
As $\vert r\vert\rightarrow +\infty$, $e^{-a/r}\rightarrow 0$, and all non-zero tensor components and invariants become inversely proportional to some polynomial function of $r$. Therefore for large $r$, all non-zero components and invariants tend to zero, consistent with the fact that asymptotic infinity models Minkowski space. Of note is the fact that as $r\rightarrow 0^{+}$, all non-zero tensor components and invariants also asymptote to zero. This is directly related to the exponentially suppressed mass at small $r$, even though we are nearing the massive object controlling the curvature of the spacetime, the exponential expression suppressing the mass dominates the components, and the spacetime tends to the flat space limit. As such we have a geometry which approaches Minkowski both near the centralised mass and at asymptotic infinity, with some maximised area of curvature located in between. We may conclude that all non-zero tensor components and invariants are most certainly globally finite, and as such the geometry possesses no curvature singularities as predicted -- we are indeed dealing with a regular black hole spacetime.

%%%%%%

\subsection{ISCO and photon sphere analysis}

Let us now calculate the location of both the photon sphere for massless particles and the ISCO for massive particles as functions of $m$ and $a$.

Consider the tangent vector to the worldline of a massive or massless particle, paramterized by some arbitrary affine parameter, $\lambda$:

\begin{equation}
    g_{\mu\nu}\frac{dx^{\mu}}{d\lambda}\frac{dx^{\nu}}{d\lambda}=-g_{tt}\left(\frac{dt}{d\lambda}\right)^{2}+g_{rr}\left(\frac{dr}{d\lambda}\right)^{2}+r^2\left\lbrace\left(\frac{d\theta}{d\lambda}\right)^{2}+\sin^{2}\theta \left(\frac{d\phi}{d\lambda}\right)^{2}\right\rbrace \ .
\end{equation}\clearpage
\noindent We may define a scalar-valued object as follows:

\begin{equation}
    \epsilon = \left\{
    \begin{array}{rl}
    -1 & \qquad\mbox{Massive particle, \emph{i.e.} timelike worldline} \ ; \\
     0 & \qquad\mbox{Massless particle, \emph{i.e.} null geodesic} \ .
    \end{array}\right. 
\end{equation}
That is, $g_{\mu\nu}\frac{dx^{\mu}}{d\lambda}\frac{dx^{\nu}}{d\lambda}=\epsilon$, and due to the metric being spherically symmetric we may fix $\theta=\frac{\pi}{2}$ arbitrarily and view the reduced equatorial problem:
\begin{equation}
    g_{\mu\nu}\frac{dx^{\mu}}{d\lambda}\frac{dx^{\nu}}{d\lambda}=-g_{tt}\left(\frac{dt}{d\lambda}\right)^{2}+g_{rr}\left(\frac{dr}{d\lambda}\right)^{2}+r^2\left(\frac{d\phi}{d\lambda}\right)^{2}=\epsilon \ .
\end{equation}
The Killing symmetries yield the following expressions for the conservation of energy $E$, and angular momentum $L$:

\begin{equation}
    \left(1-\frac{2me^{{-a/r}}}{r}\right)\left(\frac{dt}{d\lambda}\right)=E \ ; \qquad\quad r^2\left(\frac{d\phi}{d\lambda}\right)=L \ .
\end{equation}
Hence:

\begin{equation}
    \left(1-\frac{2me^{{-a/r}}}{r}\right)^{-1}\left\lbrace -E^{2}+\left(\frac{dr}{d\lambda}\right)^{2}\right\rbrace+\frac{L^{2}}{r^{2}}=\epsilon \ ;
\end{equation}
\\
\begin{equation}
    \Longrightarrow\quad\left(\frac{dr}{d\lambda}\right)^{2}=E^{2}+\left(1-\frac{2me^{{-a/r}}}{r}\right)\left\lbrace\epsilon-\frac{L^{2}}{r^{2}}\right\rbrace \ .
\end{equation}
\\

\noindent This gives `effective potentials' for geodesic orbits as follows:

\begin{equation}
    V_{\epsilon}(r)=\left(1-\frac{2me^{{-a/r}}}{r}\right)\left\lbrace -\epsilon+\frac{L^{2}}{r^{2}}\right\rbrace \ .
\end{equation}

\begin{itemize}
    \item For a photon orbit we have the massless particle case $\epsilon=0$. This corresponds to the photon sphere location, and these circular orbits occur at $V_{0}^{'}(r)=0$, hence:
    \begin{equation}
        V_{0}(r)=\left(1-\frac{2me^{{-a/r}}}{r}\right)\left(\frac{L^{2}}{r^{2}}\right) \ ,
    \end{equation}
    leading to:
    \begin{equation}
        V_{0}^{'}(r) = \frac{2mL^{2}e^{-a/r}}{r^{5}}\left\lbrace 3r-a-\frac{r^{2}e^{a/r}}{m}\right\rbrace \ .
    \end{equation}\clearpage
    \noindent When $V_{0}^{'}(r)=0$, require:
    
    \begin{eqnarray}
        && 3r-a-\frac{r^{2}e^{a/r}}{m} = 0 \ , \nonumber \\
        && \nonumber \\
        &\Longrightarrow& \quad m\left(3r-a\right) = r^{2}e^{a/r} \ , \nonumber \\
        && \nonumber \\
        &\Longrightarrow& \quad \frac{am\left(3r-a\right)}{r^{3}} = \frac{a}{r}e^{a/r} \ , \nonumber \\
        && \nonumber \\
        &\Longrightarrow& \quad \frac{a}{r} = W\left(\frac{am(3r-a)}{r^{3}}\right) \ , \nonumber \\
        && \nonumber \\
        &\Longrightarrow& \quad r = \frac{a}{W\left(\frac{am(3r-a)}{r^{3}}\right)} \ .
    \end{eqnarray}
    We have a solution for the photon sphere location \emph{implicitly} defined by the real-valued branches of the Lambert $W$ function. Subject to a choice of branch, this corresponds to a real-valued coordinate location within our spacetime, however the intractability of the implicitly defined solution renders it inefficient to pursue further analysis pertaining to the stability of the circular orbit -- we expect the orbit to be \emph{unstable} based on the known literature, \emph{i.e.} $V^{''}(r)<0$, and the coordinate location must lie strictly outside of the horizon location, $r_{H}$.

    \item For massive particles the geodesic orbit corresponds to a timelike worldline and we have the case that $\epsilon=-1$. Therefore:
    
    \begin{equation}
        V_{-1}(r)=\left(1-\frac{2me^{{-a/r}}}{r}\right)\left(1+\frac{L^{2}}{r^{2}}\right) \ ,
    \end{equation}
    and it can be shown that:
    
    \begin{equation}
        V_{-1}^{'}(r) = \frac{2mL^{2}e^{-a/r}}{r^{5}}\left\lbrace 3r-a+\frac{r^{2}\left(r-a\right)}{L^{2}}-\frac{r^{2}}{me^{-a/r}}\right\rbrace \ .
    \end{equation}
    When $V_{-1}^{'}(r)=0$, we therefore require:
    
    \begin{eqnarray}
        && 3r-a+\frac{r^{2}(r-a)}{L^{2}}-\frac{r^{2}e^{a/r}}{m} = 0 \ , \nonumber
    \end{eqnarray}\vfill\clearpage
    \begin{eqnarray}
        &\Longrightarrow& \quad \frac{a}{r}e^{a/r} = \frac{am\left(L^{2}(3r-a)+r^{2}(r-a)\right)}{L^{2}r^{3}} \ , \nonumber \\
        && \nonumber \\
        &\Longrightarrow& \quad \frac{a}{r} = W\left(\frac{am\left(L^{2}(3r-a)+r^{2}(r-a)\right)}{L^{2}r^{3}}\right) \ , \nonumber \\
        && \nonumber \\
        &\Longrightarrow& \quad r = \frac{a}{W\left(\frac{am\left(L^{2}(3r-a)+r^{2}(r-a)\right)}{L^{2}r^{3}}\right)} \ .
    \end{eqnarray}
    Once again, the solution is \emph{implicitly} defined in terms of the Lambert $W$ function. Subject to a choice of branch, we have a valid coordinate location for our ISCO in this spacetime, and expect it to be located outside of both the horizon at $r_{H}$ and the photon sphere. Verifying the stability of this circular orbit is intractable due to the difficult nature of working with the Lambert $W$ function.
\end{itemize}

%%%%%%

\subsection{Regge-Wheeler analysis}

Consider now the Regge-Wheeler equation for scalar and vector perturbations around this spacetime. Similarly to previous analyses, the subsequent analysis closely parallels the general formalism developed in~\cite{ReggeWheeler1}. We begin with the `exponentially suppressed mass' metric:

\begin{equation}
    ds^2=-\left(1-\frac{2me^{{-a/r}}}{r}\right)dt^2+\frac{dr^2}{\left(1-\frac{2me^{{-a/r}}}{r}\right)}+r^2\left(d\theta^2+\sin^2{\theta}d\phi^2\right) \ .
\end{equation}
Define a tortoise coordinate by:

\begin{equation}
    dr_{*} = \left(1-\frac{2me^{{-a/r}}}{r}\right) \, dr \ ,
\end{equation}
then the metric can be rewritten as:

\begin{equation}
    ds^2 = \left(1-\frac{2me^{{-a/r}}}{r}\right)\left\lbrace -dt^2+dr_{*}^{2}\right\rbrace +r^2\left(d\theta^2+\sin^2\theta d\phi^2\right) \ .
\end{equation}
Here $r$ is now implicitly a function of $r_*$. The Regge-Wheeler equation can be written as~\cite{ReggeWheeler1}:
\begin{equation}
 \partial_{r_{*}}^2\, \hat \phi 
+ \left\{\omega^2- \mathcal{V} \right\}
  \hat \phi = 0 \ .
\end{equation}\clearpage
\noindent For a general spherically symmetric metric with respect to curvature coordinates, the Regge-Wheeler potential for spins $S\in\lbrace 0, 1, 2\rbrace$ and angular momentum $\ell\geq S$ is~\cite{ReggeWheeler1}:

\begin{equation}
    \mathcal{V}_{S}=\left(-g_{tt}\right)\Bigg\lbrace\frac{\ell\left(\ell+1\right)}{r^2}+\frac{S\left(S-1\right)\left(g^{rr}-1\right)}{r^{2}}\Bigg\rbrace+\left(1-S\right)\frac{\partial_{r_{*}}^{2}r}{r} \ .
\end{equation}
For the `exponentially suppressed mass' metric we therefore have the following Regge-Wheeler potential:

\begin{eqnarray}
    \mathcal{V}_{S} &=& \left(1-\frac{2me^{{-a/r}}}{r}\right)\Bigg\lbrace\frac{\ell\left(\ell+1\right)}{r^{2}}-\frac{2me^{-a/r}S(S-1)}{r^{3}}\Bigg\rbrace \nonumber \\
    && \nonumber \\
    && \qquad \qquad \qquad \qquad \qquad \qquad \qquad +\left(1-S\right)\frac{\partial_{r_{*}}^{2}r}{r} \ .
\end{eqnarray}
It can be readily shown that:

\begin{equation}
    \frac{\partial_{r_{*}}^{2}r}{r} = \frac{2me^{-a/r}\left(a-r\right)}{r\left(r-2me^{-a/r}\right)^{3}} \ ,
\end{equation}
and so we may rephrase the Regge--Wheeler potential as:

\begin{eqnarray}
    \mathcal{V}_{S} &=& \left(1-\frac{2me^{{-a/r}}}{r}\right)\Bigg\lbrace\frac{\ell\left(\ell+1\right)}{r^{2}}-\frac{2me^{-a/r}S(S-1)}{r^{3}}\Bigg\rbrace \nonumber \\
    && \nonumber \\
    && \qquad \qquad \qquad \qquad +\left(1-S\right)\frac{2me^{-a/r}\left(a-r\right)}{r\left(r-2me^{-a/r}\right)^{3}} \ .
\end{eqnarray}

%%%%

\subsubsection{Spin zero}

In particular for spin zero one has:
\begin{eqnarray}
\mathcal{V}_0 &=& \left(1-\frac{2me^{{-a/r}}}{r}\right)\left\lbrace\frac{\ell\left(\ell+1\right)}{r^{2}}+\frac{2me^{-a/r}(a-r)}{\left(r-2me^{-a/r}\right)^{4}}\right\rbrace \ .
\end{eqnarray}
This result can also be readily checked by brute force computation. For scalars the $s$-wave ($\ell=0$) is particularly important:
\begin{equation}
\mathcal{V}_{0,\ell=0}  =  \frac{\partial_{r_{*}}^{2}r}{r} = \frac{2me^{-a/r}\left(a-r\right)}{r\left(r-2me^{-a/r}\right)^{3}} \ .
\end{equation}

\subsubsection{Spin one}

For the spin one vector field the $\left\{ {r^{-1}} \partial_{r_{*}}^2{ r} \right\} $ term drops out; this can ultimately be traced back to the conformal invariance of massless spin one particles in `3+1'-dimensions. We are left with the particularly simple result ($\ell\geq 1$):
\begin{equation}
\mathcal{V}_1 = \left(1-\frac{2me^{{-a/r}}}{r}\right)\left\lbrace\frac{\ell\left(\ell+1\right)}{r^{2}}\right\rbrace \ .
\end{equation}

%%%%

\subsubsection{Spin two}

For the spin two axial mode (\emph{i.e.} $S=2$) we have the following ($\ell\geq 2$):

\begin{eqnarray}
    \mathcal{V}_{S} &=& \left(1-\frac{2me^{{-a/r}}}{r}\right)\Bigg\lbrace\frac{\ell\left(\ell+1\right)}{r^{2}}-\frac{4me^{-a/r}}{r^{4}} \nonumber \\
    && \qquad \qquad -\frac{2me^{-a/r}(a-r)}{\left(r-2me^{-a/r}\right)^{4}}\Bigg\rbrace \ .
\end{eqnarray}

%%%%%%

\subsection{Stress-energy-momentum tensor}

Let us examine the Einstein field equations for this spacetime, and subsequently analyse the various energy conditions. We shall use the form of the stress-energy-momentum tensor defined by the mixed non-zero Einstein tensor components from Eq.~\ref{einsteinsuppressed}. This gives the following form of the Einstein field equations: $G^{\mu}{}_{\nu}=8\pi T^{\mu}{}_{\nu}$. Accordingly, this yields the following general form of the stress-energy-momentum tensor:

\begin{equation}\label{stresssuppressed}
    T^{\mu}{}_{\nu} = \begin{bmatrix}
    -\rho & 0 & 0 & 0 \\
    0 & p_{\parallel} & 0 & 0 \\
    0 & 0 & p_{\perp} & 0 \\
    0 & 0 & 0 & p_{\perp}
    \end{bmatrix} \ ,
\end{equation}
and we have the following specific forms for the principal pressures:

\begin{eqnarray}
    \rho &=& \frac{2mae^{-a/r}}{8\pi r^{4}} \ , \nonumber \\
    && \nonumber \\
    p_{\parallel} &=& -\frac{2mae^{-a/r}}{8\pi r^{4}} \ , \nonumber \\
    && \nonumber \\
    p_{\perp} &=& -\frac{ma\left(a-2r\right)e^{-a/r}}{8\pi r^{5}} \ .
\end{eqnarray}
Let us examine where the energy density is maximised for this spacetime (this is of specific interest due to the exponential suppression of the mass -- we wish to see how the suppression affects the distribution of energy densities in through the geometry):

\begin{equation}
    \frac{\partial{\rho}}{\partial{r}} = \frac{2mae^{-a/r}}{8\pi r^{6}}\left(a-4r\right) \ .
\end{equation}
Setting this to zero we can clearly see that $\rho$ is maximised at coordinate location $r=\frac{a}{4}$. Let us now analyse the various energy conditions and see whether they are violated in our spacetime.

\subsubsection{Null energy condition}

In order to satisfy the null energy condition, we require that both $\rho+p_{\parallel}\geq 0$ \emph{and} $\rho+p_{\perp}\geq 0$ globally in our spacetime. Let us first consider $\rho+p_{\parallel}$:

\begin{equation}\label{nullsuppressed}
    \rho+p_{\parallel} = \frac{2mae^{-a/r}}{8\pi r^{4}}-\frac{2mae^{-a/r}}{8\pi r^{4}} = 0 \ .
\end{equation}
This is manifestly zero for our spacetime, and the condition that $\rho+p_{\parallel}\geq 0$ is satisfied. Let us now consider $\rho+p_{\perp}$:

\begin{eqnarray}
    \rho+p_{\perp} &=& \frac{2mae^{-a/r}}{8\pi r^{4}}-\frac{ma\left(a-2r\right)e^{-a/r}}{8\pi r^{5}} \nonumber \\
    && \nonumber \\
    &=& \frac{mae^{-a/r}}{8\pi r^{5}}\left\lbrace 4r-a\right\rbrace \ .
\end{eqnarray}
This changes sign when $r=\frac{a}{4}$ and is negative in the region $r<\frac{a}{4}$. We therefore have the non-typical instance where the radial NEC is satisfied by the geometry whilst the transverse NEC is violated.

\subsubsection{Strong energy condition}

In order to satisfy the strong energy condition (SEC), we require that $\rho+p_{\parallel}+2p_{\perp}\geq 0$ globally in our spacetime. Evaluating:

\begin{eqnarray}
    \rho+p_{\parallel}+2p_{\perp} &=& 2p_{\perp} \nonumber \\
    && \nonumber \\
    &=& -\frac{ma\left(a-2r\right)e^{-a/r}}{4\pi r^{5}} \ ,
\end{eqnarray}
The is negative for the region $r<\frac{a}{2}$, and we may conclude that the SEC is violated for this spacetime -- this is consistent with expectations.

\subsection{Overview}

The `exponentially suppressed mass' geometry therefore accurately models a regular black hole geometry in the sense of Bardeen when the parameter $a\in\left(0, \frac{2m}{e}\right)$, and violates the strong energy condition accordingly. It also satisfies the radial NEC in the \emph{absence} of a wormhole throat, but violates the tangential NEC in the region nearest the centre of mass, when $r<\frac{a}{4}$, due to the mathematical side effects which come from exponentially suppressing the centralised mass. The energy density $\rho$ is maximised at $r=\frac{a}{4}$, and for $r<\frac{a}{4}$ the exponential expression present in the metric has the effect of suppressing the mass of our centralised object, asmpyotically heading to zero as we approach the centre of mass ($r=0$). This implies the local region near the centre of mass is asymptotically Minkowski, while for some finite positive $r$-coordinate value the curvature of the geometry is maximised (this can be found by analysing the extrema of the Kretschmann scalar, \emph{e.g.}, but the subsequent calculations are rather intractable -- the key point is to note that the curvature of the geometry is asymptotically flat at infinity, at the centre of mass, and has some maximal peak in between). We can find a valid ISCO and photon sphere location, but these results are \emph{implicitly} defined by the Lambert $W$ function, and as such are rather difficult to deal with. The intractability of dealing with the Lambert $W$ function in general means that the `exponentially suppressed mass' metric is a tricky candidate spacetime to analyse, however it is certainly a curious geometry modelling a scenario of physical interest. Accordingly further analysis of this metric, as well as the concept of `exponentially suppressed mass' in general, shall be the subject of further research.
%%%%%%%%%%%%%%%%%%%%%%%%%%%%%%%%%%%%%%%%%

%%%%%%%%%%%%%%%%%%%%%%%%%%%%%%%%%%%%%%%%%%%%%%%%%%%%%%%

\chapter{From `black-bounce' to traversable wormhole}\label{C:Black-bounce}

%%%%%%

A particularly interesting regular black hole spacetime is described by the line element:
\begin{equation}
ds^{2}=-\left(1-\frac{2m}{\sqrt{r^{2}+a^{2}}}\right)dt^{2}+\frac{dr^{2}}{1-\frac{2m}{\sqrt{r^{2}+a^{2}}}}
+\left(r^{2}+a^{2}\right)\left(d\theta^{2}+\sin^{2}\theta \;d\phi^{2}\right) \ .
\end{equation}
This spacetime neatly interpolates between the standard Schwarzschild black hole and the Morris-Thorne traversable wormhole;
at intermediate stages passing through a `black-bounce' (into a future incarnation of the universe), an extremal null-bounce (into a future incarnation of the universe), and a traversable wormhole.  As long as the parameter $a$ is non-zero the geometry is everywhere regular, so one has a somewhat unusual form of `regular black hole', where the `origin' $r=0$ can be either spacelike, null, or timelike. Thus this spacetime generalizes and broadens the class of `regular black holes' beyond those usually considered. This spacetime is carefully designed to be a minimalist modification of the ordinary Schwarzschild spacetime; when
adjusting the parameter $a$, this metric represents either: 
\begin{itemize}
\itemsep-3pt
\item 
The ordinary Schwarzschild spacetime; 
\item
A regular black hole geometry with a one-way spacelike throat;
\item
A one-way wormhole geometry with an extremal null throat (compare this case especially with reference~\cite{Cano:2018}); or
\item
A canonical traversable wormhole geometry, (in the Morris-Thorne sense~\cite{MorrisThorne, MTY, Visser:1989a, Visser:1989b, LorentzianWormholes, Visser:2003, Hochberg:1997, Poisson:1995, Barcelo:2000, Hochberg:1998, Cramer:1994, Visser:1997, Barcelo:1999, Garcia:2011, Expmetric, lobo:2004}), with a two-way timelike throat.
\end{itemize} 
In the region where the geometry represents a regular black hole the geometry is unusual in that it describes a bounce into a future incarnation of the universe, rather than a bounce back into our own universe~\cite{Barcelo:2014, Barcelo:2014b, Barcelo:2015, Barcelo:2016, Garay:2017, Rovelli:2014, Haggard:2015, Christodoulou:2016, DeLorenzo:2015, Malafarina:2017, Olmedo:2017, Barrau:2018, Malafarina:2018}. Let us conduct a standard analysis of the metric within the context of general relativity.

%=====================================================
\section{Metric analysis and Carter-Penrose diagrams}\label{sec:metric}
%=====================================================
Consider the metric:
\begin{equation}\label{RBHmetric}
    ds^{2}=-\left(1-\frac{2m}{\sqrt{r^{2}+a^{2}}}\right)dt^{2}+\frac{dr^{2}}{1-\frac{2m}{\sqrt{r^{2}+a^{2}}}}+\left(r^{2}+a^{2}\right)\left(d\theta^{2}+\sin^{2}\theta d\phi^{2}\right).
\end{equation}
Note that if $a=0$ then this is simply the Schwarzschild solution, so enforcing $a\neq 0$ is a sensible starting condition if we are to conduct an analysis concerning either regular black holes or traversable wormholes (trivially,  the Schwarz\-schild solution models a geometry which is neither). Furthermore, this spacetime geometry is manifestly static and spherically symmetric.  That is,  it admits a global, non-vanishing, timelike Killing vector field that is hypersurface orthogonal, and there are no off-diagonal components of the matrix representation of the metric tensor; fixed $r$ coordinate locations in the spacetime correspond to spherical surfaces. This metric does not correspond to a traditional regular black hole such as the Bardeen, Bergmann-Roman, Frolov, or Hayward geometries \cite{Bardeen1968, Bergmann-Roman, Hayward:2005, Bardeen:2014, Frolov:2014, Frolov:2014b, Frolov:2016, Frolov:2017, Frolov:2018, Cano:2018,Bardeen:2018}. Instead, depending on the value of the parameter $a$, it is either a regular black hole (bouncing into a future incarnation of the universe) or a traversable wormhole.

Before proceeding any further, note that the coordinate patch is global, and the coordinates have natural domains:

\begin{equation}
r\in(-\infty,+\infty);\qquad 
t\in(-\infty,+\infty);\qquad
\theta\in [0,\pi];\qquad
\phi\in(-\pi,\pi] \ .
\end{equation}
Analysis of the radial null curves in this metric yields (setting $ds^{2}=0$, $d\theta=d\phi=0$):

\begin{eqnarray}
    \frac{dr}{dt} = \pm\left(1-\frac{2m}{\sqrt{r^{2}+a^{2}}}\right) \ .
\end{eqnarray}
It is worth noting that this defines a `coordinate speed of light' for the metric (\ref{RBHmetric}):

\begin{equation}
    c(r)=\left\vert\frac{dr}{dt}\right\vert=\left(1-\frac{2m}{\sqrt{r^{2}+a^{2}}}\right) \ ,
\end{equation}
and hence an effective refractive index of:

\begin{equation}
    n(r)=\frac{1}{\left(1-\frac{2m}{\sqrt{r^{2}+a^{2}}}\right)} \ .
\end{equation}\clearpage
\noindent Let us now examine the coordinate location(s) of horizon(s) in this geometry:

\begin{itemize}
 
    \item If $a>2m$, then $\forall \ r\in(-\infty,+\infty)$ we have $\frac{dr}{dt}\neq 0$, so this geometry is in fact a (two-way) traversable wormhole~\cite{MorrisThorne, MTY, Visser:1989a, Visser:1989b, LorentzianWormholes, Visser:2003, Hochberg:1997, Poisson:1995, Barcelo:2000, Hochberg:1998, Cramer:1994, Visser:1997, Barcelo:1999, Garcia:2011, Expmetric, lobo:2004}.
      
    \item If $a=2m$, then as $r\rightarrow 0$ from either above or below, we have $\frac{dr}{dt}\rightarrow 0$. Hence we have a horizon at coordinate location $r=0$. However, this geometry is not a black hole. Rather, it is  a one-way wormhole with an extremal null throat at $r=0$.

    \item If $a<2m$, then consider the two locations $r_\pm=\pm\sqrt{(2m)^{2}-a^{2}}$; this happens when $\sqrt{r_\pm^{2}+a^{2}}=2m$.  
    Thence:
    
    \begin{eqnarray}
        \exists \ r_\pm \in\mathbb{R}; \qquad \frac{dr}{dt}=0 \ .
    \end{eqnarray}
    That is, when $a<2m$ there will be symmetrically placed $r$-coordinate values $r_\pm = \pm|r_\pm|$  which correspond to a pair of horizons.
\end{itemize}
The coordinate location $r=0$ maximises both the non-zero curvature tensor components (see  \S\ref{sec:curvature}) and the curvature invariants (see \S\ref{sec:invariants}).
\enlargethispage{40pt}

\begin{figure}[!htb]
%\vspace{-1cm}
\begin{center}
\includegraphics[scale=0.38]{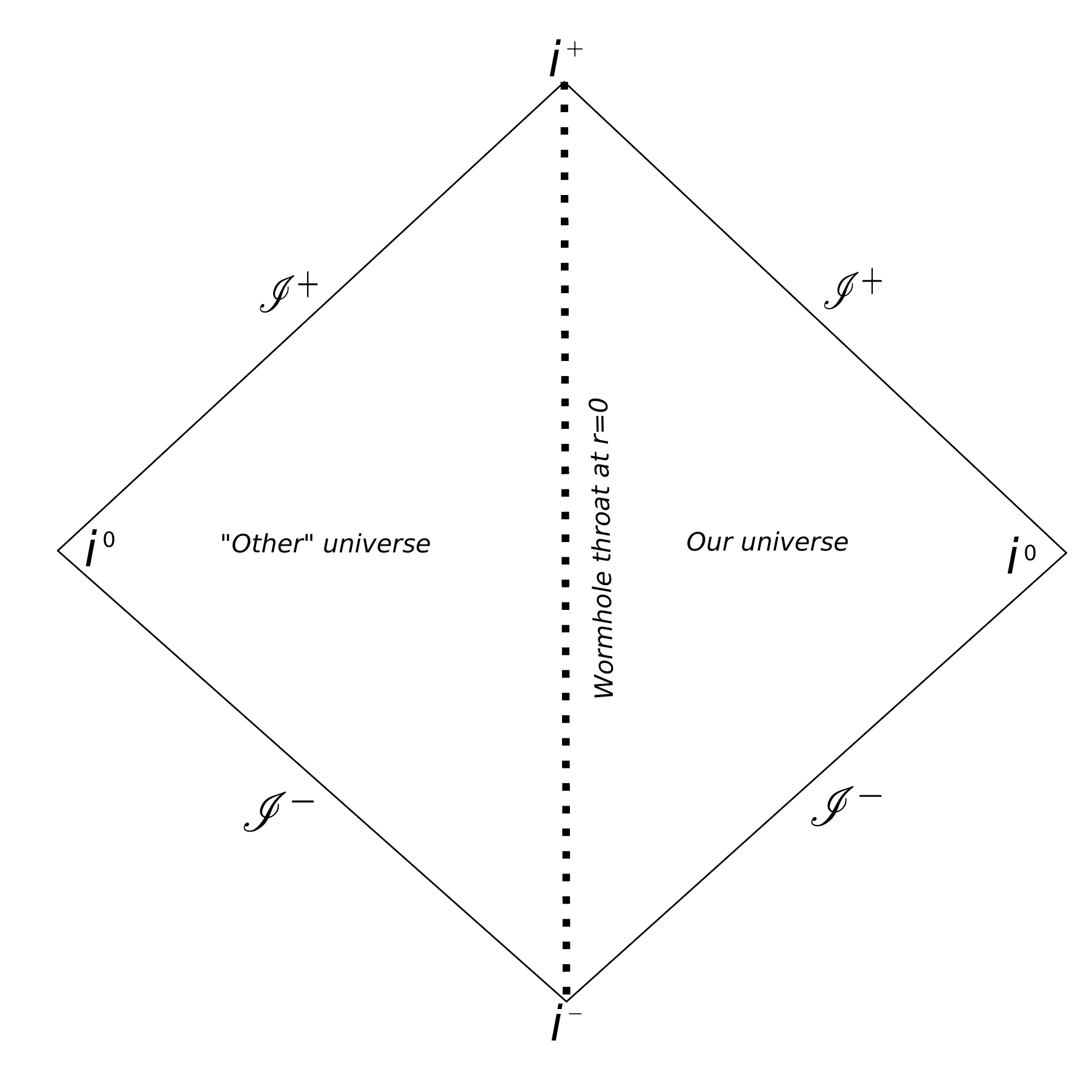}\qquad
\end{center}
{\caption[Carter-Penrose diagram for the canonical Morris-Thorne\newline traversable wormhole]{{Carter-Penrose diagram for the case when $a>2m$ and we have a traditional traversable wormhole in the Morris-Thorne sense}.}
\label{F:wormhole}}
\end{figure}

We may therefore conclude that in the case where $a>2m$ the two-way traversable wormhole throat is a timelike hypersurface located at $r=0$, and negative $r$-values correspond to the universe on the other side of the geometry from the perspective of an observer in our own universe. We then have the quite standard Carter-Penrose diagram for traversable wormholes as presented in Fig.~\ref{F:wormhole}.

%=====================================================
%=====================================================

\begin{figure}[!htb]
\vspace{-0.5cm}
\begin{center}
\includegraphics[scale=0.60]{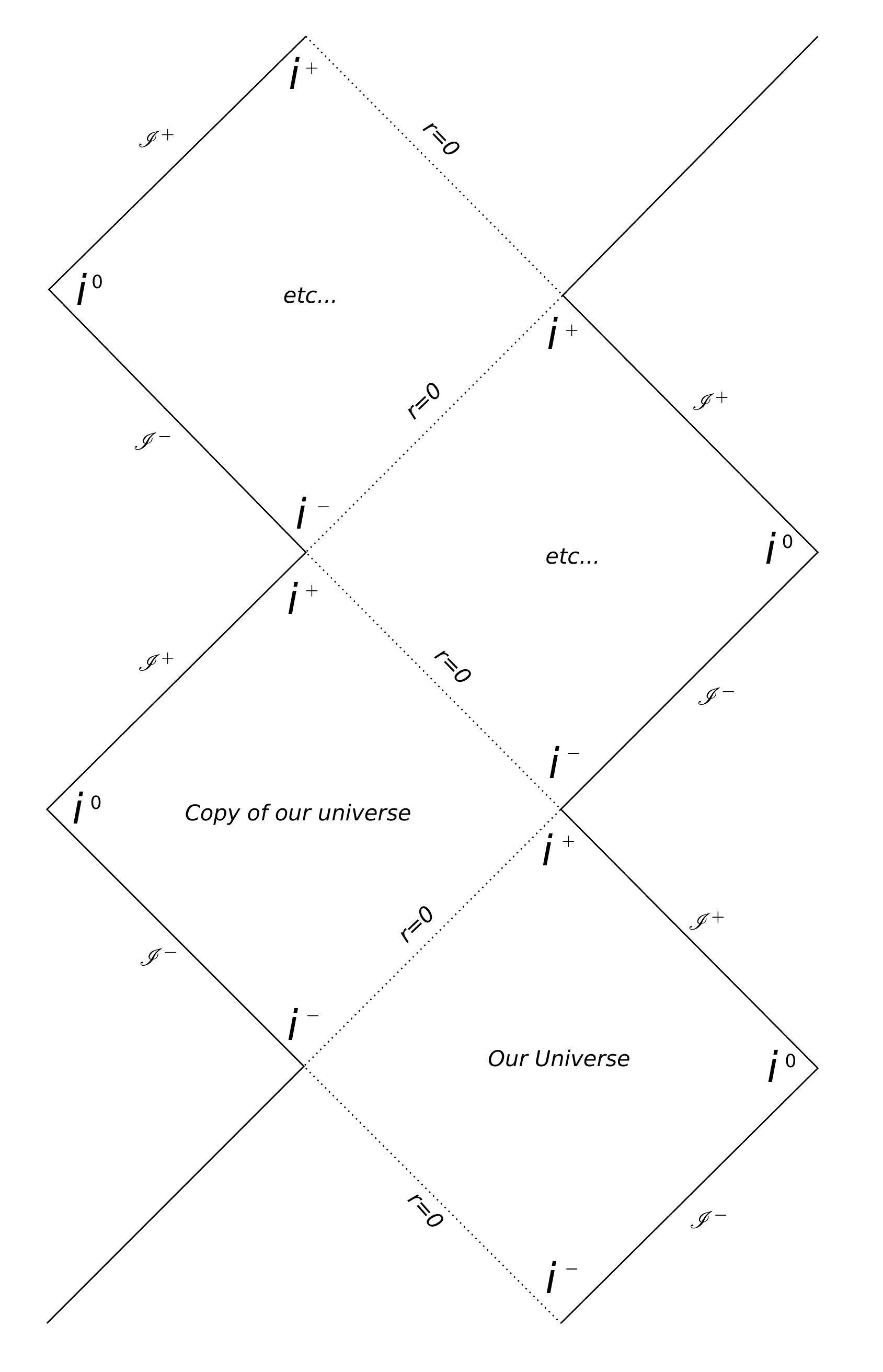}\qquad
\end{center}
{\caption[Carter-Penrose diagram for a one-way wormhole with an extremal null throat]{{Carter-Penrose diagram for the maximally extended spacetime in the case when $a=2m$. In this example we have a one-way wormhole geometry with a null throat}.}
\label{F:null-bounce-1}}
\end{figure}

Similarly for the null case $a=2m$ the null throat is located at the horizon $r=0$. Note that in this instance the wormhole geometry is only one-way traversable. 
The Carter-Penrose diagram for the maximally extended spacetime in this case is given in Fig.~\ref{F:null-bounce-1}.

%=====================================================
\null
%=====================================================
\clearpage

\begin{figure}[!htb]
\begin{center}
\includegraphics[scale=0.50]{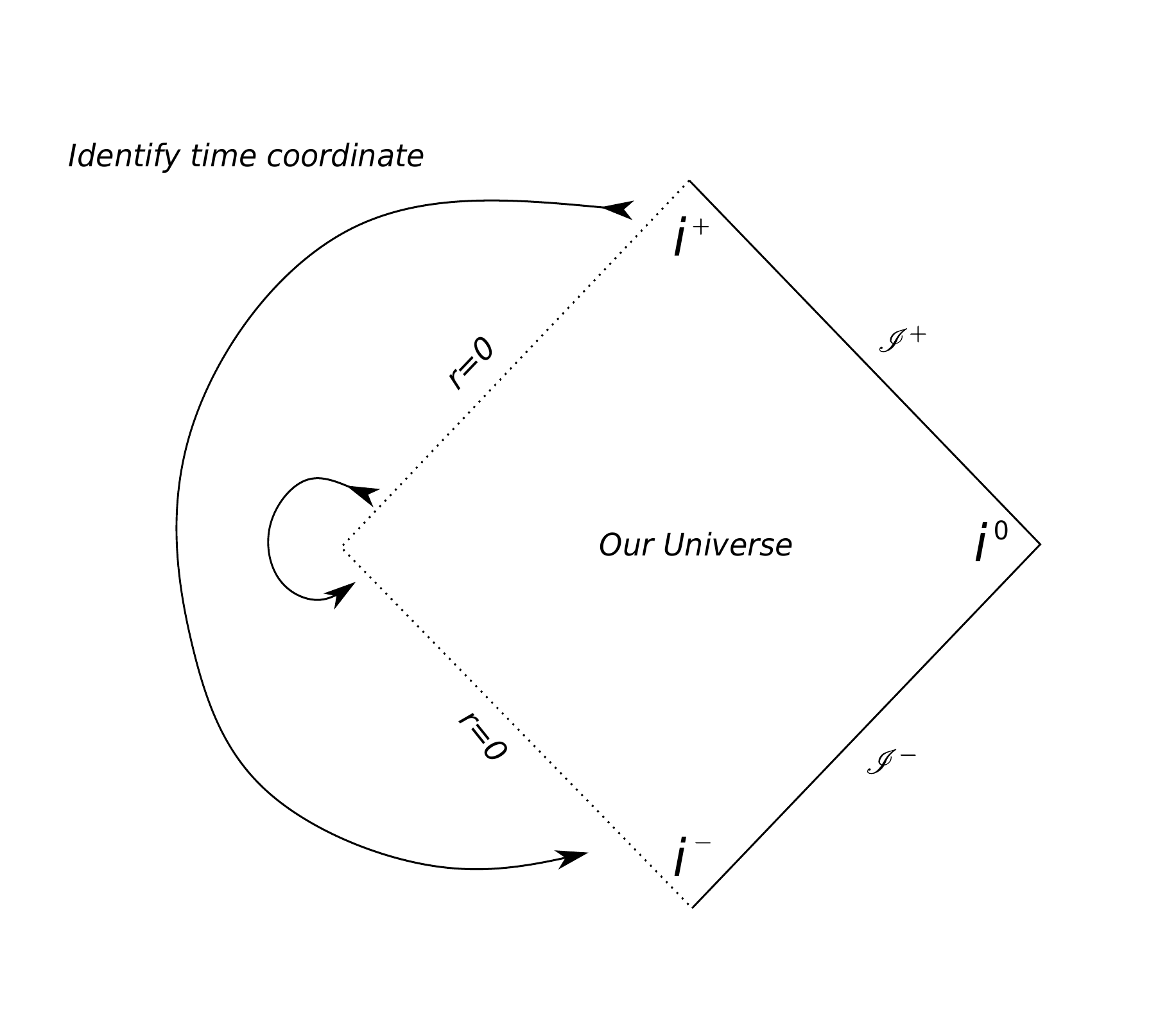}\qquad
\end{center}
{\caption[Carter-Penrose diagram for the one-way wormhole with an extremal null throat and cyclical $t$]{{Carter-Penrose diagram for the case when $a=2m$ where we have identified the future null bounce at $r=0$ with the past null bounce at $r=0$}.}
\label{F:null-bounce-2}}
\end{figure}

\noindent As an alternative construction we can identify the past null bounce at $r=0$ with the future null bounce at $r=0$, 
yielding the `looped' Carter-Penrose diagram of Fig.~\ref{F:null-bounce-2}.

%=====================================================
%=====================================================

\begin{figure}[!htb]
\begin{center}
\includegraphics[scale=0.47]{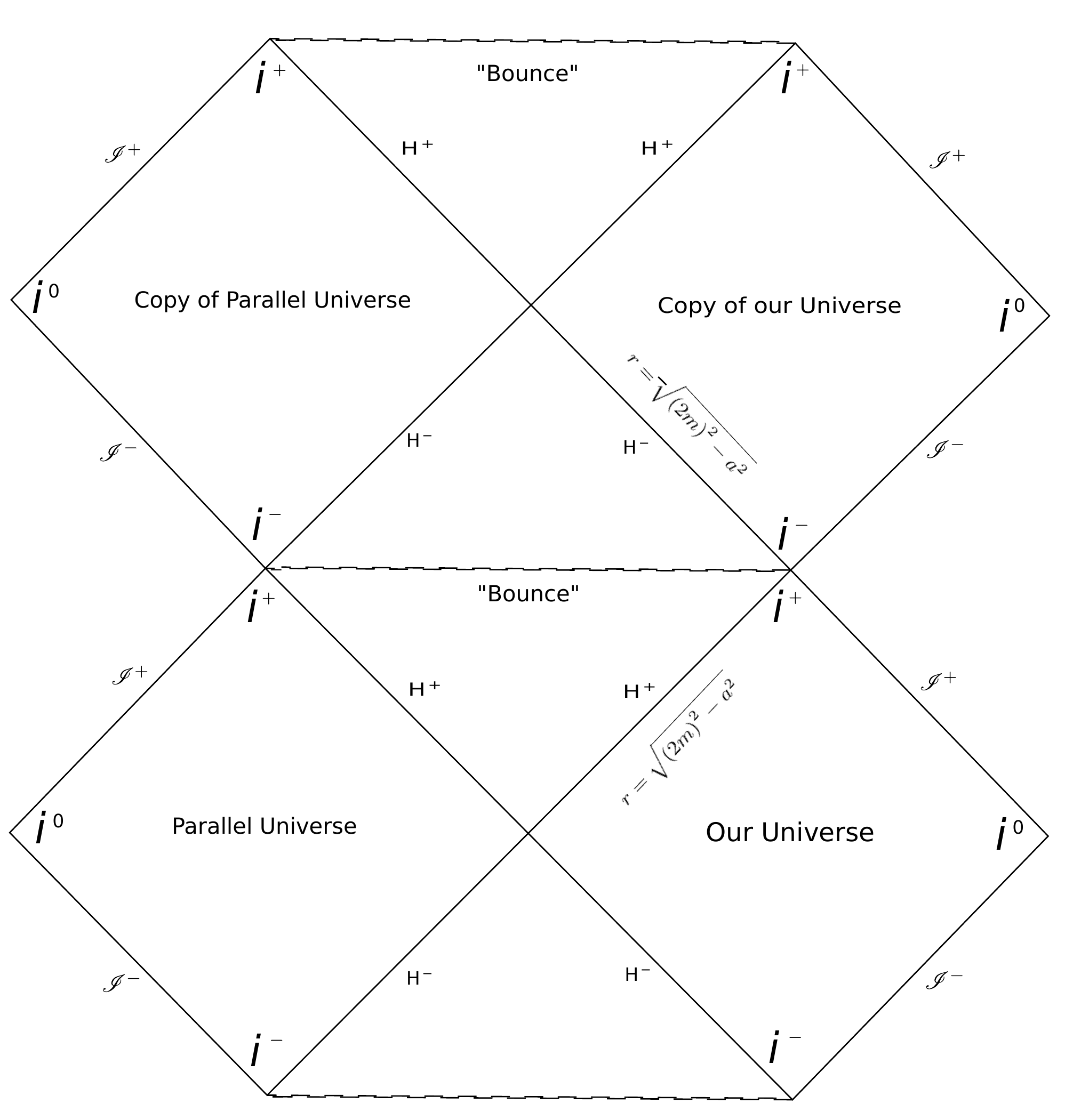}\qquad
\end{center}
{\caption[Carter-Penrose diagram for the `black-bounce' geometry]{{Carter-Penrose diagram for the maximally extended spacetime when $a\in(0,2m)$. In this example the time coordinate
runs up the page, 
`bouncing' through the $r=0$  hypersurface in each black hole region into a future copy of our own universe \emph{ad infinitum}.}}\label{F:bounce-1}}
\end{figure}

For regular black holes, 
we can restrict our attention to the interval $a\in(0,2m)$. Then the hypersurface $r=0$ is a spacelike spherical surface which marks the boundary between our universe and a bounce into a separate copy of our own universe. For negative values of $r$ we have `bounced' into another universe. See Fig.~\ref{F:bounce-1} for the  relevant Carter-Penrose diagram (contrast these Carter-Penrose diagrams with the standard one for the maximally extended Kruskal-Szekeres version of Schwarzschild -- see for instance references~\cite{Wald, telebook, largescale} -- the major difference is that the singularity has been replaced by a spacelike hypersurface representing a `bounce').

\begin{figure}[!htb]
\begin{center}
\includegraphics[scale=0.47]{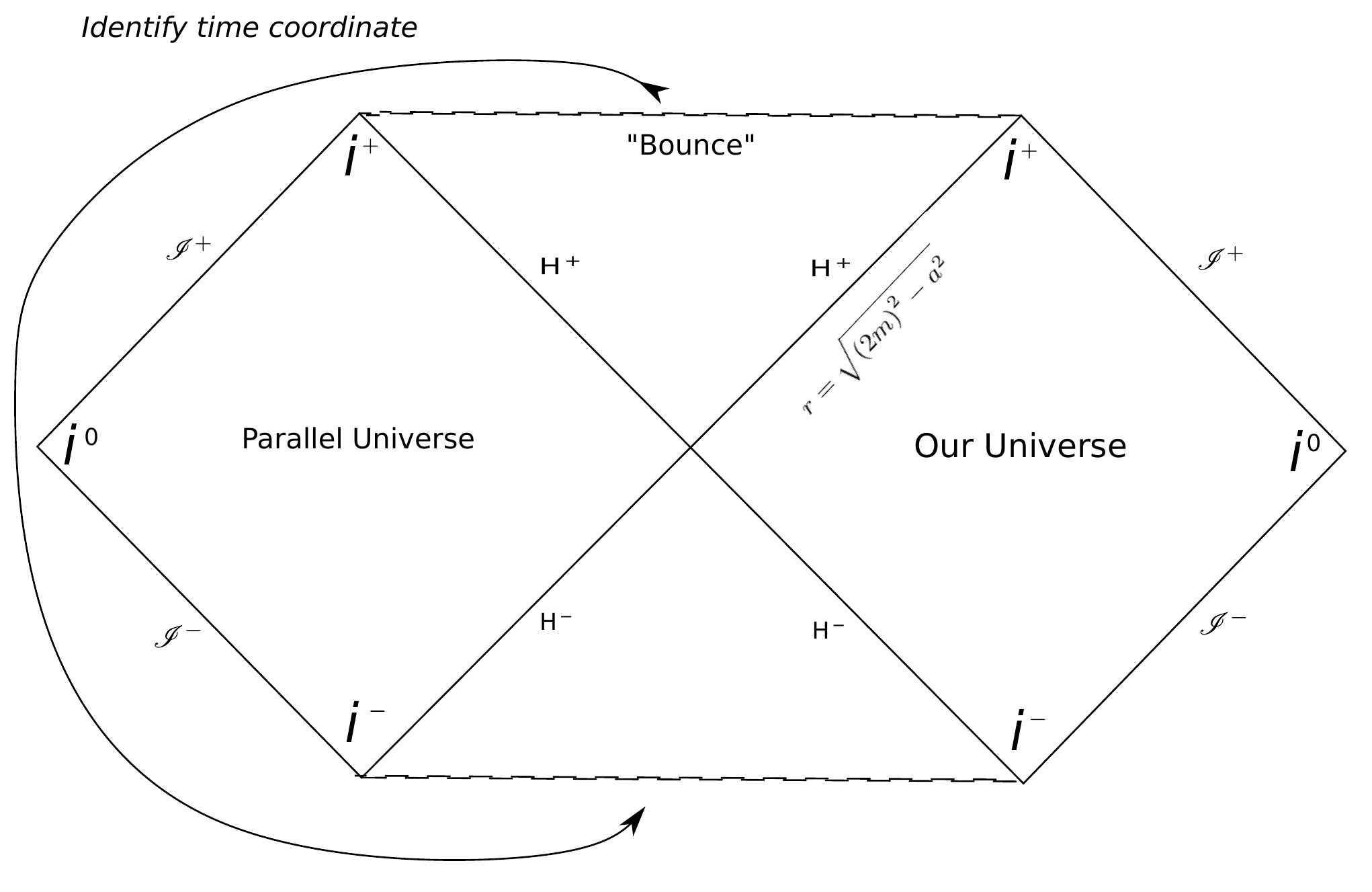}\qquad
\end{center}
{\caption[Carter-Penrose diagram for the `black-bounce' geometry\newline with cyclical $t$]{{Periodic boundary conditions in time when $a\in(0,2m)$. In this example we impose periodic boundary conditions on the time coordinate such that the future bounce is identified with the past bounce}.}
\label{F:bounce-2}}
\end{figure}

Another possibility of interest for when $a\in(0,2m)$ arises when the  $r=0$ coordinate for the `future bounce' is identified with the  $r=0$ coordinate for the `past bounce'. That is, there is still a distinct time orientation but we impose periodic boundary conditions on the time coordinate such that time is cyclical. This case yields the Carter-Penrose diagram of Fig.~\ref{F:bounce-2} (note that the global causal structure is much milder than that for the so-called `twisted' black holes~\cite{twisted}).

%=====================================================
%=====================================================

%=====================================================
%=====================================================

%====================================================
\section{Curvature tensors}\label{sec:curvature}
%====================================================

Next it is prudent to check that there are no singularities in the geometry, otherwise we do not satisfy the requirements for the regularity of our black hole. 
In view of the diagonal metric environment of Eq.~\ref{RBHmetric} we can clearly see that the chosen coordinate basis is orthogonal though not orthonormal, and it therefore follows that the (mixed) non-zero components of the Riemann tensor shall be the same with respect to this basis as to any orthonormal tetrad, ensuring that the appearance (or lack thereof) of any singularities is not simply a coordinate artefact. 

With this in mind, for simplicity we first consider the mixed $C^{\mu\nu}{}_{\alpha\beta}$ non-zero components of the Weyl tensor (these are equivalent to orthonormal components):

\begin{eqnarray}
    C^{t\theta}{}_{t\theta} = C^{t\phi}{}_{t\phi} = C^{r\theta}{}_{r\theta} = C^{r\phi}{}_{r\phi} &=&  -\frac{1}{2}C^{tr}{}_{tr} = -\frac{1}{2}C^{\theta\phi}{}_{\theta\phi} \nonumber \\
    &=&  \frac{6r^2m+a^2\left(2\sqrt{r^2+a^2}-3m\right)}{6\left(r^2+a^2\right)^{\frac{5}{2}}} \ . \quad
\end{eqnarray}
Note that as $r\to0$ these Weyl tensor components approach the finite value ${2a-3m\over 6 a^3}$.\newline

\noindent For the Riemann tensor the non-zero components are a little more complicated:
\begin{eqnarray}
    R^{tr}{}_{tr} &=& \frac{m(2r^{2}-a^{2})}{(r^{2}+a^{2})^{\frac{5}{2}}} ;
    \nonumber\\    
    R^{t\theta}{}_{t\theta} &=& R^{t\phi}{}_{t\phi}=\frac{-r^{2}m}{(r^{2}+a^{2})^{\frac{5}{2}}}; 
    \nonumber \\
    R^{r\theta}{}_{r\theta} &=& R^{r\phi}{}_{r\phi}=\frac{m\left(2a^{2}-r^{2}\right)-a^{2}\sqrt{r^{2}+a^{2}}}{(r^{2}+a^{2})^{\frac{5}{2}}}; 
    \nonumber \\
    &&\nonumber \\
    R^{\theta\phi}{}_{\theta\phi} &=& \frac{2r^{2}m+a^{2}\sqrt{r^{2}+a^{2}}}{(r^{2}+a^{2})^{\frac{5}{2}}}.
\end{eqnarray}
Provided $a\neq0$, as $\vert r\vert\rightarrow 0$ all of these Riemann tensor components approach finite limits:

\begin{eqnarray}
    && R^{tr}{}_{tr} \to -\frac{m}{a^3} \ ; \nonumber\\
    && R^{t\theta}{}_{t\theta} = R^{t\phi}{}_{t\phi} \to 0 \ ; \nonumber \\
    && R^{r\theta}{}_{r\theta} = R^{r\phi}{}_{r\phi} \to \frac{2m-a}{a^3} \ ; \nonumber\\
    &&  R^{\theta\phi}{}_{\theta\phi} \to \frac{1}{a^2} \ .
\end{eqnarray}
As $\vert r\vert$ increases, with $m$ and $a$ held fixed,  all components asymptote to multiples of $m/r^3$, hence as $\vert r\vert\rightarrow+\infty$, all components tend to $0$ (this is synonymous with the fact that for large $\vert r\vert$ this geometry models weak field general relativity).
Hence $\forall \ r\in(-\infty,+\infty)$ the components of the Riemann tensor are strictly finite. We may conclude that on the interval $a\in(0,2m]$ there is a horizon, but no singularity, and the metric really does represent the geometry of a regular black hole. In the case when $a>2m$ and we have a traversable wormhole, trivially there are also no singularities.\newline

\noindent The Ricci tensor has non-zero (mixed) components:

\begin{eqnarray}
-2R^{t}{}_{t} = R^{\theta}{}_{\theta} &=& R^{\phi}{}_{\phi} = \ \frac{2a^{2}m}{\left(r^{2}+a^{2}\right)^{\frac{5}{2}}} \ , \nonumber \\
&& \nonumber \\
R^{r}{}_{r} &=& \frac{a^{2}\left(3m-2\sqrt{r^{2}+a^{2}}\right)}{\left(r^{2}+a^{2}\right)^{\frac{5}{2}}} \ .
\end{eqnarray}
The Einstein tensor has non-zero (mixed) components:

\begin{eqnarray}
G^{t}{}_{t} &=& \frac{a^{2}\left(\sqrt{r^{2}+a^{2}}-4m\right)}{\left(r^{2}+a^{2}\right)^{\frac{5}{2}}} \ ,
\qquad
G^{r}{}_{r} = \frac{-a^{2}}{\left(r^{2}+a^{2}\right)^{2}}\; \ , \nonumber \\
G^{\theta}{}_{\theta} &=& G^{\phi}{}_{\phi} = \frac{a^{2}\left(\sqrt{r^{2}+a^{2}}-m\right)}{\left(r^{2}+a^{2}\right)^{\frac{5}{2}}} \ .
\end{eqnarray}

%====================================================
\section{Curvature invariants}\label{sec:invariants}
%====================================================

The Ricci scalar is:

\begin{equation}
    R=\frac{2a^{2}\left(3m-\sqrt{r^{2}+a^{2}}\right)}{\left(r^{2}+a^{2}\right)^{\frac{5}{2}}} \ .
\end{equation}
The Ricci contraction $R_{\mu\nu}R^{a\mu\nu}$ is:

\begin{equation}\label{3.10}
    R_{\mu\nu}R^{\mu\nu} = \frac{a^4\left[4\left(\sqrt{r^2+a^2}-\frac{3}{2}m\right)^{2}+(3m)^2\right]}{\left(r^2+a^2\right)^{5}} \ .
\end{equation}
Note that this is a sum of squares and so automatically non-negative (and finite).\newline

\noindent The Weyl contraction $C_{\mu\nu\alpha\beta}C^{\mu\nu\alpha\beta}$:

\begin{equation}
    C_{\mu\nu\alpha\beta}C^{\mu\nu\alpha\beta} = \frac{4}{3\left(r^2+a^2\right)^{5}}\Bigg\lbrace 3m\left(2r^2-a^2\right)+2a^2\sqrt{r^2+a^2}\Bigg\rbrace^{2} \ .
\end{equation}
Note that this is a perfect square and so is automatically non-negative (and finite).\newline

\noindent The Kretschmann scalar is:

\begin{equation}
R_{\mu\nu\alpha\beta} \, R^{\mu\nu\alpha\beta} = C_{\mu\nu\alpha\beta} \, C^{\mu\nu\alpha\beta} +2 R_{\mu\nu}\, R^{\mu\nu} - \frac{1}{3}R^2 \ ,
\end{equation}
and so (in view of the above) is guaranteed finite without further calculation. Explicitly:

\begin{eqnarray}
R_{\mu\nu\alpha\beta}\,R^{\mu\nu\alpha\beta} &=& \frac{4}{\left(r^{2}+a^{2}\right)^{5}}\bigg\lbrace\sqrt{r^{2}+a^{2}}\left[8a^2m\left(r^2-a^2\right)\right] \nonumber \\
&& +3a^{4}\left(r^2+a^2\right)+3m^{2}\left(3a^4-4a^2r^2+4r^4\right)\bigg\rbrace \ .
\end{eqnarray}

%===========================================
\section{Stress-energy-momentum tensor}
%===========================================

Let us examine the Einstein field equations for this spacetime. 
We first note that for $\sqrt{r^2+a^2}>2m$, that is, \emph{outside} any horizon that may potentially be present, one has $\rho = - T_t{}^t$ while $p_\parallel = T_r{}^r$ and $p_\perp =  T_\theta{}^\theta = T_\phi{}^\phi$. Using the mixed components $G^{\mu}_{\ \nu}=8\pi \; T^{\mu}_{\ \nu}$, this
yields the following form of the stress-energy-momentum tensor:

\begin{eqnarray}
\rho &=& -\frac{a^{2}\left(\sqrt{r^{2}+a^{2}}-4m\right)}{8\pi \left(r^{2}+a^{2}\right)^{\frac{5}{2}}} \ ; \nonumber \\
p_\parallel &=& \frac{-a^{2}}{8\pi \left(r^{2}+a^{2}\right)^{2}} \ ; \nonumber \\
p_{\perp} &=& \frac{a^{2}\left(\sqrt{r^{2}+a^{2}}-m\right)}{8 \pi \left(r^{2}+a^{2}\right)^{\frac{5}{2}}} \ .
\end{eqnarray}
Now a necessary condition for the NEC (null energy condition) to hold is that both $\rho + p_\parallel\geq 0$ and $\rho+p_\perp\geq 0$  for all $r$, $a$, $m$.  
It is sufficient to consider:

\begin{eqnarray}
        \rho+p_\parallel  &=& \quad \frac{1}{8\pi }
        \left\lbrace - \frac{a^2\left(\sqrt{r^2+a^2}-4m\right)}{\left(r^2+a^2\right)^{\frac{5}{2}}}-\frac{a^2}{\left(r^2+a^2\right)^{2}}\right\rbrace 
        \nonumber \\
        &=&  \frac{-a^2 (\sqrt{r^2+a^2}-2m) }{4\pi \left(r^2+a^2\right)^{\frac{5}{2}}} \ .
\end{eqnarray}
Assuming $\sqrt{r^2+a^2}>2m$, this is manifestly negative for all values of $a$ and $m$ in our domain, and the NEC is clearly violated.

Note that for $\sqrt{r^2+a^2}<2m$, that is, \emph{inside} any horizon that may potentially be present, the $t$ and $r$ coordinates swap their timelike/spacelike characters and one has $\rho = - T_r{}^r$ while $p_\parallel = T_t{}^t$ and $p_\perp =  T_\theta{}^\theta = T_\phi{}^\phi$. So inside the horizon:

\begin{eqnarray}
\rho &=& \frac{a^{2}}{8\pi G_{N}\left(r^{2}+a^{2}\right)^{2}} \ ; \nonumber \\
p_\parallel &=&  \frac{a^{2}\left(\sqrt{r^{2}+a^{2}}-4m\right)}{8\pi G_{N}\left(r^{2}+a^{2}\right)^{\frac{5}{2}}} \ ,
\end{eqnarray}
and:

\begin{eqnarray}
        \rho+p_\parallel &=&  \frac{a^2 (\sqrt{r^2+a^2}-2m) }{4\pi G_{N}\left(r^2+a^2\right)^{\frac{5}{2}}} \ .
\end{eqnarray}
But since we are now working in the region $\sqrt{r^2+a^2}<2m$ this is again negative, and the NEC is again clearly violated. We can summarize this by stating:

\begin{eqnarray}
        \rho+p_\parallel &=&  - \frac{a^2\; |\sqrt{r^2+a^2}-2m| }{4\pi G_{N}\left(r^2+a^2\right)^{\frac{5}{2}}}\ ,
\end{eqnarray}
which now holds for all values of $r$ and is negative everywhere except \emph{on} any horizon that may potentially be present. 

Demonstrating that the NEC is violated is sufficient to conclude that the weak, strong, and dominant energy conditions shall also be violated~\cite{LorentzianWormholes}. We therefore have a spacetime geometry that accurately models that of a regular black hole or a traversable wormhole depending on the value of $a$, but clearly violates all of the classical energy conditions associated with the stress-energy-momentum tensor~\cite{Kar:2004, Molina-Paris:1998, Visser:cosmo1999, Barcelo:2000b, Visser:1999-super, Visser:1998-super, Abreu:2008, Abreu:2010, LNP, Martin-Moruno:2013a, Martin-Moruno:2013b,Martin-Moruno:2015, Martin-Moruno:2015b, Visser:1994, Visser:1996a, Visser:1996b, Visser:1997-ec}.

%==============================
\section{Surface gravity and Hawking temperature}\label{sec:3+1}
%==============================

Let us now calculate the surface gravity at the event horizon for the regular black hole case when $a\in(0,2m]$. The Killing vector which is null at the event horizon is $\xi^{\mu}=\partial_{t}$. This yields the following norm:

\begin{equation}
    \xi^{\mu}\xi_{\mu} = g_{\mu\nu}\xi^{\mu}\xi^{\nu} = g_{tt} = -\left(1-\frac{2m}{\sqrt{r^2+a^2}}\right) \ .
\end{equation}\clearpage
\noindent Then we have the following relation for the surface gravity $\kappa$ (see for instance~\cite{Wald, telebook, largescale}):

\begin{equation}
    \nabla_{\nu}\left(-\xi^{\mu}\xi_{\mu}\right) = 2\kappa\xi_{\nu} \ .
\end{equation}
That is:

\begin{equation}
 \nabla_{\nu}\left(1-\frac{2m}{\sqrt{r^2+a^2}}\right) = 2\kappa\xi_{\nu} \ ;
\end{equation}
Keeping in mind that the event horizon is located at radial coordinate $r=\sqrt{(2m)^2-a^2}$ we see:

\begin{equation}
 \kappa = \frac{\partial_{r}}{2}\left(1-\frac{2m}{\sqrt{r^2+a^2}}\right)\Bigg\vert_{r=\sqrt{(2m)^2-a^2}} = \frac{\sqrt{(2m)^2-a^2}}{8m^2} 
 = \kappa_\mathrm{Sch}\; \sqrt{1- {a^2\over(2m)^2}} \ .
\end{equation}
As a consistency check it is easily observed that for the Schwarzschild case when $a=0$, we have $\kappa=\frac{1}{4m}$, which is the expected surface gravity for the Schwarzschild black hole. For $a=2m$ the null horizon (one-way throat) is seen to be extremal.  It now follows that the temperature of Hawking radiation for our regular black hole is as follows (see for instance~\cite{Wald, telebook, largescale}):

\begin{eqnarray}
&& T_{H} = \frac{\hslash\kappa}{2\pi k_{B}}  = \frac{\hslash\sqrt{(2m)^2-a^2}}{16\pi k_{B}m^2} = T_{H,\mathrm{Sch}} \; \sqrt{1- {a^2\over(2m)^2}} \ .
\end{eqnarray}

%==============================
\section{ISCO and photon sphere analysis}\label{sec:isco}
%==============================

Let us now find the location of both the photon sphere for massless particles and the ISCO for massive particles as functions of $m$ and $a$.

Consider the tangent vector to the worldline of a massive or massless particle, parameterized by some arbitrary affine parameter, $\lambda$:

\begin{equation}
    g_{\mu\nu}\frac{dx^{\mu}}{d\lambda}\frac{dx^{\nu}}{d\lambda}=-g_{tt}\left(\frac{dt}{d\lambda}\right)^{2}+g_{rr}\left(\frac{dr}{d\lambda}\right)^{2}+\left(r^{2}+a^{2}\right)\left\lbrace\left(\frac{d\theta}{d\lambda}\right)^{2}+\sin^{2}\theta \left(\frac{d\phi}{d\lambda}\right)^{2}\right\rbrace \ .
\end{equation}
We may, without loss of generality, separate the two physically interesting cases (timelike and null) by defining:

\begin{equation}
    \epsilon = \left\{
    \begin{array}{rl}
    -1 & \qquad\mbox{massive particle, \emph{i.e.} timelike worldline} \ ; \\
     0 & \qquad\mbox{massless particle, \emph{i.e.} null worldline} \ .
    \end{array}\right. 
\end{equation}
That is, $ds^{2}/d\lambda^2=\epsilon$. Due to the metric being spherically symmetric we may fix $\theta=\frac{\pi}{2}$ arbitrarily and view the reduced equatorial problem:

\begin{equation}
    g_{\mu\nu}\frac{dx^{\mu}}{d\lambda}\frac{dx^{\nu}}{d\lambda}=-g_{tt}\left(\frac{dt}{d\lambda}\right)^{2}+g_{rr}\left(\frac{dr}{d\lambda}\right)^{2}+\left(r^{2}+a^{2}\right)\left(\frac{d\phi}{d\lambda}\right)^{2}=\epsilon \ .
\end{equation}
The Killing symmetries yield the following expressions for the conserved energy $E$ per unit mass and angular momentum $L$ per unit mass (see for instance~\cite{Wald, telebook, largescale}):

\begin{equation}
    \left(1-\frac{2m}{\sqrt{r^{2}+a^{2}}}\right)\left(\frac{dt}{d\lambda}\right)=E \ ; \qquad\quad \left(r^{2}+a^{2}\right)\left(\frac{d\phi}{d\lambda}\right)=L \ .
\end{equation}
Hence:

\begin{equation}
    \left(1-\frac{2m}{\sqrt{r^{2}+a^{2}}}\right)^{-1}\left\lbrace -E^{2}+\left(\frac{dr}{d\lambda}\right)^{2}\right\rbrace+\frac{L^{2}}{r^{2}+a^{2}}=\epsilon \ ;
\end{equation}
implying:

\begin{equation}
\left(\frac{dr}{d\lambda}\right)^{2}=E^{2}+\left(1-\frac{2m}{\sqrt{r^{2}+a^{2}}}\right)\left\lbrace\epsilon-\frac{L^{2}}{r^{2}+a^{2}}\right\rbrace \ .
\end{equation}
This gives `effective potentials' for geodesic orbits as follows:

\begin{equation}
    V_{\epsilon}(r)=\left(1-\frac{2m}{\sqrt{r^{2}+a^{2}}}\right)\left\lbrace -\epsilon+\frac{L^{2}}{r^{2}+a^{2}}\right\rbrace \ .
\end{equation}

\begin{itemize}
    \item For a photon orbit we have the massless particle case $\epsilon=0$. Since we are in a spherically symmetric environment, solving for the locations of such orbits amounts to finding the coordinate location of the `photon sphere'. That is, the value of the $r$-coordinate sufficiently close to our central mass such that photons are forced to propagate along circular geodesic orbits. These circular orbits occur at $V_{0}^{'}(r)=0$.  That is:
    
    \begin{equation}
        V_{0}(r)=\left(1-\frac{2m}{\sqrt{r^{2}+a^{2}}}\right)\left(\frac{L^{2}}{r^{2}+a^{2}}\right) \ ,
    \end{equation}
    leading to:
    
    \begin{equation}
        V_{0}^{'}(r)=\frac{2rL^{2}}{\left(r^{2}+a^{2}\right)^{\frac{5}{2}}}\bigg\lbrace 3m-\sqrt{r^{2}+a^{2}}\bigg\rbrace \ .
    \end{equation}
    When $V_{0}^{'}(r)=0$, if we discount the solution $r=0$ (as this spherical surface is clearly invalid for the location of the photon sphere), this gives the location of these circular orbits as $r=\pm\sqrt{(3m)^{2}-a^{2}}$. Firstly note that if $a\in(0,2m]$, $(3m)^{2}>a^{2}\ \forall \ a$, hence this solution does in fact correspond to a real-valued $r$-coordinate within our domain. Hence the photon sphere in our universe (\emph{i.e.} taking positive solution) for the case when the geometry is a regular black hole has coordinate location $r=\sqrt{(3m)^{2}-a^{2}}$. It also follows that in the case when $a>2m$ and we have a traversable wormhole, since we have strictly defined our $r$-coordinate to take on real values, there exists a photon sphere location in our universe only for the case when $2m<a<3m$. When $a>3m$ we have no photon sphere.
    To verify stability, check the sign of $V_{0}^{''}(r)$:
    
    \begin{equation}
        V_{0}^{''}(r)=\frac{2L^{2}}{\left(r^{2}+a^{2}\right)^{\frac{7}{2}}}\Bigg\lbrace \sqrt{r^{2}+a^{2}}\left(3r^2-a^2\right)-3m\left(4r^2-a^2\right)\Bigg\rbrace \ .
    \end{equation}
    For ease of notation let us first establish that when $r=\sqrt{(3m)^{2}-a^{2}}$, then $r^{2}+a^{2}=(3m)^{2}$, hence it can be shown that:
    
    \begin{equation}
        V_{0}^{''}\left(r=\sqrt{(3m)^{2}-a^{2}}\right)=\frac{-2L^{2}}{(3m)^{6}} \left((3m)^{2}-a^{2}\right)<0 \ .
    \end{equation}
    
    Now $V^{''}_{0}<0$ implies instability, hence there is an unstable photon sphere at $r=\sqrt{(3m)^{2}-a^{2}}$ as presumed. For the Schwarzschild solution the location of the unstable photon sphere is at $r=3m$; which provides a useful consistency check.
      
    \item For massive particles the geodesic orbit corresponds to a timelike worldline and we have the case that $\epsilon=-1$. Therefore:
    
    \begin{equation}
        V_{-1}(r)=\left(1-\frac{2m}{\sqrt{r^{2}+a^{2}}}\right)\left(1+\frac{L^{2}}{r^{2}+a^{2}}\right) \ ,
    \end{equation}
    and it is easily verified that this leads to:
    
    \begin{equation}
        V_{-1}^{'}(r)=\frac{2r}{\left(r^{2}+a^{2}\right)^{\frac{5}{2}}}\bigg\lbrace L^{2}\left(3m-\sqrt{r^{2}+a^{2}}\right)+m\left(r^{2}+a^{2}\right)\bigg\rbrace \ .
    \end{equation}
    Equating this to zero and rearranging for $r$ gives a messy solution for $r$ as a function of $L$, $m$ and $a$. Instead it is preferable to assume a fixed circular orbit at some $r=r_{c}$, and rearrange the required angular momentum $L_{c}$ to be a function of $r_{c}$, $m$, and $a$. It then follows that the innermost circular orbit shall be the value of $r_{c}$ for which $L_{c}$ is minimised.\clearpage % (since $L_{c}\sim r_{c}$ \red {(ref)}).
    
    \noindent Hence if $V_{-1}^{'}(r_{c})=0$, we have:  
    
    \begin{equation}
        L_{c}^{2}\left(3m-\sqrt{r_{c}^{2}+a^{2}}\right)+m\left(r_{c}^{2}+a^{2}\right)=0 \ ,
    \end{equation}
    implying:
    
    \begin{equation}
         L_{c}\left(r_{c}, m, a\right)=\sqrt{\frac{m\left(r_{c}^{2}+a^{2}\right)}{\sqrt{r_{c}^{2}+a^{2}}-3m}} \ ,
    \end{equation}
    As a consistency check, for large $r_{c}$ (\emph{i.e.} $r_{c}>>a$) we observe that $L_{c}\approx\sqrt{mr_{c}}$, which is consistent with the expected value when considering circular orbits in weak-field general relativity.
    
    \noindent It is then easily obtained that:
    
    \begin{equation}
        \frac{\partial L_{c}}{\partial r_{c}}=\left(\frac{\sqrt{m}r_{c}}{2\sqrt{\sqrt{r_{c}^{2}+a^{2}}-3m}}\right)\left(\frac{2}{\sqrt{r_{c}^{2}+a^{2}}}-\frac{1}{\sqrt{r_{c}^{2}+a^{2}}-3m}\right) \ .
    \end{equation}
    Solving for stationary points, and excluding $r_{c}=0$ (as this lies within the photon sphere, which is clearly an invalid solution for the ISCO of a massive particle):
    
    \begin{equation}
        \sqrt{r_{c}^{2}+a^{2}}-6m=0 \ ; \qquad \Longrightarrow\quad r_{c}=\sqrt{(6m)^{2}-a^{2}} \ ,
    \end{equation}
    (once again, discounting the negative solution for $r_{c}$ in the interests of remaining in our own universe). We therefore have a coordinate ISCO location at $r_{c}=\sqrt{(6m)^{2}-a^{2}}$. This is consistent with the expected value ($r=6m$) for Schwarzschild, when $a=0$. For our traversable wormhole geometry, provided $2m<a<6m$ we will have a valid ISCO location in our coordinate domain. When $a>6m$, we have a traversable wormhole with no ISCO.
\end{itemize}

\enlargethispage{10pt}
Denoting $r_{H}$ as the location of the horizon, $r_{\scriptscriptstyle{Photon}}$ as the location of the photon sphere, and $r_{\scriptscriptstyle{ISCO}}$ as the location of the ISCO, we have the following summary:

\begin{itemize}
    \item $r_{\scriptscriptstyle{H}}=\sqrt{(2m)^{2}-a^2}$ ;
    \item $r_{\scriptscriptstyle{Photon}}=\sqrt{(3m)^{2}-a^2}$ ;
    \item $r_{\scriptscriptstyle{ISCO}}=\sqrt{(6m)^{2}-a^2}$ .
\end{itemize}\vfill

%==============================
\section{Regge-Wheeler analysis}\label{sec:regge-wheeler}
%==============================

Considering the Regge-Wheeler Equation in view of the formalism developed in~\cite{ReggeWheeler1}, (see also reference~\cite{Expmetric}), we may explicitly evaluate the Regge-Wheeler potentials for particles of spin $S\in\lbrace 0,1\rbrace$ in our spacetime. 
Firstly define a tortoise coordinate as follows:

\begin{equation}
dr_{*} = \frac{dr}{\left(1-\frac{2m}{\sqrt{r^2+a^2}}\right)} \ ,
\end{equation}
which gives the following expression for the metric Eq.~\ref{RBHmetric}:

\begin{equation}
    ds^2 = \left(1-\frac{2m}{\sqrt{r^2+a^2}}\right)\bigg\lbrace -dt^2+dr_{*}^2\bigg\rbrace+\left(r^2+a^2\right)\left(d\theta^2+\sin^2\theta\;\ d\phi^2\right) \ .
\end{equation}
It is convenient to write this as:

\begin{equation}
    ds^2 = A(r_*)^2\bigg\lbrace -dt^2+dr_{*}^2\bigg\rbrace+B(r_*)^2\left(d\theta^2+\sin^2\theta \;d\phi^2\right) \ .
\end{equation}
The Regge--Wheeler equation is \cite{ReggeWheeler1}:

\begin{equation}
    \partial_{r_{*}}^{2}\hat{\phi}+\lbrace \omega^2-\mathcal{V}_S\rbrace\hat\phi = 0 \ ,
\end{equation}
where $\hat\phi$ is the scalar or vector field, $\mathcal{V}$ the spin-dependent Regge-Wheeler potential for our test particle, and $\omega$ is some temporal frequency component in the Fourier domain.
For a scalar field ($S=0$) examination of the d'Alembertian equation quickly yields:

\begin{equation}
\mathcal{V}_{S=0} =   \left\lbrace{A^2 \over B^2} \right\rbrace \ell(\ell+1)
+ {\partial_{r_{*}}^2 B \over B} \ .
\end{equation}
For a vector field ($S=1$) conformal invariance in `3+1'-dimensions guarantees that the Regge-Wheeler potential can depend only on the ratio $A/B$, whence normalizing to known results implies:

\begin{equation}
\mathcal{V}_{S=1} =   \left\lbrace{A^2 \over B^2} \right\rbrace \ell(\ell+1)\ .
\end{equation}
Collecting results, for $S\in\{0,1\}$ we have:

\begin{equation}
\mathcal{V}_{S} =   \left\lbrace{A^2 \over B^2} \right\rbrace \ell(\ell+1)
+ (1-S) {\partial_{r_{*}}^2 B \over B} \ .
\end{equation}
The spin 2 axial mode is somewhat messier, and not of immediate interest.\clearpage

Noting that for our metric: $\partial_{r_{*}}=\left(1-\frac{2m}{\sqrt{r^2+a^2}}\right)\partial_{r}$, and $B=\sqrt{r^2+a^2}$,  we have:

\begin{equation}
    \frac{\partial_{r_{*}}^2 B}{B}=\left\lbrace 1-\frac{2m}{\sqrt{r^2+a^2}}\right\rbrace \; \left(2m(r^2-a^2) +a^2\sqrt{r^2+a^2} \over (r^2+a^2)^{5/2}\right) \ .
\end{equation}
Therefore:

\begin{eqnarray}
\mathcal{V}_{S\in\{0,1\}} &=&  \left(1-\frac{2m}{\sqrt{r^2+a^2}}\right) \Bigg\lbrace{\ell(\ell+1)\over r^2+a^2} \nonumber \\
&& \nonumber \\
&& \qquad \qquad \qquad \qquad + (1-S) \left(2m(r^2-a^2) +a^2\sqrt{r^2+a^2} \over (r^2+a^2)^{5/2}\right)\Bigg\rbrace \ . \nonumber \\
&& \nonumber \\
&&
\end{eqnarray}
This has the correct behaviour as $a\to0$. Note that this Regge-Wheeler potential is symmetric about $r=0$. 
For $a<2m$ the situation is qualitatively similar to the usual Schwarzschild case (the tortoise coordinate diverges at either horizon, and $\mathcal{V}_{S\in\{0,1\}} \to 0$ at either horizon).  For $a=2m$ the tortoise coordinate diverges at the extremal horizon (one-way null throat), while we still have $\mathcal{V}_{S\in\{0,1\}} \to 0$. For $a>2m$ the tortoise coordinate converges at the wormhole throat, while we now have have $\mathcal{V}_{S\in\{0,1\}}$ nonzero and positive at the throat:

\begin{equation}
\mathcal{V}_{S\in\{0,1\}} \to  \left(1-\frac{2m}{a}\right) \left\lbrace{\ell(\ell+1)\over a^2} 
+ (1-S) \left(a-2m \over a^3\right) \right\rbrace \ .
\end{equation}

%%%%%%

\section{Overview}

We have analysed a candidate spacetime which models qualitatively different geometries in an $a$-dependent manner with minimal perversion to the Schwarzschild solution (with $a$ being the newly introduced scalar parameter in the metric). When $a\in\left(0, 2m\right)$ the geometry models a regular black hole (in the sense of Bardeen) possessing a spacelike `bounce' at $r=0$, when $a=2m$ we have an extremal null bounce corresponding to a one-way traversable wormhole, and for $a>2m$ we have a standard two-way traversable wormhole in the sense of Morris and Thorne. This metric candidate therefore neatly interpolates between the classes of spacetime which are of primary interest in this thesis, and it does so in a neat and tractable manner, broadening the class of `regular black holes' and `traversable wormholes' beyond those typically considered. Accordingly, further analysis of this candidate spacetime is prudent, and we immediately look to extend the discussion by imposing a time dependence on the metric in \S\ref{C:Vaidya}.

%%%%%%%%%%%%%%%%%%%%%%%%%%%%%%%%%%%%%%%%%%%%%%%%%%%%%%

%%%%%%%%%%%%%%%%%%%%%%%%%%%%%%%%%%%%%%%%%%%%%%%%%%%%%%%

\chapter {Beyond the static case: Introducing Vaidya spacetimes}\label{C:Vaidya}

%=====================================================
\section{Introduction}
%====================================================
\enlargethispage{20pt}
In \S\ref{C:Black-bounce} (also see reference~\cite{black-bounce}) the following static spacetime was considered:

\begin{equation}
    ds^{2}=-\left(1-\frac{2m}{\sqrt{r^{2}+a^{2}}}\right)dt^{2}+\frac{dr^{2}}{1-\frac{2m}{\sqrt{r^{2}+a^{2}}}}+\left(r^{2}+a^{2}\right)\left(d\theta^{2}+\sin^{2}\theta d\phi^{2}\right).
\end{equation}
Adjusting the parameter $a$, assuming without loss of generality that $a>0$, following the analysis of \S\ref{C:Black-bounce}, this metric represents either:

\begin{itemize}
\itemsep-3pt
\item 
The ordinary Schwarzschild spacetime ($a=0$) ; 
\item
A `black-bounce' with a one-way spacelike throat  ($a<2m$) ;
\item
A one-way wormhole with a null throat ($a=2m$); or
\item
A traversable wormhole in the Morris-Thorne sense ($a>2m$) .
\end{itemize}

Now let us look to explore a (relatively) \emph{tractable} way of adding time-dependence to this spacetime.
We start by re-writing the static spacetime in Eddington--Finkelstein coordinates:

\begin{equation}
ds^{2}=-\left(1-\frac{2m}{\sqrt{r^{2}+a^{2}}}\right)dw^{2}-( \pm 2 \,dw \,dr)
+\left(r^{2}+a^{2}\right)\left(d\theta^{2}+\sin^{2}\theta \;d\phi^{2}\right) \ .
\end{equation}
Here $w=\{u,v\}$ is the $\{outgoing,ingoing\}$ null time coordinate, that is it represents $\{retarded,advanced\}$ time. 
Here the upper + sign corresponds to $u$, and the lower $-$ sign corresponds to $v$.
We now invoke a Vaidya-like trick~\cite{Vaidya:1951a,Vaidya:1951b,Vaidya:1999a,Vaidya:1970, Wang:1998, Parikh:1998}, by allowing the mass parameter $m(w)$ to depend on the null time coordinate.\clearpage

\noindent That is we consider the spacetime described by the metric:

\begin{equation}\label{E:dmetric}
ds^{2}=-\left(1-\frac{2m(w)}{\sqrt{r^{2}+a^{2}}}\right)dw^{2}-(\pm 2 \,dw \,dr)
+\left(r^{2}+a^{2}\right)\left(d\theta^{2}+\sin^{2}\theta \;d\phi^{2}\right).
\end{equation}
When $a\to0$ this is just the standard Vaidya spacetime~\cite{Vaidya:1951a,Vaidya:1951b,Vaidya:1999a,Vaidya:1970, Wang:1998, Parikh:1998} (either a `shining star' or a star accreting a flux of infalling null dust).
This metric can be used either to study the collapse of null dust, or the semiclassical evaporation of black holes.
When the parameter $m(w)\to m$ is a constant we just have the static `black-bounce'/traversable wormhole of \S\ref{C:Black-bounce}. 
The point of introducing time-dependence in this precise manner is to keep calculations algebraically tractable; and so provide a simple model of an evolving regular black hole. Another considerably less tractable option, which will not be explored in this thesis, would consist of promoting the parameter $a$ to $a(w)$, with $m$ either kept constant or not.

So it is natural to argue that, on one hand, for an increasing function $m(v)$ crossing the $a/2$ limit, the spacetime metric Eq.~\ref{E:dmetric} describes the conversion of a wormhole into a regular black hole by the accretion of null dust. On the other hand, for a decreasing  function $m(u)$ crossing the $a/2$ limit, the situation will correspond to the evaporation of a regular black hole leaving a wormhole remnant. 
Moreover, this may be related to the more-or-less equivalent process of phantom energy accretion onto black holes, which should, however, be studied considering negative energy and using the ingoing null coordinate $v$ (for related discussion see~\cite{Babichev:2004yx,Babichev:2004qp,MartinMoruno:2006mi,GonzalezDiaz:2007gt,Martin-Moruno:2007,Madrid:2010}). Finally, it is worth noticing that one can describe the transmutation of a regular black hole into a wormhole and \emph{vice versa} in this classical description only because the black hole is regular and, therefore, there is no topology change. 
It should be noted that `black-bounce' models have recently become quite popular, though more typically for bounces back into our own universe (see for instance references~\cite{Barcelo:2014, Barcelo:2014b, Barcelo:2015, Barcelo:2016, Garay:2017, Rovelli:2014, Haggard:2015, Christodoulou:2016, DeLorenzo:2015, Malafarina:2017, Olmedo:2017, Barrau:2018, Malafarina:2018}). Not all of these bounce models are entirely equivalent, either to each other or to the bounce scenarios of this current analysis. Let us now investigate whether the above mentioned physical scenarios can actually be described by the metric Eq.~\ref{E:dmetric}, and subsequently analyse interesting physical characteristics of this geometry.

%=====================================================
\section{Geometric basics}\label{S:basics}
%=====================================================
Consider the metric:

\begin{equation}\label{E:VBBmetric}
ds^{2}=-\left(1-\frac{2m(w)}{\sqrt{r^{2}+a^{2}}}\right)dw^{2}-(\pm 2 \,dw \,dr)
+\left(r^{2}+a^{2}\right)\left(d\theta^{2}+\sin^{2}\theta \;d\phi^{2}\right) \ ,
\end{equation}
where the coordinates have natural domains:

\begin{equation}
w\in(-\infty,+\infty);\qquad
r\in(-\infty,+\infty);\qquad 
\theta\in [0,\pi];\qquad
\phi\in(-\pi,\pi] \ .
\end{equation}
Here $w=\{u,v\}$ denotes the $\{outgoing,ingoing\}$ null time coordinate, and $\pm\to+$ for $u$, while $\pm\to-$ for $v$.\newline

\noindent The radial null curves are found by setting:

\begin{equation}
0=ds^2 = dw \left[ \left(1-\frac{2m(w)}{\sqrt{r^{2}+a^{2}}}\right)dw \pm 2\,dr\right] \ , 
\end{equation}
corresponding to:

\begin{equation}
dw=0 \qquad \hbox{and} \qquad  dr = \mp {1\over 2} \left(1-\frac{2m(w)}{\sqrt{r^{2}+a^{2}}}\right)dw \ ,
\end{equation}
and the associated radial null vectors are proportional to:

\begin{equation}
k^\mu = (0,1,0,0) \qquad \hbox{and} \qquad k^\mu = \left( 1, \mp{1\over2}\left(1-\frac{2m(w)}{\sqrt{r^{2}+a^{2}}}\right) , 0,0\right) \ ,
\end{equation}
respectively. 
For tangential null curves (that is, $dr=0$) we can without any loss of generality set $\phi=0$ and concentrate on:

\begin{equation}
0=ds^2 = - \left(1-\frac{2m(w)}{\sqrt{r^{2}+a^{2}}}\right)dw^2 +  (r^2+a^2) d\theta^2 \ , 
\end{equation}
for which the associated tangential null vectors (defined only for \newline $\sqrt{r^2+a^2}\geq 2 m(w)$) are proportional to:

\begin{equation}
 k^\mu = \left( \sqrt{r^2+a^2},0, \sqrt{1-\frac{2m(w)}{\sqrt{r^{2}+a^{2}}}} ,0\right) \ .
\end{equation}
Similarly to the static case analysed in \S\ref{C:Black-bounce}, we can define a `coordinate speed of light' that is equal to ${\rm d}r/{\rm d}w$. If  $2m(w)>a$, this quantity vanishes at:

\begin{equation}
r_{AH}(w)=\pm\sqrt{(2m(w))^2-a^2} \ ,
\end{equation}
so we have a dynamical apparent horizon. The existence of a future/past event horizon depends on the presence or absence of an apparent horizon in the limit $t\rightarrow\pm\infty$.
We already know from the static case that there is a throat/bounce hypersurface at $r=0$. At this hypersurface the induced three-metric is:

\begin{equation}\label{E:VBBthroat}
ds|_\Sigma^{2}=-\left(1-\frac{2m(w)}{a}\right)dw^{2}
+a^{2}\left(d\theta^{2}+\sin^{2}\theta \;d\phi^{2}\right) \ .
\end{equation}
Geometrically, this induced three-geometry is always a cylinder, though potentially of variable signature.
Specifically this $r=0$ hypersurface is timelike if $2m(w)/a<1$, null (lightlike) if $2m(w)/a=1$, and spacelike if $2m(w)/a>1$. These correspond to a traversable wormhole throat, a one-way null throat, or a `black-bounce' respectively, where now (as opposed to the static discussion of \S\ref{C:Black-bounce}) the nature of the throat can change in a $w$-dependent manner.
Because of this feature, the relevant Carter-Penrose diagrams will thus depend on the entire history of the ratio $2m(w)/a$ over the entire domain $w\in(-\infty,+\infty)$. Since the Carter-Penrose diagrams are constructed to exhibit intrinsically global causal structure, to determine them one needs global information regarding $2m(w)/a$.

%====================================================
\section{Einstein tensor and energy conditions}\label{S:einstein}
%====================================================

In Eddington-Finkelstein coordinates, as long as $a\neq 0$, both the metric $g_{\mu\nu}$ and the inverse metric $g^{\mu\nu}$ have finite components for all values of $r$.
Moreover, as it was shown in detail for the static case, and as we shall analyse for the dynamical case, all the curvature tensors (Riemann, Weyl, Ricci, Einstein) have finite components for all values of $r$.
Consequently, even for a time-dependent $m(w)$ one still has a regular spacetime geometry -- there are no curvature singularities.

We discuss here in some detail the results for the Einstein tensor, since it is strongly related with the stress-energy tensor in general relativity. 
The Einstein tensor has non-zero components:

\begin{eqnarray}
G_{ww} &=& \mp{2r\;\dot m(w)\over (r^2+a^2)^{3/2}} -
{a^2\left\{1-{2m(w)\over\sqrt{r^2+a^2}}\right\} \left\{ 1-{4m(w)\over\sqrt{r^2+a^2}}\right\} \over(r^2+a^2)^{2} } \ ; \nonumber \\
&& \nonumber \\
G_{wr} &=&
\mp a^2 {\sqrt{r^2+a^2}-4m(w)\over(r^2+a^2)^{5/2}} \ ; \nonumber \\
&& \nonumber \\
G_{rr} &=& {-2a^2 \over(r^2+a^2)^2} \ ; \nonumber \\
&& \nonumber \\
G_{\theta\theta} &=& 
+{a^2(\sqrt{r^2+a^2}-m(w))\over (r^2+a^2)^{3/2}}
= {G_{\phi\phi}\over \sin^2\theta} \ .
\end{eqnarray}
with $\dot m(w)=dm/dw$. Note that the derivative term $\dot m(w)$  only shows up linearly, and only in a very restricted way. 
In fact we can write:

\begin{equation}\label{Gs}
G_{\mu\nu} = G_{\mu\nu}^{nonderivative} \mp{2r\;\dot m(w)\over (r^2+a^2)^{3/2}} \; (dw)_\mu (dw)_\nu \ ,
\end{equation}
where we remind the reader that the upper $-$ sign corresponds to the outgoing coordinate $u$ and the lower $+$ sign
to the ingoing coordinate $v$.
It is interesting to underline that the derivative term is precisely the only term present in the pure Vaidya case where $a=0$. Note that $G_{\mu\nu}\propto T_{\mu\nu}$. So, it is like we were considering a flux equivalent to that of the Vaidya geometry on top of the (now dynamical) fluid that generates the static spacetime. 
It is in this sense that discussion ensues on the existence of a null flux proportional to $\dot m(w)$ in the dynamical region of the geometry in  \S\ref{S:models}.

Now, let us consider the nature of the matter content generating these geometries. 
We already know that the material supporting the static geometry, with $m(w)=m$, violates the null energy condition (NEC). 
This condition is a necessary requirement for forcing all timelike observers to see non-negative energy densities.
As the NEC is used in the singularity theorems to assure convergence of geodesics in general relativity, one should already expect to have some violations in wormholes, where the throat has to flare out, or in `black bounces', which avoid the formation of singularities~\cite{twisted,Kar:2004, Molina-Paris:1998, Visser:cosmo1999, Barcelo:2000b, Visser:1999-super, Visser:1998-super, Abreu:2008, Abreu:2010, LNP, Martin-Moruno:2013a, Martin-Moruno:2013b,Martin-Moruno:2015, Visser:1994, Visser:1996a, Visser:1996b, Visser:1997-ec}.

In the dynamical case, some results of the static geometry will be recovered, but there will also be some crucial differences. For the specific radial null vector \,$k^\mu = (0,1,0,0)$\, we have:

\begin{equation}\label{NEC1}
T_{\mu\nu} k^\mu k^\nu \; \propto \; G_{\mu\nu} k^\mu k^\nu = G_{rr} = - {2a^2\over(r^2+a^2)^2} \ . 
\end{equation}
This implies that in general relativity the stress-energy-momentum tensor is always NEC violating.
Although the result above is already enough to conclude the violation of the NEC, let us study other contractions in order to figure out the effect of having a non-constant mass. For the other radial null vector $k^\mu = \left( 1, \mp {1\over2}\left(1-\frac{2m(w)}{\sqrt{r^{2}+a^{2}}}\right) , 0,0\right)$, where the minus sign corresponds to $u$ and the plus sign to $v$, we have:

\begin{equation}
G_{\mu\nu} k^\mu k^\nu = - {a^2\left(\sqrt{r^2+a^2}-2m(w)\right)^2\over2(r^2+a^2)^3} \mp {2r\dot m(w)\over(r^2+a^2)^{3/2}} \ . 
\end{equation}
The non-derivative term is always NEC violating. The derivative term $\dot m(w)$ might or might not be NEC violating depending on sign. 
When considering ingoing radiation (described by $v$) the stress-energy-momentum tensor that can be constructed considering only the derivative term satisfies the NEC for non-decreasing $m(v)$. For outgoing radiation the situation is the opposite, so the NEC is satisfied by that flux for $\dot m(u)<0$.
Overall NEC violation in this particular direction would depend on relative magnitudes and signs. 

In contrast, for the transverse null vector:

\begin{equation}
    k^\mu = \left( \sqrt{r^2+a^2},0, \sqrt{1-\frac{2m(w)}{\sqrt{r^{2}+a^{2}}}} ,0\right) \ ,
\end{equation}
we have:

\begin{equation}
G_{\mu\nu} k^\mu k^\nu =   {3m(w)a^2(\sqrt{r^2+a^2}-2m(w))\over(r^2+a^2)^2}\mp {2r\dot m(w)\over\sqrt{r^2+a^2}}.
\end{equation}
The non-derivative term is now NEC satisfying for wormholes and outside the horizon of regular black holes. The derivative term $\dot m(w)$ might or might not be NEC violating depending on sign. Overall NEC violation in this particular direction would depend on relative magnitudes and signs. 
However, we emphasise that to violate the NEC it is sufficient  to have even one direction in which we have non-positive contraction $G_{\mu\nu} k^\mu k^\nu$. This certainly occurs for the radial direction; see Eq.~\ref{NEC1}.\newline

Summarising, we can write:

\begin{equation}
T_{\mu\nu} = T_{\mu\nu}^{nonderivative}+ T_{\mu\nu}^{derivative},\quad {\rm with}\quad T_{\mu\nu}^{derivative}\propto \mp\dot m(w)\; (dw)_\mu (dw)_\nu.
\end{equation}
Whereas $T_{\mu\nu}^{nonderivative}$ always violates the NEC in the radial direction; the flux described by $T_{\mu\nu}^{derivative}$ satisfies the NEC for ingoing radiation with $\dot m(v)\geq0$ and for outgoing radiation with  $\dot m(u)\leq 0$.

%%%%%%%%%%%%%%%%%%%%%%%%%%%%%%%%%%%%%%%%%%%%%%
\section{Physical models}\label{S:models}
%%%%%%%%%%%%%%%%%%%%%%%%%%%%%%%%%%%%%%%%%%%%%
Let us now analyse some particular evolutionary scenarios that can be described by the spacetime metric Eq.~\ref{E:dmetric}. In particular, focus is placed on several situations of direct physical interest first taking ingoing Eddington-Finkelstein coordinates and later outgoing Eddington-Finkelstein coordinates. We classify those scenarios as having ingoing or outgoing radiation, respectively, focusing attention on the $T_{\mu\nu}^{derivative}$ part of the stress-energy-momentum tensor, which is not present in the static case.

\subsection{Models with ingoing radiation}
%%%%%%%%%%%%%%%%%%%%%%%%%%%%%%%%%%%%%%%%%%%%%
Let us now focus on the metric with ingoing (advanced) Eddington-Finkel\-stein coordinates. That is:

\begin{equation}\label{E:dmetricv}
ds^{2}=-\left(1-\frac{2m(v)}{\sqrt{r^{2}+a^{2}}}\right)dv^{2}+ 2 \,dv \,dr
+\left(r^{2}+a^{2}\right)\left(d\theta^{2}+\sin^{2}\theta \;d\phi^{2}\right) \ .
\end{equation}
As is well known, in the standard Vaidya situation~\cite{Vaidya:1951a,Vaidya:1951b,Vaidya:1999a,Vaidya:1970, Wang:1998, Parikh:1998} (that is for $a=0$), this metric describes an ingoing null flux with $T_{vv}\propto 2\dot m(v)/r^2$. So, the black hole mass increases as a result of an ingoing flux with positive energy. 
When $a\neq0$, the geometry is generated by a non-vanishing stress-energy-momentum tensor even in the static case, $m(v)=m$. But, 
as we have discussed in the previous section, when one allows $m(v)$ to be a dynamical quantity, then an extra null flux term will appear in that tensor. That is:

\begin{equation}
T_{\mu\nu}^{derivative}\propto \dot m(v)\; (dv)_\mu (dv)_\nu \ .
\end{equation}
So, the derivative contribution to the null flux is positive for $\dot m(v)>0$ and negative for $\dot m(v)<0$.
In this case we can distinguish three different physically relevant situations. Denoting $m_0$ as the initial mass, two of them are characterised by $\dot m(v)>0$ and the last one by $\dot m(v)<0$. These three scenarios are:

%%%%%%%%%%%%%%%%%%%%%%%%%%%
\subsubsection{Growing `black-bounce' ($a<2m_0$)} 
%%%%%%%%%%%%%%%%%%%%%%%%%%%

\begin{figure}[!htb]
%\vspace{-1cm}
\begin{center}
\includegraphics[scale=0.90]{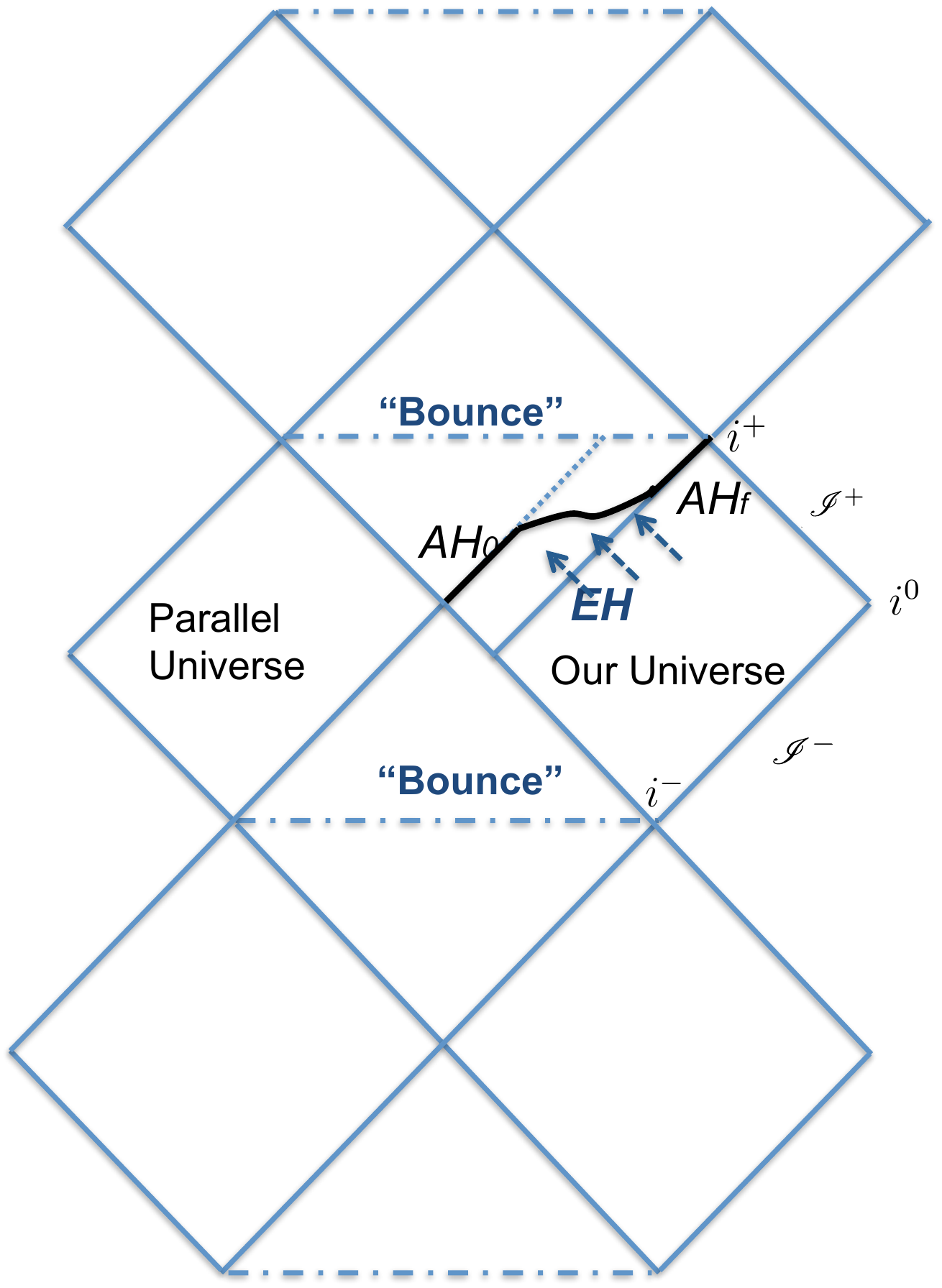}
\end{center}
\caption[Asymmetric Carter-Penrose diagram for a growing `black-bounce']{Carter-Penrose diagram for a growing `black-bounce'. There is positive radiation being accreted by the `black-bounce' for $0 < v < v_f$ (shown by arrows in the diagram). The apparent horizon evolves from $AH_0$ to $AH_f$. Note that before the influx of this radiation the diagram is symmetric; 
however, during accretion of the fluid by the `black-bounce' the diagram is asymmetric, and after the subsequent post-accretion bounce the diagram is again symmetric but shifted to the right.}
\label{F:1}
\end{figure}

\begin{figure}[!htb]
%\vspace{-1cm}
\begin{center}
\includegraphics[scale=0.90]{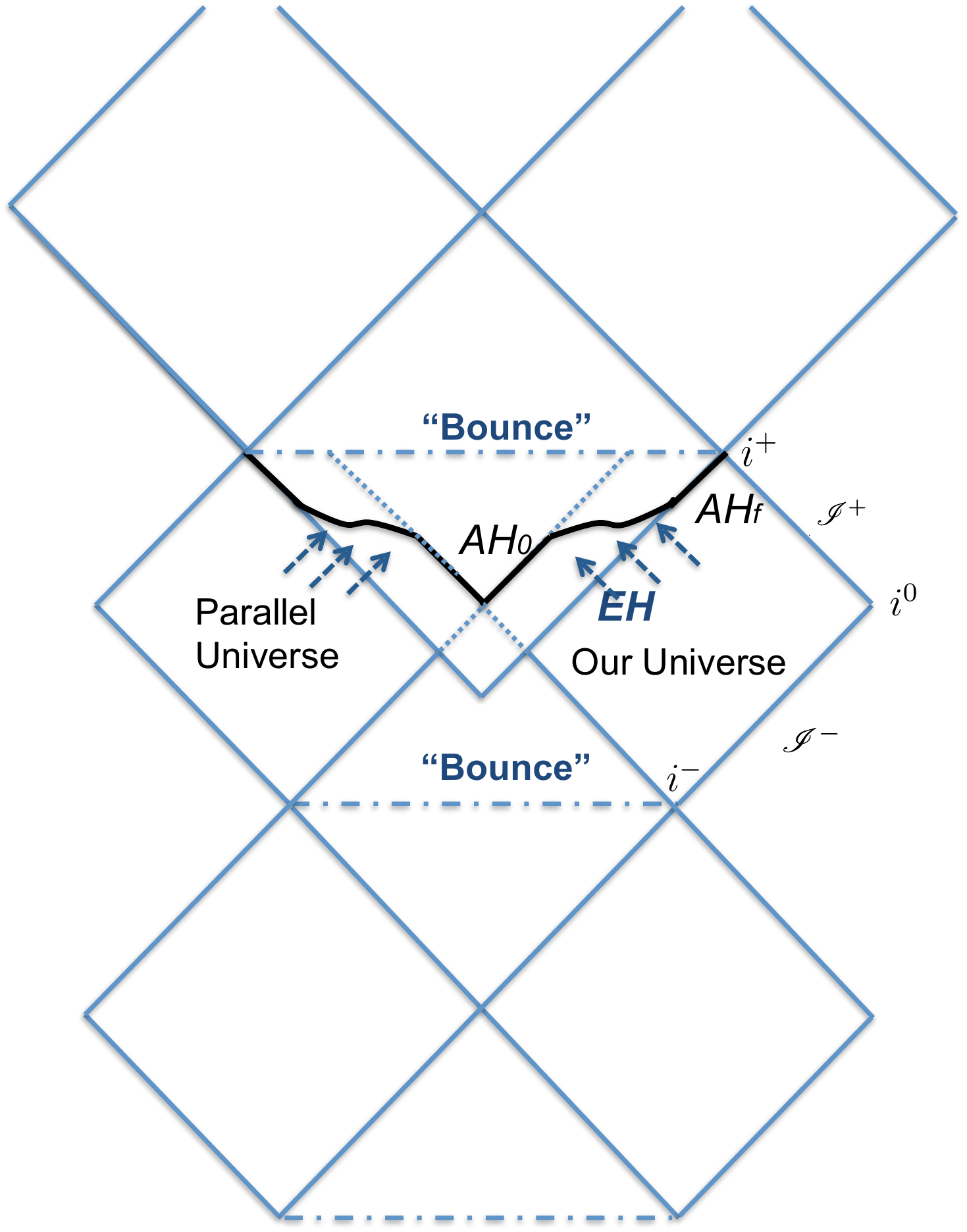}
\end{center}
\caption[Symmetric Carter-Penrose diagram for a growing `black-\newline bounce']{Carter-Penrose diagram for a growing `black-bounce'. We now restore the symmetric character of the diagram by assuming that, for some reason, there is also positive radiation being accreted by the `black-bounce' of the parallel universe for $0 < v < v_f$.}
\label{F:2}
\end{figure}

For an outside observer in our universe the initial situation will be similar to that for a black hole with an apparent horizon given by $r_{+0}=\sqrt{(2m_0)^2-a^2}$; however, the interior region will instead describe a bounce into another universe. Now, turn on an additional positive ingoing null flux by considering a non-constant increasing function $m(v)$. With the increase of $m(v)$, the radius of the apparent horizon will also increase, $r_+(v)$, leading to a bigger black object.
A particularly simple example is that of linear growth, given by:

\begin{equation}
m(v)= \left\{ \begin{array}{ll}
             m_0>a/2, &\qquad   v\leq 0 \ ; \\
             m_0+\alpha v, &\qquad 0 < v < v_f \ ; \\
             m_f=m_0+\alpha\, v_f, &\qquad v\geq v_f \ ,
             \end{array}
   \right.
\end{equation}
with $\alpha>0$. In this case, there is an apparent horizon at:

\begin{equation}
r_{+}(v)=\sqrt{(2m(v))^2-a^2} \ ,
\end{equation}
and an event horizon, which partially overlaps with the final apparent horizon, located at:

\begin{equation}
r_{+f}=\sqrt{4m_f^2-a^2} = \sqrt{(2m_0+2\alpha\, v_f)^2-a^2} \ .
\end{equation}
The Carter-Penrose diagram for this scenario can be seen in Fig.~\ref{F:1}, whereas in Fig.~\ref{F:2} we show the resulting spacetime if one considers that a similar flux is turned on in the parallel universe.
%=====================================================
%=====================================================
%=====================================================
\clearpage

%%%%%%%%%%%%%%%%%%%%%%%%%%%%%%%%%%
\subsubsection{Wormhole to `black-bounce' transition ($a>2m_0$)} 
%%%%%%%%%%%%%%%%%%%%%%%%%%%%%%%%%%

\begin{figure}[!htb]
%\vspace{-1cm}
\begin{center}
\includegraphics[scale=0.9]{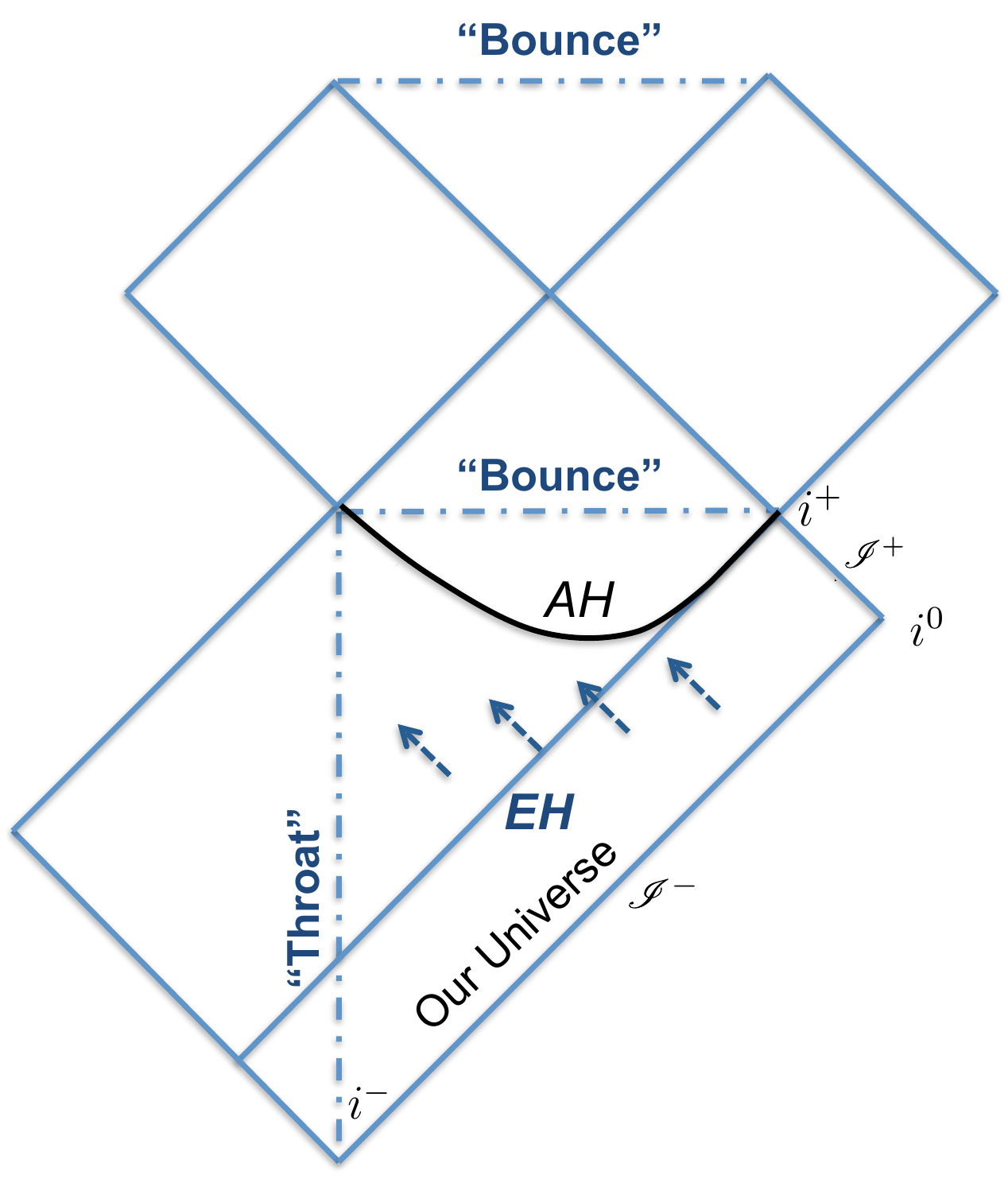}
\end{center}
\caption[Carter-Penrose diagram for a wormhole to `black-bounce' transition]{Carter-Penrose diagram for a wormhole to `black-bounce' transition. There is an incoming flux of positive radiation into the wormhole (depicted by arrows) that causes its transmutation into a `black-bounce'. That is, the timelike wormhole throat hypersurface becomes a spacelike `black-bounce' hypersurface, passing through being null at the point from which the apparent horizon emerges. Since there is a final apparent horizon, our universe would have an event horizon 
which cannot end at the throat (which is not a boundary of the spacetime) and, therefore, continues through the other universe.}
\label{F:3}
\end{figure}

In this case, the initial scenario will be that of a traversable Morris--Thorne wormhole (which could even have $m_0=0$). 
Now, we again turn on an additional ingoing flux with positive energy, by taking a non-constant increasing function $m(v)$. At first, this will have no effect in the causal properties of the geometry. But, if the increasing function $m(v)$ crosses the critical value $a/2$, then we will momentarily have a one-way wormhole, and then a regular black hole will form.  So sufficiently large ingoing positive null flux will lead to the transition from a wormhole to a regular black hole. As in the previous case, we could consider:

\begin{equation}
m(v)= \left\{ \begin{array}{ll}
             m_0<a/2, &\qquad   v\leq 0 \ ; \\
             m_0+\alpha v, &\qquad 0 < v < v_f \ ; \\
             m_f=m_0+\alpha\, v_f>a/2, &\qquad v\geq v_f \ .
             \end{array}
   \right.
\end{equation}
The Carter-Penrose diagram of this scenario can be seen in Fig.~\ref{F:3}. This situation can be interpreted as the accretion of energy satisfying the NEC onto a wormhole. When the mass of the hole reaches the value $2m(v)=a$, its causal character changes from timelike to spacelike, momentarily passing through null. At that point, an apparent horizon forms to hide the spacelike bounce. The event horizon of our space, which partially overlaps with the final apparent horizon, is placed at:

\begin{equation}
r_{+f}=\sqrt{4m_f^2-a^2} =\sqrt{(2m_0+2\alpha\, v_f)^2-a^2} \ .
\end{equation} 
%=====================================================
%=====================================================
\clearpage

%%%%%%%%%%%%%%%%%%%%%%%%%%%%%%%%%%%%%%%%%%%%%
\subsubsection{Phantom energy accretion onto a `black-bounce'}

\begin{figure}[!htb]
%\vspace{-1cm}
\begin{center}
\includegraphics[scale=0.90]{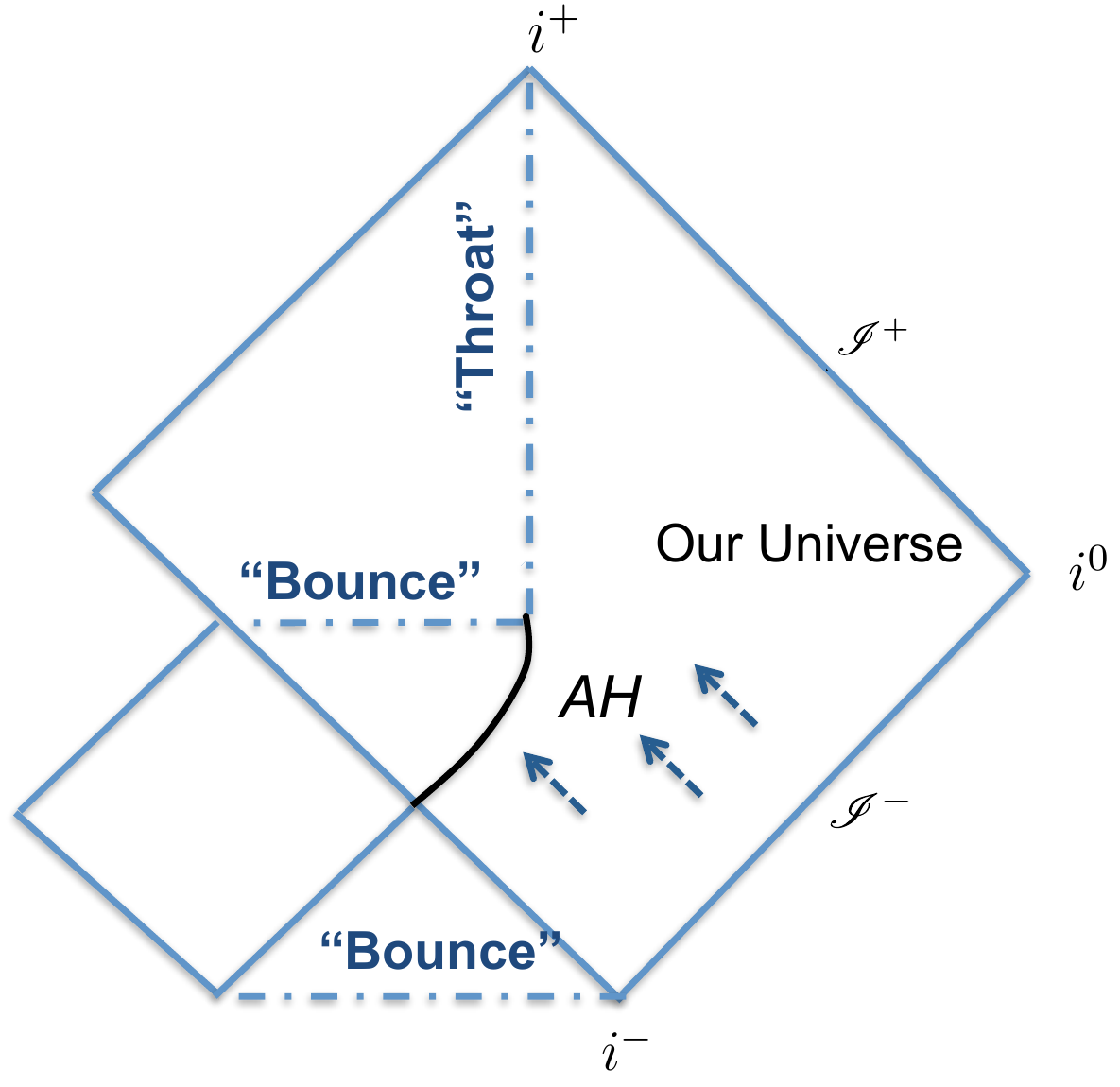}
\qquad
\end{center}
\caption[Carter-Penrose diagram for a `black-bounce' to wormhole transition due to phantom energy accretion]{Carter-Penrose diagram for a `black-bounce' to wormhole transition due to the accretion of phantom energy. The arrows indicate the region where the phantom fluid is being accreted. There is a `black-bounce' in our universe, characterised by an apparent horizon, that converts into a wormhole. Therefore, there is no event horizon in our universe.}
\label{F:4}
\end{figure}

One could also consider the case in which the additional ingoing flux that we turn on when allowing $m(v)$ to vary is characterised by a negative energy density. This type of exotic fluid is called phantom energy in a cosmological setting. The accretion of phantom energy into black holes has been studied in the test-fluid regime~\cite{Babichev:2004yx,Babichev:2004qp,MartinMoruno:2006mi,GonzalezDiaz:2007gt,Martin-Moruno:2007,Madrid:2010}, predicting a decrease of the black hole mass. With the present formalism we could take into account the back-reaction of this process, by using the advanced metric Eq.~\ref{E:dmetricv}, but considering $\dot m(v)<0$. 
However, an important difference with that picture is that our static geometry is a non-vacuum solution of the Einstein equations. We consider again for simplicity a finite region of linear evolution, that is now:

\begin{equation}
m(v)= \left\{ \begin{array}{ll}
             m_0>a/2, &\qquad   v\leq 0 \ ; \\
             m_0-\alpha v, &\qquad 0 < v < v_f \ ; \\
             m_f=m_0-\alpha\, v_f<a/2, &\qquad v\geq v_f \ .
             \end{array}
   \right.
\end{equation}
The apparent horizon of the regular black hole decreases due to the accretion of phantom energy. At $2m(v)=a$, this horizon disappears and the bounce surface is null, becoming then timelike. So, an ideal observer in this universe will see a black hole that is converted into a wormhole.
The Carter-Penrose diagram of this scenario can be seen in Fig.~\ref{F:4}.

%=====================================================
\clearpage
%%%%%%%%%%%%%%%%%%%%%%%%%%%%%%%%%%%%%%%%%%%%%%%%%%%%%%%%%%%%%%
%%%%%%%%%%%%%%%%%%%%%%%%%%%%%%%%%%%%%%%%%%%%%%%%%%%%%%%%%%%%%%
\subsection{Model with outgoing radiation}
%%============================================================================================

It is also interesting to consider the spacetime metric with outgoing (retarded) Eddington-Finkelstein coordinates. That is:

\begin{equation}\label{E:dmetricu}
ds^{2}=-\left(1-\frac{2m(u)}{\sqrt{r^{2}+a^{2}}}\right)du^{2}- 2 \,du \,dr
+\left(r^{2}+a^{2}\right)\left(d\theta^{2}+\sin^{2}\theta \;d\phi^{2}\right).
\end{equation}
For $a=0$, this is the standard retarded Vaidya metric that describes an outgoing null flux with $T_{uu}\propto -2\dot m(u)/r^2$. 
This scenario can be used to describe classically the back reaction of the semi-classical Hawking radiation by a black hole, in which case there is a positive outgoing flux of radiation that corresponds to a decrease of the black hole mass.
For our case, $a\neq0$ 
and we have a non-vacuum solution even for $m(u)=m$. So, when $m(u)$ varies, an extra null flux term will appear in that tensor (see   \S\ref{S:einstein}), with:

\begin{equation}
T_{\mu\nu}^{derivative}\propto -\dot m(u)\; (du)_\mu (du)_\nu \ .
\end{equation}
Therefore, we have a positive outgoing flux for $\dot m(u)<0$.

%%%%%%%%%%%%%%%%%%%%%%%%%%%%%%%%%%%%
\subsubsection{Classical effective description of black hole radiation} 
%%%%%%%%%%%%%%%%%%%%%%%%%%%%%%%%%%%%

\begin{figure}[!htb]
%\vspace{-1cm}
\begin{center}
\includegraphics[scale=0.90]{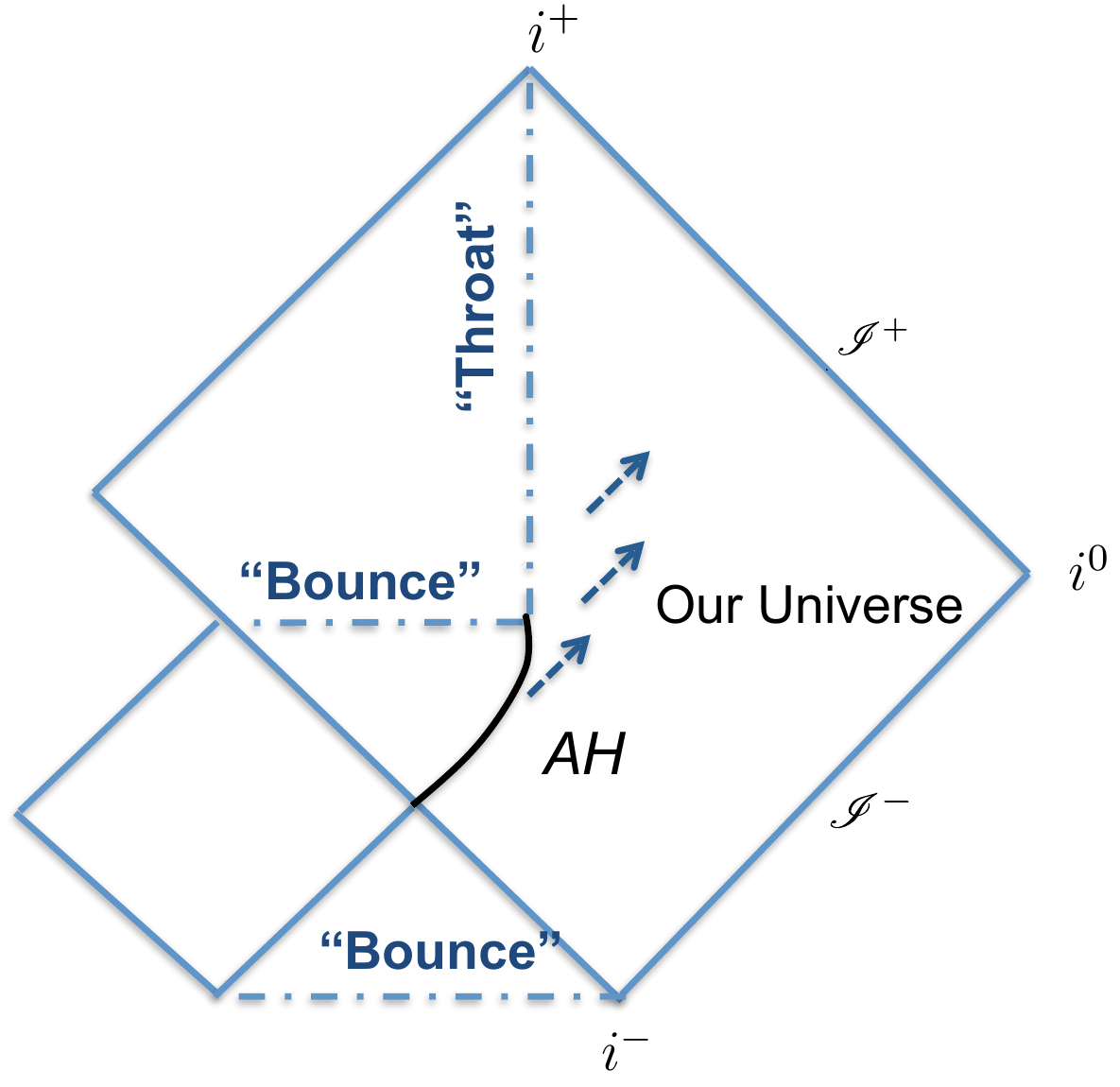}
\qquad
\end{center}
\caption[Carter-Penrose diagram for a `black-bounce' to wormhole transition due to positive energy emission]{Carter-Penrose diagram for a `black-bounce' to wormhole transition due to the emission of positive energy. This diagram is very similar to that shown in Figure~\ref{F:4}, however, now there is a (positive) flux being emitted by the `black-bounce' and wormhole.}
\label{F:5}
\end{figure}

Of course, one should first study carefully the semi-classical properties of this solution to interpret the outgoing flux as semi-classical~\cite{sparsity, sparsity2}. However, it is interesting to consider this scenario as we may have a `black-bounce' to wormhole transition similar to that already considered in the previous subsection. In this case it would be interesting to emphasise that the remnant of the `black-bounce' would be a wormhole.
The Carter-Penrose diagram of this scenario can be seen in Fig.~\ref{F:5}.

%=====================================================
%%%%%%%%%%%%%%%%%%%%
\clearpage

%========================================================

%=============================================================
\section{Curvature tensors and curvature invariants}\label{S:appendix}
%=============================================================
%=============================================================
%\def\thesection{A} %%% to get section numbering to work
%=============================================================
The key point is that in Eddington-Finkelstein coordinates, as long as $a\neq 0$, both the metric $g_{\mu\nu}$ and the inverse metric $g^{\mu\nu}$ have finite components for all values of $r$.
Specifically (taking upper sign for $u$, lower sign for $v$):

\begin{equation}
g_{\mu\nu} = \left[ \begin{array}{crcc} -\left(1-{2m(w)\over\sqrt{r^2+a^2}} \right)&  \mp1 & 0 & 0\\
\mp1 &\;0 & 0 & 0\\
0 & 0 & (r^2+a^2) & 0 \\
0 & 0 & 0 & (r^2+a^2)\sin^2\theta\\
  \end{array}\right] \ ,
\end{equation}
and:

\begin{equation}
g^{\mu\nu} = \left[ \begin{array}{rccc}0&  \mp1 & 0 & 0\\
\mp1 &  \;+\left(1-{2m(w)\over\sqrt{r^2+a^2}} \right) & 0 & 0\\
0 & 0 & {1\over(r^2+a^2)} & 0 \\
0 & 0 & 0 & {1\over(r^2+a^2)\sin^2\theta}\\
  \end{array}\right] \ .
\end{equation}
Similarly we shall soon see that the curvature tensors (Riemann, Weyl, Ricci, Einstein) have finite components for all values of $r$.
Consequently, even for a time-dependent $m(w)$ one still has a regular spacetime geometry -- there are no curvature singularities.

With this in mind, for simplicity we first consider the non-zero components of the Weyl tensor:
\begin{eqnarray}
C_{wrwr} &=& {m(w) ( a^2-2r^2) \over(r^2+a^2)^{5/2}} - {2a^2 \over3(r^2+a^2)^2}\nonumber\\
&& \nonumber \\
 &=& \mp{2C_{w\theta r\theta} \over r^2+a^2}   =  \mp{2  C_{w\phi r\phi} \over (r^2+a^2)\sin^2\theta } 
 =  -{C_{\theta\phi \theta\phi}\over (r^2+a^2)^2\sin^2\theta } \; \ ;\qquad \nonumber
 \\
 && \nonumber \\
 C_{w\theta w\theta} &=&
    -{(2r^2-a^2) m(w)^2\over (r^2+a^2)^2}  + {(6r^2-7a^2)m(w)\over6(r^2+a^2)^{3/2}} + {a^2\over 3(r^2+a^2) } 
    = {C_{w\phi w\phi} \over \sin^2\theta} \ . \nonumber \\
    &&
\end{eqnarray}
Note that there are no derivative contributions (no $\dot m(w) = dm(w)/dw$ contributions) to the Weyl tensor, and that the Weyl tensor components are finite at all values of $r$.\clearpage

\noindent For the Riemann tensor the non-zero components are a little more complicated:

\begin{eqnarray}
R_{wrwr} &=&-  {(2r^2-a^2)m(w)\over(r^2+a^2)^{5/2}} \ ; \nonumber \\
&& \nonumber \\
R_{w\theta r\theta} &=&\pm {r^2m(w)\over(r^2+a^2)^{3/2}} = {R_{w\phi r\phi}\over \sin^2\theta} \ ; \nonumber \\
&& \nonumber \\
R_{r\theta r\theta} &=&- {a^2\over(r^2+a^2)} = {R_{r\phi r\phi}\over \sin^2\theta} \ ; \nonumber \\
&& \nonumber \\
 R_{\theta\phi\theta\phi} &=& \left({2r^2 m(w)\over \sqrt{r^2+a^2}} + a^2 \right) \sin^2\theta \ ; \nonumber \\
 && \nonumber \\
 R_{w\theta w\theta} &=&  \mp {r \; \dot m(w)\over \sqrt{r^2+a^2}} + {r^2 m(w)\over (r^2+a^2)^{3/2}} - {2 r^2 m(w)^2 \over (r^2+a^2)^2} 
 = {R_{w\phi w\phi} \over \sin^2\theta} \ .
\end{eqnarray}
Note that the derivative term $\dot m(w)$  only shows up linearly, and only in a very restricted way.\newline 

\noindent The Ricci tensor has non-zero  components:

\begin{eqnarray}
R_{uu} &=& \mp{2r\;\dot m(w)\over (r^2+a^2)^{3/2}} + {a^2m(w)\; \left\{1-{2m(w)\over\sqrt{r^2+a^2}}\right\} \over(r^2+a^2)^{5/2}} \ ; \nonumber \\
&& \nonumber \\
R_{ur} &=&\pm{a^2 m(w)\over (r^2+a^2)^{5/2}} \ ; \nonumber \\
&& \nonumber \\
R_{rr} &=& {-2a^2 \over(r^2+a^2)^2} \ ; \nonumber \\
&& \nonumber \\
R_{\theta\theta} &=& {2 a^2 m(w)\over (r^2+a^2)^{3/2}} = {R_{\phi\phi}\over \sin^2\theta} \ .
\end{eqnarray}
Note that the derivative term $\dot m(w)$  only shows up linearly, and only in a very restricted way. 
In fact for the outgoing $u$ coordinate we we can write:

\begin{equation}
R_{\mu\nu} = R_{\mu\nu}^{nonderivative} -{2r\;\dot m(u)\over (r^2+a^2)^{3/2}} \; (du)_\mu (du)_\nu \ .
\end{equation}
On the other hand, if we had taken instead the ingoing coordinate $v$, we would have obtain $R_{vr}=-R_{ur}$ and a sign flip in the derivative term of $R_{vv}$ with respect that of $R_{uu}$. That is:

\begin{equation}
R_{\mu\nu} = R_{\mu\nu}^{nonderivative} +{2r\;\dot m(v)\over (r^2+a^2)^{3/2}} \; (dv)_\mu (dv)_\nu \ .
\end{equation}
The Einstein tensor has been discussed in \S\ref{S:einstein}, and those formulae will not be repeated here.

Note that all of these curvature tensor components are finite at all values of $r$. From the discussion above, it is already clear that all of the (polynomial) curvature invariants are all finite for all values of $r$.
For instance, the Ricci scalar is:

\begin{equation}
    R = - {2a^2\over(r^2+a^2)^2} \left\{ 1 - {3\,m(w)\over\sqrt{r^2+a^2}}\right\} \ .
\end{equation}
Note this is independent of the derivative term $\dot m(w)$.\newline  

\noindent Furthermore, the Ricci contraction $R_{\mu\nu}R^{\mu\nu}$ is:

\begin{eqnarray}
R_{\mu\nu}R^{\mu\nu} &=& \pm{8a^2r \; \dot m(w) \over(r^2+a^2)^{7/2} } 
+ {4a^4\over (r^2+a^2)^4} \left\{1 - {3m(w)\over\sqrt{r^2+a^2}} + {9 m(w)^2\over2(r^2+a^2)}\right\} \ ; \nonumber \\
&& \nonumber \\
&=& \pm{8a^2r \; \dot m(w) \over(r^2+a^2)^{7/2} } 
+ {4a^4\over (r^2+a^2)^4} \left\{ \left(1 - {3m(w)\over2\sqrt{r^2+a^2}}\right)^2 + {9 m(w)^2\over4(r^2+a^2)}\right\} \ . \nonumber \\
&& 
\end{eqnarray}
Note that the derivative term $\dot m(w)$  only shows up linearly. Note that the non-derivative contribution is a sum of squares and so automatically non-negative. In `3+1'-dimensions $G_{\mu\nu} G^{\mu\nu} = R_{\mu\nu} R^{\mu\nu}$, so the  $G_{\mu\nu} G^{\mu\nu}$ contraction provides nothing new.\newline 

\noindent The Weyl contraction $C_{\mu\nu\alpha\beta}C^{\mu\nu\alpha\beta}$ is a perfect square:

\begin{equation}
    C_{\mu\nu\alpha\beta}C^{\mu\nu\alpha\beta} = {16a^4\over(r^2+a^2)^4} \left\{1-{3m(w)\over2\sqrt{r^2+a^2}} + {3m(w)r^2\over a^2\sqrt{r^2+a^2}} \right\}^2 \ .
\end{equation}

\noindent The Kretschmann scalar is:

\begin{equation}
R_{\mu\nu\alpha\beta} \, R^{\mu\nu\alpha\beta} = C_{\mu\nu\alpha\beta} \, C^{\mu\nu\alpha\beta} +2 R_{\mu\nu}\, R^{\mu\nu} - \frac{1}{3}R^2 \ ,
\end{equation}
and so (in view of the above) without further calculation we have:

\begin{eqnarray}
R_{\mu\nu\alpha\beta}\,R^{\mu\nu\alpha\beta} &=& \pm{16a^2r \; \dot m(w) \over(r^2+a^2)^{7/2}} + {12a^4\over (r^2+a^2)^4} 
\Bigg\lbrace 1+ {8m(w)(r^2-a^2)\over3a^2\sqrt{r^2+a^2}} + \nonumber \\
&& \nonumber \\
&& \qquad \qquad \qquad \qquad {m(w)^2(4r^4-4a^2r^2+3a^4)\over a^4(r^2+a^2)}\Bigg\rbrace \ .
\end{eqnarray}
Note that the derivative term $\dot m(w)$  only shows up linearly.
All the curvature invariants are well-behaved everywhere throughout the spacetime.

%%%%%%

\section{Overview}

We have analysed a simple method of adding time-dependence to the `black-bounce' to traversable wormhole geometry presented in \S\ref{C:Black-bounce}, by imposing a Vaidya-like time-dependence on the metric. Subsequently, various physical models of interest have been analysed in a tractable fashion, providing an amenable framework for discussing the time-dependent transition between `black-bounce' and traversable wormhole geometries. Further analysis of interest could potentially involve extending the time dependence to the parameter $a$, such that \emph{both} $a(w)$ and $m(w)$ are functions of the outgoing/ingoing null time coordinate, however this analysis will be extremely algebraically involved and is relegated to the domain of future research.

%========================================================

%%%%%%%%%%%%%%%%%%%%%%%%%%%%%%%%%%%%%%%%%%%%%%%%%%%%%%%%%%%%%%%%

%%%%%%%%%%%%%%%%%%%%%%%%%%%%%%%%%%%%%%%%%%%%%%%%%%%%%%%

%\include{thestyle}
%\include{library}
%\include{binding}
%\include{lipsum}
\chapter{Conclusions}\label{C:con}

%%%%%%

The various analyses and discussions present in this thesis are primarily focused on nonsingular model spacetimes. Numerous geometries modelling either traversable wormholes or regular black holes (or the transition between the two) have been the subject of thorough general relativistic analysis, and several specific candidate spacetimes have been presented which extend the prior discussion beyond the class of singularity-free spacetimes typically considered. Let us review the spacetimes of primary interest individually, before presenting a holistic conclusion of the key findings.

\section{The exponential metric}\label{S:Exp-Discussion}

Regardless of one's views regarding the merits of some of the `justifications' used for advocating the exponential metric, the exponential metric can simply be viewed as a phenomenological model that can be studied in its own right.  Viewed in this way the exponential metric has a number of interesting features:

\begin{itemize}
\item It is a traversable wormhole, with time slowed down for stationary observers on the other side of the wormhole throat. 
\item Strong field lensing phenomena are markedly different from the\newline Schwarzs\-child geometry.
\item ISCOs and unstable photon orbits still exist, and they are moderately shifted from where they would be located in Schwarzschild spacetime.
\item  Regge--Wheeler potentials can still be extracted,  and are moderately different from what they would be in Schwarzschild spacetime.
\end{itemize}
Many of the proponents of the exponential metric are arguing for using it as a replacement for the Schwarzschild geometry of general relativity -- however typically without any detailed assessment of the phenomenology. It is strongly felt that if one wishes to replace all the black hole candidates astronomers have identified with traversable wormholes, then certainly a careful phenomenological analysis of this quite radical proposal (somewhat along the lines of \S\ref{C:Exponential}) should be carried out.
Perhaps most ironically, despite the fact that many of the proponents of the exponential metric reject general relativity, the exponential metric has a quite natural interpretation in terms of general relativity coupled to a phantom scalar field.
\enlargethispage{10pt}
%%%%%%%%%%%%%%%%%%%%%%%%%%%%%%%%%%%%%%%%%%%%%%%%%%%%%%%%%%%%%%%%%

\section{`Black-bounce' to traversable wormhole}\label{sec:dis}
%========================================================

The regular black hole presented in \S\ref{C:Black-bounce} in some sense represents minimal violence to the standard Schwarzschild solution. Indeed for $a=0$ it \emph{is} the standard Schwarzschild solution.  For $a\in(0,2m)$ the Carter-Penrose diagram is in some sense `as close as possible' to that for the maximally extended Kruskal-Szekeres version of Schwarzschild, except that the singularity is converted into a spacelike hypersurface representing a `bounce' into a future incarnation of the universe. This is qualitatively different from the picture where the collapsing regular black hole `bounces' back into our own universe~\cite{Barcelo:2014, Barcelo:2014b, Barcelo:2015, Barcelo:2016, Garay:2017, Rovelli:2014, Haggard:2015, Christodoulou:2016, DeLorenzo:2015, Malafarina:2017, Olmedo:2017, Barrau:2018, Malafarina:2018},  and is a scenario that deserves some attention in its own right. The specific model introduced also has the very nice feature that it analytically interpolates between black holes and traversable wormholes in a particularly clear and tractable manner. The `one-way' wormhole at $a=2m$, where the throat becomes null and extremal, is particularly interesting and novel.

%%%%%%%%%%%%%%%%%%%%%%%%%%%%%%%%%%%%%%%%%%%%%%%%%%%%%%%%%%%%%%%%%

\section{`Black-bounces' in Vaidya spacetimes}\label{S:discussion}
%========================================================

In \S\ref{C:Vaidya} several simple and tractable scenarios were presented for the time evolution of the regular `black-bounce'/traversable wormhole spacetime considered in \S\ref{C:Black-bounce}. These models provide a good framework for considering `black-bounce' $\longleftrightarrow$ traversable wormhole transitions. However, despite the generality of the simple models shown, it should be noted that in this framework a `black-bounce' cannot be formed by gravitational collapse from an ordinary stellar object. This is because in the limit $m\rightarrow0$, we have a traversable wormhole instead of Minkowski spacetime. So, in order to describe the physically relevant situation of stellar collapse one should go beyond our simple treatment above and consider \emph{both} $a(w)$ and $m(w)$ appropriately. Note that computations would then be significantly more complex, and more importantly that there would then be a \emph{qualitative} difference between the cases $a=0$ and $a\neq 0$. Such considerations are left for future work.

%%%%%%%%%%%%%%%%%%%%%%%%%%%%%%%%%%%%%%%%%%%%%%%%%%%%%%%%%%%%%%%%%

\section{Overall findings}\label{C:mainconc}

All candidate spacetimes analysed have required \emph{some} violation of one or more of the classical energy conditions associated with the stress-energy-momentum tensor in general relativity. For traversable wormhole geometries, the `flare-out' at the wormhole throat has consistently led to the violation of the null energy condition (NEC), and consequently the remaining energy conditions are also violated. For regular black hole spacetimes we observe the specific violation of the strong energy condition (SEC), and in general the satisfaction of the radial NEC, in the \emph{absence} of a wormhole throat. This is consistent with the pre-existing discussion on the viability of nonsingular geometries in the context of general relativity, and we may conclude that a lack of singularities certainly requires violation of one or more of the classical energy conditions on $T_{\mu\nu}$.

The traversable wormhole geometry induced by the exponential metric of \S\ref{C:Exponential} has been thoroughly analysed and conclusively found to have a standard interpretation in general relativity -- it is the opinion of the author that this should be sufficient evidence to stop using it as a `prop' for alternative theories of gravity (such as Yilmaz gravity, vector theory of gravity, \emph{etc.}).

A result of primary interest is the establishment of a tractable framework in which to discuss regular black hole/`black-bounce' to traversable wormhole transitions; this is a corollary of the analyses in \S\ref{C:Black-bounce} and \S\ref{C:Vaidya}. These spacetimes are of significant physical interest and extend the pre-existing class of nonsingular geometries beyond those usually considered. There are many potential avenues for further research in this specific framework, and the conversion of regular black hole regions to wormhole geometries is a subject of particular interest to those who are interested in the potential fate of regular black holes, as it presents a qualitatively different discussion to the canonical discourse pertaining to black hole evaporation.

The `exponentially suppressed mass' metric analysed in \S\ref{C:suppressed} also pre\-sents opportunities for further research, defining a physically interesting scenario where the effective mass of the central object is $r$-dependent, asymptotically heading to zero as one nears the centre of mass. This is a novel method of imposing geodesic completeness on the manifold, and further analysis could include adding a time-dependence to the metric, or thoroughly analysing the ISCO and photon sphere locations by unpacking the solutions which are \emph{implicitly} defined by the Lambert $W$ function.

Holistically, it is hoped that the research presented in this thesis has made a valuable contribution to some of the more qualitatively interesting questions which remain open in general relativity, and once again I would like to acknowledge and thank all those who have assisted me in any way over the course of its construction. Special thanks goes to Professor Matt Visser -- your time and wisdom has been invaluable.\newpage

%%%%%%

\appendix
\addcontentsline{toc}{chapter}{Appendix of Publications}
\section*{Appendix of Publications}

Publications in Journals:

\begin{itemize}
    \item \textbf{Exponential metric represents a traversable wormhole}\newline P. Boonserm, T. Ngampitipan, A. Simpson, and M. Visser\newline Phys. Rev. D \textbf{98}, 084048 (2018)\newline [arXiv:gr-qc/1805.03781]
    \item \textbf{Black-bounce to traversable wormhole}\newline A. Simpson and M. Visser\newline Journal of Cosmology and Astroparticle Physics \textbf{Feb 2019, 02, 042}\newline doi:10.1088/1475-7516/2019/02/042 \newline [arXiv:gr-qc/1812.07114]
\end{itemize}

\noindent Article accepted for publication in Classical Quantum Gravity:

\begin{itemize}
    \item \textbf{Vaidya spacetimes, black-bounces, and traversable wormholes}\newline A. Simpson, P. Mart\'in-Moruno, and M. Visser\newline [arXiv:gr-qc/1902.04232] .
\end{itemize}

%%%%%%%%%%%%%%%%%%%%%%%%%%%%%%%%%%%%%%%%%%%%%%%%%%%%%%%%%%%%%%%%%

%%%%%%%%%%%%%%%%%%%%%%%%%%%%%%%%%%%%%%%%%%%%%%%%%%%%%%%

% and of course book style knows about backmatter
% \backmatter caused problems with appendices :-(
% and of course report style doesn't
%%%%%%%%%%%%%%%%%%%%%%%%%%%%%%%%%%%%%%%%%%%%%%%%%%%%%%%

%\bibliographystyle{ieeetr}
\bibliographystyle{acm}
\addcontentsline{toc}{chapter}{Bibliography}
\bibliography{10-myrefs.bib}

\begin{thebibliography}{100}

\bibitem{LIGO}
{\sc Abbott, B.~P., Abbott, R., Abbott, T., Abernathy, M., Acernese, F.,
  Ackley, K., Adams, C., Adams, T., Addesso, P., Adhikari, R., et~al.}
\newblock Observation of gravitational waves from a binary black hole merger.
\newblock {\em Physical {R}eview {L}etters 116}, 6 (2016), 061102.

\bibitem{Abreu:2008}
{\sc Abreu, G., and Visser, M.}
\newblock Quantum interest in (3+1)-dimensional minkowski space.
\newblock {\em Physical Review D 79}, 6 (2009), 065004.

\bibitem{Abreu:2010}
{\sc Abreu, G., and Visser, M.}
\newblock The quantum interest conjecture in (3+ 1)-dimensional minkowski
  space.
\newblock In {\em The Twelfth Marcel Grossmann Meeting: On Recent Developments
  in Theoretical and Experimental General Relativity, Astrophysics and
  Relativistic Field Theories (In 3 Volumes)\/} (2012), World Scientific,
  pp.~2371--2373.

\bibitem{Introtopology}
{\sc Adams, C.~C., and Franzosa, R.~D.}
\newblock {\em Introduction to topology: pure and applied}.
\newblock No.~i9780131848696. Pearson Prentice Hall, Upper Saddle River, 2008.

\bibitem{Aldama:2015}
{\sc Aldama, M.~E.}
\newblock The gravity apple tree.
\newblock In {\em Journal of Physics: Conference Series\/} (2015), vol.~600,
  IOP Publishing, p.~012050.

\bibitem{Alley:1995}
{\sc Alley, C.~O., Aschan, P.~K., and Yilmaz, H.}
\newblock Refutation of {C}.{W}. {M}isner's claims in his article ``{Y}ilmaz
  cancels {N}ewton''.
\newblock {\em arXiv preprint gr-qc/9506082\/} (1995).

\bibitem{Arias}
{\sc Arias, R.~E., Cantcheff, M.~B., and Silva, G.~A.}
\newblock Lorentzian {AdS} geometries, wormholes, and holography.
\newblock {\em Physical Review D 83}, 6 (2011), 066015.

\bibitem{Babichev:2004yx}
{\sc Babichev, E., Dokuchaev, V., and Eroshenko, Y.}
\newblock Black hole mass decreasing due to phantom energy accretion.
\newblock {\em Physical Review Letters 93}, 2 (2004), 021102.

\bibitem{Babichev:2004qp}
{\sc Babichev, E., Dokuchaev, V., and Eroshenko, Y.}
\newblock Dark energy cosmology with generalized linear equation of state.
\newblock {\em Classical and Quantum Gravity 22}, 1 (2004), 143.

\bibitem{Big-O}
{\sc Bachman, P.}
\newblock Die analytische zahlentheorie.
\newblock {\em Teubner, Leipzig\/} (1894).

\bibitem{Barcelo:2014b}
{\sc Barcel{\'o}, C., Carballo-Rubio, R., and Garay, L.~J.}
\newblock Mutiny at the white-hole district.
\newblock {\em International Journal of Modern Physics D 23}, 12 (2014),
  1442022.

\bibitem{Barcelo:2016}
{\sc Barcel{\'o}, C., Carballo-Rubio, R., and Garay, L.~J.}
\newblock Exponential fading to white of black holes in quantum gravity.
\newblock {\em Classical and Quantum Gravity 34}, 10 (2017), 105007.

\bibitem{Barcelo:2015}
{\sc Barcel{\'o}, C., Carballo-Rubio, R., Garay, L.~J., and Jannes, G.}
\newblock Do transient white holes have a place in nature?
\newblock In {\em Journal of Physics: Conference Series\/} (2015), vol.~600,
  IOP Publishing, p.~012033.

\bibitem{Barcelo:2014}
{\sc Barcel{\'o}, C., Carballo-Rubio, R., Garay, L.~J., and Jannes, G.}
\newblock The lifetime problem of evaporating black holes: mutiny or
  resignation.
\newblock {\em Classical and Quantum Gravity 32}, 3 (2015), 035012.

\bibitem{LRR}
{\sc Barcel{\'o}, C., Liberati, S., and Visser, M.}
\newblock Analogue gravity.
\newblock {\em Living reviews in relativity 14}, 1 (2011), 3.

\bibitem{Barcelo:1999}
{\sc Barcel{\'o}, C., and Visser, M.}
\newblock Traversable wormholes from massless conformally coupled scalar
  fields.
\newblock {\em Physics Letters B 466}, 2-4 (1999), 127--134.

\bibitem{Barcelo:2000b}
{\sc Barcel{\'o}, C., and Visser, M.}
\newblock Brane surgery: energy conditions, traversable wormholes, and voids.
\newblock {\em Nuclear Physics B 584}, 1-2 (2000), 415--435.

\bibitem{Barcelo:2000}
{\sc Barcel{\'o}, C., and Visser, M.}
\newblock Scalar fields, energy conditions and traversable wormholes.
\newblock {\em Classical and Quantum Gravity 17}, 18 (2000), 3843.

\bibitem{Bardeen1968}
{\sc Bardeen, J.~M.}
\newblock Non-singular general-relativistic gravitational collapse.
\newblock {\em In Proceedings of of International Conference GR5, Tbilisi,
  USSR, page 174, 1968\/} (1968).

\bibitem{Bardeen:2014}
{\sc Bardeen, J.~M.}
\newblock Black hole evaporation without an event horizon.
\newblock {\em arXiv preprint arXiv:1406.4098\/} (2014).

\bibitem{Bardeen:2018}
{\sc Bardeen, J.~M.}
\newblock Models for the nonsingular transition of an evaporating black hole
  into a white hole.
\newblock {\em arXiv preprint arXiv:1811.06683\/} (2018).

\bibitem{Barrau:2018}
{\sc Barrau, A., Martineau, K., and Moulin, F.}
\newblock A status report on the phenomenology of black holes in loop quantum
  gravity: Evaporation, tunneling to white holes, dark matter and gravitational
  waves.
\newblock {\em Universe 4}, 10 (2018), 102.

\bibitem{BenAmots:2007}
{\sc Ben-Amots, N.}
\newblock Relativistic exponential gravitation and exponential potential of
  electric charge.
\newblock {\em Foundations of Physics 37}, 4 (2007), 773--787.

\bibitem{BenAmots:2011}
{\sc Ben-Amots, N.}
\newblock Some features and implications of exponential gravitation.
\newblock In {\em Journal of Physics: Conference Series\/} (2011), vol.~330,
  IOP Publishing, p.~012017.

\bibitem{Bhawal}
{\sc Bhawal, B., and Kar, S.}
\newblock Lorentzian wormholes in {E}instein-{G}auss-{B}onnet theory.
\newblock {\em Physical Review D 46}, 6 (1992), 2464.

\bibitem{Eotvos}
{\sc Bod, L., Fischbach, E., Marx, G., and N{\'a}ray-Ziegler, M.}
\newblock One hundred years of the {E}{\"o}tv{\"o}s experiment.
\newblock {\em Acta Physica Hungarica 69}, 3-4 (1991), 335--355.

\bibitem{Vaidya:1970}
{\sc Bonnor, W., and Vaidya, P.}
\newblock Spherically symmetric radiation of charge in {E}instein-{M}axwell
  theory.
\newblock {\em General Relativity and Gravitation 1}, 2 (1970), 127--130.

\bibitem{refractive}
{\sc Boonserm, P., Cattoen, C., Faber, T., Visser, M., and Weinfurtner, S.}
\newblock Effective refractive index tensor for weak-field gravity.
\newblock {\em Classical and Quantum Gravity 22}, 11 (2005), 1905.

\bibitem{Expmetric}
{\sc Boonserm, P., Ngampitipan, T., Simpson, A., and Visser, M.}
\newblock Exponential metric represents a traversable wormhole.
\newblock {\em Physical Review D 98}, 8 (2018), 084048.

\bibitem{ReggeWheeler1}
{\sc Boonserm, P., Ngampitipan, T., and Visser, M.}
\newblock {Regge-Wheeler equation, linear stability, and greybody factors for
  dirty black holes}.
\newblock {\em Phys. Rev. D 88\/} (2013), 041502.

\bibitem{tortoise}
{\sc Boonserm, P., and Visser, M.}
\newblock Bounding the greybody factors for {S}chwarzschild black holes.
\newblock {\em Physical Review D 78}, 10 (2008), 101502.

\bibitem{Cano:2018}
{\sc Cano, P.~A., Chimento, S., Ortin, T., and Ruiperez, A.}
\newblock Regular stringy black holes?
\newblock {\em arXiv preprint arXiv:1806.08377\/} (2018).

\bibitem{viability}
{\sc Carballo-Rubio, R., Di~Filippo, F., Liberati, S., Pacilio, C., and Visser,
  M.}
\newblock On the viability of regular black holes.
\newblock {\em Journal of High Energy Physics 2018}, 7 (2018), 23.

\bibitem{beyond}
{\sc Carballo-Rubio, R., Di~Filippo, F., Liberati, S., and Visser, M.}
\newblock Phenomenological aspects of black holes beyond general relativity.
\newblock {\em Physical Review D 98}, 12 (2018), 124009.

\bibitem{Carroll}
{\sc Carroll, S.~M.}
\newblock An introduction to general relativity: {S}pacetime and geometry.
\newblock {\em Addison Wesley 101\/} (2004), 102.

\bibitem{LIGO2}
{\sc Castelvecchi, D., and Witze, A.}
\newblock Einstein's gravitational waves found at last.
\newblock {\em Nature News\/} (2016).

\bibitem{Poincare}
{\sc Cervantes-Cota, J.~L., Galindo-Uribarri, S., and Smoot, G.~F.}
\newblock A brief history of gravitational waves.
\newblock {\em Universe 2}, 3 (2016), 22.

\bibitem{Christodoulou:2016}
{\sc Christodoulou, M., Rovelli, C., Speziale, S., and Vilensky, I.}
\newblock Planck star tunneling time: An astrophysically relevant observable
  from background-free quantum gravity.
\newblock {\em Physical Review D 94}, 8 (2016), 084035.

\bibitem{Clapp:1973}
{\sc Clapp, R.~E.}
\newblock Preliminary quasar model based on the {Y}ilmaz exponential metric.
\newblock {\em Physical Review D 7}, 2 (1973), 345.

\bibitem{Claudel:2000}
{\sc Claudel, C.-M., Virbhadra, K.~S., and Ellis, G.~F.}
\newblock The geometry of photon surfaces.
\newblock {\em Journal of Mathematical Physics 42}, 2 (2001), 818--838.

\bibitem{corben}
{\sc Corben, H.~C., Stehle, P., and Chako, N.}
\newblock Classical mechanics.
\newblock {\em Physics Today 14\/} (1961), 62.

\bibitem{Corless}
{\sc Corless, R.~M., Gonnet, G.~H., Hare, D.~E., Jeffrey, D.~J., and Knuth,
  D.~E.}
\newblock On the {L}ambert {$W$} function.
\newblock {\em Advances in Computational mathematics 5}, 1 (1996), 329--359.

\bibitem{Cramer:1994}
{\sc Cramer, J.~G., Forward, R.~L., Morris, M.~S., Visser, M., Benford, G., and
  Landis, G.~A.}
\newblock Natural wormholes as gravitational lenses.
\newblock {\em Physical Review D 51}, 6 (1995), 3117.

\bibitem{Dadhich:2001}
{\sc Dadhich, N., Kar, S., Mukherjee, S., and Visser, M.}
\newblock ${R}=0$ spacetimes and self-dual {L}orentzian wormholes.
\newblock {\em Physical Review D 65}, 6 (2002), 064004.

\bibitem{Weylcurv}
{\sc Danehkar, A.}
\newblock On the significance of the {W}eyl curvature in a relativistic
  cosmological model.
\newblock {\em Modern Physics Letters A 24}, 38 (2009), 3113--3127.

\bibitem{DeLorenzo:2015}
{\sc De~Lorenzo, T., and Perez, A.}
\newblock Improved black hole fireworks: Asymmetric black-hole-to-white-hole
  tunneling scenario.
\newblock {\em Physical Review D 93}, 12 (2016), 124018.

\bibitem{dewitt}
{\sc DeWitt, B.~S., Hawking, S., and Israel, W.}
\newblock {\em {General relativity: an Einstein centenary survey}}.
\newblock {C}ambridge {U}niversity {P}ress, {E}ngland (re-print in 2010), 1979.

\bibitem{einsteinrosen}
{\sc Einstein, A., and Rosen, N.}
\newblock The particle problem in the general theory of relativity.
\newblock {\em Physical Review 48}, 1 (1935), 73.

\bibitem{Faraoni:2016}
{\sc Faraoni, V., Hammad, F., and Belknap-Keet, S.~D.}
\newblock Revisiting the {B}rans solutions of scalar-tensor gravity.
\newblock {\em Physical Review D 94}, 10 (2016), 104019.

\bibitem{Faraoni:2018}
{\sc Faraoni, V., Hammad, F., Cardini, A.~M., and Gobeil, T.}
\newblock Revisiting the analogue of the {J}ebsen-{B}irkhoff theorem in
  {B}rans-{D}icke gravity.
\newblock {\em Physical Review D 97}, 8 (2018), 084033.

\bibitem{Fennelly:1976}
{\sc Fennelly, A., and Pavelle, R.}
\newblock Nonviability of {Y}ilmaz' gravitation theories and his criticisms of
  {R}osen's gravitation theory.
\newblock Tech. rep., 1976.

\bibitem{conformal}
{\sc Frauendiener, J.}
\newblock Conformal infinity.
\newblock {\em Living Reviews in Relativity 7}, 1 (2004), 1.

\bibitem{Frolov:2014b}
{\sc Frolov, V.~P.}
\newblock Do black holes exist?
\newblock {\em arXiv preprint arXiv:1411.6981\/} (2014).

\bibitem{Frolov:2014}
{\sc Frolov, V.~P.}
\newblock Information loss problem and a {`black hole'} model with a closed
  apparent horizon.
\newblock {\em Journal of High Energy Physics 2014}, 5 (2014), 49.

\bibitem{Frolov:2016}
{\sc Frolov, V.~P.}
\newblock Notes on nonsingular models of black holes.
\newblock {\em Physical Review D 94}, 10 (2016), 104056.

\bibitem{Frolov:2018}
{\sc Frolov, V.~P.}
\newblock Remarks on non-singular black holes.
\newblock In {\em EPJ Web of Conferences\/} (2018), vol.~168, EDP Sciences,
  p.~01001.

\bibitem{Frolov:2017}
{\sc Frolov, V.~P., and Zelnikov, A.}
\newblock Quantum radiation from an evaporating nonsingular black hole.
\newblock {\em Physical Review D 95}, 12 (2017), 124028.

\bibitem{Gao}
{\sc Gao, P., Jafferis, D.~L., and Wall, A.~C.}
\newblock Traversable wormholes via a double trace deformation.
\newblock {\em Journal of High Energy Physics 2017}, 12 (2017), 151.

\bibitem{Garay:2017}
{\sc Garay, L.~J., Barcel{\'o}, C., Carballo-Rubio, R., and Jannes, G.}
\newblock Do stars die too long?
\newblock In {\em The Fourteenth Marcel Grossmann Meeting; World Scientific
  Press: Singapore\/} (2017), vol.~2, World Scientific, pp.~1718--1723.

\bibitem{Garcia:2011}
{\sc Garcia, N.~M., Lobo, F.~S., and Visser, M.}
\newblock Generic spherically symmetric dynamic thin-shell traversable
  wormholes in standard general relativity.
\newblock {\em Physical Review D 86}, 4 (2012), 044026.

\bibitem{variational}
{\sc Gelfand, I., and Fomin, S.}
\newblock Calculus of variations. {R}evised {E}nglish edition translated and
  edited by {R}ichard {A}. {S}ilverman.
\newblock {\em Prentice Hall, Englewood Clis, NJ 7\/} (1963), 10--11.

\bibitem{GonzalezDiaz:2007gt}
{\sc Gonzalez-Diaz, P.~F., and Martin-Moruno, P.}
\newblock Wormholes in the accelerating universe.
\newblock In {\em The Eleventh Marcel Grossmann Meeting: On Recent Developments
  in Theoretical and Experimental General Relativity, Gravitation and
  Relativistic Field Theories (In 3 Volumes)\/} (2008), World Scientific,
  pp.~2190--2192.

\bibitem{twisted}
{\sc Gray, F., Santiago, J., Schuster, S., and Visser, M.}
\newblock {``Twisted''} black holes are unphysical.
\newblock {\em Modern Physics Letters A 32}, 18 (2017), 1771001.

\bibitem{sparsity}
{\sc Gray, F., Schuster, S., Van-Brunt, A., and Visser, M.}
\newblock The {H}awking cascade from a black hole is extremely sparse.
\newblock {\em Classical and Quantum Gravity 33}, 11 (2016), 115003.

\bibitem{mechequilibrium}
{\sc Griffith, B.~A., and Synge, J.~L.}
\newblock {\em Principles of mechanics}.
\newblock McGraw-Hill, 1970.

\bibitem{Griffiths}
{\sc Griffiths, D.~J.}
\newblock {\em Introduction to electrodynamics}.
\newblock AAPT, 2005.

\bibitem{Haggard:2015}
{\sc Haggard, H.~M., and Rovelli, C.}
\newblock Black to white hole tunneling: An exact classical solution.
\newblock {\em International Journal of Modern Physics A 30}, 28n29 (2015),
  1545015.

\bibitem{Hartle}
{\sc Hartle, J.~B.}
\newblock {\em Gravity: An introduction to Einstein{'}s general relativity}.
\newblock {P}earson {E}ducation {I}nc. publishing as {A}ddison {W}esley, 2003.

\bibitem{largescale}
{\sc Hawking, S., and Ellis, G.}
\newblock The large scale structure of spacetime, {C}ambridge {U}niversity
  {P}ress.
\newblock {\em Cambridge, England\/} (1973).

\bibitem{Hayward:2005}
{\sc Hayward, S.~A.}
\newblock Formation and evaporation of nonsingular black holes.
\newblock {\em Physical {R}eview {L}etters 96}, 3 (2006), 031103.

\bibitem{Hochberg:1997}
{\sc Hochberg, D., and Visser, M.}
\newblock Geometric structure of the generic static traversable wormhole
  throat.
\newblock {\em Physical Review D 56}, 8 (1997), 4745.

\bibitem{Hochberg:1998}
{\sc Hochberg, D., and Visser, M.}
\newblock Dynamic wormholes, antitrapped surfaces, and energy conditions.
\newblock {\em Physical Review D 58}, 4 (1998), 044021.

\bibitem{Hochberg:1998b}
{\sc Hochberg, D., and Visser, M.}
\newblock Null energy condition in dynamic wormholes.
\newblock {\em Physical {R}eview {L}etters 81}, 4 (1998), 746.

\bibitem{Hunt}
{\sc Hunt, B.~J.}
\newblock {\em The Maxwellians}.
\newblock Cornell University Press, 2005.

\bibitem{Ibison:2006b}
{\sc Ibison, M.}
\newblock Cosmological test of the {Y}ilmaz theory of gravity.
\newblock {\em Classical and Quantum Gravity 23}, 3 (2006), 577.

\bibitem{Ibison:2006a}
{\sc Ibison, M.}
\newblock The {Y}ilmaz cosmology.
\newblock In {\em AIP Conference Proceedings\/} (2006), vol.~822, AIP,
  pp.~181--186.

\bibitem{Kar:2004}
{\sc Kar, S., Dadhich, N., and Visser, M.}
\newblock Quantifying energy condition violations in traversable wormholes.
\newblock {\em Pramana 63}, 4 (2004), 859--864.

\bibitem{Raychad1}
{\sc Kar, S., and Sengupta, S.}
\newblock The {R}aychaudhuri equations: A brief review.
\newblock {\em Pramana 69}, 1 (2007), 49--76.

\bibitem{speedofthought}
{\sc Kennefick, D.}
\newblock {\em Travelling at the speed of thought}.
\newblock Princeton University Press, Princeton, 2007.

\bibitem{Kepler}
{\sc Kepler, J.}
\newblock {\em Harmonices mundi libri V}.
\newblock 1969.

\bibitem{universeofGR}
{\sc Kox, A.~J., Eisenstaedt, J., et~al.}
\newblock {\em The Universe of General Relativity}.
\newblock Springer, 2005.

\bibitem{little-o}
{\sc Landau, E.}
\newblock {\em Handbuch der Lehre von der Verteilung der Primzahlen. 2
  B{\"a}nde}.
\newblock Chelsea Publishing Co., New York, 1953.

\bibitem{tensors}
{\sc Landsberg, J.~M.}
\newblock Tensors: geometry and applications.
\newblock {\em Representation theory 381\/} (2012), 402.

\bibitem{Lemos:2003}
{\sc Lemos, J.~P., Lobo, F.~S., and de~Oliveira, S.~Q.}
\newblock Morris-{T}horne wormholes with a cosmological constant.
\newblock {\em Physical Review D 68}, 6 (2003), 064004.

\bibitem{lobo:2004}
{\sc Lobo, F.~S.}
\newblock Thin shells around traversable wormholes.
\newblock {\em arXiv preprint gr-qc/0401083\/} (2004).

\bibitem{Lobo:2005}
{\sc Lobo, F.~S.}
\newblock Phantom energy traversable wormholes.
\newblock {\em Physical Review D 71}, 8 (2005), 084011.

\bibitem{solarmass}
{\sc Luzum, B., Capitaine, N., Fienga, A., Folkner, W., Fukushima, T., Hilton,
  J., Hohenkerk, C., Krasinsky, G., Petit, G., Pitjeva, E., et~al.}
\newblock The {IAU} 2009 system of astronomical constants: the report of the
  {IAU} working group on numerical standards for fundamental astronomy.
\newblock {\em Celestial Mechanics and Dynamical Astronomy 110}, 4 (2011), 293.

\bibitem{Madrid:2010}
{\sc Madrid, J. A.~J., and Mart{\'\i}n-Moruno, P.}
\newblock On accretion of dark energy onto black-and worm-holes.
\newblock {\em arXiv preprint arXiv:1004.1428\/} (2010).

\bibitem{Malafarina:2017}
{\sc Malafarina, D.}
\newblock Classical collapse to black holes and quantum bounces: A review.
\newblock {\em Universe 3}, 2 (2017), 48.

\bibitem{Malafarina:2018}
{\sc Malafarina, D.}
\newblock Black hole bounces on the road to quantum gravity.
\newblock {\em Universe 4}, 9 (2018), 92.

\bibitem{Maldacena}
{\sc Maldacena, J., Stanford, D., and Yang, Z.}
\newblock Diving into traversable wormholes.
\newblock {\em Fortschritte der Physik 65}, 5 (2017), 1700034.

\bibitem{marchtorsion}
{\sc March, R., Bellettini, G., Tauraso, R., and Dell{'}Agnello, S.}
\newblock Constraining spacetime torsion with the {M}oon and {M}ercury.
\newblock {\em Physical Review D 83}, 10 (2011), 104008.

\bibitem{Martin-Moruno:2007}
{\sc Mart{\'i}n-Moruno, P.}
\newblock On the formalism of dark energy accretion onto black-and worm-holes.
\newblock {\em Physics Letters B 659}, 1-2 (2008), 40--44.

\bibitem{MartinMoruno:2006mi}
{\sc Mart{\'\i}n-Moruno, P., Madrid, J. A.~J., and Gonz{\'a}lez-D{\'\i}az,
  P.~F.}
\newblock Will black holes eventually engulf the universe?
\newblock {\em Physics Letters B 640}, 4 (2006), 117--120.

\bibitem{Martin-Moruno:2013b}
{\sc Mart{\'\i}n-Moruno, P., and Visser, M.}
\newblock Classical and quantum flux energy conditions for quantum vacuum
  states.
\newblock {\em Physical Review D 88}, 6 (2013), 061701.

\bibitem{Martin-Moruno:2015}
{\sc Mart{\'i}n-Moruno, P., and Visser, M.}
\newblock Semiclassical energy conditions for quantum vacuum states.
\newblock {\em Journal of High Energy Physics 2013}, 9 (2013), 50.

\bibitem{Martin-Moruno:2015b}
{\sc Mart{\'i}n-Moruno, P., and Visser, M.}
\newblock Semi-classical and nonlinear energy conditions.
\newblock In {\em 14th Marcel Grossmann Meeting on Recent Developments in
  Theoretical and Experimental General Relativity, Astrophysics, and
  Relativistic Field Theories (MG14) Rome, Italy\/} (2015), World Scientific.

\bibitem{Martin-Moruno:2013a}
{\sc Mart{\'\i}n-Moruno, P., and Visser, M.}
\newblock Classical and semi-classical energy conditions.
\newblock In {\em Wormholes, Warp Drives and Energy Conditions}. Springer,
  2017, pp.~193--213.

\bibitem{Martinis:2010}
{\sc Martinis, M., and Perkovi{\'c}, N.}
\newblock Is exponential metric a natural space-time metric of {N}ewtonian
  gravity?
\newblock {\em arXiv preprint arXiv:1009.6017\/} (2010).

\bibitem{supermassive}
{\sc McConnell, N.~J., Ma, C.-P., Gebhardt, K., Wright, S.~A., Murphy, J.~D.,
  Lauer, T.~R., Graham, J.~R., and Richstone, D.~O.}
\newblock Two ten-billion-solar-mass black holes at the centres of giant
  elliptical galaxies.
\newblock {\em Nature 480}, 7376 (2011), 215.

\bibitem{Misner:1995}
{\sc Misner, C.~W.}
\newblock Yilmaz cancels {N}ewton.
\newblock {\em arXiv preprint gr-qc/9504050\/} (1995).

\bibitem{telebook}
{\sc Misner, C.~W., Thorne, K.~S., and Wheeler, J.~A.}
\newblock {\em Gravitation}.
\newblock Princeton University Press, 2017.

\bibitem{plancklength}
{\sc Mohr, P.~J., Newell, D.~B., and Taylor, B.~N.}
\newblock Codata recommended values of the fundamental physical constants:
  2014.
\newblock {\em Journal of Physical and Chemical Reference Data 45}, 4 (2016),
  043102.

\bibitem{Molina-Paris:1998}
{\sc Molina-Par{\'i}s, C., and Visser, M.}
\newblock Minimal conditions for the creation of a
  {F}riedmann--{R}obertson--{W}alker universe from a {``bounce''}.
\newblock {\em Physics Letters B 455}, 1-4 (1999), 90--95.

\bibitem{MorrisThorne}
{\sc Morris, M.~S., and Thorne, K.~S.}
\newblock Wormholes in spacetime and their use for interstellar travel: A tool
  for teaching general relativity.
\newblock {\em American Journal of Physics 56}, 5 (1988), 395--412.

\bibitem{MTY}
{\sc Morris, M.~S., Thorne, K.~S., and Yurtsever, U.}
\newblock Wormholes, time machines, and the weak energy condition.
\newblock {\em Physical Review Letters 61}, 13 (1988), 1446.

\bibitem{delmatt}
{\sc Nawarajan, D., and Visser, M.}
\newblock Global properties of physically interesting {L}orentzian spacetimes.
\newblock {\em International Journal of Modern Physics D 25}, 14 (2016),
  1650106.

\bibitem{Newton}
{\sc Newton, I.}
\newblock {\em Philosophiae naturalis principia mathematica}, vol.~1.
\newblock G. Brookman, 1833.

\bibitem{Olmedo:2017}
{\sc Olmedo, J., Saini, S., and Singh, P.}
\newblock From black holes to white holes: a quantum gravitational, symmetric
  bounce.
\newblock {\em Classical and Quantum Gravity 34}, 22 (2017), 225011.

\bibitem{ONeill}
{\sc O'neill, B.}
\newblock {\em Semi-Riemannian geometry with applications to relativity},
  vol.~103.
\newblock Academic press, 1983.

\bibitem{Parikh:1998}
{\sc Parikh, M.~K., and Wilczek, F.}
\newblock Global structure of evaporating black holes.
\newblock {\em Physics Letters B 449}, 1-2 (1999), 24--29.

\bibitem{EddFink}
{\sc Penrose, R.}
\newblock Gravitational collapse and space-time singularities.
\newblock {\em Physical Review Letters 14}, 3 (1965), 57.

\bibitem{Poisson:1995}
{\sc Poisson, E., and Visser, M.}
\newblock Thin-shell wormholes: Linearization stability.
\newblock {\em Physical Review D 52}, 12 (1995), 7318.

\bibitem{Rastall:1975}
{\sc Rastall, P.}
\newblock Gravity without geometry.
\newblock {\em American Journal of Physics 43}, 7 (1975), 591--595.

\bibitem{pathways}
{\sc Renn, J., and Sauer, T.}
\newblock Pathways out of classical physics.
\newblock In {\em The genesis of general relativity}. Springer, 2007,
  pp.~113--312.

\bibitem{Robertson:1999}
{\sc Robertson, S.~L.}
\newblock Bigger bursts from merging neutron stars.
\newblock {\em The Astrophysical Journal Letters 517}, 2 (1999), L117.

\bibitem{MECO:1999}
{\sc Robertson, S.~L.}
\newblock X-ray novae, event horizons, and the exponential metric.
\newblock {\em The Astrophysical Journal 515}, 1 (1999), 365.

\bibitem{MECO:2016}
{\sc Robertson, S.~L.}
\newblock {MECO} in an exponential metric.
\newblock {\em arXiv preprint arXiv:1606.01417\/} (2016).

\bibitem{Rodrigues2018Bardeen}
{\sc Rodrigues, M.~E., and Silva, M. V. d.~S.}
\newblock Bardeen regular black hole with an electric source.
\newblock {\em arXiv preprint arXiv:1802.05095\/} (2018).

\bibitem{Bergmann-Roman}
{\sc Roman, T.~A., and Bergmann, P.~G.}
\newblock Stellar collapse without singularities?
\newblock {\em Physical Review D 28}, 6 (1983), 1265.

\bibitem{Rovelli:2014}
{\sc Rovelli, C., and Vidotto, F.}
\newblock Planck stars.
\newblock {\em International Journal of Modern Physics D 23}, 12 (2014),
  1442026.

\bibitem{Sahoo}
{\sc Sahoo, P., Moraes, P., Sahoo, P., and Ribeiro, G.}
\newblock Phantom fluid supporting traversable wormholes in alternative gravity
  with extra material terms.
\newblock {\em arXiv preprint arXiv:1802.02465\/} (2018).

\bibitem{Schuller}
{\sc Schuller, F.~P.}
\newblock The {WE-H}eraeus international winter schoold on gravity and light.
\newblock {\em Perimeter Institute course notes\/} (2015).

\bibitem{black-bounce}
{\sc Simpson, A., and Visser, M.}
\newblock Black-bounce to traversable wormhole.
\newblock {\em {J}ournal of {C}osmology and {A}stroparticle {P}hysics Feb
  2019}, 02 (2019), 042--042.

\bibitem{Stewart:11}
{\sc Stewart, S.~M.}
\newblock Wien peaks and the {L}ambert {$W$} function.
\newblock {\em Revista Brasileira de Ensino de F{\'\i}sica 33}, 3 (2011), 1--6.

\bibitem{Stewart:12}
{\sc Stewart, S.~M.}
\newblock Spectral peaks and {W}ien's displacement law.
\newblock {\em Journal of Thermophysics and Heat Transfer 26}, 4 (2012),
  689--692.

\bibitem{Sushkov:2005}
{\sc Sushkov, S.}
\newblock Wormholes supported by a phantom energy.
\newblock {\em Physical Review D 71}, 4 (2005), 043520.

\bibitem{Svidzinsky:2009}
{\sc Svidzinsky, A.~A.}
\newblock Vector theory of gravity in {M}inkowski space-time: Flat universe
  without black holes.
\newblock {\em arXiv preprint arXiv:0904.3155\/} (2009).

\bibitem{Svidzinsky:2015}
{\sc Svidzinsky, A.~A.}
\newblock Vector theory of gravity: Universe without black holes and solution
  of dark energy problem.
\newblock {\em Physica Scripta 92}, 12 (2017), 125001.

\bibitem{Vaidya:1951b}
{\sc Vaidya, P.}
\newblock Nonstatic solutions of {E}instein's field equations for spheres of
  fluids radiating energy.
\newblock {\em Physical Review 83}, 1 (1951), 10.

\bibitem{Vaidya:1999a}
{\sc Vaidya, P.}
\newblock The external field of a radiating star in general relativity.
\newblock {\em General Relativity and Gravitation 31}, 1 (1999), 119--120.

\bibitem{Vaidya:1951a}
{\sc Vaidya, P.~C.}
\newblock The gravitational field of a radiating star.
\newblock In {\em Proceedings of the Indian Academy of Sciences-Section A\/}
  (1951), vol.~33, Springer, p.~264.

\bibitem{Valluri:09}
{\sc Valluri, S.~R., Gil, M., Jeffrey, D., and Basu, S.}
\newblock The {L}ambert {$W$} function and quantum statistics.
\newblock {\em Journal of Mathematical Physics 50}, 10 (2009), 102103.

\bibitem{Valluri:00}
{\sc Valluri, S.~R., Jeffrey, D.~J., and Corless, R.~M.}
\newblock Some applications of the {L}ambert {$W$} function to physics.
\newblock {\em Canadian Journal of Physics 78}, 9 (2000), 823--831.

\bibitem{Vial:12}
{\sc Vial, A.}
\newblock Fall with linear drag and {W}ien's displacement law: approximate
  solution and {L}ambert function.
\newblock {\em European Journal of Physics 33}, 4 (2012), 751.

\bibitem{Virbhadra:2008}
{\sc Virbhadra, K.}
\newblock Relativistic images of {S}chwarzschild black hole lensing.
\newblock {\em Physical Review D 79}, 8 (2009), 083004.

\bibitem{Virbhadra:2002}
{\sc Virbhadra, K., and Ellis, G.~F.}
\newblock Gravitational lensing by naked singularities.
\newblock {\em Physical Review D 65}, 10 (2002), 103004.

\bibitem{Virbhadra:2007}
{\sc Virbhadra, K., and Keeton, C.}
\newblock Time delay and magnification centroid due to gravitational lensing by
  black holes and naked singularities.
\newblock {\em Physical Review D 77}, 12 (2008), 124014.

\bibitem{Virbhadra:1998}
{\sc Virbhadra, K., Narasimha, D., and Chitre, S.}
\newblock Role of the scalar field in gravitational lensing.
\newblock {\em {A}stronomy and {A}strophysics {B}erlin 337\/} (1998), 1--8.

\bibitem{Virbhadra:1999}
{\sc Virbhadra, K.~S., and Ellis, G.~F.}
\newblock Schwarzschild black hole lensing.
\newblock {\em Physical Review D 62}, 8 (2000), 084003.

\bibitem{Visser:1989b}
{\sc Visser, M.}
\newblock Traversable wormholes from surgically modified {S}chwarzschild
  spacetimes.
\newblock {\em Nuclear Physics B 328}, 1 (1989), 203--212.

\bibitem{Visser:1989a}
{\sc Visser, M.}
\newblock Traversable wormholes: Some simple examples.
\newblock {\em Physical Review D 39}, 10 (1989), 3182.

\bibitem{LorentzianWormholes}
{\sc Visser, M.}
\newblock {\em Lorentzian wormholes: from Einstein to Hawking}.
\newblock Springer-Verlag New York inc., 1995.

\bibitem{Visser:1994}
{\sc Visser, M.}
\newblock Scale anomalies imply violation of the averaged null energy
  condition.
\newblock {\em Physics Letters B 349}, 4 (1995), 443--447.

\bibitem{Visser:1996a}
{\sc Visser, M.}
\newblock Gravitational vacuum polarization. {I}. energy conditions in the
  {H}artle-{H}awking vacuum.
\newblock {\em Physical Review D 54}, 8 (1996), 5103.

\bibitem{Visser:1996b}
{\sc Visser, M.}
\newblock Gravitational vacuum polarization. {II}. energy conditions in the
  {B}oulware vacuum.
\newblock {\em Physical Review D 54}, 8 (1996), 5116.

\bibitem{Visser:1997-ec}
{\sc Visser, M.}
\newblock Gravitational vacuum polarization. {IV}. energy conditions in the
  {U}nruh vacuum.
\newblock {\em Physical Review D 56}, 2 (1997), 936.

\bibitem{LNP}
{\sc Visser, M.}
\newblock Survey of analogue spacetimes.
\newblock In {\em Analogue Gravity Phenomenology}. Springer, 2013, pp.~31--50.

\bibitem{Visser:14}
{\sc Visser, M.}
\newblock Physical observability of horizons.
\newblock {\em Physical Review D 90}, 12 (2014), 127502.

\bibitem{MATH465}
{\sc Visser, M.}
\newblock Notes on general relativity.
\newblock {\em Victoria University course notes\/} (2017).

\bibitem{primes}
{\sc Visser, M.}
\newblock Primes and the {L}ambert {$W$} function.
\newblock {\em Mathematics 6}, 4 (2018), 56.

\bibitem{Visser:cosmo1999}
{\sc Visser, M., and Barcelo, C.}
\newblock Energy conditions and their cosmological implications.
\newblock In {\em Cosmo-99}. World Scientific, 2000, pp.~98--112.

\bibitem{Visser:1999-super}
{\sc Visser, M., Bassett, B., and Liberati, S.}
\newblock Perturbative superluminal censorship and the null energy condition.
\newblock In {\em AIP Conference Proceedings\/} (1999), vol.~493, AIP,
  pp.~301--305.

\bibitem{Visser:1998-super}
{\sc Visser, M., Bassett, B., and Liberati, S.}
\newblock Superluminal censorship.
\newblock {\em Nuclear Physics B-Proceedings Supplements 88}, 1-3 (2000),
  267--270.

\bibitem{sparsity2}
{\sc Visser, M., Gray, F., Schuster, S., and Van-Brunt, A.}
\newblock Sparsity of the {H}awking flux.
\newblock In {\em Proceedings of the MG14 Meeting on General Relativity\/}
  (2017), World Scientific, pp.~1724--1729.

\bibitem{Visser:1997}
{\sc Visser, M., and Hochberg, D.}
\newblock Generic wormhole throats.
\newblock {\em Annals of the Israeli Physics Society 13\/} (1997), 249--295.

\bibitem{Visser:2003}
{\sc Visser, M., Kar, S., and Dadhich, N.}
\newblock Traversable wormholes with arbitrarily small energy condition
  violations.
\newblock {\em Physical {R}eview {L}etters 90}, 20 (2003), 201102.

\bibitem{Wald}
{\sc Wald, R.~M.}
\newblock {\em General relativity}.
\newblock University of Chicago Press, 2007.

\bibitem{Wang:1998}
{\sc Wang, A., and Wu, Y.}
\newblock Generalized {V}aidya solutions.
\newblock {\em General Relativity and Gravitation 31}, 1 (1999), 107--114.

\bibitem{Willenborg}
{\sc Willenborg, F., Grunau, S., Kleihaus, B., and Kunz, J.}
\newblock Geodesic motion around traversable wormholes supported by a massless
  conformally coupled scalar field.
\newblock {\em Physical Review D 97}, 12 (2018), 124002.

\bibitem{Kerr}
{\sc Wiltshire, D.~L., Visser, M., and Scott, S.~M.}
\newblock {\em The {K}err Spacetime}.
\newblock Cambridge, UK: Cambridge University Press, 2009, 2009.

\bibitem{Worthington}
{\sc Worthington, A.~M.}
\newblock {\em Dynamics of rotation: an elementary introduction to rigid
  dynamics}.
\newblock Longmans, Green, and Company, 1910.

\bibitem{Yilmaz:1958}
{\sc Yilmaz, H.}
\newblock New approach to general relativity.
\newblock {\em Physical Review 111}, 5 (1958), 1417.

\bibitem{Yilmaz:1971}
{\sc Yilmaz, H.}
\newblock New theory of gravitation.
\newblock {\em Physical Review Letters 27}, 20 (1971), 1399.

\bibitem{Yilmaz:1973}
{\sc Yilmaz, H.}
\newblock New approach to relativity and gravitation.
\newblock {\em Annals of Physics 81}, 1 (1973), 179--200.

\end{thebibliography}

\end{document}